\documentclass[12pt]{report}

\pdfoutput=1

\usepackage{xcolor}
\usepackage{appendix}
\usepackage{hyperref}

\usepackage{epsfig}
\usepackage{amssymb}
\usepackage{amsfonts}
\usepackage{amsmath}
\usepackage{euscript}
\usepackage{verbatim}
\usepackage{latexsym}
\usepackage{graphicx}
\usepackage{slashed}
\usepackage[utf8]{inputenc}
\usepackage{graphicx}	
\usepackage{subcaption}
\usepackage{wrapfig}
\usepackage[T1]{fontenc}
\usepackage{mathtools}
\usepackage{caption}
\usepackage{float}
\usepackage{subfloat}
\usepackage{calligra}
\usepackage{braket}
\usepackage{caption}
\usepackage{url}

\usepackage{tikz}
\usetikzlibrary{shapes.geometric, arrows,patterns,snakes}
\tikzstyle{ellip} = [ellipse, minimum width=3cm, minimum height=1cm,text centered, draw=black]

\newcommand{\beq}{\begin{equation}}
\newcommand{\eeq}{\end{equation}}
\newcommand{\bea}{\begin{eqnarray}}
\newcommand{\eea}{\end{eqnarray}}

\def\e{\epsilon}

\def\e{\epsilon}

\def\Tr{ \hbox{\rm Tr}}

\def\const{\hbox {\rm const.}}
\def\im{\hbox{\rm Im}}
\def\re{\hbox{\rm Re}}
\def\bra{\langle}
\def\ket{\rangle}
\def\Arg{\hbox {\rm Arg}}
\def\dag{{}^{\dagger}}

\def\N{{\cal N}}

\def\diag{\mathop{\mathrm{diag}}}

\def\hsp{,\hspace{.7cm}}

\def\br{\nonumber \\}

\newcommand{\sgm}[1]{\sigma_{#1}}
\newcommand{\idd}{\mathbf{1}}
\newcommand{\psiin}{\psi_{0}}
\newcommand{\phiin}{\phi_{1}}
\newcommand{\hin}{h_{0}}
\newcommand{\rh}{r_{h}}
\newcommand{\rb}{r_{b}}
\newcommand{\psibnd}{\psi_{0}^{b}}
\newcommand{\psibndp}{\psi_{1}^{b}}
\newcommand{\phibnd}{\phi_{0}^{b}}
\newcommand{\phibndp}{\phi_{1}^{b}}
\newcommand{\gbnd}{g_{0}^{b}}
\newcommand{\hbnd}{h_{0}^{b}}
\newcommand{\zh}{z_{h}}
\newcommand{\zb}{z_{b}}
\newcommand{\psizero}{\psi_{0}}
\newcommand{\phizero}{\phi_{0}}
\newcommand{\hzero}{h_{0}}
\newcommand{\man}{\mathcal{M}}
\newcommand{\hbr}{\bar{h}}

\newcommand{\Z}{\mathbb{Z}}
\newcommand{\non}{\nonumber}
\newcommand{\lag}{\langle}
\newcommand{\rag}{\rangle}

\newcommand{\ee}{\mathbf{e}}
\newcommand{\real}{\text{Re}}
\newcommand{\imag}{\text{Im}}
\newcommand{\caln}{\mathcal{N}}
\newcommand{\f}{\phi}
\newcommand{\m}{\mu}
\newcommand{\n}{\nu}
\newcommand{\z}{\zeta}
\newcommand{\F}{\Phi}
\newcommand{\g}{\gamma}
\newcommand{\G}{\Gamma}
\newcommand{\ci}{\mathcal{I}}
\newcommand{\cj}{\mathcal{J}}
\newcommand{\ck}{\mathcal{K}}
\newcommand{\bphi}{b^{\Phi}}
\newcommand{\cor}{c_{\mathcal{R}}^{\Omega}}
\newcommand{\boi}{b_{\mathcal{I}}^{\Omega}}
\newcommand{\calg}{\mathcal{G}}

\newcommand{\calw}{\mathcal{W}}

\newcommand{\dtn}{_{10}}
\newcommand{\calv}{\mathcal{V}}
\newcommand{\calo}{\mathcal{O}}
\newcommand{\zz}{$ SU(2)\times SU(2)\times\Z_2 \ $}
\newcommand{\lr}[1]{\big\langle #1 \big\rangle}
\newcommand{\exval}[1]{\big\langle \mathcal{O}_{#1}\big\rangle}
\newcommand{\exvalb}[2]{\big\langle \overline{\mathcal{O}}^{#1}_{#2} \big\rangle}
\newcommand{\ol}[1]{\hat{#1}}
\newcommand{\dep}{\delta_{\epsilon}}
\newcommand{\des}{\delta_{\sigma}}
\newcommand{\depm}{\delta_{\epsilon^{-}}}
\newcommand{\hzd}[1]{\hat{\zeta}_{\ol#1}}
\newcommand{\qft}{_{\text{\tiny QFT}}}
\newcommand{\cals}{\mathcal{S}}
\newcommand{\la}{\log a z\,}
\newcommand{\gs}{g_{s}}
\newcommand{\hks}{h_{\tx{KS}}}

\newcommand{\hzdb}[1]{\overline{\hat{\zeta}}_{\ol#1}}
\newcommand{\tx}{\text}
\newcommand{\de}{\partial}
\newcommand{\R}{\mathbb{R}}

\hypersetup{colorlinks=false, linkcolor=blue, citecolor=red}

\begin{document}
	
	\pagenumbering{gobble}

\begin{titlepage}
\pagenumbering{gobble}

\begin{center}

	\null %\vspace{1.6cm}
	
	{\LARGE \emph{Applications of Holography}}
	
	\vskip 1cm

	\large{A thesis submitted for the degree of \\ \textbf{Doctor of Philosophy} \\ in the Faculty of Sciences}
	
	\vspace{3cm}
	by\\
	{\Large\textbf{\textsf{{P N Bala Subramanian}}}}
	
	\vspace{3cm}
	under the guidance of\\
	{\Large\textbf{\textsf{{Dr. Chethan Krishnan}}}}
	
	\vspace{1.7cm}
	\begin{figure}[h]
		\centering
		\includegraphics[scale=0.25]{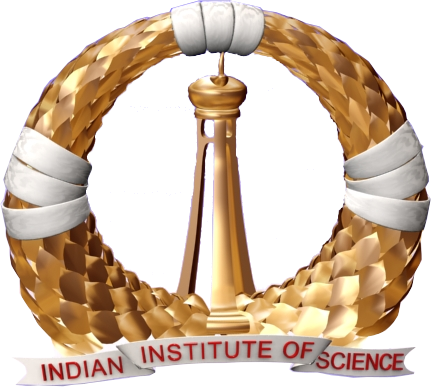}
	\end{figure}
	
	\vspace{.5cm}
	\textit{Centre for High Energy Physics\\
		Indian Institute of Science \\
		\ Bangalore - 560012. India. \\}

\end{center}

\end{titlepage}

\pagebreak

\vskip 2cm

\begin{flushleft}
\textsf{\textbf{\LARGE Declaration}}
\end{flushleft}

\vskip 2cm

{\large I hereby declare that this thesis "Applications of Holography" is based on my own research work, that I carried out with my collaborators in the Centre for High Energy Physics, Indian Institute of Science, during my tenure as a PhD student under the supervision of Dr. Chethan Krishnan. The work presented here does not feature as the work of someone else in obtaining a degree in this or any other institute. Any other work that may have been quoted or used in writing this thesis has been duly cited and acknowledged.

\vskip 2cm

 \ \\
Date: 10/04/2018   \hspace{7.5cm} P N Bala Subramanian

\vskip 1.5cm

 \ \\
Certified by:

\vskip 1.5cm

 \ \\
Dr. Chethan Krishnan\\
Centre for High Energy Physics\\
Indian Institute of Science\\
Bangalore: 560012\\
India}

\pagebreak

\pagebreak

\vskip 0.5cm

\begin{center}
\textsf{\textbf{\LARGE Acknowledgement}}
\end{center}

The set of events that has lead me to present this thesis was shaped by a lot of people, and I would like to acknowledge them for the non-trivial effect they had in my life. This effort is bound to be an under-representation of the gratitude I have towards them, but I will try anyway.\\

My adviser Chethan Krishnan has been an inspiration for me since the day I started working with him. He has been a constant source of energy and motivation throughout the last four-and-half years. The tenacious way he goes about doing research has been fascinating, and it is something I will strive to emulate down the road. I have been fortunate to spend an inordinate amount of time discussing topics in Physics, philosophy, psychology and various other topics, and his restructured my world-view for the better. Honestly, I am delighted to have been his student, and any words I can put together to thank him would be inadequate to portray my true gratitude. \\

Next, I would like to convey my gratitude to all my collaborators, in particular Pallab Basu who continually helped me throughout the duration of my PhD, A P Balachandran, Sachindeo Vaidya, Avinash, Pavan Kumar, Veronica and Nirmalendu. I also extend my gratitude to Prof. Ananthanarayan for all the support as Chairman, and to all the faculty in CHEP, who helped me broaden my perspective in the subject. It is with great pleasure I thank Saravana, Mallika, Aruna and most importantly Keshava for making sure that the Department functions so smoothly. At this point, I would also like to thank my school teachers Joseph Cherian and Maria Savio, my college mentor Prof. A Varadarajan, my teachers in Univ of Madras Dr. Ranabir Chakrabarty, Dr. Rita John and Dr. Vaitheeswaran and my Master's thesis adviser Dr. D Indumathi.\\

It is thanks to the awesome friends around me, that I have memories to smile about. I send my regards to my friends since school Ajay, Pranab, Ankit, Sravan and Delon for over a decade of crazy hangouts, Aparna, Keziah and Sarath for the beautiful times in Cochin, Kalpana, Kanchana, Siva, Gayatri for the Besant Nagar sunsets, Susmith and Mayur for the cycle trips and late night hangouts, Priya and Myna for making IISc feel like home, Madhu, Deby, Pranjal, Abhi, Andreas and Meltem for the fun-soaked times at ICTP, Subbu and Arghya for trusting me with their lives on a Pune roadtrip and much needed company in Pilani(as there was no internet), Suchetha, Amrutha for the banters during schools, Prateek, Vaishak, Lijo et. al. for the inordinate amount of badminton, Vishakh, Kiran, Vimal and the friends at Grand Cafe for entertaining dinner conversations, and to Anusree for the wonderful evenings at Calicut. I thank  Vaishak V, Anitta and my seniors and friends at CHEP, for their help, support and company. \\

I am deeply grateful to my grandparents for their unconditional love and to my uncles Ramakrishnan, Swaminathan and Sreenivasan. It goes without saying, its all thanks to the love, care and support of Amma, Appa and Keechu, that any of this has been possible.

\pagebreak

\begin{center}
\textbf{\large \underline{ABSTRACT}}
\end{center}

\vspace{0.5cm}

\noindent This thesis consists of four parts. In the first part of the thesis, we investigate the phase structure of Einstein-Maxwell-Scalar system with a negative cosmological constant. For the conformally coupled scalar, an intricate phase diagram is charted out between the four relevant solutions: global AdS, boson star, Reissner-Nordstrom black hole and the hairy black hole. The nature of the phase diagram undergoes qualitative changes as the charge of the scalar is changed, which we discuss. We also discuss the new features that arise in the extremal limit.

In the second part, we do a systematic study of the phases of gravity coupled to an electromagnetic field and charged scalar in flat space, with box boundary conditions. The scalar-less box has previously been investigated by Braden, Brown, Whiting and York (and others) before AdS/CFT and we elaborate and extend their results in a language more familiar from holography. The phase diagram of the system is analogous to that of AdS black holes, but we emphasize the differences and explain their origin. Once the scalar is added, we show that the system admits both boson stars as well as hairy black holes as solutions, providing yet another way to evade flat space no-hair theorems. Furthermore both these solutions can exist as stable phases in regions of the phase diagram. The final picture of the phases that emerges is strikingly similar to that of holographic superconductors in global AdS, discussed in part one. We also point out previously unnoticed subtleties associated to the definition quasi-local charges for gravitating scalar fields in finite regions. 

In part three, we investigate a class of tensor models which were recently outlined as potentially calculable examples of holography, as their perturbative large-$N$ behavior is similar to the Sachdev-Ye-Kitaev (SYK) model, but they are fully quantum mechanical (in the sense that there is no quenched disorder averaging). We explicitly diagonalize the simplest non-trivial Gurau-Witten tensor model and study its spectral and late-time properties. We find parallels to (a single sample of) SYK where some of these features were recently attributed to random matrix behavior and quantum chaos. In particular, after a running time average, the spectral form factor exhibits striking qualitative similarities to SYK. But we also observe that even though the spectrum has a unique ground state, it has a huge (quasi-?)degeneracy of intermediate energy states, not seen in SYK. If one ignores the delta function due to the degeneracies however, there is level repulsion in the unfolded spacing distribution hinting chaos. Furthermore, the spectrum has gaps and is not (linearly) rigid. The system also has a spectral mirror symmetry which we trace back to the presence of a unitary operator with which the Hamiltonian anticommutes. We use it to argue that to the extent that the model exhibits random matrix behavior, it is controlled not by the Dyson ensembles, but by the BDI (chiral orthogonal) class in the Altland-Zirnbauer classification.

In part four, we construct general asymptotically Klebanov-Strassler solutions of a five dimensional $SU(2) \times SU(2) \times \Z_2\times \Z_{2R}$ truncation of IIB supergravity on $ T^{1,1} $, that break supersymmetry. This generalizes results in the literature for the $SU(2) \times SU(2) \times \Z_2\times U(1)_R$ case, to a truncation that is general enough to capture the deformation of the conifold in the IR. We observe that there are only two SUSY-breaking modes even in this generalized set up, and by holographically computing Ward identities, we confirm that only one of them corresponds to spontaneous breaking: this is the mode triggered by smeared anti-D3 branes at the tip of the warped throat. Along the way, we address some aspects of the holographic computation of one-point functions of marginal and relevant operators in the cascading gauge theory. Our results strengthen the evidence that {\it if} the KKLT construction is meta-stable, it is indeed a spontaneously SUSY-broken (and therefore bona fide) vacuum of string theory.

\pagebreak

\vspace*{7cm}

\begin{center}
	\textit{Dedicated to Appa, Amma, Keechu}\\[20pt]
	\textit{In Memory of} \\
	\textit{My Grandparents and my uncle Ramakrishnan}
\end{center}

\pagebreak

\pagenumbering{arabic}

\tableofcontents

\newpage

\leavevmode\thispagestyle{empty}\newpage

\newpage

\chapter*{\textbf{{ List of Publications}}}
\addcontentsline{toc}{chapter}{{{ List of Publications}}}

\begin{flushleft}
	\textsf{\textbf{\large This thesis is based on the following articles}}
\end{flushleft}

\begin{enumerate}

%%CITATION = doi:10.1007/JHEP03(2017)056;%%
%69 citations counted in INSPIRE as of 21 Feb 2018
%\cite{Basu:2016srp}

%%CITATION = doi:10.1007/JHEP11(2016)041;%%
%10 citations counted in INSPIRE as of 21 Feb 2018

%\cite{Basu:2016mol}
	\item
P.~Basu, C.~Krishnan and P.~N.~Bala Subramanian,
``Phases of Global AdS Black Holes,''
JHEP {\bf 1606}, 139 (2016),
doi:10.1007/JHEP06(2016)139,
[arXiv:1602.07211 [hep-th]].
%%CITATION = doi:10.1007/JHEP06(2016)139;%%
%8 citations counted in INSPIRE as of 21 Feb 2018

	\item
	P.~Basu, C.~Krishnan and P.~N.~Bala Subramanian,
	``Hairy Black Holes in a Box,''
	JHEP {\bf 1611}, 041 (2016),
	doi:10.1007/JHEP11(2016)041,
	[arXiv:1609.01208 [hep-th]].

	%\cite{Krishnan:2016bvg}
	\item C.~Krishnan, S.~Sanyal and P.~N.~Bala Subramanian,
	``Quantum Chaos and Holographic Tensor Models,''
	JHEP {\bf 1703}, 056 (2017),
	doi:10.1007/JHEP03(2017)056,
	[arXiv:1612.06330 [hep-th]].
	
	\item   C.~Krishnan, H.~Raj and P.~N.~Bala Subramanian,
	``On the KKLT Goldstino,''
	JHEP {\bf 1806}, 092 (2018),
	doi:10.1007/JHEP06(2018)092,
	arXiv:1803.04905 [hep-th].

\end{enumerate}

\begin{flushleft}
	\textsf{\textbf{\large Other publications during  PhD, not included in thesis}}
\end{flushleft}

\begin{enumerate}
\item	P.~Basu, C.~Krishnan and P.~N.~Bala Subramanian,
		``AdS (In)stability: Lessons From The Scalar Field,''
		Phys.\ Lett.\ B {\bf 746}, 261 (2015),
		doi:10.1016/j.physletb.2015.05.009,
		[arXiv:1501.07499 [hep-th]].
		%%CITATION = doi:10.1016/j.physletb.2015.05.009;%%
		%19 citations counted in INSPIRE as of 21 Feb 2018
\item	N.~Acharyya, A.~P.~Balachandran, V.~Errasti Díez, P.~N.~Bala Subramanian and S.~Vaidya,
``BRST Symmetry: Boundary Conditions and Edge States in QED,''
Phys.\ Rev.\ D {\bf 94}, no. 8, 085026 (2016),
doi:10.1103/PhysRevD.94.085026,
[arXiv:1604.03696 [hep-th]].
%%CITATION = doi:10.1103/PhysRevD.94.085026;%%
%1 citations counted in INSPIRE as of 21 Feb 2018
			
			%\cite{Krishnan:2016dgy}
\item C.~Krishnan, A.~Raju and P.~N.~B.~Subramanian,
			``Dynamical boundary for anti–de Sitter space,''
			Phys.\ Rev.\ D {\bf 94}, no. 12, 126011 (2016),
			doi:10.1103/PhysRevD.94.126011,
			[arXiv:1609.06300 [hep-th]].
			%%CITATION = doi:10.1103/PhysRevD.94.126011;%%
			%12 citations counted in INSPIRE as of 21 Feb 2018
			%\cite{Acharyya:2016xaq}

\item	C.~Krishnan, S.~Maheshwari and P.~N.~Bala Subramanian,
``Robin Gravity,''
J.\ Phys.\ Conf.\ Ser.\  {\bf 883}, no. 1, 012011 (2017),
doi:10.1088/1742-6596/883/1/012011, [arXiv: 1702.01429 [gr-qc]].
%%CITATION = doi:10.1088/1742-6596/883/1/012011;%%
%4 citations counted in INSPIRE as of 21 Feb 2018
			%\cite{Basu:2015efa}
\item		C.~Krishnan, K.~V.~Pavan Kumar and P.~N.~Bala Subramanian,
``(Anti-)Symmetrizing Wave Functions,''
arXiv:1711.09811 [hep-th].
%%CITATION = ARXIV:1711.09811;%%
%1 citations counted in INSPIRE as of 21 Feb 2018
%\cite{Krishnan:2017bte}

\end{enumerate}

\pagebreak

\leavevmode\thispagestyle{empty}\newpage

\chapter{Introduction}
In this thesis, we address some questions related to quantum gravity by applying techniques learned from Holography. The puzzle of finding a \emph{Quantum Theory of Gravity} has occupied the minds of theorists for almost a century, and the closest we have got to such a theory is \emph{String Theory}. String theory in itself is only understood as a perturbation theory, and a full non-perturbative understanding of the theory is still an open question. In the last couple of decades, thanks to String Theory which lead to to the AdS/CFT conjecture\cite{Mal}, we have been able to uncover much about gravity using a Holographic description. The idea of \emph{Holography} was first put forward by 't Hooft\cite{tHooft:1993dmi}, followed by Susskind\cite{Susskind:1994vu}, in which they suggested that the degrees of freedom of a theory with gravity in a region should be the same as the degrees of freedom living on the boundary of the region. The AdS/CFT conjecture put forward by Maldacena\cite{Mal} is a concrete realization of the Holographic principle, and it opened up new avenues to understand gravity. In the last two decades, the duality picture has been expanded to numerous other examples. Many of these examples are dualities between certain AdS backgrounds and boundary CFTs. However, some examples have also been found which do not have a bulk (super)gravity picture\cite{Losev:1997hx,Banks:1996vh,Fradkin:1987ks,Sezgin:2003pt,Klebanov:2002ja}, or even a continuous bulk geometry\cite{Gubser:2016guj}. The AdS/CFT correspondence has led to a paradigm, termed as \emph{Gauge/Gravity duality}. Additionally, using the \emph{Strong-Weak} duality nature of Gauge/Gravity duality has led to important discoveries in the previously inaccessible regimes of strongly-coupled gauge theories, by mapping them to more tractable gravity problems. It also provides a test-bed for ideas in strongly coupled regimes of gravity, as we can check them against the weakly coupled field theories in the boundary.

The thesis is organized as follows. In the rest of this Chapter, we will briefly review the AdS/CFT conjecture. In Chapter \ref{gads}, we will work with the idea of \emph{Holographic Superconductors}, which is used to model a theory of superconductor using a bulk gravity theory with matter, and explore the phase structure of the theory. The aim of this Chapter is to provide a road-map to explore a system presumed to be very similar, namely a Flat Box, which we investigate in Chapter \ref{hbbox}. In Appendix \ref{app_ch2}, we have provided some computational details required for this chapter and the next. Although \emph{No-Hair Theorems} exist in asymptotically flat space, we find in Chapter \ref{hbbox} that the Flat Box allows for Hairy Black Holes to exist. Moreover, the phase structure of the system is non-trivial and interesting, as they allow for thermodynamically stable configurations of (Hairy) Black Holes. We contrast the Flat Box case against the AdS case and also note important subtleties previously not noticed. In Chapter \ref{tensormodel} we investigate the Gurau-Witten(GW) Model, which is a potentially holographic quantum mechanical model that is very closely related to the Sachdev-Ye-Kitaev(SYK) Model. The GW Model is a fermionic tensor model, which in the large-N limit has the same diagrammatics as the SYK Model, suggesting saturation of the chaos bound\cite{bound}. We look into the smallest non-trivial $ N $, the Hamiltonian of which can be numerically diagonalized, and find that even in the finite $ N $ case the system indicates quantum chaos. In Chapter \ref{KKLT}, we consider $ SU(2)\times SU(2)\times \Z_2\times \Z_{2R} $ truncation of Type IIB supergravity in search of mode(s) that are responsible for spontaneous breaking of supersymmetry. We find two SUSY breaking modes and using techniques of Holography we find that one of the modes breaks SUSY spontaneously. This is precisely the mode that corresponds to the anti-D3 branes in KKLT. Some computational tools for studying this system are relegated to Appendix \ref{app_ch4}. 

In the rest of this Chapter, we will provide a brief review of the AdS/CFT correspondence.

\section{Introduction to the AdS/CFT Conjecture}
The first clues to $ AdS/CFT $ correspondence came in \cite{Brown:1986nw}, where the authors showed that the asymptotic symmetry algebra of $ AdS_{3} $ under Brown-Henneaux boundary conditions consists of two copies of Virasoro Algebra. The Hilbert space should form a representation of this algebra. In retrospect, we can understand this as a hint of the $ AdS_{3}/CFT_{2} $ correspondence. However, at the time it did not lead to the AdS/CFT correspondence, which had to wait until Maldacena's seminal work \cite{Mal}. We will now briefly outline the logic that led to the AdS/CFT conjecture. For a more detailed analysis, see \cite{Aharony:1999ti,Witten1,Natsuume:2014sfa}.

The $ AdS/CFT $ conjecture was first proposed in the context of string theory in $ AdS_{5}\times S^{5} $ being dual to $ \caln=4 $ Super Yang-Mills(SYM) with gauge group $ U(N) $ in 3+1 dimensions. To illustrate this, consider IIB string theory in $ \R^{9,1} $ with a stack of $ N $ parallel $ D3 $-branes. The precise meaning of stack is that we will be considering the string length $ l_{s}\rightarrow 0 $ limit later. The theory has two kinds of excitations: closed and open strings. The closed string excitations capture the gravitational perturbations around $ \R^{9,1} $ while the open strings capture those of the $ D3 $-branes. We can write down the low energy effective action, by looking at only the low energy excitations of the theory. If we are interested in energies lower than the string scale $ 1/\alpha' $ ($\sim 1/l_{s}^{2} $), then only the massless states are of interest. The full action can be written as
\begin{align}
	S= S_{open}+ S_{closed} + S_{int},
\end{align}
where $ S_{open/closed} $ corresponds to the low energy effective actions of the open/closed string excitations and $ S_{int} $ refers to the open-closed string interactions. $ S_{closed} $ corresponds to the action of IIB supergravity in 9+1-d, plus higher-derivative terms. $ S_{open} $ corresponds to the world-volume theory living on the 3+1-d branes, which consists of $ \caln=4 \; U(N)$ SYM plus higher-derivative corrections. $ S_{int} $, which denotes the open-closed interactions, can be understood as the interactions of the modes in the branes with those of the bulk. 

The bulk-brane interactions can be safely omitted in the low-energy limit. To see this, we can take the $ l_{s}\rightarrow 0 $ $ (\alpha'\rightarrow 0) $ limit keeping the energies and dimensionless parameters fixed, whence the terms in $ S_{int} $ drop out\cite{Aharony:1999ti}. In this limit, the higher derivative terms in the $ S_{open/closed} $ also drop out, leaving us with pure $ \caln=4 \; U(N)$ SYM, which is also conformal. and the free bulk supergravity. The bulk and brane modes are completely decoupled.

There is another perspective to look at the same system, where one starts by solving the supergravity with backreacted $ D3 $-branes (by taking $ \gs N $ to be large, where $ \gs $ is string coupling and $ N $ is the number of $ D3 $-branes). The system has two types of low energy excitations as seen by the observer at infinity. One of them corresponds to the very large wavelength excitations of the bulk, which spans the gravitational size of the brane (and hence is decoupled with respect to the branes), and the ones of arbitrarily high energy but very close to the stack of branes(which forms a horizon) which get infinitely red-shifted at asymptotic infinity. These two decoupled pieces can be described as the free bulk supergravity plus the excitations of the Near-Horizon geometry of the stack, which is $ AdS_{5}\times S^{5} $. The fact that the system is protected by supersymmetry is what allows one to freely tune the quantity $ \gs N $, as the string coupling $ \gs $ is just a modulus.

The two perspectives are descriptions of the same system, both of which has a decoupled piece corresponding to the free bulk supergravity. This leads to identifying string theory in $ AdS_{5}\times S^{5} $ as a dual description of the $ \caln=4 \; U(N) $ SYM theory. The duality is stated with $ \caln=4 \; SU(N) $ or $ U(N) $ SYM, depending on whether one is interested in including the $ U(1) $ vector multiplet which corresponds to the center of mass motion of the branes\footnote{See \cite{Aharony:1999ti} for more details}. The perturbative analysis in SYM is trustworthy in the limit
\bea\label{stronglim}
g_{YM}^{2}N \sim \gs N \sim \dfrac{R^{4}}{l_{s}^{4}} \ll 1,
\eea
whereas the classical supergravity description is valid when the $ AdS $ radius $ R $ is large comparable to string length $ l_{s} $
\bea\label{weaklim}
\dfrac{R^{4}}{l_{s}^{4}} \sim g_{YM}^{2}N \sim \gs N \gg 1.
\eea
The beauty of this duality is that when the gravity theory is classical, the SYM is strongly coupled and vice versa. The duality is hence a \emph{strong-weak duality}. It is only one of the regimes we can compute, hence, it is extremely difficult to show that the duality is precisely met. However, assuming the conjecture we can probe the non-calculable sectors of string theory or the SYM using the dual description. The stronger version of the conjecture is that the duality is valid for all values of $ \gs N $\cite{Mal,Aharony:1999ti}.

Although the duality cannot be exactly shown, we can gather evidence that reinforces the confidence in the conjecture. To start with, the spacetime and supersymmetries of both the descriptions match. The group of spacetime isometry group of $ AdS_{5} $ is given by $ SO(4,2) $, which in fact is the 3+1-d conformal group. For the case of supersymmetries, both the $ \caln=4 \; SU(N) $ SYM and $ AdS_{5}\times S^{5} $ have 32 supercharges, enhanced due to conformal symmetry, and the R-symmetry group $ SU(4)_{R} $ of the field theory can be identified with the $ SO(6) $ rotation group of the spacetime $ S^{5} $. Furthermore, the correlations functions that are protected by anomalies and the spectrum of chiral primaries can be matched\cite{Aharony:1999ti}, among many other tests, none of which we will go into. 

\section{Field-Operator Correspondence and Correlation Functions}
We have seen how in \cite{Mal}, it was argued that the string theory on $ AdS_{5}\times S^{5} $ was dual to $ \caln =4 \; U(N) $ SYM in 3+1-d. A precise statement of the duality would require a dictionary that illustrates the correspondence at the level of observables from one description to the other. This was done in \cite{Witten1}, which we will discuss in some detail below. The discussion is aimed at a more general $ AdS_{d+1}/CFT_{d} $ duality.
 
To start with, consider a massless scalar field $ \phi $ in $ AdS_{d+1} $, satisfying the condition $ \phi|_{\de AdS} =\phi_{0} $, then $ \phi_{0} $ should couple to a conformal field $ \calo $ as $ \int_{\de AdS} \phi_{0}\calo $ \cite{Witten1}. The AdS/CFT correspondence can be then stated as
\bea
\lag e^{\int d^{d}x \phi_{0}(x)\calo(x)}\rag_{CFT} = {\cal Z}_{String}[\phi_{0}],
\eea
where $ {\cal Z}_{String}[\phi_{0}] $ is the string partition function with the boundary condition for the scalar $ \phi|_{\de AdS} =\phi_{0} $. As we can see, the LHS is the generating function of the $ n $-point correlation functions of the CFT operator $ \calo(x) $. It is to be noted that we have only mentioned the scalar, as it is easier to deal with. The correspondence could be stated in more generality, with bulk gauge fields and metric fluctuations, and we would get 
\bea
\lag e^{\int d^{d}x (\phi_{0}(x)\calo(x) + J_{\m}(x)A^{\m}_{0}(x) + T_{\m\n}(x) g^{\m\n}_{0}(x) ) }\rag_{CFT} = {\cal Z}_{String}[\phi_{0},A^{\mu}_{0}(x),g^{\m\n}_{0}(x)],
\eea
where $ J_{\m}(x) $ corresponds to conserved current in the CFT and $ T_{\m\n}(x) $ is the CFT stress-tensor. The boundary values of the gauge field is $ A^{\m}_{0}(x) $ and of metric is $ g^{\m\n}_{0}(x) $. This relation can be used to find the correlation function consisting of $ \calo, J_{\m} $ and $ T_{\m\n} $. The boundary values $ \phi_{0} ,A^{\m}_{0}$ and $ g^{\m\n}_{0} $ act as sources. For the case of the scalar, the $ n $-point correlation function is
\bea\label{nptcorr}
\lag \calo(x_{1})\calo(x_{n})\dots \calo(x_{n}) \rag = \dfrac{\delta^{n}}{\delta\phi_{0}(x_{1})\delta\phi_{0}(x_{2})\dots \delta\phi_{0}(x_{n})}  {\cal Z}_{String}[\phi_{0}] \bigg|_{\phi_{0}(x) = 0}.
\eea

In the limit where the gauge theory is strongly coupled \eqref{weaklim}, where the supergravity approximation is valid, the string partition function can be approximated using the saddle point approximation to to be
\bea
{\cal Z}_{String}[\phi_{0}] \simeq e^{-I_{sugra}[\phi_{0}]},
\eea
where $ I_{sugra} $ is the classical supergravity action. 

We will now look at the massive scalar in $ AdS_{d+1} $ to see that we can indeed compute the 2-point function using this dictionary, and match with the result expected from $ CFT_{d} $. The action for the bulk theory is given by
\bea
S = \dfrac{1}{2}\int d^{d+1}x \sqrt{-g} \Big( g^{\m\nu} \de_{\m}\phi \de_{\n}\phi +m^{2} \phi^{2} \Big),
\eea
where $ g_{\m\n} $ is the $ AdS_{d+1} $ metric in the Poincar\'e coordinates, given by
\bea
ds^{2} = g_{\m\n} dx^{\m}dx^{\n} = \dfrac{R^{2}}{z^{2}}\Big(dz^{2}+ \eta_{ij}dx^{i}dx^{j}\Big),
\eea
with the AdS radius $ R $ and $ x^{i} $ denoting the $ \R^{d-1,1} $ boundary coordinates. In these coordinates $ z =0 $ represents the AdS boundary. Near the boundary, we can solve the differential equation, by setting $ \phi = z^{\Delta} $, which yields the relation
\bea
\Delta(\Delta-d) = m^{2}R^{2}\qquad \Rightarrow \ \ \Delta_{\pm} = \dfrac{d}{2}\pm \sqrt{\frac{d^{2}}{4}+m^{2}R^{2}}.
\eea
It can be easily seen that $ \Delta_{+} \geq \Delta_{-}$ and $ \Delta_{+}\geq \frac{d}{2} $, which means that the $ z^{\Delta_{+}} $ always decays as $ z \rightarrow 0 $. Hence the boundary condition is imposed as
\bea
\phi(z=0,x) = \lim\limits_{z\rightarrow 0 } z^{\Delta_{-}} \phi_{0}(x),
\eea
where $ \phi_{0}(x) $ is an arbitrary function of the boundary coordinates. One important point to notice is that the scaling dimensions $ \Delta_{\pm} $ are real for $ m^{2}R^{2} \geq -\frac{d^{2}}{4} $, which is called the Breitenlohner-Freedman bound. For the choice of mass such that
\bea
-\dfrac{d^{2}}{4}\leq m^{2}R^{2}< -\dfrac{d^{2}}{4}+1,
\eea
both $ \Delta_{\pm} $ can be interpreted as a source for the other, and both of these choices lead to consistent CFTs in the boundary. This mass range, where two choices of quantizations exist, is called the Breitenlohner-Freedman window.

In order to find the correlation functions using the relation \eqref{nptcorr}, first we have to define the bulk to boundary propagator. To this end, we first have to solve for the Green's function
\bea
(\Box -m^{2})K(z,x,x') = 0,\qquad\quad K(0,x,x') = \lim\limits_{z\rightarrow 0} z^{\Delta_{-} }\delta(x-x'), 
\eea
the solution to which is given by \cite{Aharony:1999ti,Witten1,GKP}
\bea
K(z,x,x') = \dfrac{\Gamma(\Delta_{+})}{\pi^{2}\Gamma(\Delta_{+}-2)} \left(\dfrac{z}{z^{2}+(x-y)^{2}}\right)^{\Delta_{+}}.
\eea
Using this bulk to boundary propagator, we can write the solution to the wave equation as
\bea
\phi(z,x) = \dfrac{\Gamma(\Delta_{+})}{\pi^{2}\Gamma(\Delta_{+}-2)} \int d^{d+1}x' \left(\dfrac{z}{z^{2}+(x-y)^{2}}\right)^{\Delta_{+}}\phi_{0}(x'),
\eea
where $ \phi_{0}(x') $ is the boundary condition we have set for the scalar field. Plugging this solution into the on-shell action and taking two functional derivatives and removing the divergence coming from the contact-term, one can find the 2-point function to be
\bea
\lag\calo(x_{1})\calo(x_{2})\rag = \dfrac{1}{|x-x'|^{2\Delta_{+}}},
\eea
which indeed is the 2-point function of the CFT operator $ \calo(x) $ of dimension $ \Delta_{+} $.\footnote{See \cite{Aharony:1999ti} for more details on the computation 2-point, 3-point and 4-pont functions.}

\section{Holography}
The AdS/CFT duality has revived the concept of holography with a concrete realization. The holographic principle was introduced after the following paradox was encountered. In \cite{Bekenstein:1993dz,Bekenstein:1972tm,Bekenstein:1973ur,Hawking:1976ra}, it was found that the maximum entropy enclosed by a region of spacetime is given by
\bea
S_{Bek} = \dfrac{A}{4G_{N}},
\eea
where $ G_{N} $ is the Newton's constant and $ A $ is the area of the surface enclosing the region of volume $ V $. This entropy is associated with a black hole filling the volume. Suppose there exists another state of larger entropy than $ S_{Bek} $, then we could add more matter along with it until it collapses into a black hole, whose entropy will be given by $ S_{Bek} $. This means that the entropy would have to decrease, violating the second law of thermodynamics. Furthermore, quantum field theories generally have the entropy scaling as volume, and not as area. The scaling however is consistent with a quantum theory living on the surface enclosing the region\cite{Shomer:2007vq}. The interior can be then thought of as a \emph{Hologram} of the quantum theory in the boundary of the region.

In the case of AdS/CFT this proposal is hard to check, owing to the infinite degrees of freedom of the CFT and the infinite volume of AdS bulk. We could however introduce a cutoff at $ z= \epsilon $ for the AdS ($ z=0 $ is the boundary of AdS and $ \epsilon $ is small), and the CFT will have a UV cutoff at distance $ \epsilon $. For $ \caln=4 \; SU(N)$ SYM living on $ S^{3} $ of unit radius and a UV cutoff at $ \epsilon $, the number of degrees of freedom is given by 
\bea
S\sim N^{2} \epsilon^{-3}.
\eea
In the gravity side, the entropy is given by 
\bea
S = \dfrac{A}{4G_{N}} = \dfrac{Vol(S^{5})R^{3}\epsilon^{-3}}{4G_{N}} \sim N^{2}\epsilon^{-3}.
\eea
This matching suggests that we can understand the AdS/CFT correspondence as a Holographic theory. What this suggests is that the entire AdS bulk can be reconstructed using the boundary CFT, in a process called bulk reconstruction (see \cite{Balasubramanian:1999ri,Giddings:1999qu} for details).

\section{Gauge/Gravity Duality}
The realization of the AdS$ _5 $/CFT$ _4 $ duality by Maldacena\cite{Mal} has led to numerous generalizations in the years that followed. The fundamental idea that one can study D-branes in string theory, and similarly M-branes in 11-d supergravity, has allowed us to explore the existence of such dualities. One of the earliest such studies was to look at Type IIB supergravity on $ T^{1,1} $ with $ N $ D3-branes \cite{KW,Morrison:1998cs}, where the bulk $ AdS_{5}\times T^{1,1} $ geometry was dual to conformal $ \caln=1\, SU(N)\times SU(N)$ gauge theory. Extending this picture with the addition of $ M $ fractional D3-branes (D5 branes with 2 directions wrapping the $ S^2 $ of $ T^{1,1} \sim S^2\times S^3$), a duality was established between a non-AdS bulk geometry and a boundary non-conformal $ SU(N+M) \times SU(N)$ gauge theory\cite{KT}. This example is an illustration that one does not need AdS in the bulk and CFT in the boundary in order to have a duality. This leads to the idea of \emph{Gauge/Gravity duality}. Many such non-AdS/non-CFT dualities are known, and are interesting for the fact that they provide insights into QCD-like theories(see \cite{Zaffaroni} for more details). Analogous to D-branes in String Theory, one could study M-branes in 11-d supergravity. Two different cases to look are that of M2 and M5 branes. With M2-branes, the most interesting one is the duality between M-theory on $ AdS_{4}\times S^{7}/ \Z_{k} $ and $ \caln=6 \; U(N)\times U(N)$  superconformal Chern-Simons-matter theories at level $ k $\cite{Aharony:2008ug}, and M5-branes play a role in understanding 6d $ \caln=(2,0) $ SCFTs(see \cite{Heckman:2018jxk} for a brief recent review). This is by no means a complete list of the known dualities that can come from studying branes in String/M-/F-theories. The general idea one can get is that by studying branes, a plethora of systems can be constructed by a top-down approach.

There is another avenue that Gauge/Gravity duality has opened up, which is a bottom-up approach. In this kind of a setup, one starts with a gravitational system coupled to Abelian gauge fields and various types of scalars. The essential difficulty is in knowing whether the system can be embedded in a UV-complete theory like String theory. Modulo such questions, it allows for the construction of theories which are of particular interest. For eg., studying Einstein-Maxwell-Scalar system with a negative cosmological constant lead to the construction of Holographic Superconductors \cite{HHH2,hhh0810}. Another such avenue is where we can start with a boundary theory at strong coupling which is solvable like SYK Model \cite{SYK,Kitaev}, Gurau-Witten Model\cite{WittenSYK,Gurau} or Klebanov-Tarnopolsky Model\cite{klebanov} and then try to understand the bulk gravity theory. Another system of interest is the duality between Vasiliev Higher spin theories \cite{Fradkin:1987ks} in 3+1-d and $ O(N) $ vector models in 2+1-d \cite{Sezgin:2003pt,Klebanov:2002ja}), which is not borne out of a String theory construction, but inspired by AdS/CFT. The Guage/Gravity picture has also lead to research in dS/CFT\cite{Strominger:2001pn} and Flat-Space Holography\cite{Bagchi:2010eg}.

\section{A Brief Introduction to the Thesis}

In this chapter, we have gone into some detail of the original AdS/CFT conjecture, as it is one of the simplest cases to study, and  gives the necessary intuition for understanding more general cases of Gauge/Gravity duality. In the rest of the thesis, our goals would be to use the ideas in Holography to study some interesting systems. Although the systems of interest in the thesis are superficially different, the questions we address with them are of interest from the viewpoint of Holography. The study of Einstein-Maxwell-Scalar theory in asymptotically global AdS in Chapter \ref{gads} illustrate the phase structures present in the system, and in the dual description describes a Insulator/Superconductor phases of the CFT. In Chapter \ref{hbbox}, we will use the intuition gained by studying global AdS to describe the phase structure in a Flat Box, although its implications for a (possible) boundary Holographic dual is not clear. In Chapter \ref{tensormodel}, we study the Gurau-Witten Model, which has the same large-$ N $ behaviour as the SYK Model, while not being plagued by the problems of taking disorder averaging. It is a strongly coupled finite-$ N $ theory that is numerically solvable (for the case that we consider). The final system that we consider, in Chapter \ref{KKLT} is the Klebanov-Strassler theory, where we look for spontaneous supersymmetry breaking modes. The core of the problem is to identify the spontaneous SUSY-breaking mode using the techniques of Holography, to address the problem of SUSY-breaking in KKLT, while not delving into the details of moduli stabilization of the string compactification.

%In what followed the original AdS/CFT correspondence, the proposal has been extended to numerous examples of AdS/CFT dualities. It has also been extended to cases where the bulk theory is not AdS, and the boundary theory a non-CFT. In addition, dualities have been established for bulk theories that do not have a supergravity interpretation (for eg, Vasiliev Higher spin theories \cite{Fradkin:1987ks} in 3+1-d are conjectured to be dual to $ O(N) $ vector models in 2+1-d \cite{Klebanov:2002ja,Sezgin:2003pt}). This has led to a more general terminology of \emph{Gauge/Gravity duality}. It is beyond the scope of this discussion to illustrate the success of AdS/CFT and the spin-offs.

\chapter{Phases of Global AdS Black Holes}\label{gads}

In the past decade we witnessed many interesting applications of holography in condensed matter inspired systems. Ones which are of immediate relevance to us, are a series of works \cite{Gubser,HHH2,hhh0810}, where it has been argued numerically that an Einstein-Maxwell-Scalar (EMS) system in AdS goes through a second order phase transition, in suitable parametric regimes, from a RN black hole to a hairy black hole. The resulting hairy black hole has been identified with a superconducting state (with a non-zero condensate) in the dual theory. 

In \cite{Gubser}, it was suggested that the black hole hair in AdS spacetime (in Einstein-Maxwell-Scalar theory with negative cosmological constant) will lead to a charged scalar condensate very close to the outside of the horizon, and this can be interpreted as the spontaneous breaking of the gauge symmetry in the theory, leading to superconductivity of black holes. This idea was used to construct the holographic superconductor \cite{HHH2,hhh0810}, and the physics of the phenomenon can be summed up in the following way. The presence of a non-trivial gauge field will drive the effective mass of the scalar negative in the region close to the outside of the horizon. If the effective scalar mass is sufficiently negative, this forces the scalar to take up a non-trivial profile (i.e. the scalar solution $ \psi = 0 $ becomes unstable). This leads to the formation of a condensate. The system takes the form of a superconductor when the scalar profile is such that it has a boundary value of zero, while the condensate is non-zero. To note, the formation of a condensate leads to a second order phase transition, as the entropy change in this case would be continuous. In terms of the boundary CFT, this means that the one-point function is non-zero with the source turned off. The EMS system in the bulk leads to a conserved boundary stress-tensor, a conserved $ U(1) $ current in the boundary and spontaneous symmetry breaking (due to formation of a non-zero condensate with the source turned off), whereby the boundary theory can be interpreted as describing a superconductor. In certain range of values for the parameters, these will form a thermodynamically stable solution. 

In these works the background considered was the Poincar\'e patch of the AdS geometry. It is natural to ask what happens if we look at the phases of the EMS system in an asymptotically global AdS spacetime. The aim of this chapter is to investigate this question in some detail. Unlike previous works in this direction (see \cite{Min1, Min2,Markeviciute:2016ivy,Bhattacharyya:2010yg} ), which considered the problem in the fixed charge\&mass ensemble, we consider this problem in the (grand) canonical ensemble\footnote{Also, we work with a conformally coupled scalar field in AdS$_4$, unlike the massless cases in AdS$_5$ considered elsewhere.}. 

Even without a scalar field, the phase structure of EMS system in global AdS is more interesting than in the Poincar\'e patch. Unlike the Poincar\'e patch case where the dual field theory lives in a flat space, here the dual theory lives on a sphere and has a mass gap coming from the curvature of the sphere. Because of this, at any finite value of the temperature $T$ and chemical potential $\mu$ there is only one phase in the Poincar\'e patch, corresponding to the RN (or Schwarzschild, if $\mu=0$) black hole. The situation changes in global AdS and we have (generalizations of) the Hawking-Page transition \cite{HP, Hawking-Reall,catastrophic}. 

When we add a scalar, in addition to global AdS and RN black hole, we have two new hairy saddle points of gravity: one is called the boson star (see eg., \cite{Brihaye, Gentle}) and the other is the hairy black hole. Depending on the boundary chemical potential and temperature, one of the four solutions dominates the free energy landscape, giving rise to an intricate phase diagram. To have a quick idea about the phase diagram, the reader may consult Figs. \ref{gq53}, \ref{gql1}. Our phase diagrams bears a rough similarity with the phase diagram of EMS in another gapped geometry, the AdS soliton \cite{HorWay2010,Takayanagi,RaamBasu}.

This Chapter is based on \cite{Basu:2016mol}.

\section{The Setup}
\noindent The action for a Maxwell field and a charged scalar coupled to gravity \footnote{From now on we refer this system by EMS, i.e. Einstein-Maxwell-Scalar system. EM would stand for a similar system without the scalar field.}, is given by\cite{hhh0810}
\bea\label{gaction}
S=  -\frac{1}{16 \pi G}\int d^{4}x \sqrt{-g}\left( R + \dfrac{6}{L^{2}} -\dfrac{1}{4} F_{\mu\nu} F^{\mu\nu} -|\nabla\psi - i q A \psi|^{2}  - V(|\psi|)\right).
\eea
We will set $G=1$ in what follows. We would like to work with a time independent ansatz, which is also spherically symmetric. For the metric we pick 
\bea
ds^{2} = -g(r) h(r) dt^{2} + \dfrac{dr^{2}}{g(r)} + r^{2}\, d\Omega_{2}^{2},
\eea
and for the Maxwell and scalar fields
\bea
A = \phi(r)dt,\ \ \text{and}\ \psi = \psi(r).
\eea
The scalar field can be taken to be real, using a gauge transformation that fixes this phase \cite{hhh0810}. We also work with the potential for the scalar field of the form $V(|\psi|)= -2 M^{2}\psi(r)^{2}/L^{2}$. %In the analysis, we will set the scalar mass with $M=1$. 
With the above ansatz, we get the equations of motion
\begin{align}
\label{geoms}
&\psi ''(r) + \frac{g'(r) \psi '(r)}{g(r)}+\frac{q^2 \psi (r) \phi (r)^2}{g(r)^2 h(r)}-\frac{V'(\psi)}{2 g(r)}+\frac{h'(r) \psi '(r)}{2 h(r)}+\frac{2 \psi '(r)}{r}= 0 ,\\
&\phi ''(r) -\frac{2 q^2 \psi (r)^2 \phi (r)}{g(r)}-\frac{h'(r) \phi '(r)}{2 h(r)}+ \frac{2 \phi '(r)}{r} = 0 ,\\
&\frac{1}{2} \psi '(r)^2 + \frac{g'(r)}{r g(r)}+\frac{q^2 \psi (r)^2 \phi (r)^2}{2 g(r)^2 h(r)}+\frac{\phi '(r)^2}{4 g(r) h(r)}- \dfrac{3}{L^{2} g(r)}-\frac{1}{r^2 g(r)}+\frac{V(\psi)}{2 g(r)}+\frac{1}{r^2} = 0 ,\\
&h'(r)-r h(r) \psi '(r)^2 -\frac{q^2 r \psi (r)^2 \phi (r)^2}{g(r)^2} = 0.
\end{align}
What we have to look for are solutions, with the asymptotic behaviour of global AdS. The set of equations have the following scaling symmetries for the functions
\begin{itemize}
\item $r \rightarrow a r, \ \ q\rightarrow \frac{q}{a}, \ {\rm and } \ L\rightarrow a L $. With this rescaling, one could set $L =1$.
\item $h \rightarrow \bar{h}= a^{2} h, \ \phi \rightarrow \bar{\phi} =  a \phi, \ {\rm and }\ t \rightarrow \bar{t}= \frac{t}{a}$, so that the time part of the metric becomes
\bea
-g h dt^{2} = - g \dfrac{\bar{h}}{a^{2}}dt^{2} = -g \bar{h} d\bar{t}^{2}.
\eea
Now, since we need the space to be asymptotically global AdS, we need to have\footnote{We will suppress the \emph{bar} on $\bar{h}$ in what follows with the understanding that we are always using the rescaled metric.}
\bea
\lim_{r\rightarrow\infty}\bar{h}(r) = 1,\ \text{i.e} \ \ \bar{h}(\rb)= a^{2} h(\rb)=1 \Rightarrow a = \dfrac{1}{\sqrt{h(r\rightarrow\infty)}}.
\eea
\end{itemize} 
We will be looking at conformally coupled scalar, which sets $M =1$. The asymptotic expansion for such a scalar for any asymptotically-$AdS_{4}$ space has the falloff
\bea
\lim_{r\rightarrow\infty} \psi(r) \simeq \dfrac{\psi_{(1)}}{r} + \dfrac{\psi_{(2)}}{r^{2}} + \dots \; .
\eea
We will consider the boundary conditions for the systems at $r\rightarrow\infty$ of the following form,
\bea
g(r\rightarrow\infty) \simeq 1 +r^{2}, \ h(r\rightarrow\infty)= 1,\ \psi_{(1)}= 0,\ \text{and} \  \phi(r\rightarrow\infty) = 2 \mu.
\eea
(See footnote 4 for a discussion on the origin of the factor of 2 in the definition of $\mu$.)
Depending on whether we are looking for solutions with or without the horizon, we have to specify the boundary condition at $r=0$, if there is no horizon, or at $r=\rh$ which is the location of the horizon. This boundary condition will be discussed separately, depending on the solutions that we will be looking at.\\
We use the scaling symmetry to set $L=1$, for all the numerical solutions in the following.

\section{The different solutions}
As we will see in the following discussion in detail, there exist four different (classes of) solutions for the action in \eqref{gaction}. Of them, global AdS and RNAdS are solutions where the scalar is zero all throughout and will be discussed first. For non-trivial scalar profiles the solutions are called boson star and the hairy black hole, which will be discussed in more detail. The thermodynamic favourability of these four solutions are based on the free energy of the respective solution. 
\subsection{Global AdS}
The vacuum solution of the action in \eqref{gaction}, is the global AdS, given by
\bea
g(r) = 1+ \dfrac{r^{2}}{L^{2}}, \ h(r) =1, \ \phi'(r) =0,\ \text{and} \ \psi(r) = 0, 
\eea
We can choose $\phi(r)=2 \mu$, a constant. This solution exists for any chemical potential and temperature, but it is not necessarily the global minimum of the free energy. At $T=\mu=0$ the only possible solution of \eqref{gaction} is global AdS.
 
\subsection{RNAdS}
\begin{figure}[h!]
\begin{center}
\includegraphics[width=8.2cm]{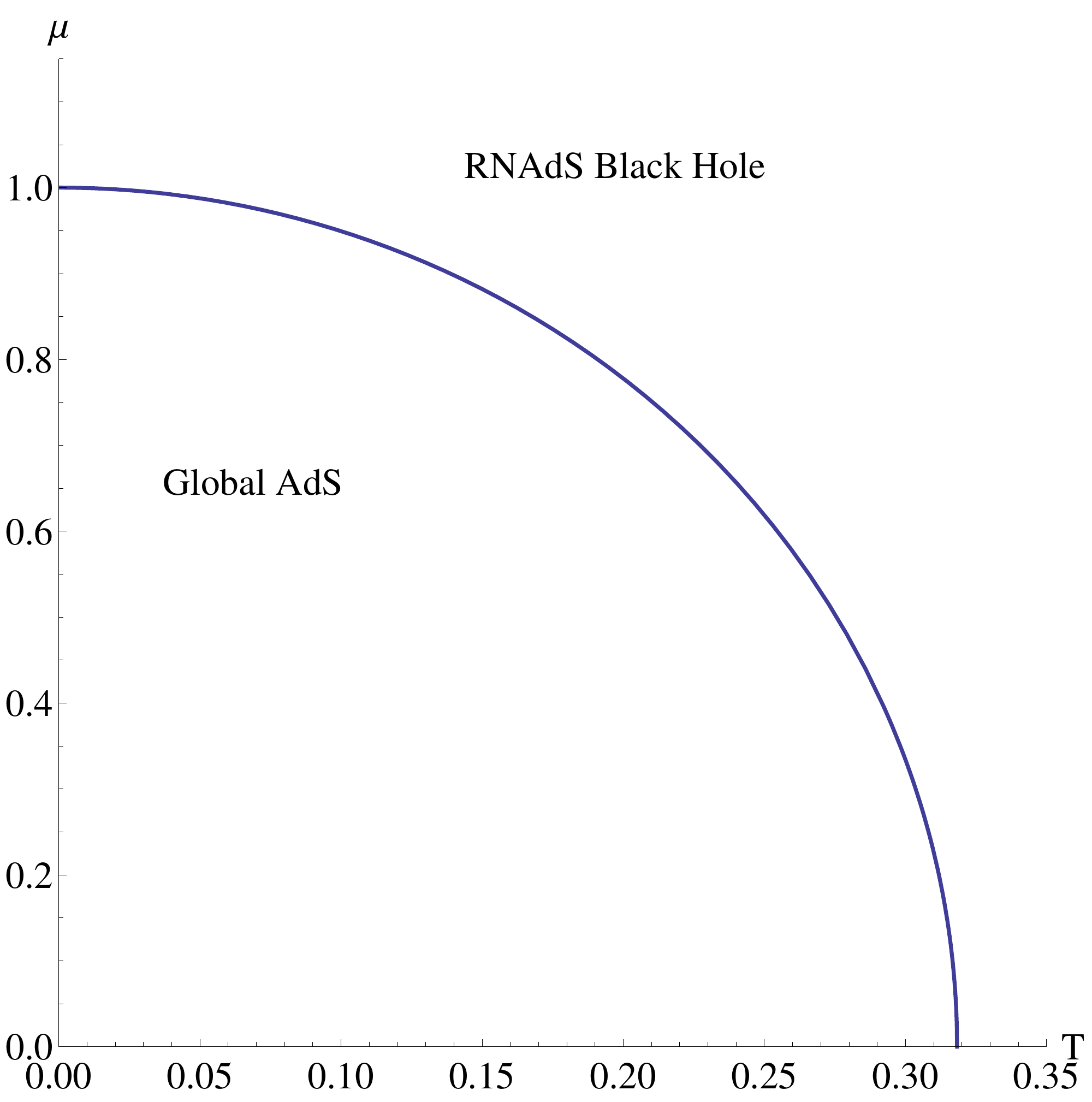}
\end{center}
\caption{Phase diagram for RNAdS black hole}
\label{grnadsphasedig} 
\end{figure}

With the scalar in \eqref{gaction} turned off, we impose the boundary conditions at the horizon for the metric
\bea
g(\rh)=0,\ \text{and} \ \phi(\rh) =0,
\eea
which is required to ensure the regularity of the gauge connection. This is enough to completely fix the solution, which is given by
\bea
g(r) = 1 + \dfrac{\rho^2}{4 r^2} +\dfrac{r^2}{L^2}- \dfrac{L^2 \rho^2 + 4 L^2 \rh^2 + 4 \rh^4}{4 L^2 \rh \, r}, \, h(r) =1, \, \phi(r) =\rho\left(\dfrac{1}{\rh} - \dfrac{1}{r}\right).
\eea
The number of scaling symmetries in the case of global AdS is one less than that in the Poincar\'e patch\cite{hhh0810}. This means that one cannot set the horizon and $L$ to be 1 at the same time. This means that for different horizon radii, the solutions are different, for a given value of $L$. The temperature for the metric ansatz that we have, by imposing that there is no conical singularity after Wick rotation, is given by 
\bea
T = \dfrac{1}{4\pi} \dfrac{g'(\rh)}{\sqrt{h(r\rightarrow \infty)}}.
\eea
For the RNAdS system, we get the temperature and chemical potential to be
\bea 
T =\frac{1}{4 \pi}\left( \dfrac{1}{\rh} - \dfrac{\rho^2}{4 \rh^3} + \dfrac{3 \rh}{L^2}\right),\ \text{and} \ \mu = \dfrac{\rho}{2 \rh}.
\eea
The free energy of the system can be computed by evaluating the classical action for the solution ($I=\beta F$), and subtracting out the same of global AdS \cite{Witten2}, for details see Appendix \ref{app_ch2}. It can alternately be evaluated using the ADM mass and charge of the system, as
\bea\label{gtherm}
F = E -T \, S -\mu \, Q,
\eea
where\footnote{We are following the conventions of \cite{catastrophic}, except for the fact that we have an extra factor of $1/4$ in the Maxwell piece in the action as in \cite{HorWay2010}. This is the source of the extra factor of 2 between $\rho$ and $Q$, as well as between $\phi(r\rightarrow \infty)$ and $\mu$.} $Q=\rho/2$. %We The unusual factor of $4$ is included due to the $\frac{1}{4}$ term multiplying the Maxwell part in the action \eqref{action}. 
Both approaches lead to the free energy
\bea
F = -\dfrac{4 \rh^4 + L^2 (\rho^2 - 4 \rh^2)}{16 \rh}
\eea
for the RN black hole. From the free energy one may chart down the phase diagram. The RNAdS black hole becomes the dominant phase when the free energy goes negative. Imposing this condition, we can solve for $T$ in terms of $\mu$ or vice versa, which demarcates the two phases. These two phases are separated by a line of first order transition know as Hawking-Page transition \cite{HP}(see Fig.\ref{grnadsphasedig}). 

\subsection{The different instabilities and hairy solutions}
We will now look into the hairy solutions, by turning on the scalar field and finding the instability of global AdS and RNAdS for forming scalar hair. The corresponding solutions are boson star and hairy black hole respectively. Also, we will look into the properties of the hairy solutions, and investigate the phase structure of the full system.
\subsubsection{Boson star instability}
As the chemical potential increases to a critical value, say $\mu =\mu_{c 1}$, the scalar field will develop a zero mode, i.e. we will have  $\psi_{(1)}=0$ (scalar condensate), without the formation of a horizon. This configuration is called Boson star \cite{Brihaye}. Here, boundary conditions at $r=0$ is given by $\phi'(0)=0,h'(0)=0,\psi'(0)=0,g'(0)=0$ which ensure that there is no kink at $r=0$. Also, we set $g(0)=1$ and $h(0)$ is set to an arbitrary value, as we will rescale it to make sure that the asymptotic boundary conditions are satisfied. \\

To understand the onset of the formation of the scalar zero mode, we look at the equation of motion of scalar in global AdS. A probe computation is enough to specify the instability because for very small scalar profile, the back-reaction is negligible. By a probe computation here, we mean taking $\psi(r)\rightarrow \epsilon \psi(r)$ with $\epsilon\ll 1$. Now, looking only upto linear order terms in $\epsilon$, the two equations coming from the Einstein equation and the Maxwell equation decouple from the scalar, giving the RNAdS solution. The scalar equation of motion then becomes a homogeneous equation in this background, and is given by
\bea
\psi ''(r) + \left(\frac{2 r}{L^2 \left(\frac{r^2}{L^2}+1\right)}+\frac{2}{r}\right) \psi '(r) + \left(\frac{2 M^2}{L^2 \left(\frac{r^2}{L^2}+1\right)}+\frac{4\mu ^2 q^2}{\left(\frac{r^2}{L^2}+1\right)^2}\right) \psi (r) = 0.
\eea
 The solution of this equation with the above mentioned boundary conditions can be analytically determined and is given by
 \bea
\psi(r) = C \ \dfrac{\sin(2 q\mu \tan^{-1}r)}{r}, \ \text{where} \ q\mu =n, \ n\in \mathbf{Z},
\eea
where $C$ is arbitrary, and signifies the overall scaling freedom we get in the probe limit. The quantization in $n$  happens because AdS is like a box, as far as the scalar is considered. 

We will look at the first non trivial solution, which is $n=1$, for the boson star phase, as there are no nodes in the scalar profile. For the $n=1$ mode, we have the relation $\mu \, q =1$, which gives the chemical potential ($\mu_{c1}$) at which the global AdS phase goes to the boson star phase, for a given $q$. It can be seen that for large $q$ the condensate can be formed by a very small $\mu$ and vice-versa.
 
\begin{figure}[h!]
    \centering
    \begin{subfigure}[b]{8cm}
        \includegraphics[width=\linewidth]{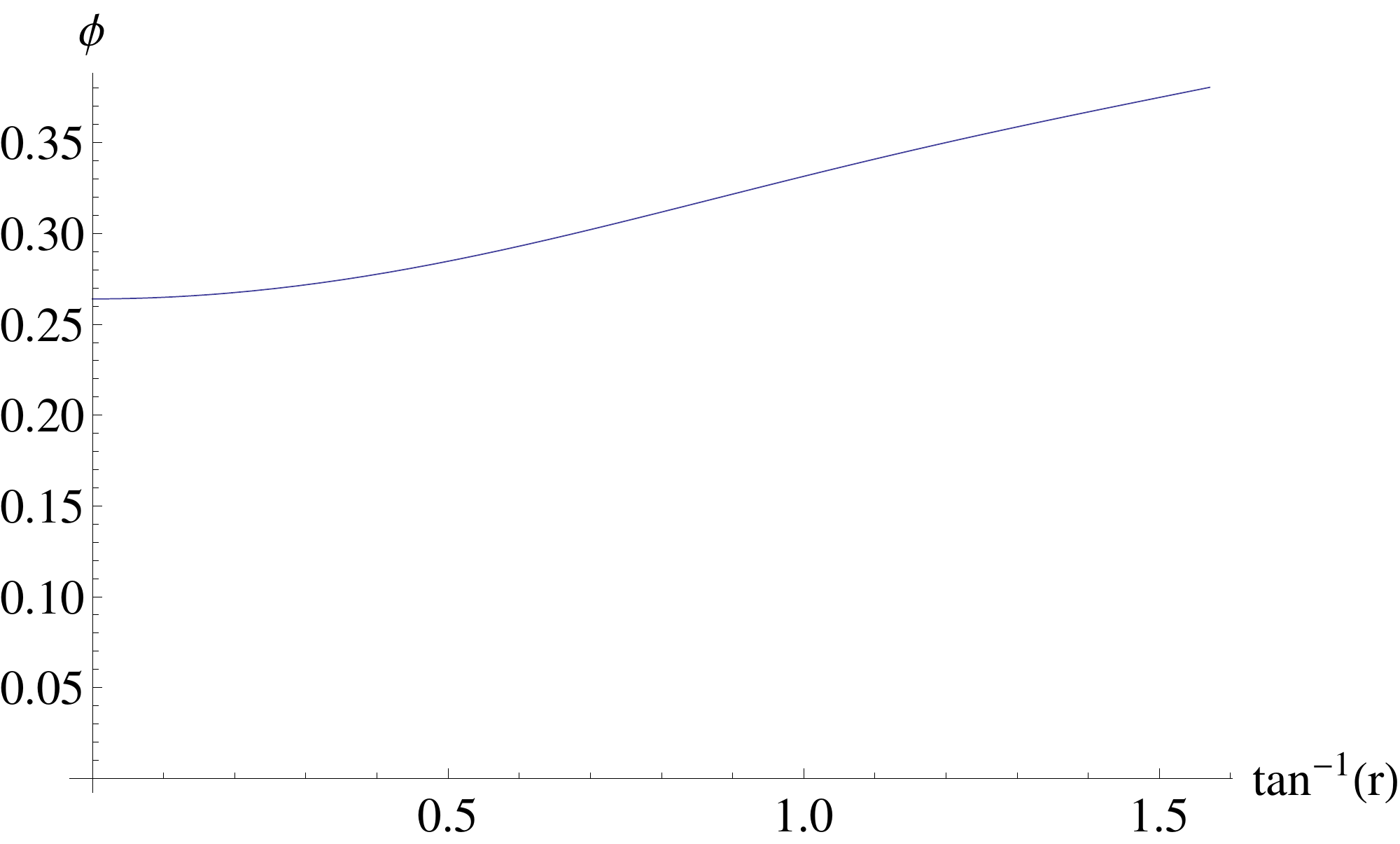}
        \caption{}
     \end{subfigure}
    \begin{subfigure}[b]{8cm}
        \includegraphics[width=\linewidth]{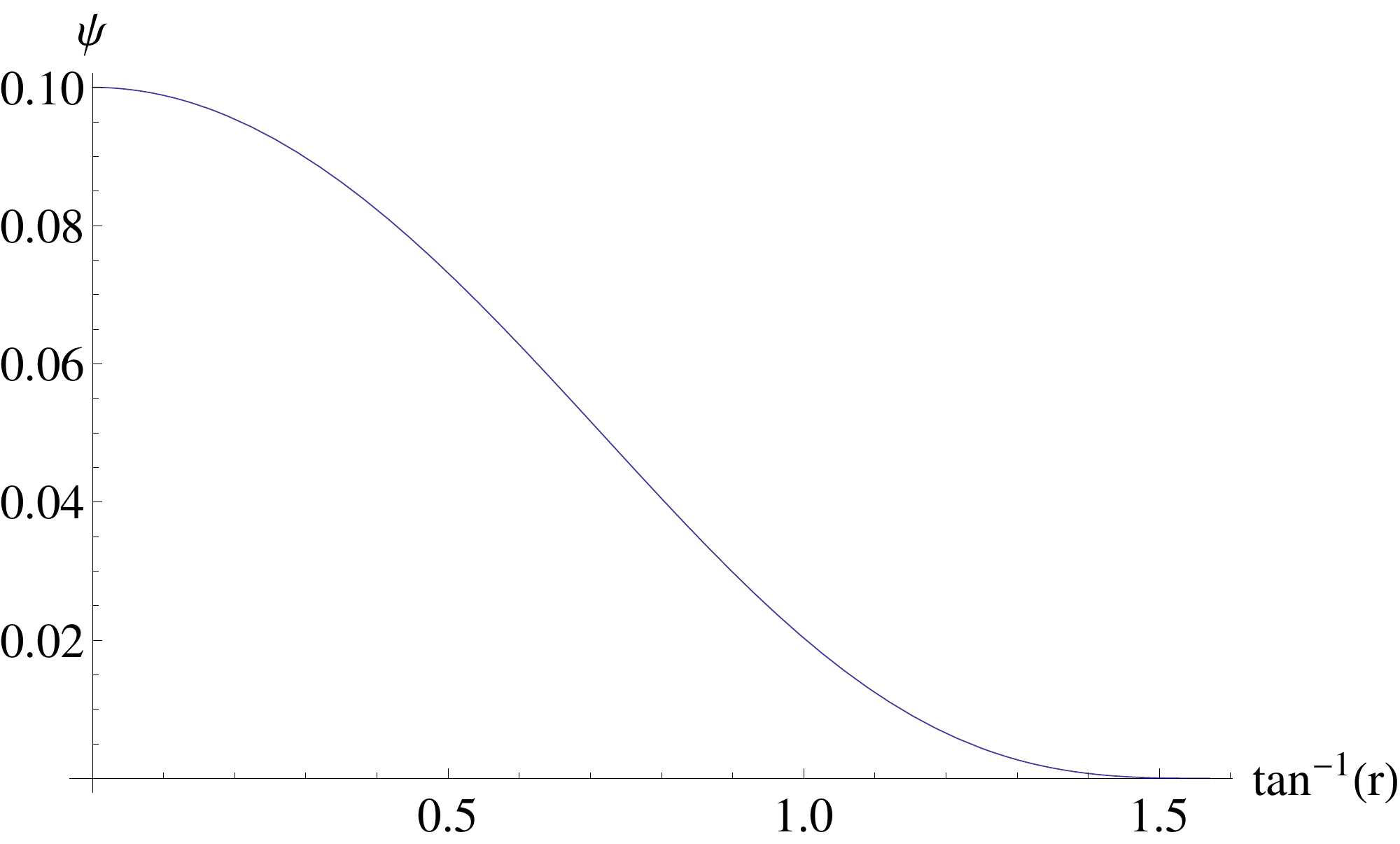}
        \caption{}
        \end{subfigure}
     \caption{The profiles of $\psi(r)$ and $\phi(r)$ in the fully backreacted solution for boson star.}\label{gbsprofiles}
\end{figure}

As we move away from the probe case, the solutions can be found numerically\footnote{Except for determining the point of onset of the instability, we do not use the probe computation anywhere.}. Here we have plotted the fully back reacted solution for $q=10$ and $2 \mu=0.3804$, see Fig.(\ref{gbsprofiles}). Since it is the fully back-reacted solution, the relation $q\mu =1$ doesn't hold.
 
\subsubsection{Hairy Black Hole instability}
A similar analysis can be done in a RNAdS background, for a probe scalar for the formation of a hair. The scalar equation of motion is solved in the RNAdS background, with the boundary conditions at the horizon $r =\rh $ for the scalar given by $\psi(\rh )= \psi_{0}$, and the first derivative $\psi'(\rh )$ fixed by the consistency of the series expansion of the equations of motion around the horizon. For the asymptotic boundary, $\psi_{(1)} = 0$. The scalar equation of motion in the RNAdS background is given by,
\bea
\psi ''(r) +\left(\frac{\rho^2 (r-2\rh)+4 r \rh \left(2 r^3+\rh^3+\rh\right)}{r (r-\rh) \left(4 r \rh \left(r^2+r \rh+\rh^2+1\right)-\rho^2\right)}+\frac{2}{r}\right) \psi '(r) \qquad\qquad \nonumber\\ \qquad \qquad +\left(\frac{8r^{2} \left[2q^{2} \rho^{2} - M^{2}\rh (\rho^{2} - 4r\rh (1+r^{2}+ r \rh +\rh^{2}))\right]}{(r-\rh )(\rho^{2} - 4r\rh (1+r^{2}+ r \rh +\rh^{2}))^{2}} \right) \psi(r) = 0.
\eea
For a given value of $q$ and $\rh $, we take an arbitrary value for $\psi_{0} $ (because the probe equation is homogeneous, the solutions are rescalable) and find %$\phi'(\rh) $ 
$\rho$ such that the scalar develops a zero mode $\psi_{(1)}=0$, by a numerical shooting method. This fully determines the RNAdS black hole background as we know $\rh$ and $\mu=\rho/2 \rh$, from which we can also find the temperature $T$. Repeating this for different values of $\rh$ keeping $q$ fixed will give the instability curve\footnote{Where this line features will be discussed while considering the full phase diagram}. This instability indicates a second order phase transition from the RNAdS to hairy black hole phase. For a fixed $q$, and varying $\rh $ we get a curve in the $\mu-T$ plane, which demarcates the two phases. As in the case of the boson star, here too, the larger the value of $q$, the lower the chemical potential $\mu$ required to form a condensate. 

\begin{figure}[h!]
    \centering
    \begin{subfigure}[b]{0.4\linewidth}
        \includegraphics[width=\linewidth]{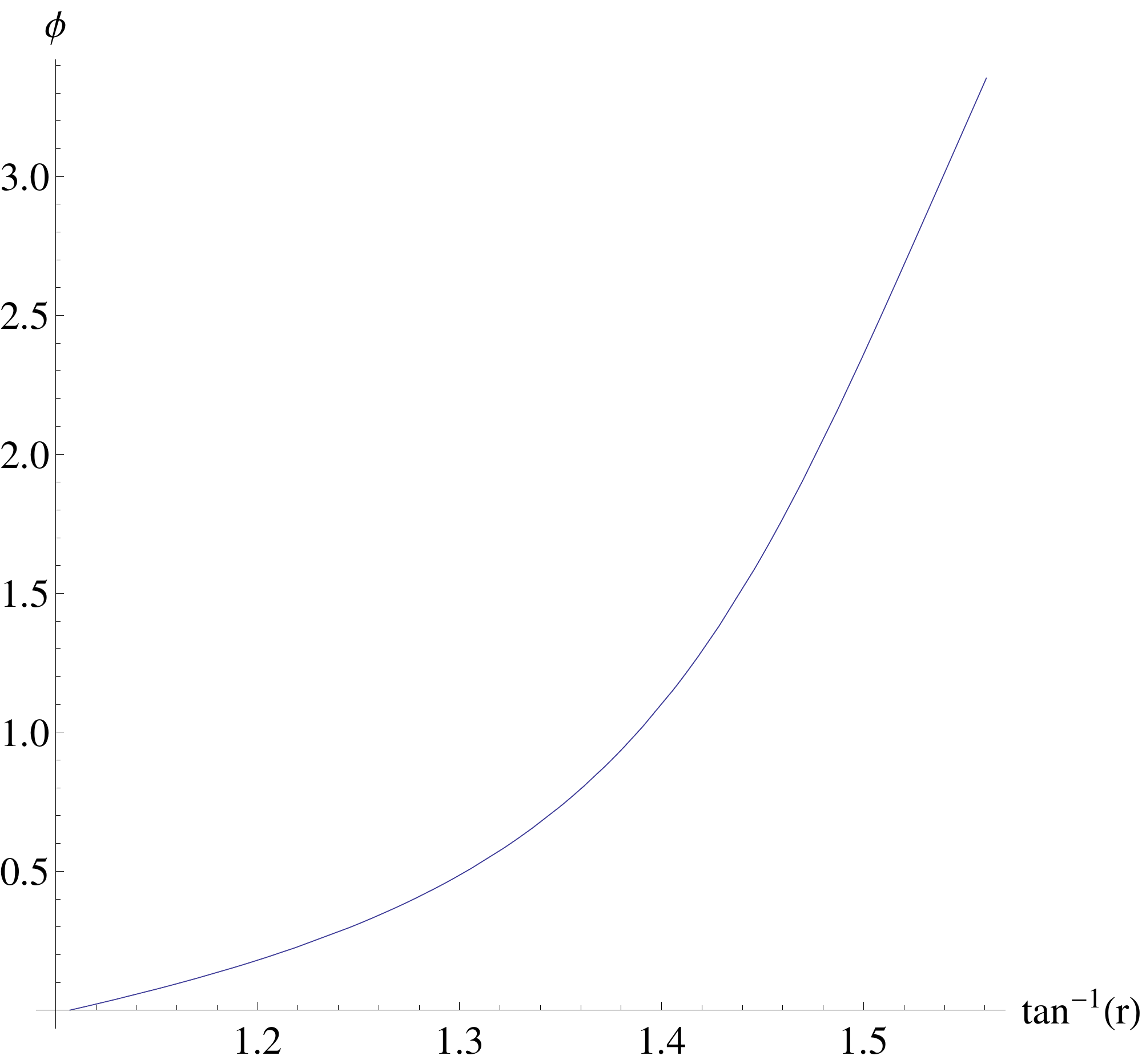}
        \caption{}
         \end{subfigure}
    \begin{subfigure}[b]{0.4\linewidth}
        \includegraphics[width=\linewidth]{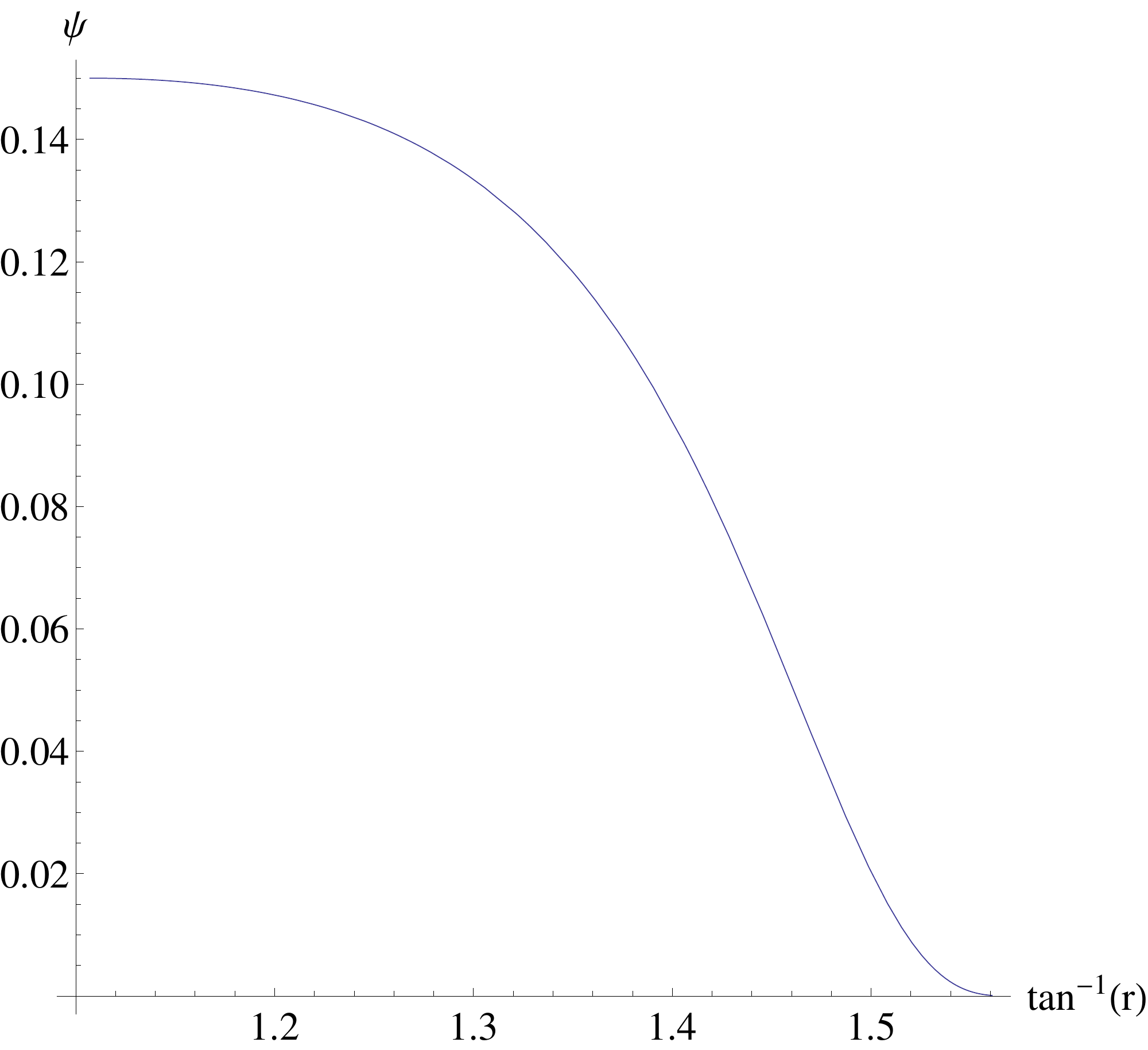}
        \caption{}
         \end{subfigure}
     \caption{Sample profiles of $\psi(r)$ and $\phi(r)$ in the fully backreacted solution for hairy black hole.}\label{gbhprofiles}
\end{figure}

We also can construct the fully back-reacted hairy solution numerically, see Fig.(\ref{gbhprofiles}).

%
%\section{Instability and Hairy Solutions}
%Now we can add a scalar field into the theory
%\bea
%S_{scalar} = \dfrac{1}{16\pi G}\int d^{4}x \sqrt{-g}\left(-|\nabla\psi - i q A \psi|^{2}  - V(|\psi|)\right)
%\eea
%A scalar field in global AdS has eigenfrequencies given by $e$. If the $\mu>\mu_c=\lambda_0$, then the first eigenmode of the scalar field condenses without forming an horizon. In short we would like to find the condition for $Q$ for a given $\rh$ such that the scalar field has a zero mode.  This configuration is called Boson star \cite{keylist}. 
%
%Similar instability exists for RN-AdS black holes also. 
%
%\begin{figure}[H]
%\begin{center}
%\includegraphics[width=.7\linewidth]{instability_gads.pdf}
%\end{center}
%\caption{Phase diagram of RNAdS along with instability curve}
%\label{instablty}
%\end{figure}
%The phase diagram of hairy AdS black hole is more complicated than the EYM system. In EYM system there are only two possible radially symmetric solutions, i.e. global AdS and a charged BH. Whereas EYMH system has two more type of solutions with non-zero scalar condensate: boson stars and hairy black holes. Depending on the temperature and the chemical potential one of the four type of solutions may be the dominant phase of the theory. First we would go through a short introduction to the different kind of solutions. Then we would discuss the details of the phase structure in in section\ref{}.

\section{Phase diagram}
The phase diagram of the Einstein-Maxwell-Scalar (EMS) system is more complicated than the EM system. In EM system there are only two possible radially symmetric solutions\cite{Hawking-Reall}, i.e. global AdS and a charged BH. Whereas, as we discussed, EMS system has two more types of solutions, which are boson stars and hairy black holes. Depending on the temperature and the chemical potential, one of the four possible solutions becomes the dominant phase (i.e. the phase with least free energy) of the theory. The intricateness of the phase diagram depends on the value of $q$.

As we saw in the boson star instability discussion, the chemical potential that sets up this instability is given by $\frac{1}{q}$. In Fig.\ref{grnadsphasedig}, we can see that $\mu =1$ for $T=0$, above which the global AdS is not the stable solution. Since the boson star is a second order transition from global AdS, it does not exist once the chemical potential required for this instability is not within the global AdS phase, i.e. if $\mu>1$, or in other words, when $q<1$. Hence, there will be a qualitative difference in the phase diagram for $q>1$ and $q<1$, which we will deal with separately.

\subsection*{$ \infty >q>1$}
\begin{figure}[h!]
\begin{center}
\begin{tikzpicture}
\coordinate (a) at (0,0) {};
\coordinate (b) at (8,0) {};
\coordinate (c) at (0,8) {};
\coordinate (d) at (0,2.5) {};
\coordinate (e) at (4.5,0) {};
\coordinate (f) at (4,2.5) {};
\coordinate (g) at (3,4) {};
\coordinate (h) at (2.5,7.5) {};
\coordinate (i) at (7,7) {};

\draw[ultra thick] (a) -- (b);
\draw[ultra thick] (a) -- (c);
\draw[ultra thick,dashed] (d) --node[midway, below right] {S-1} (f);
\draw[blue,ultra thick] (e) to [out=90,in=300] node[midway, below right] {F-1} (f);
\draw[green,ultra thick] (f) to [out=120,in=280] node[midway, below right] {  \;\;F-2} (g);
\draw[purple,ultra thick] (g) to [out=100,in=280] node[midway, below right] {F-3}(h);
\draw[red,ultra thick,dashed] (g) --node[midway, below right] {S-2} (i);
\node at (7.5,-0.5) {$T$};
\node at (-0.5,7.5) {$\mu$};
\node at (6,2.5) {$\framebox{RNAdS}$};
\node at (1.5,1) {$\framebox{global AdS}$};
\node at (1.2,6) {$\framebox{Boson Star}$};
\node at (5,7) {$\framebox{Hairy BH}$};

\end{tikzpicture}
\end{center}
\caption{Schematic phase diagram for $\infty  >q >1$.}
\label{gfig:phase1}
\end{figure}
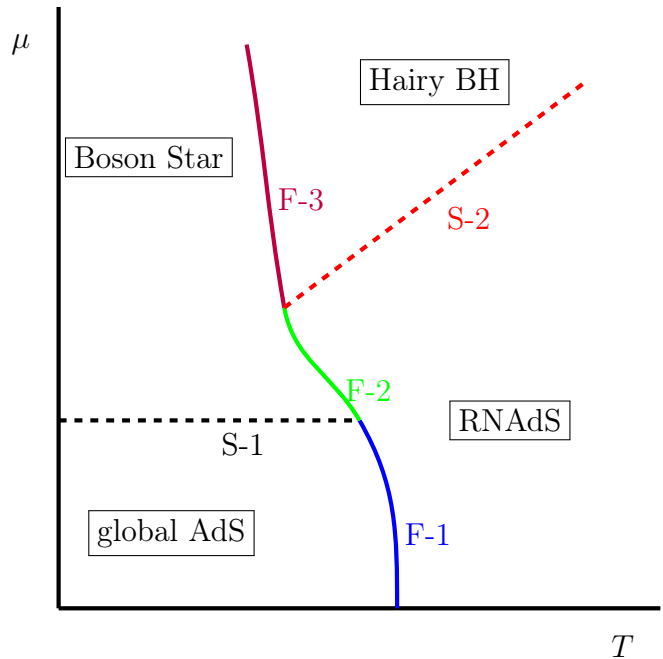

In the case of $q>1$, the boson star instability happens at $\mu<1$. The schematic phase diagram is given in Fig.\ref{gfig:phase1}. These phase diagrams are found for a fixed $q$, by a mixture of analytic and numerical methods where appropriate.

For sufficiently small $\mu$, which is less than required for the boson star instability ($\mu_{c1}$), we will have global AdS and RNAdS as the phases in the theory, demarcated by a first order phase transition, which is the Hawking-Page transition, given by the curve \emph{F-1} in the figure. As we increase $\mu$, beyond $\mu_{c1}$, there is a second order phase transition \emph{S-1}, from global AdS to boson star. These two phase boundaries are analytically tractable, as we saw. The Hairy black hole instability happens at a larger value of $\mu$, say $\mu_{c2}$, for any given temperature. The two phases in the region $\mu_{c1}<\mu<\mu_{c2}$, will be separated by a first order transition, given by the curve \emph{F-2}. We can find the curve \emph{F-2}, semi-analytically, as follows. First we numerically find the fully back-reacted boson star solutions, for some value of $\psi(0)$, above the line \emph{S-1}. A given value of $\psi(0)$ corresponds to a unique boson star solution, so that fixes its $\mu$ and $F$. It is now straightforward to analytically compute the temperature $T$ of the RNAdS black hole with this same $\mu$ and $F$ and that gives us a point on the \emph{F-2} curve in the $T-\mu$ plane. Changing the value of $\psi(0)$ and repeating the process produces the full \emph{F-2}.

As we keep increasing the value of $\mu$ further, the RNAdS black hole is unstable towards formation of scalar hair. The RNAdS and hairy black hole phases are separated by a second order phase transition, given by the curve \emph{S-2}. This curve is found numerically, by looking at the hairy black hole instability, as discussed in the previous section, for different $\rh$. Now, the hairy black hole and boson star phases will also have a first order phase boundary given by the curve \emph{F-3}, which is again found numerically as we discuss presently.

The way in which the phase diagram is computed for the global AdS is slightly different from the way in which it is done for the case of the Poincar\'e patch\cite{HorWay2010}. For the Poincar\'e patch, since there is an additional coordinate rescaling freedom, what one does is take a value of $\mu$, rescale all the solutions to have the same value of $\mu$, and then evaluate the free energies for different $T$, thus finding the phase boundaries. The lack of that rescaling in global AdS does not become an issue while evaluating the curves \emph{F-1} and \emph{F-2} because they are (semi-)analytic. Determining \emph{S-1,S-2} is also straightforward because they are probe computations as we explained, which rely on the fact the phases with condensates have lesser free energy than phases with condensate.

\begin{figure}[h!]
	\centering
	\begin{subfigure}[b]{0.45\linewidth}
		\includegraphics[width=\linewidth]{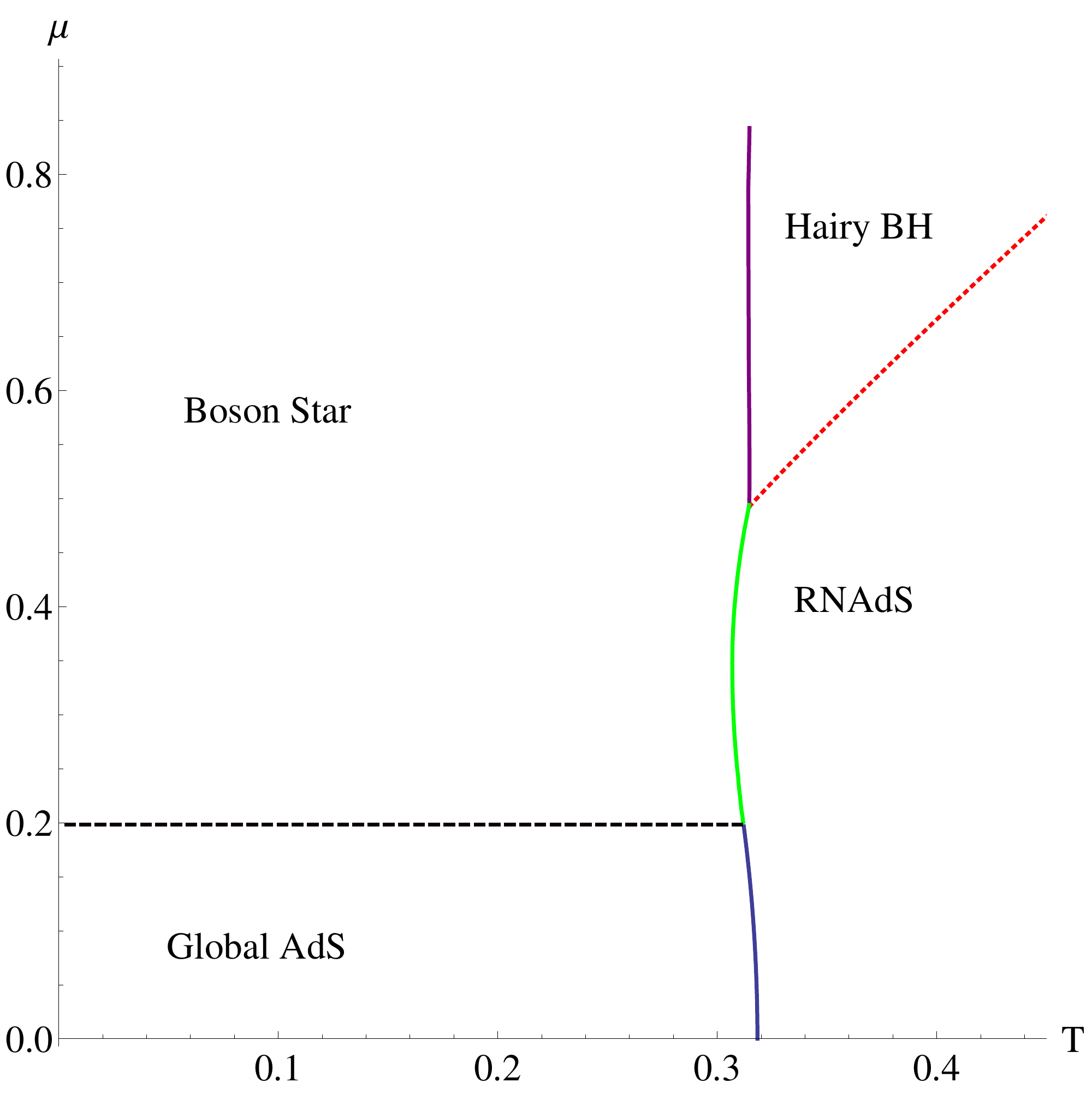}
		\caption{}
	\end{subfigure}
	\begin{subfigure}[b]{0.45\linewidth}
		\includegraphics[width=\linewidth]{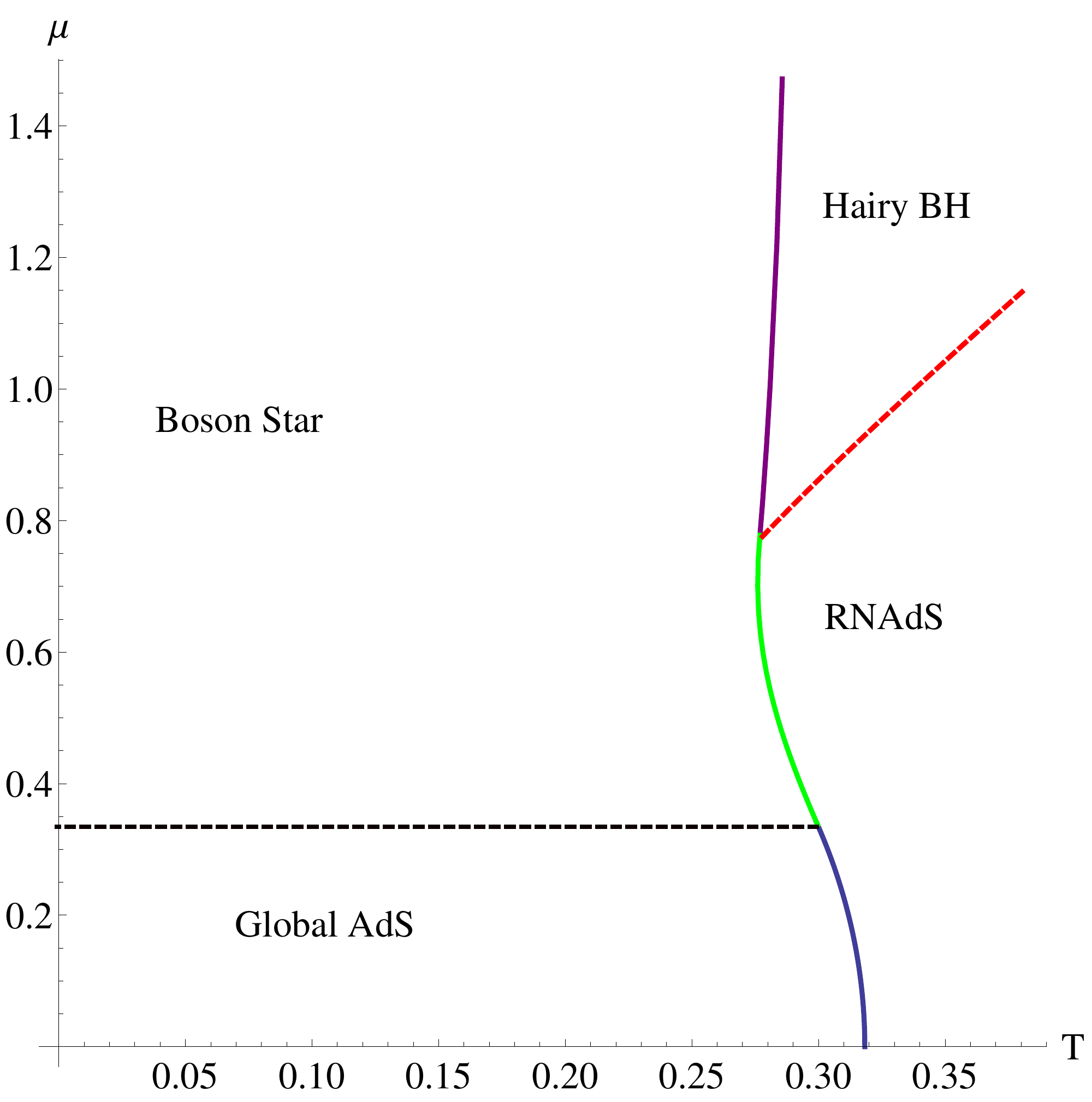}
		\caption{}
	\end{subfigure}
	\begin{subfigure}[b]{0.45\linewidth}
		\includegraphics[width=\linewidth]{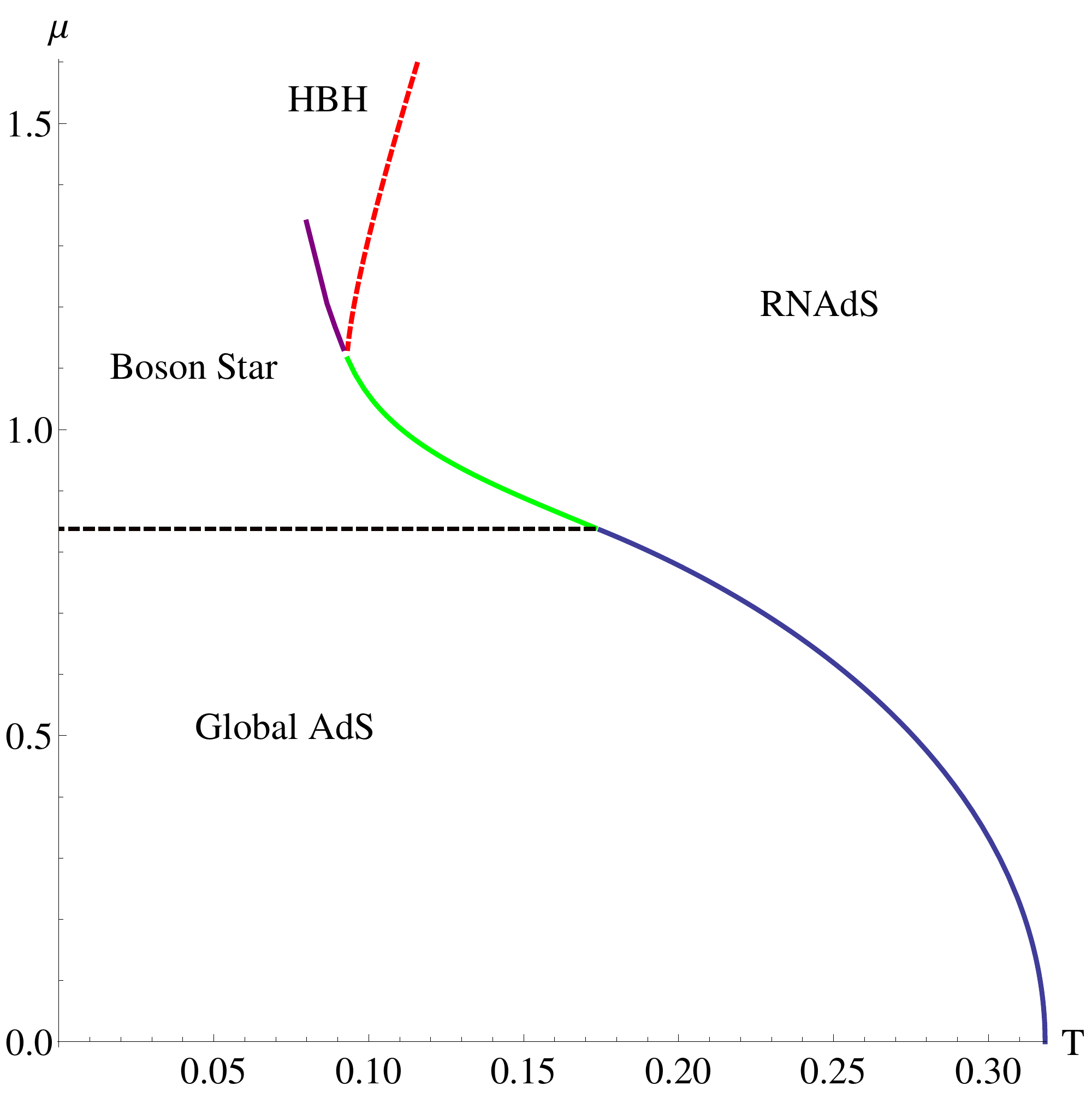}
		\caption{}
	\end{subfigure}
	\caption{Phase diagram calculated for $q=5,3,1.2$.}\label{gq53}
\end{figure}

The curve \emph{F-3} is more difficult to find, precisely because of the missing coordinate rescaling freedom, and the fact that both the phases to be compared are fully back-reacted solutions, hence, completely numerical. What we do first is to find the free energy ($F$) of the boson star for different values of $\mu$, and use a numerical fit to get $F$ as a function of $\mu$. One useful fact to keep in mind for the hairy black hole solutions is that for a fixed $\rh $, as we increase the value of $\psi_{0}$, the chemical potential and the temperature both increase, the latter very slowly. Thus, to obtain the curve \emph{F-3}, we take horizon radius slightly less than that which corresponds to the black hole %given the $(T,\mu)$ value of the 
at the intersection of the curves \emph{F-2} and \emph{S-1} (we will call this critical horizon radius, $r_{h22}$), and keep increasing $\psi_{0}$, simultaneously evaluating $\mu$, $T$ and the free energy. The curve \emph{F-3} is gotten by finding the $(T,\mu)$ values for which the free energies of the hairy black hole and the boson star match, by repeating the process mentioned above for smaller and smaller values of $\rh$ compared to $r_{h22}$.

One thing to be noted is that for large $\mu$, the black hole is very big and the line of hair-forming instability asymptotes to a linear curve $\mu=\alpha T$, where the parameter $\alpha$ is $\frac{\mu}{T}$ in the Poincar\'e patch-AdS, and is dependent on $q$. %({\bf WHY IS THIS EXPECTED?})

The exact phase diagrams evaluated for $q=5,3,1.2$, are given in Fig.\ref{gq53}.

\begin{figure}[h!]
\begin{center}
\begin{tikzpicture}
\coordinate (a) at (0,0) {};
\coordinate (b) at (8,0) {};
\coordinate (c) at (0,8) {};
\coordinate (d) at (0,4.5) {};
\coordinate (e) at (4.5,0) {};
\coordinate (f) at (0,5) {};
\coordinate (g) at (4,7){};

\draw[ultra thick] (a) -- (b);
\draw[ultra thick] (a) -- (c);
\draw[blue,ultra thick] (e) to [out=90,in=0] node[midway, below right] {\;\;F-1} (d);

\draw[red,ultra thick,dashed] (f) to [out=0,in=235] node[midway, below right] {S-2} (g);
\node at (7.5,-0.5) {$T$};
\node at (-0.5,7.5) {$\mu$};
\node at (6,2.5) {$\framebox{RNAdS}$};
\node at (1.5,1) {$\framebox{global AdS}$};
\node at (1.2,6) {$\framebox{Hairy BH}$};

\end{tikzpicture}
\end{center}
\caption{Schematic phase diagram for $q\leq 1$.}
\label{gfig:phase2}
\end{figure}
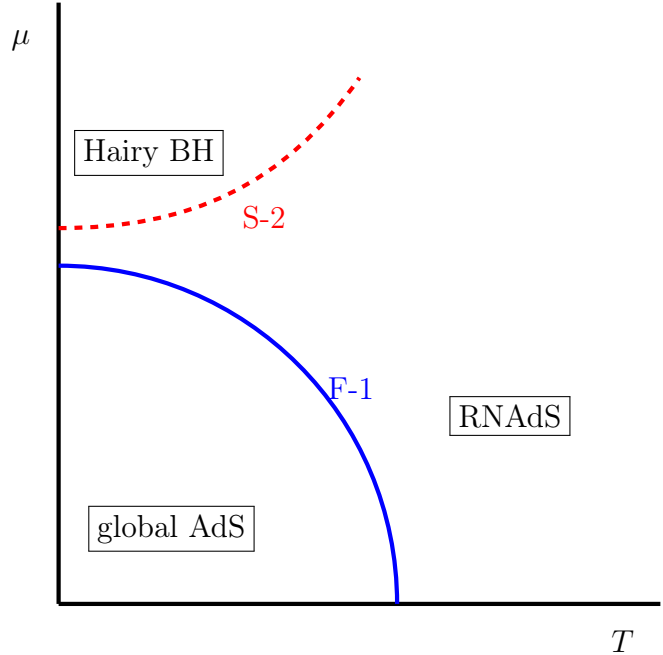

\subsection*{$ q \le 1 $}

\begin{figure}[h!]
    \centering
    \begin{subfigure}[b]{0.45\linewidth}
        \includegraphics[width=\linewidth]{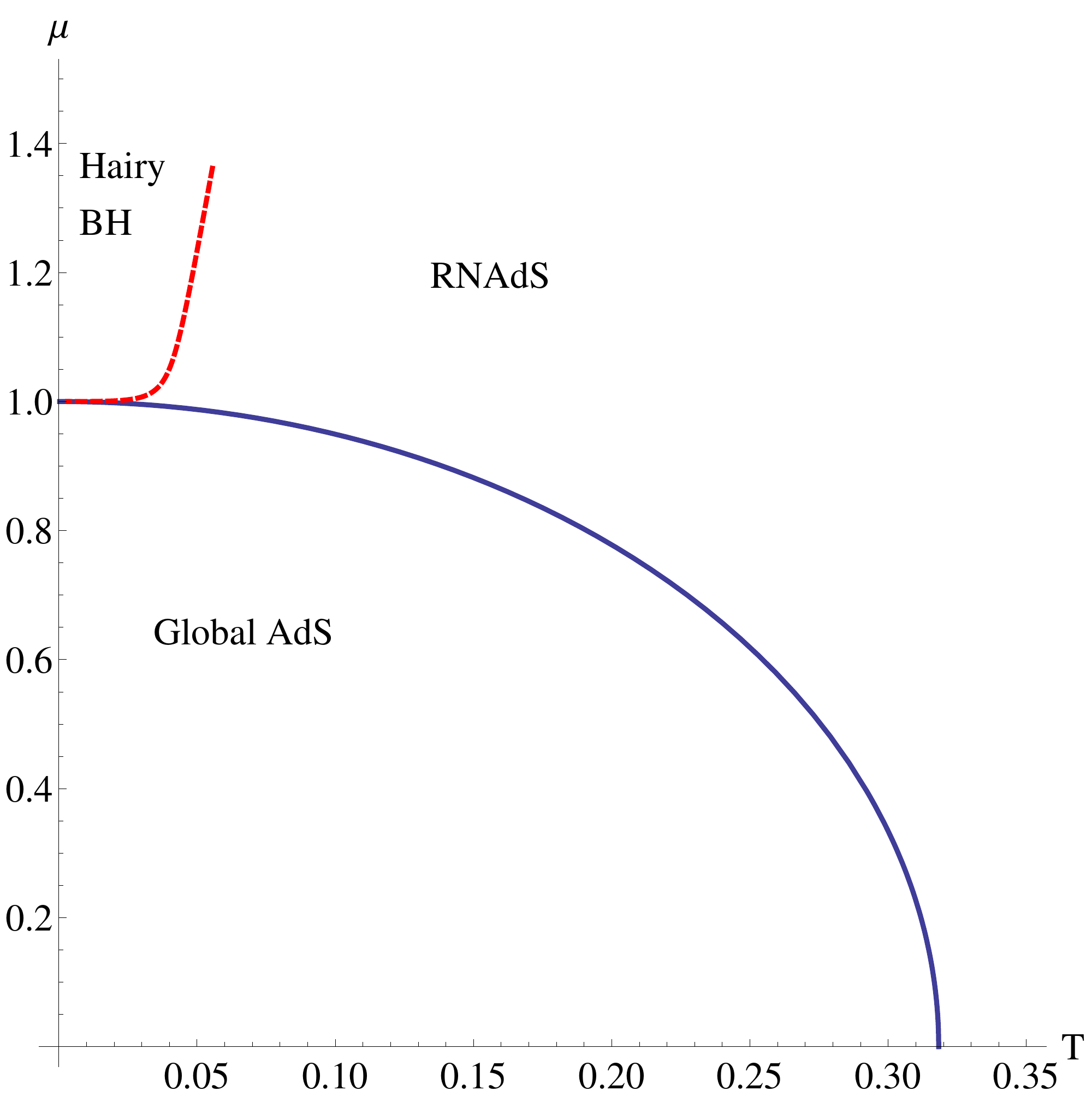}
        \caption{}
        \label{gq1}
        \end{subfigure}
    \begin{subfigure}[b]{0.45\linewidth}
        \includegraphics[width=\linewidth]{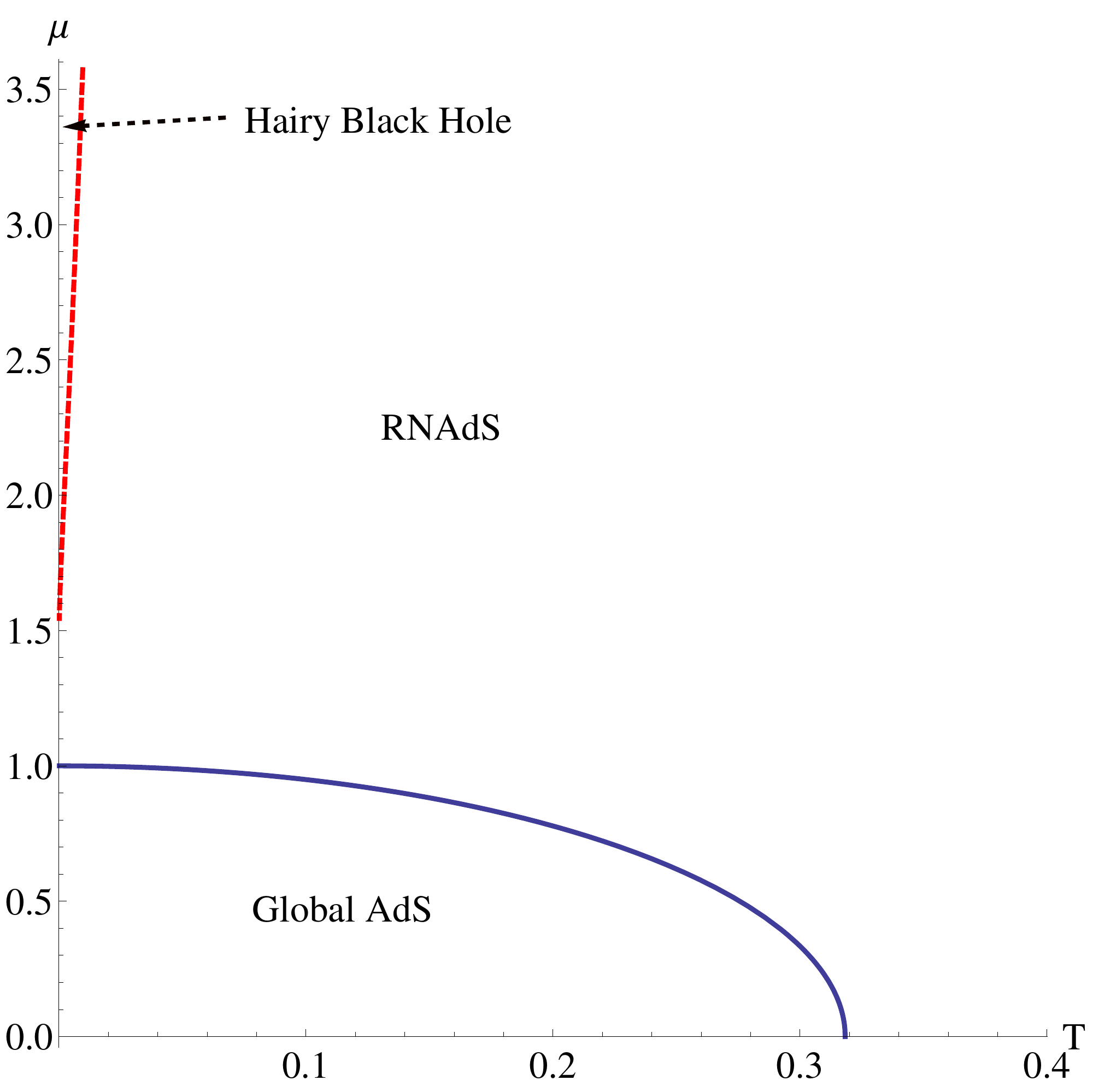}
        \caption{}
          \end{subfigure}
     \caption{Phase diagram calculated for $q=1,\frac{1}{2}$.}\label{gql1}
\end{figure}

At $q=1$, it can easily be seen from our analytic results that boson star instability happens at $\mu=1$. As the extremal RN black hole near $\mu=1$ has a radius approaching zero, the geometry is almost identical to the global AdS. Hence it is no surprise that line of hairy black hole instability and boson star instability coincides for $T=0,\mu=1$ and $q=1$ in Fig.\ref{gq1}. 
 
We now look at $q\leq 1$, where, as we had discussed earlier, the boson star instability happens at $\mu\geq 1$. The transition from global AdS to boson star is a second order transition. So as we increase the chemical potential to $\mu\geq 1$, the system undergoes a first order transition to RNAdS, the Hawking-Page transition (the curve \emph{F-1} in Fig.\ref{gfig:phase2}), before the boson star instability could set in. Since the system is now in the black hole phase, the possible second order transition is the one in which the RNAdS develops a hair. This transition can be found as in the earlier case, numerically. The starting point of the instability curve (named here also as \emph{S-2}) is at the $T=0$ axis. %, i.e. this is an instability also of the extremal black hole. ({\bf Check THIS stenetence})

The phase diagram in this case has only three different phases. The exact phase diagrams for $q=1,1/2$ are given in Fig.\ref{gql1}. In the case of $q=1$, the boson star and hairy black hole instability seem to happen at $\mu =1$. Here too, the number of phases remain three.

A Chandrasekhar-like instability for AdS black holes was found in \cite{Min1, Min2} when the $q$ was below some bound. In our case, that bound corresponds to $q=1$. We have checked that the for values of $q$ that are less than 1, the curvatures diverge at $r=0$ as we increase the condensate of the boson star. (But note that in our case, in this regime the boson star is not the dominant phase anyway.)

It is interesting to note that in our case we form a hairy block hole for arbitrarily small values of $q$ (see also, \cite{Sumit}). In the Appendix \ref{app_ch2}, we discuss the extremal case analytically.

\subsection{Comments on Condensate Plots}

%We also look at the plots for the condensates by looking at invariant quantities. 

\begin{figure}[h!]
\centering
\includegraphics[width=0.5\linewidth]{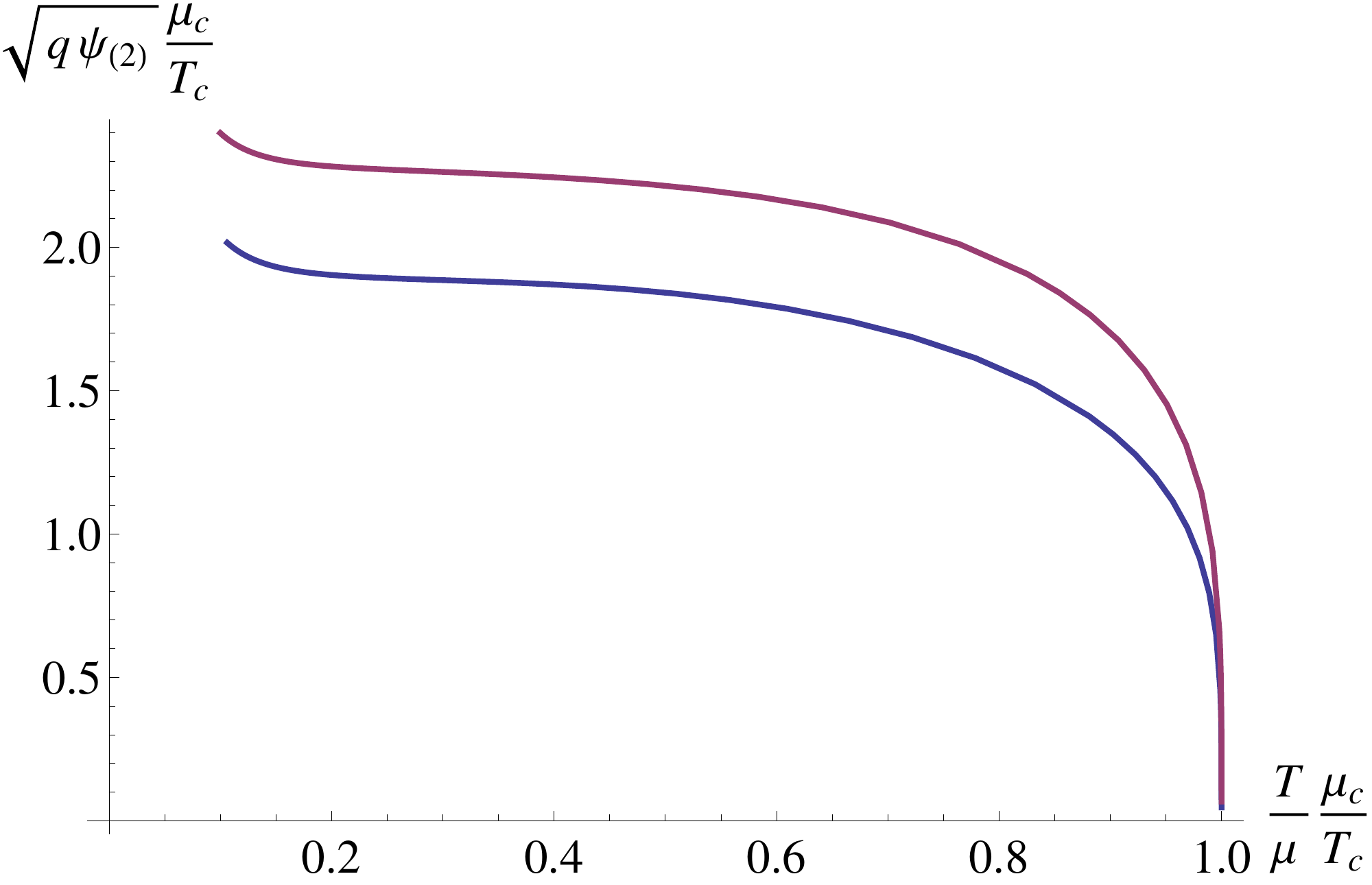}
\caption{Condensate plots for $q=10$ and $\rh = \tan^{-1}(0.9)$  [purple], $\tan^{-1}(1)$ [blue].}
\label{gcondensate}
\end{figure}

In the case of the Poincar\'e patch, the condensate plots were made for different values of $q$ alone, as there were two coordinate rescalings available. Here, however, one could plot for various values of $q$ and for different values of $\rh$. We have given the plot for $q=10$, for $\rh = \tan^{-1}(0.9),\tan^{-1}(1)$ in Fig.\ref{gcondensate}. Here $\mu_{c}$ and $T_{c}$ are defined to be the $\mu,T$ for the smallest value of $\psi_{0}$ considered -- this is the onset of the phase transition line (numerically in our computation it is $\approx 10^{-5}$). It can be seen that for larger $\rh$ the condensate decreases and the curve profile is slightly smaller. %, and is very similar to that of the Poincar\'e patch.
It should be noted that because we have two dimensionful quantities available, we have some freedom in choosing how to plot the condensate plots. 

\section{Concluding Comments}

In this chapter we have investigated the phases of black holes in global AdS spacetime for large and moderate values of the scalar charge $q$. We have not investigated the phase structure for very small values of $q$ but we expect that it should not change qualitatively from the behavior we found for $q \leq 1$. It will be interesting to see if this expectation is correct, by doing a more exhaustive scan of the low charge regime. 

The instabilities/phases that we have uncovered are to be understood in the context of phases of thermal partition functions in the dual gauge theory. Recently there has been a lot of interest in various questions regarding non-linear dynamical instability of classical gravity in AdS \cite{instability,Basu:2014sia,Basu:2015efa,Evnin:2015gma}, which should be understood in the context of thermalization in the gauge theory. It will be interesting to try and fit these two perspectives into a coherent whole.  Another obvious line of development is to consider, in analogy with the work on Poncar\'e patch superfluids in \cite{superfluid}, adding a spatial component to the gauge field to the configurations considered here.

\chapter{Hairy Black Holes in a Box}\label{hbbox}
%We do a systematic study of the phases of gravity coupled to an electromagnetic field and charged scalar in flat space, with box boundary conditions. The scalar-less box has previously been investigated by Braden, Brown, Whiting and York (and others) before AdS/CFT and we elaborate and extend their results in a language more familiar from holography.   The phase diagram of the system is analogous to that of AdS black holes, but we emphasize the differences and explain their origin. Once the scalar is added, we show that the system admits both boson stars as well as hairy black holes as solutions, providing yet another way to evade flat space no-hair theorems. Furthermore both these solutions can exist as stable phases in regions of the phase diagram. The final picture of the phases that emerges is strikingly similar to that found recently for holographic superconductors in global AdS, arXiv: 1602.07211. Our construction lays bare certain previously unnoticed subtleties associated to the definition quasi-local charges for gravitating scalar fields in finite regions. 

%
%%%%%%%%%%%%%%%%%%%%%%%%%%%%%%%%%%%%%%%%%%%%%%%%%%%%%%%%%%%%%%%%%%%%%%%%%%%%%%%%%%%%%%%%%%%%%%
%\section{Introduction}

Schwarzschild black holes in flat space have negative specific heat, which means that they heat up by Hawking radiating and cool down by absorbing radiation. Therefore, they cannot be in equilibrium with thermal radiation in asymptotically flat space. As is well known, one way to bypass this problem is to put the black hole in a (small enough) box, and to study the phases of the black hole + radiation system. 

A natural gravitational box for the black hole is provided by Anti-de Sitter space, where the phase structure of pure gravity was studied for the the first time by Hawking and Page \cite{HP}. With the advent of the AdS/CFT correspondence \cite{Mal,Witten1,Witten2,GKP}, it became clear that this is more than just a curiosity and that the physics of the black hole in the AdS box is dual to that of a (de-)confined gauge theory. 

AdS/CFT correspondence triggered an avalanche of interest, and black holes in (asymptotically) AdS geometries have been studied from various angles. In particular it has been noted that adding a charged scalar to the Einstein-Maxwell system in AdS gives a way to evade the no-hair theorems of flat space\footnote{Black holes are essentially uniquely determined by their global charges: this is the basic message of the no-hair theorems in classical general relativity. The {\em spirit} of these theorems is unlikely to be evaded, and forms one of the cornerstones of the modern lore on black holes: we need a quantum count of the microstates of black holes to account for the Bekenstein-Hawking entropy, classical hair simply is not numerous enough. But the {\em letter} of the no-hair theorems have been evaded in many ways, and holographic superconductors in AdS we saw in the previous Chapter are an example. Our box black holes will serve as another.}: in the AdS/CFT literature such black holes are called holographic superconductors \cite{HHH2,hhh0810}. The detailed study of the phase structure and other properties of this system and its numerous generalizations have given rise to an industry in itself \cite{BKA,BasuVector,IIB,HorIntro}. 

However, the original question that motivated Hawking and Page to consider AdS space in the first place, namely the black hole in a box, has {\em not} been investigated much in the context of the added luxury of a charged scalar. In particular, the phase diagram of the Einstein-Maxwell-scalar system in a box is not known, to the best of our knowledge. Our goal in this chapter is to take a first step in that direction and to chart out the phase diagram of this system. We will find that apart from the known Reissner-Nordstrom black hole, the system also allows boson stars and hairy black holes as classical solutions. We will furthermore demonstrate that these solutions are more than a curiosity: they exist as thermodynamically stable phases in appropriate regions of the $T-\mu$ phase diagram ($T$ is the temperature and $\mu$ is the chemical potential of the system). Our construction of hairy solutions is a constructive proof for yet another way to evade the no-hair theorems of asymptotically flat space. 

The phase diagram that we uncover bears striking resemblance to that of the Einstein-Maxwell-scalar system in global AdS studied in the previous chapter (see also \cite{Arias}). This is re-assuring because our expectation is that AdS should really be viewed as a box. We work with the specific case of the massless scalar for concreteness, and our comparison will be with the conformally coupled scalar studied in the previous chapter.

In what follows, we first start out by considering the Einstein-Maxwell system (without a scalar) in a box. This system (as well as the pure Einstein system \cite{PHut,GibbonsPerry}) have been studied before and our results will overlap with those of Braden, Brown, Whiting and York \cite{BBWY}. Our approach will however be decidedly holography-inspired and somewhat more complete. We will also emphasize the definitions of the charges etc., which will have to be reconsidered when we add the charged scalar. Adding the scalar brings in a few different subtleties, related to the fact that no-hair theorems are in effect when the box size is taken to infinity. This also introduces difficulty in defining quasi-local charges directly, as we will discuss. But the free energy is well-defined and computable and gives rise to a phase diagram that matches with our qualitative expectations from global AdS as well as reduces to the hairless case when the scalar is turned off. We will conclude with some comments and possible future directions for work.

Hairy solutions in a cavity have been constructed before, most notably \cite{Win1, Win2, Win3}.
Isolated special examples of hairy solutions were shown to
arise even earlier as the final states of super-radiant instabilities
in \cite{Herdiero}, see also \cite{Bosch}\footnote{Some of the papers in \cite{Herdiero,HerdeiroRadu,Degollado:2013bha,Sanchis-Gual:2015lje,Sanchis-Gual:2016tcm} were looking at the growth of the scalar field in the linear regime only, where it grows exponentially. So these were not true solutions to the field equations, but indicative.}. A relevant conjecture here
is that of \cite{HerdeiroRadu}. These observations indicate that these
solutions can arise as the endpoints of dynamical processes,
suggesting that they can be stable and physical. This is
satisfying, in light of the results of our work: we deal with the stability of thermal phases, while these papers dealt with dynamical aspects.

The work of \cite{Win1, Win2, Win3} offers a nice complementary
discussion\footnote{The solutions they find seem identical to ours modulo
	notations and conventions, except for one caveat: we have had to be
	somewhat more careful with boundary issues than \cite{Win1, Win2,
		Win3} for various reasons. To make the box boundary fully well-defined as
	a variational problem, one needs to add
	a boundary term to the action (the Gibbons-Hawking-York term, see our
	discussion in Section 3). This means that the problem is well-defined
	only with a fixed boundary metric, which we take as our ``box"
	(\ref{bbndmet}), and we write down all our bulk solutions in the {\em same}
	gauge for the boundary metric, namely (\ref{bbndmet}). Instead, \cite{Win1,Win2, Win3} hold $h(r)$ fixed to unity at the {\em horizon}, which means that they will need a further {\em solution-dependent} rescaling of the time coordinate that brings the boundary value of $g(r) h(r)$ to some fixed value (say unity) to bring all their solutions into the same gauge.  The $ g(r) $ and $ h(r) $ here are defined in eg. \eqref{bmetric}.} of these solutions: our focus is on thermodynamic stability, they focus on perturbative stability. Taking these results together, it seems evident that these solutions are bonafide solutions of gravity in a box.

This Chapter is based on \cite{Basu:2016srp}.

\section{The Setup}

We will consider a spacetime manifold $\man$, with a time-like boundary $\partial\man$, which we will refer to as a {\em box} henceforth. We will first look at gravity, with no cosmological constant, coupled to Maxwell field: the intuition we get by studying this system will be useful when we add the scalar in later sections. The action is given by
\bea\label{baction}
S= - \frac{1}{16 \pi}\int_{\man} d^{4}x \sqrt{-g}\left( R - F_{\mu\nu} F^{\mu\nu}\right) - \dfrac{1}{8\pi }\oint_{\partial \mathcal{M}} \sqrt{-\gamma} \; \mathcal{K} \, ,
\eea
where $g_{\mu\nu}$ gives the metric in the bulk, $\gamma$ is the metric in the boundary, and $\mathcal{K}$ is the extrinsic curvature. The boundary piece in the action is called the Gibbons-Hawking-York term, and we will briefly comment about it in the next section. We have set $G=1$. Note that the normalization of our Maxwell piece follows the conventions of \cite{catastrophic}.

We would like to work with a time independent ansatz, which is also spherically symmetric.  We will be looking at a space where the boundary is at $ r = \rb $. The metric is chosen to be of the form
\bea \label{bmetric}
ds^{2} = -g(r) h(r) dt^{2} + \dfrac{dr^{2}}{g(r)} + r^{2}\, d\Omega_{2}^{2},
\eea
and for the Maxwell field (see similar constructions in eg. \cite{HHH2, BKA})
\bea
A = \phi(r)dt.
\eea
With the above ansatz, we get the equations of motion
\begin{align}
\label{beoms}
&\frac{g'(r)}{r g(r)}+\frac{\phi '(r)^2}{g(r) h(r)}-\frac{1}{r^2 g(r)}+\frac{1}{r^2} =0, \\
&h'(r) = 0, \\
&\phi ''(r)+\frac{2 \phi '(r)}{r} -\frac{h'(r) \phi '(r)}{2 h(r)} = 0.
\end{align}
The second of the above equations is solved by $h(r)$ a constant, but we will phrase the discussion below at the level of the equations of motion without setting $h(r)$ to constant. The reason for doing this is that when one adds the scalar, the $h$-equation of motion will become non-trivial (see later sections), but the discussion below will still hold. 
From the equations of motion, we can see the existence of the following two scaling symmetries.
\begin{itemize}
	\item $r \rightarrow a r$. With this rescaling, one could set $\rb  =1$.
	\item $h \rightarrow \bar{h}= a^{2} h, \ \phi \rightarrow \bar{\phi} =  a \phi, \ {\rm and }\ t \rightarrow \bar{t}= \frac{t}{a}$. This scaling symmetry can be used to set the $g_{tt}$ coefficient of the metric to be unity at $r=\rb $. The boundary metric will thus be $\mathbf{R} \times \mathbf{S}^{2}$, which will ensure that the metric of any geometry matches with the flat space metric at the boundary. This gives
	\bea
	-g h dt^{2}|_{\rb} = - g \dfrac{\bar{h}}{a^{2}}dt^{2}|_{\rb} = -g \bar{h} d\bar{t}^{2}|_{\rb} = -d\bar{t}^{2}.
	\eea
	For this, we choose rescaling in the following way\footnote{In the rest of this chapter, we will drop the {\em bar} on the variables $\bar{h}$ and $\bar{t}$, with the understanding that the boundary conditions are met}.
	\bea
	\lim_{r\rightarrow\rb }\bar{h}(r) = \dfrac{1}{g(\rb)}, \ \text{i.e} \ \ \bar{h}(\rb)= a^{2} h(\rb)=\dfrac{1}{g(\rb)} \Rightarrow a = \dfrac{1}{\sqrt{g(\rb )h(\rb )}},
	\eea
\end{itemize} 
We will be interested in looking at the Schwarzschild and Reissner-Nordstr\"om solutions, inside the box. 

\section{Gravity in a Box}

Because gravity is that mysterious force that causes spacetime itself to be dynamical, putting gravity in a box strikes can be worrisome. A natural question to ask at this point is: What if gravitational waves leak out uncontrollably beyond the box boundary? So let us start by making a few comments to alleviate such worries. First of all, a dynamical spacetime does {\em not} mean that there is  any violence done to the manifold structure: it means merely that metric is the dynamical variable. Secondly, by putting gravity in a box, what one operationally does, is to set an appropriate boundary condition for the metric. And despite the fact that it affects our notions of distance and is therefore sacred to us, on a manifold the metric is just {\em some} field. This means that at least classically, the ``box boundary condition'' is perfectly well-defined as a boundary condition for the metric, as long as the metric equations of motion can arise from a well-posed variational problem on the manifold, with the said ``box boundary condition''. In particular, gravitational waves cannot do anything illegal if this is the case, because gravitational waves are solutions of the metric equations of motion that arise from such a variational problem, and therefore {\em by construction} have to respect those boundary conditions.

So in order for the ``box boundary condition'' to be physically acceptable, what we need to make sure is that they lead to a well-defined variational formulation for the metric. Now the most natural ``box boundary condition" is to hold the metric at the boundary fixed\footnote{A ``box" is nothing but a Dirichlet boundary condition for the fields.}, but it is known ever since the work of Gibbons-Hawking \cite{GH} and York \cite{York} that Einstein gravity indeed allows a perfectly well-defined variational problem of this Dirichlet type, when one adds the so-called Gibbons-Hawking-York boundary term to it. So this is the reason why we work with an action of the form \eqref{baction} in this work (and its generalization to include a charged scalar which we will consider a bit later). We also note that the standard scalar and Maxwell pieces in the Lagrangian automatically are well-defined Dirichlet problems, so we do not need to add any boundary terms for them.

In what follows, for all geometries with or without a horizon, we will take the boundary metric to be of the form 
\bea\label{bbndmet}
ds^{2}|_{\partial\man} = -dt^{2} + \rb^{2} d\Omega_{2}^{2}.
\eea
This is our definition of the box. This means in particular that $g_{tt}^{black hole}|_{\rb } = g_{tt}^{flat space}|_{\rb } $, so  the effective temperatures defined for the two systems will be equal at $r= \rb$. This will be relevant if/when we do background subtraction of the classical action of non-trivial geometries with that of flat space.  As the manifold does not have an asymptotic region, the definitions of the energy, charge and temperature require a bit of explanation for those who are used to asymptotically flat/AdS boundary conditions. 

The Hawking temperature computation is unaffected because it relies only on the horizon and not the boundary. When there is a horizon at $r=\rh$, the line element \ref{bmetric} can be expanded via $r = \rh + \delta r$ and after the usual \cite{QFTBH} demand that there are no conical singularities in the Euclidean metric, one ends up with  
%\bea ds^{2}|_{NH} =  - g'(\rh)\, h(\rh)\; \delta r \;dt^{2} + \dfrac{dr^{2}}{g'(\rh)\delta r} + \rh^{2}\; d\Omega_{2}^{2}. \eea
%Dropping the $S^{2}$ part of the metric and after the redefinition %$\dfrac{dr}{\sqrt{g'(\rh)\delta r}} = d\rho \ \Rightarrow \ 
%$\rho = \dfrac{2\, \delta r^{1/2}}{\sqrt{g'(\rh)}}$,
%we get in the near-horizon region, 
%$ ds^{2}|_{NH} = - g'(\rh)^{2} h(\rh) \dfrac{\rho^{2}}{4} dt^{2} + d\rho^{2}$. As usual, demanding the absence of conical singularities  $\tilde{t} =\dfrac{g'(\rh)\, h(\rh)^{1/2}}{2} \; t $ and demanding that the we have a periodicity in the new coordinate, i.e. $\tilde{t} \rightarrow \tilde{t} + \beta$, we get
\bea\label{btbh}
\dfrac{1}{\beta} = T = \dfrac{1}{4\pi} g'(\rh)\,h(\rh)^{1/2}
\eea
as the Hawking temperature. The entropy of the geometry is also a horizon quantity, and is just given by the a quarter of the area of the horizon as usual:
\bea
S=\pi \rh^2.
\eea

In order to define the gravitational mass in an asymptotically flat spacetime, we can use the ADM construction. For defining the ADM mass, the spacelike slices ($\Sigma_{t}$) of the geometry are set up in such a way that they asymptotically coincide with a constant time surface of Minkowski space. The spacelike slices $\Sigma_{t}$ are bounded by closed two-surfaces $S_{t}$.  The mass is then defined as the value of the ADM Hamiltonian when the two-surface is a two-sphere at spatial infinity, for a specific choice of lapse and shift\footnote{See Sec.4.3 of \cite{Poisson} for details, the ADM mass is obtained when the lapse is taken to be unity and the shift is taken as zero. This identifies the ADM mass as the generator of boundary time translations.}, as
\bea
M = -\dfrac{1}{8\pi} \lim_{S_{t}\rightarrow\infty} \oint_{S_{t}} (k - k_{0})\sqrt{\sigma}d^{2}\theta,
\eea
where $\sigma_{AB}$ is the metric on $S_{t}$, $k$ is the extrinsic curvature of $S_{t}$ embedded in $\Sigma_{t}$ and $k_{0}$ is the extrinsic curvature of $S_{t}$ embedded in flat space. Using this definition, we find that the ADM mass of a black hole is the mass parameter $M$ of the Schwarzschild metric, and it has the interpretation of energy in the thermodynamics of the system, i.e. $E = M $.

In our construction, the space does not have an asymptotic region, instead, we set the spacelike slices $\bar{\Sigma}_{t}$ to be in such a way that the boundary metric coincides with a constant time slice of Minkowski metric with a boundary at $r=\rb$. The quasilocal energy density can be defined as \cite{BrownYork}
\bea\label{benergy}
E \equiv -\dfrac{1}{8\pi} \lim_{r\rightarrow\rb } \oint_{\mathbf{S}^{2}} (k - k_{0})\; \sqrt{\sigma} \; d^{2}\theta.
\eea
where $\sigma_{AB} = \rb^{2} d\Omega^{2}_{2}$ is the metric of the boundary 2-sphere, and the unit normal is $n_{\mu} = (0,g(r_{b})^{-1/2},0,0)$. The extrinsic curvature of the boundary $2$-sphere, of the geometry embedded in the spacelike slices $\bar{\Sigma}_{t}$ is given by $k$, and $k_{0}$ is the extrinsic of the boundary $2$-sphere embedded in flat space. Also, $k = k_{AB}\sigma^{AB}$, and 
\bea
k_{AB} = \dfrac{1}{2} (\nabla_{\mu} n_{\nu}+\nabla_{\nu} n_{\mu}) e^{\mu}_{A} e^{\nu}_{B},
\eea
where $e^{\mu}_{A}= \partial x^{\mu}/ \partial \theta^{A}$ denote the basis vectors of the $2$-sphere.

The definition of chemical potential is taken to be the value of the gauge field potential at the boundary, $\mu = \phi(\rb) $. The electric charge of the system is defined as 
\bea
Q = \lim_{r\rightarrow \rb} \dfrac{1}{4\pi}\int_{S^{2}} F_{\mu\nu} t^{\mu}n^{\nu} \sqrt{\sigma} d^{2}\theta,
\eea
where $t^{\mu}$ is the unit time-like normal at the boundary, and $n^{\nu}$ is the unit outward drawn normal at the $r=\rb $ hypersurface. 

To evaluate the classical action for the geometries directly, we have to know the boundary term or the Gibbons-Hawking term. For the metric ansatz we have chosen, the outward unit normal to $\partial\man$ is given by $n_{\mu} = ( 0 ,g(\rb)^{-1/2},0,0)$. The metric at the boundary, after appropriate rescaling, is $\gamma_{IJ}\,dy^{I}\,dy^{J} = -dt^{2} +\rb^{2}\,d\Omega_{2}^{2} $, where $y^{I} = (t,\theta,\phi)$. The extrinsic curvature of the boundary embedded in the full geometry is given by $\mathcal{K} = \mathcal{K}_{IJ}\gamma^{IJ}$, where
\bea
\mathcal{K}_{IJ} = \dfrac{1}{2}(\nabla_{\mu}\eta_{\nu} + \nabla_{\nu}\eta_{\mu}) e^{\mu}_{I}e^{\nu}_{J}
\eea
and $e^{\mu}_{I}= \partial x^{\mu}/ \partial y^{I}$ are the basis vectors at $\partial\man$. Evaluating this for our metric ansatz, we get
\bea
\mathcal{K} = \frac{1}{2 \sqrt{g(\rb)}}  g'(\rb)+\frac{1}{2} g(\rb)^{3/2} \bar{h}'(\rb)+\frac{2 \sqrt{g(\rb)}}{\rb}.
\eea
The extrinsic curvature for the Minkowski box will be denoted as $\mathcal{K}_{0}$, and will be used to do a background subtraction, which sets the free energy of the Minkowski box to zero. The background subtraction is strictly not necessary if we are looking at the box, as there are no divergences. However, doing a background subtraction makes the comparison of quantities more straightforward when we want to take the limit when the boundary goes to infinity and we hope to reproduce the known results in asymptotically flat space\footnote{When there is a non-trivial scalar profile in the problem, that there is no such smooth asymptotically flat limit, is one of the observations we make.}. 

\section{Schwarzschild in the Box}

The simplest non-trivial solution for the equations of motion in \eqref{beoms} is given by,
\bea
h(r) = C_{1},\ \ \phi(r) = \mu, \ \text{and}\ g(r) = 1- \dfrac{\rh}{r},
\eea
where $C_{1}$ is some constant, which we will set to be $1/g(\rb)$, and $\mu$ is a constant chemical potential, which is arbitrary for the Schwarzschild solution. We will set it to zero for convenience, because it does not affect the following discussion.

\begin{figure}[h!]
	\centering
	\includegraphics[width=0.5\textwidth]{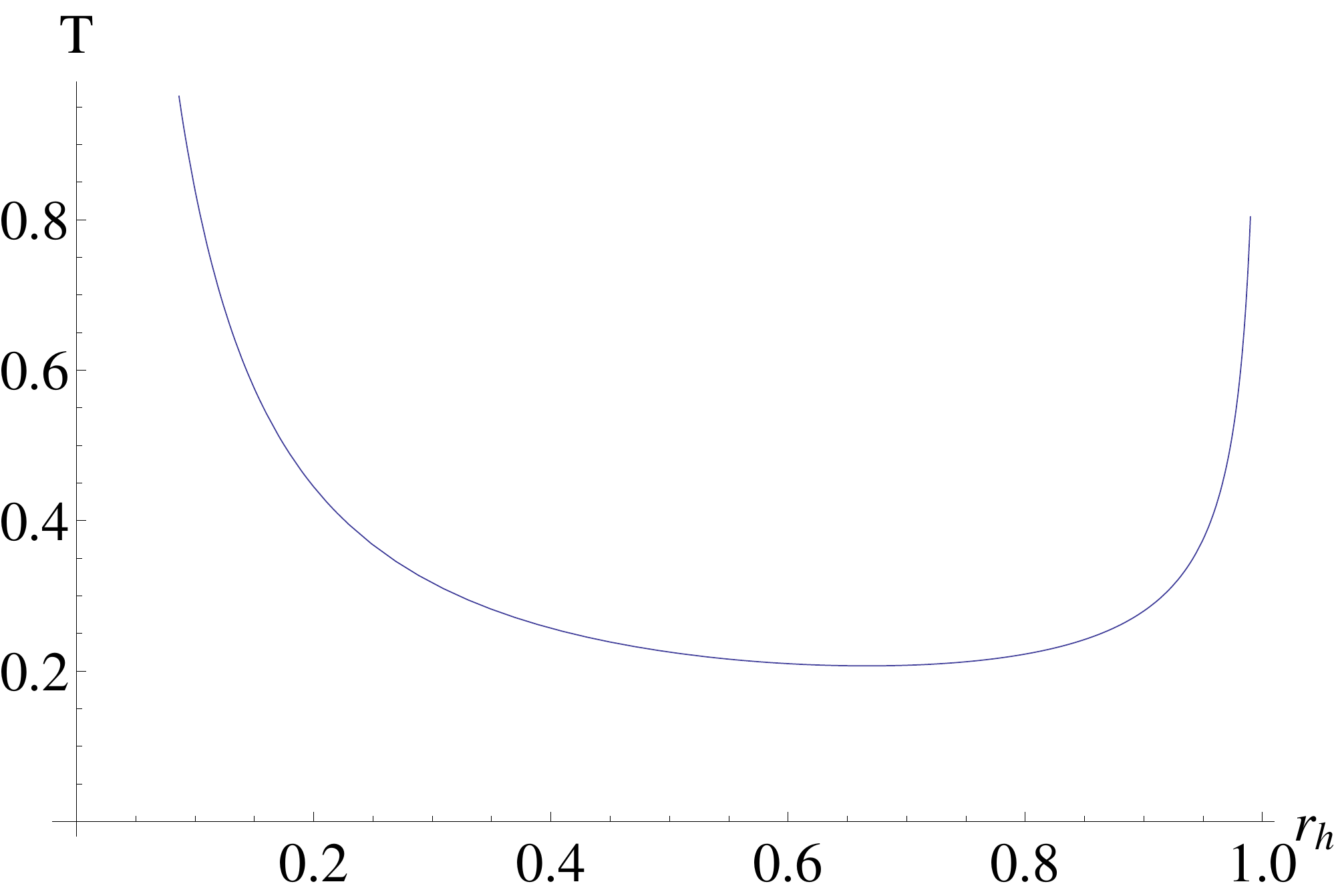}
	\caption{Temperature as a function of $\rh $, with $\rb =1 $.}
	\label{btempsch}
\end{figure}

The quasilocal energy and the temperature of the Schwarzschild solution can be computed as described above:
\bea
&E = \rb -\rb  \sqrt{1-\dfrac{\rh}{\rb}},\\
&T =\dfrac{1}{4\pi } \dfrac{1}{\rh \sqrt{g(\rb)}} = \dfrac{1}{4\pi \,\rh }\sqrt{\dfrac{\rb }{\rb -\rh }}
\eea
The temperature is plotted as a function of $\rh $, after setting $\rb = 1$, in Fig.\ref{btempsch}. It can be seen from the figure that the temp for a very small black hole and a black hole approaching the size of the box go off to infinity, and for any temperature above $T_{min}$ there are two black hole solutions.
The free energy of the system is given by 
\bea
F &= E - T S = \rb -\rb  \sqrt{1-\dfrac{\rh}{\rb}} -\dfrac{1}{4\pi}\dfrac{\rb }{\rh \sqrt{\rb - \rh }} \pi \rh ^{2} \nonumber\\
&= \rb -\rb  \sqrt{1-\dfrac{\rh}{\rb}} -\dfrac{\rb \rh }{4 \sqrt{\rb - \rh }} .
\eea

The free energy can also be computed directly from the classical action. The only term that will contribute is the surface term, because the Ricci scalar $R=0$ and the gauge field is not turned on.
\bea
F = T S_{cl} = - \dfrac{T}{8\pi} \int_{0}^{1/T}d\tau \int d\theta \, d\phi \, \sin^{2}\theta \, r^{2} \left( \mathcal{K} - \mathcal{K}_{0}\right)\biggr|_{\rb },
\eea
The free energy computed using this formula yields the same result. We can also verify that the identity $\frac{\partial E}{\partial S} = T$ holds.

\begin{figure}[h!]
	\centering
	\includegraphics[width=0.5\textwidth]{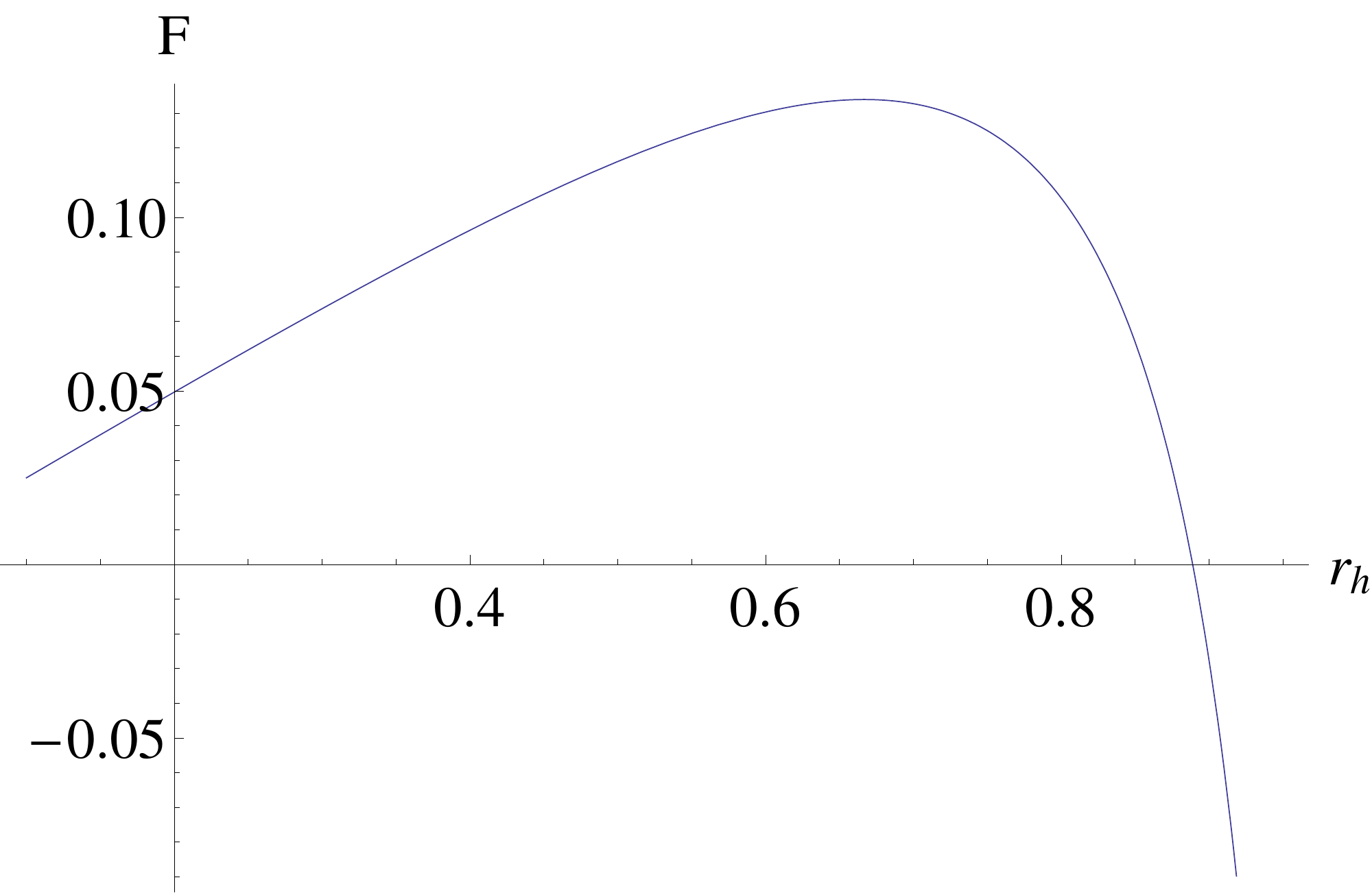}
	\caption{Free energy as a function of $\rh $, with $\rb =1 $.}
	\label{bschferg}
\end{figure}

\begin{figure}[h!]
	\centering
	\includegraphics[width=0.5\textwidth]{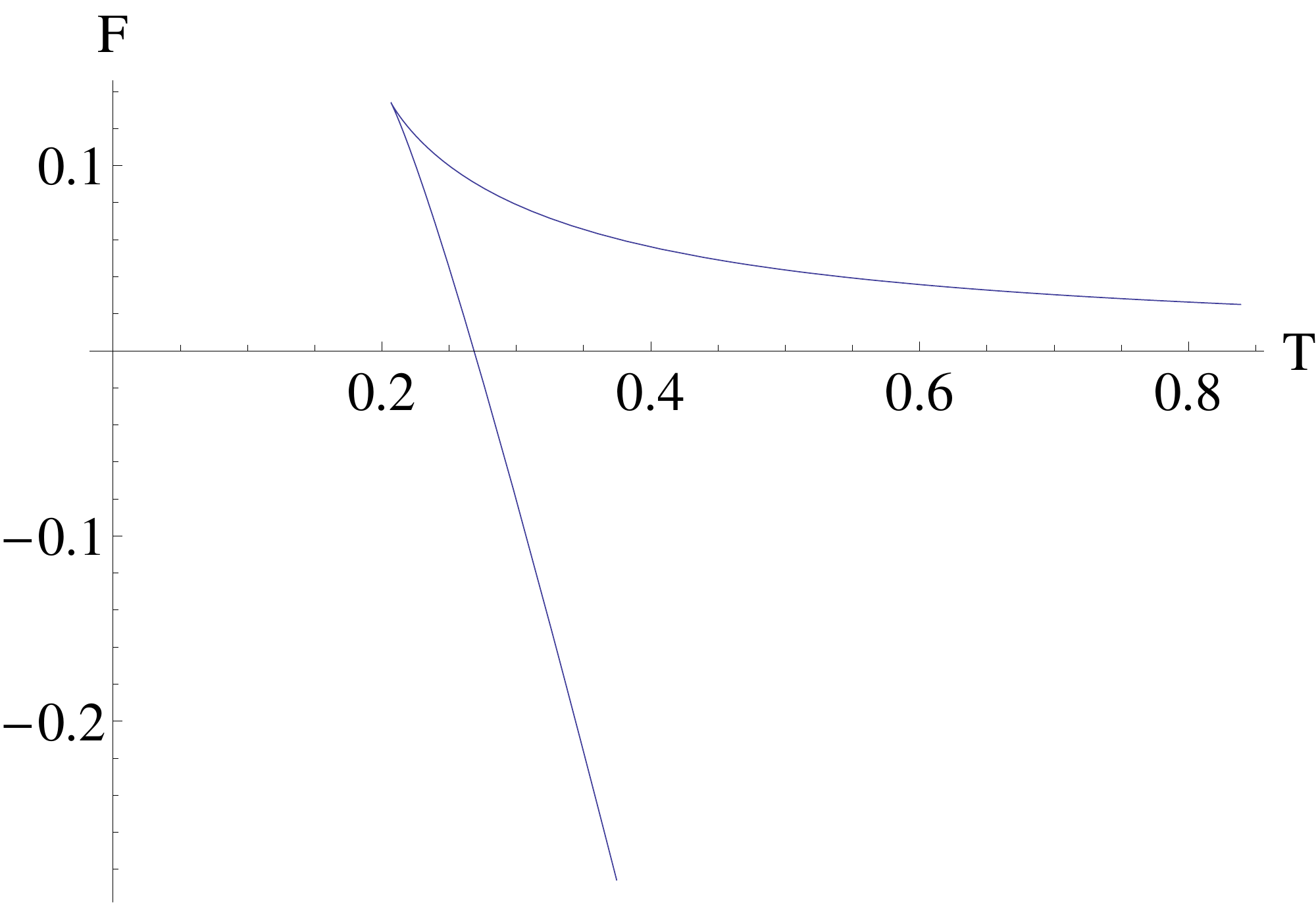}
	\caption{Free energy plotted against $T$, with $\rb =1 $.}
	\label{bschfergt}
\end{figure}

In Fig.\ref{bschferg}, we have plotted the free energy of the system against $\rh $, and against $T$ in Fig.\ref{bschfergt}, with $\rb =1$. The free energy is positive for a small black hole, and goes negative for black hole larger than $\rh =\frac{8}{9}\rb $, which is the box analogue of the AdS Hawking-Page transition.

The plots in Fig.\ref{bschferg},\ref{bschfergt} looks similar to the Schwarzschild black hole in global AdS, and so do the Penrose diagrams of Schwarzschild black hole in global AdS and in a box (with no cosmological constant)\footnote{See http://www.iopb.res.in/$\sim$mukherji/THESIS/tanay.pdf figures 2.1, 2.2, 2.3 for the Schwarzschild-AdS black hole}. 

\section{Reissner-Nordstr\"om in the Box}

Now, we will look at the RN solution in the box. We solve for $g(r),h(r),\phi(r)$ for the equations of motion in \eqref{beoms} and appropriately rescale them to get
\bea
\begin{split}
	g(r) &= 1-\dfrac{Q^2+\rh ^2}{\rh \, r }+\dfrac{Q^2}{r^2} = 1 - \dfrac{(1 + \epsilon) \rh }{r} + \dfrac{\epsilon \rh^{2}}{r^{2} }, \\ \hbr(r) &= \dfrac{1}{g(\rb)}, \ \text{and} \ \bar{\phi}(r) = \dfrac{Q}{\sqrt{g(\rb)}}\left(\dfrac{1}{\rh }- \dfrac{1}{r}\right) = \dfrac{\sqrt{\epsilon}\, \rh }{\sqrt{g(\rb)}}\left(\dfrac{1}{\rh }- \dfrac{1}{r}\right),
\end{split}
\eea
where we have parametrized the inner horizon as 
\bea\label{brinner}
r_{inner} = \frac{Q^{2}}{\rh} = \epsilon \rh 
\eea
with $0\leqslant \epsilon \leqslant 1$.
The energy and temperature of the system can again be computed:
\bea
E = \rb - \rb  \sqrt{1 -\frac{(1 + \epsilon )\rh  }{\rb } + \frac{\epsilon  \,\rh^{2}}{\rb ^2}} ,\\
T= \dfrac{1}{4\pi} \dfrac{(1 - \epsilon^{2})}{\rh }\left(1 -\frac{(1 + \epsilon )\rh  }{\rb } + \frac{\epsilon \, \rh^{2}}{\rb ^2}\right)^{-1/2}
\eea
The chemical potential of the system is 
\bea
\mu = \bar{\phi}(\rb) = \dfrac{\sqrt{\epsilon}\, \rh}{\sqrt{g(\rb)}} \left(\dfrac{1}{\rh }- \dfrac{1}{\rb }\right).
\eea

The thermodynamic relations $T= \frac{\partial E}{\partial S }\bigr|_{Q }$ and $\mu = \frac{\partial E}{\partial Q }\bigr|_{\rh }$ can be checked to hold from these. Putting all this together we get the free energy
\bea
F = E - T\, S - \mu \, Q \hspace{10.5cm}\nonumber \\
= \left(\rb\sqrt{1 -\frac{(1 + \epsilon )\rh  }{\rb } + \frac{\epsilon \rh^{2}}{\rb ^2}} - \rb +\dfrac{\epsilon  \,\rh }{4 } +\dfrac{3\rh }{4} \right) \left(1 -\frac{(1 + \epsilon )\rh  }{\rb } + \frac{\epsilon \rh^{2}}{\rb ^2}\right)^{-1/2}
\eea
The expressions for $E, T$ and $\mu$ that we obtain are the same as that in \cite{BBWY}, and also of the on-shell action. In \cite{BBWY}, the analysis is centered around finding configurations that are locally stable, although, they point out that certain configurations give a global minima for the on-shell action. We will systematically analyze the phase structure of the RN black hole in a box, using the free energy to characterize the thermodynamic stability, which is the language that is familiar from AdS-CFT.

For $\epsilon=0$, we must get the Schwarzschild case, and the black hole will be extremal when $\epsilon =1$. The free energy set to zero gives the transition curve between flat space and the RN black hole. This can be computed fully analytically, and we get the solutions
\bea
\epsilon = 1 , \; \dfrac{9 \rh - 8\rb }{\rh }.
\eea

For the (not-so-interesting) case with $\epsilon =1$, the black hole is extremal and will remain so for any value of $\rh < \rb$. The chemical potential and temperature are given by 
\bea
\mu =1  \ \text{and} \ T = 0.
\eea

\begin{figure}[h!]
	\centering
	\includegraphics[width=0.5\textwidth]{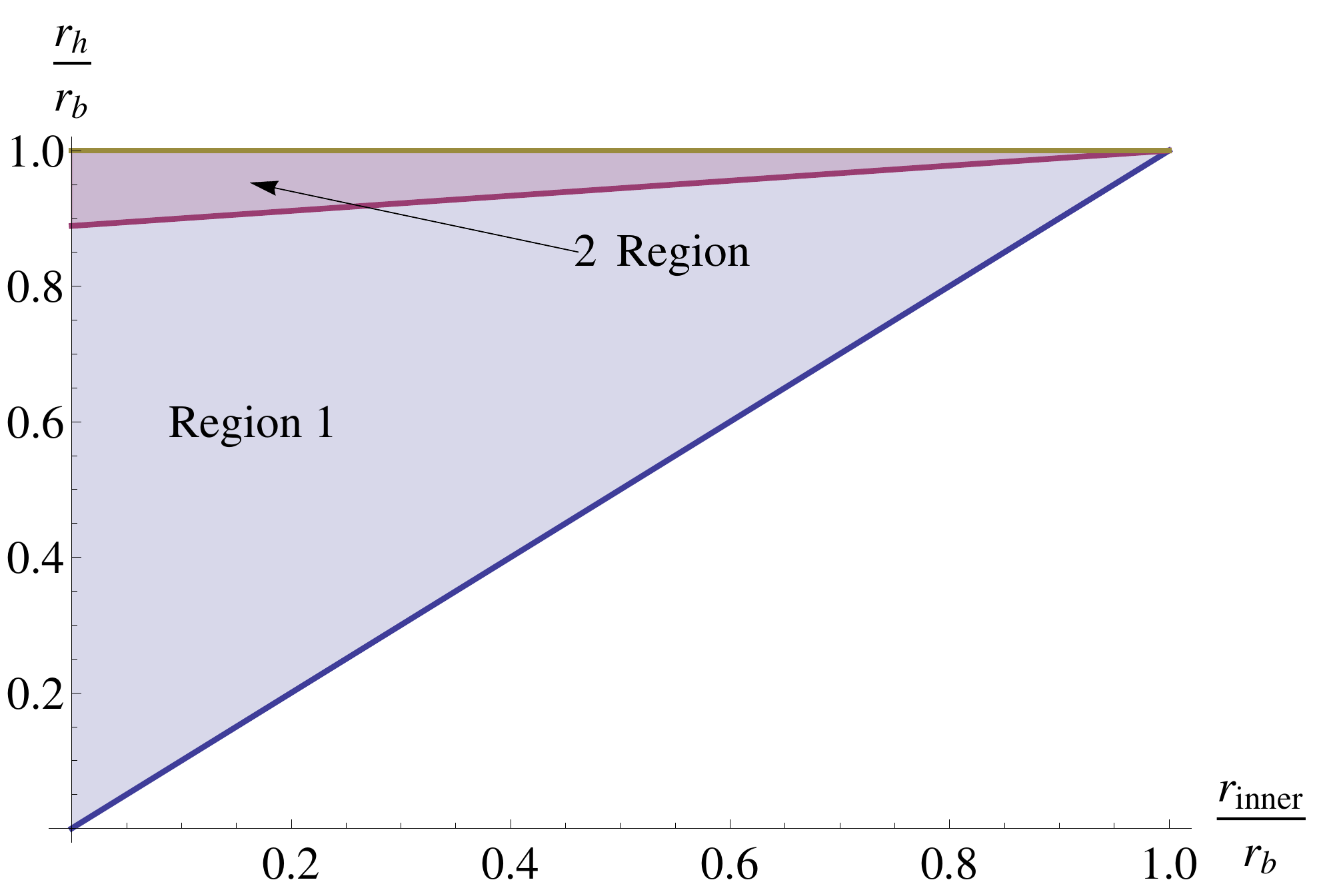}
	\caption{Region 1 and Region 2 together indicate where black holes can be formed and Region 2 is where they are thermodynamically favourable. See \eqref{brinner} for the definition of $r_{inner}$.}
	\label{brnbhregion}
\end{figure}

For the case $\epsilon = \frac{9 \rh - 8\rb }{\rh }$, the chemical potential and temperature are given by
\bea
\mu =\sqrt{1-\frac{8 r_b}{9 r_h}}\ \ \text{and} \  T= \dfrac{2\, \rb }{3\pi \rh^{2}}.
\eea
In the case of global AdS, as we look at larger values of $\mu$ along the Hawking-Page transition curve, the horizon continues to shrink, and intersects the $T=0$ axis at $\mu=1$(see Chapter \ref{gads}). However, in the case of the RN black hole in the box, the black hole gets bigger and bigger as we go up in $\mu$, and gets closer to extremality as the black hole becomes almost the size of the box itself, as can be seen from the following relation
\bea
\dfrac{\rh }{\rb } = \dfrac{8 }{9(1-\mu^{2})} \leqslant 1.
\eea
In Fig.\ref{brnbhregion}, we have shown the the regions where the RN black hole can exist in the box, and also on where it becomes thermodynamically favorable. 

\begin{figure}[h!]
	\centering
	\includegraphics[width=0.5\textwidth]{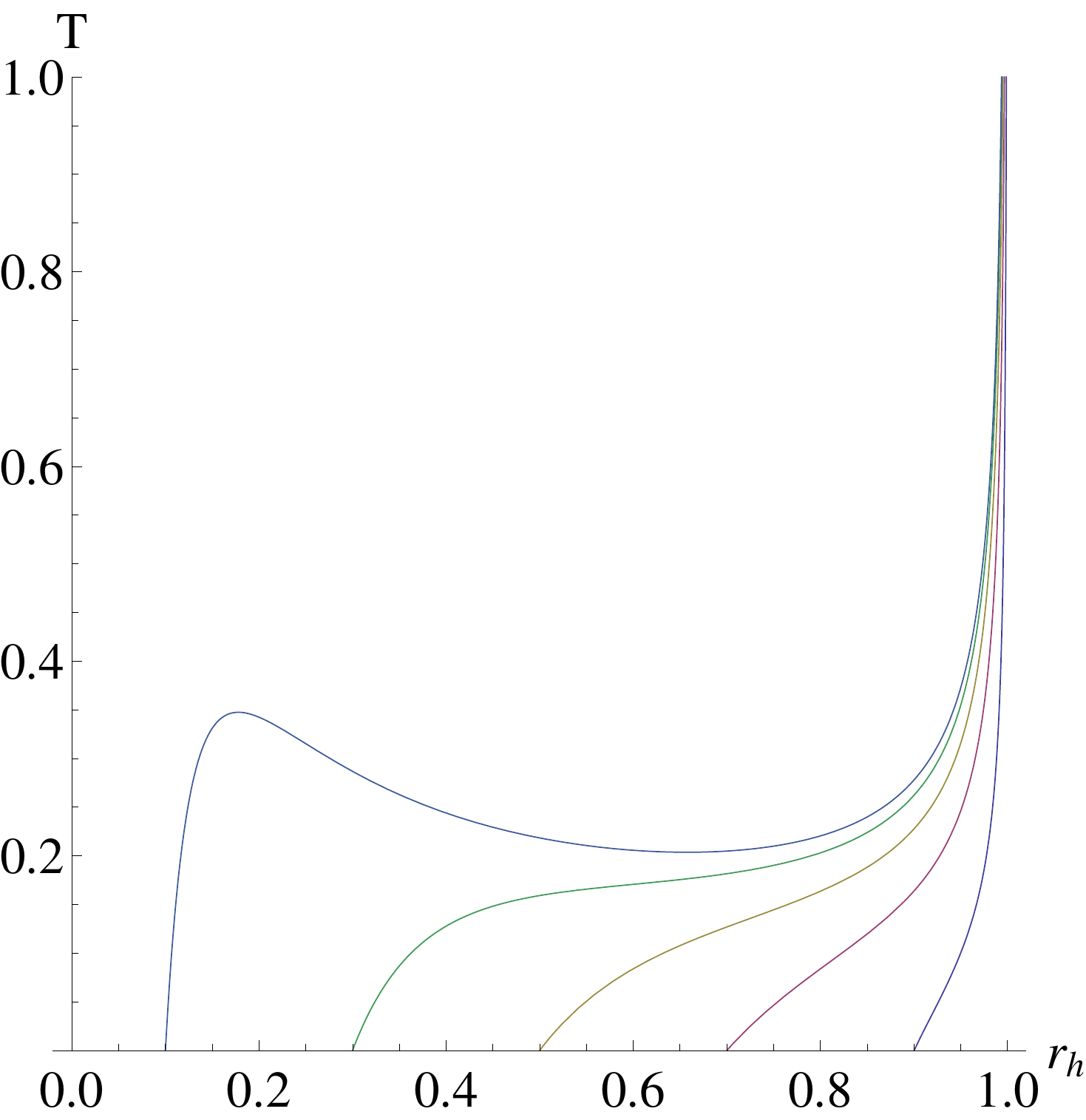}
	\caption{Temperature of RNBH with different $Q$ against $\rh $, with $\rb =1 $, for $Q=0.1,0.3,0.5,0.7,0.9$ (in that order from left to right).}
	\label{bcompare}
\end{figure}

At $\rh = \frac{8}{9}\rb $, we will get $\epsilon = 0$ in the second case, which corresponds to $\mu =0$, and $T = \frac{27}{32\pi \rb}$, which is the Schwarzschild case. The more interesting limit happens at $\rh \rightarrow \rb$. The chemical potential becomes $\mu=\frac{1}{3}$, and temperature $T = \frac{2}{3\pi}$ along with $\epsilon \rightarrow 1$. This means the phase diagram will have an abrupt ending at some finite $T$. The reason this happens can be understood as follows. As the outer horizon of the black hole is very near the boundary, the temperature diverges, see Fig.\ref{bcompare}. However, along with that, to make the free energy zero, the inner horizon is approaching the outer horizon, making it almost extremal, and it tries to take the temperature to almost zero. The existence of a finite limit is a balancing of this competition.

\begin{figure}[h!]
	\centering
	\includegraphics[width=0.6\textwidth]{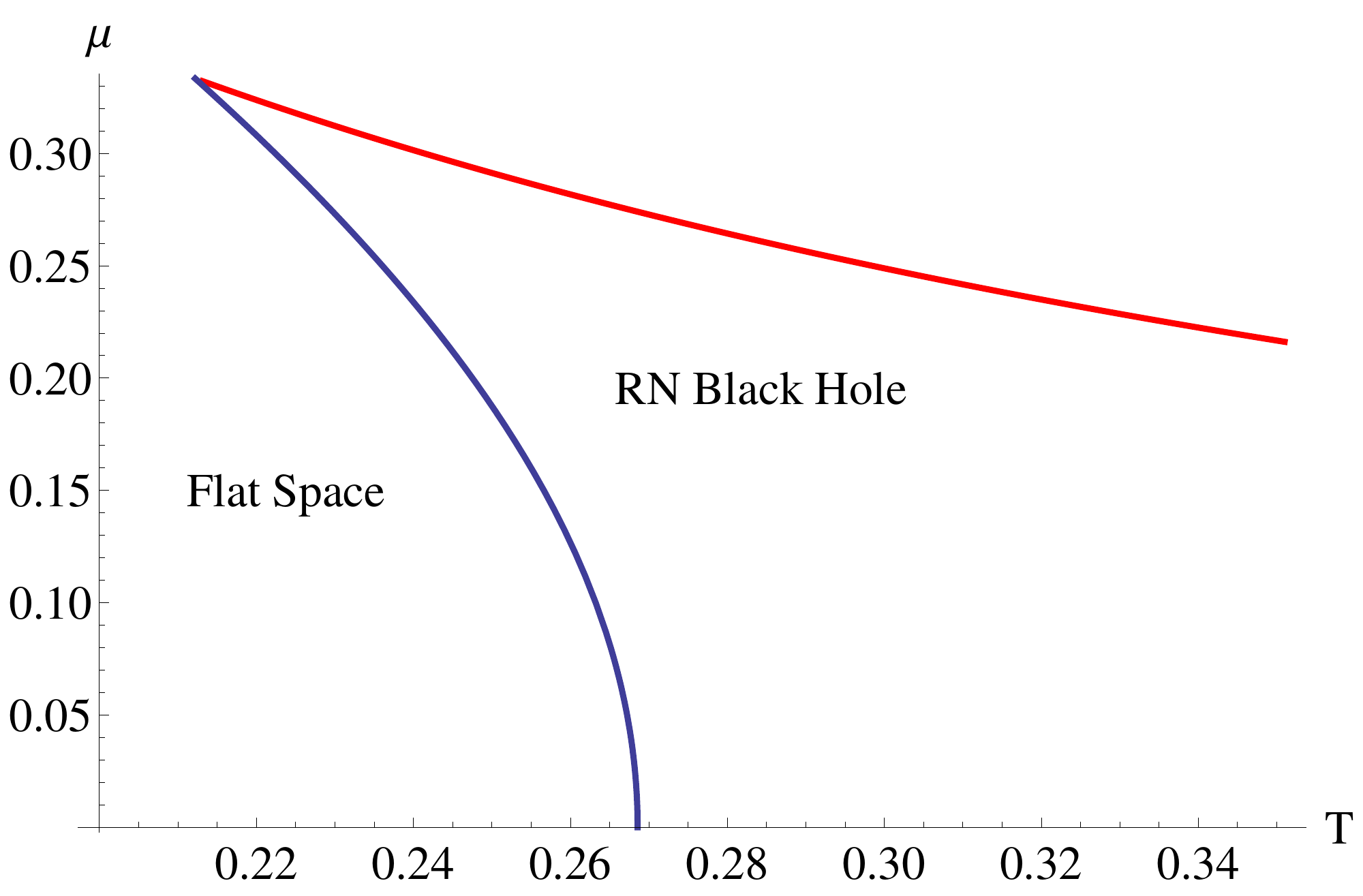}
	\caption{Phase diagram of RN BH in a box.}
	\label{brnphasedig}
\end{figure}

In Fig.\ref{brnphasedig}, this curve is shown in blue. This curve, as it can be seen has an abrupt ending, at $T=\frac{2}{3\pi} = 0.2122$. However, the big RN black hole phase has another phase boundary, which comes from the saturation of the box itself, i.e. the black hole horizon approaching the size of the full box. At the $\mu = 0$ axis, this will be at $T \rightarrow \infty$, and it is a Schwarzschild black hole limit.

To understand the behaviour of a box-sized near-extremal black hole, let us look at the expressions for $T$ and $\mu$ in this limit. First, we will parametrize $\epsilon = 1 - \delta $, where $\delta \ll 1$. Now, in this limit we get
\bea
T = \dfrac{1}{4\pi \rh } \left(\left[\dfrac{\rh \delta}{1-\rh}\right] + \dfrac{1}{2}\left[\dfrac{\rh \delta}{1-\rh}\right]^{2} + \dfrac{3}{8}\left[\dfrac{\rh \delta}{1-\rh}\right]^{3} +\dots\right),\\
\mu = 1 - \dfrac{1}{2} \left(\dfrac{\delta}{1-\rh }\right) -\dfrac{1-4\rh }{8}\left(\dfrac{\delta}{1-\rh }\right)^{2} -\dfrac{1-4\rh +8\rh^{2}}{16} \left(\dfrac{\delta}{1-\rh }\right)^{3} + \dots.
\eea
From here one can see that if $\rh$ not close to $1$, then in the $\delta\rightarrow 0$ limit, or $\epsilon \rightarrow 1$, the temperature will go to $0$, and $\mu = 1$. However, if $\rh\rightarrow 1$ at the same time then what we will end up is a limit of the form
\bea
\lim_{\delta\rightarrow 0 }\lim_{\rh\rightarrow 1} \dfrac{\delta}{1-\rh}. \nonumber
\eea
This appears in both $T$ and $\mu$ expansions and is what gives the finite temperature limit for the almost box-sized near-extremal black holes. The red curve in Fig.\ref{brnphasedig} is the plot for a almost-extremal black hole that is infinitesimally smaller than the box itself.

\begin{figure}[h!]
	\centering
	\includegraphics[width=0.5\textwidth]{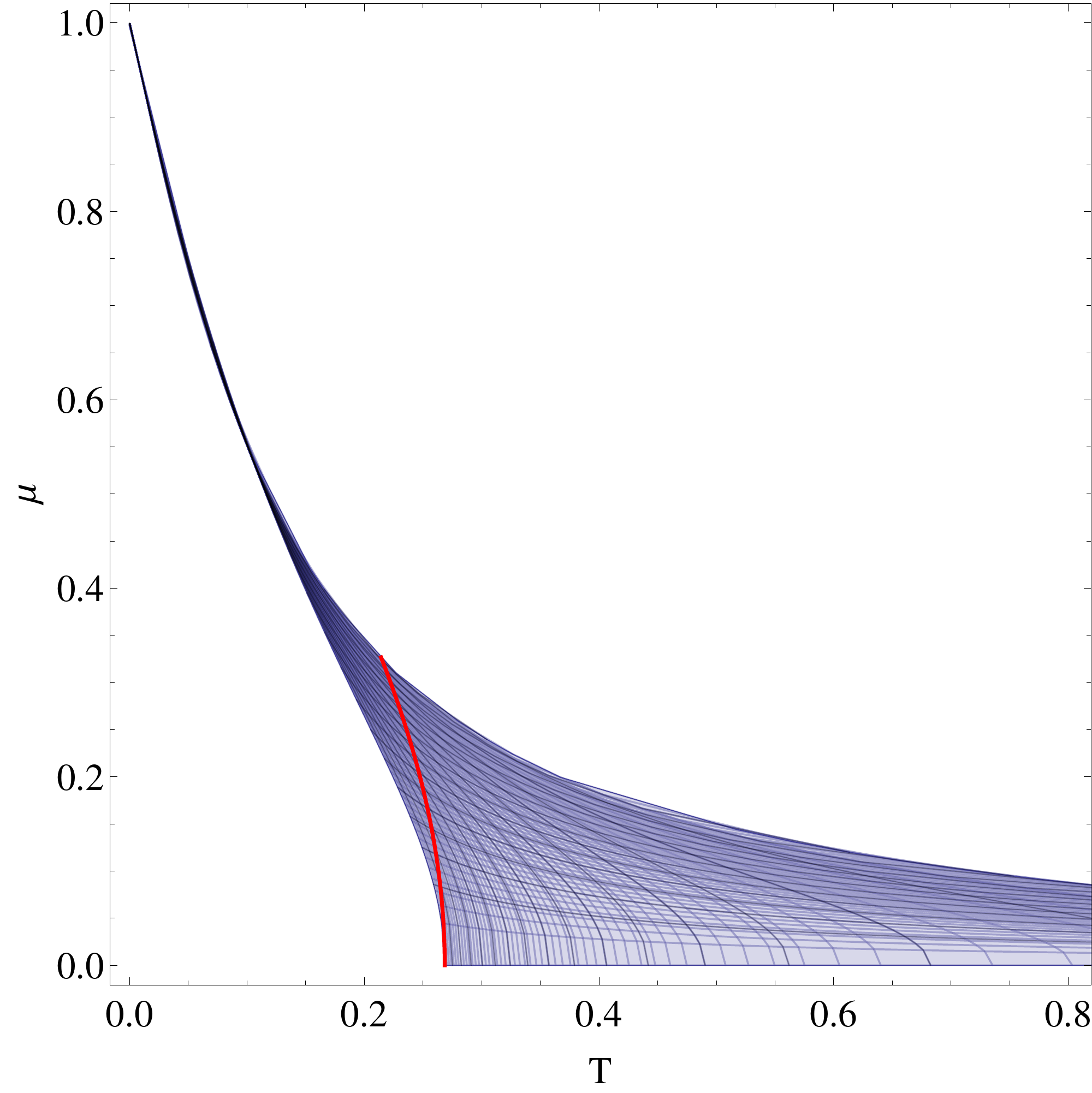}
	\caption{Blue region indicates the $(T,\mu)$ values for which large black hole solutions exist.}
	\label{bverify}
\end{figure}

The reason this will be a phase boundary may not be intuitive, therefore, let us look at it in more detail. We can invert the relations of $T$ and $\mu$ in terms of $\rh, Q$, which can be solved only numerically. Using this we can verify that for a given $(T,\mu) $ there could be upto three solutions, of which at most only one could have $\rh>\frac{8}{9}\rb $. This means there are values of $(T,\mu) $ for which there are no solutions that correspond to large black holes (with $\rh>\frac{8}{9}\rb $). In Fig.\ref{bverify}, we have the region in the $(T,\mu)$-plane which has a large black hole solution marked in blue. The Hawking-Page like curve is plotted in red, which, as one can see, falls and ends within the region marked in blue. The upper boundary of the blue region is the red curve marked in Fig.\ref{brnphasedig}.

\section{Turning on the Scalar: Hairy Solutions}

We will now add a charged scalar to this system. The result of \cite{NoHair} shows that stationary charged black hole in asymptotically flat space is completely characterized by the mass, angular momentum and charges (of the Maxwell fields), this is the \emph{no-hair theorem}. This means that asymptotically flat spaces will not support any non-trivial scalar profile. To be more concrete, let us add a scalar piece to the action \eqref{baction},
\bea\label{bscaction}
S_{scalar} = \dfrac{1}{16\pi}\int_{\man}d^{4}x \sqrt{-g}\; |\nabla\psi - i q A \psi|^{2}.
\eea
The metric and gauge field have the same functional forms as that without the scalar. Using the fact that the $r$-component of Maxwell field equation forces the phase of the scalar to be a constant, which can then be absorbed by a gauge transformation (see eg. \cite{hhh0810, BKA}), we take the scalar to be real, $\psi=\psi(r)$. The equations of motion\footnote{We have checked that these equations are equivalent to Eq. (2.15) in \cite{Win1}, as can be seen by mapping our variables $ \{g, \psi, q, \phi\} $ onto $ \{f, \phi, \sqrt{2} q, A_0/\sqrt{2}\} $ in \cite{Win1}.} for this choice of the fields is given by
\bea
\frac{1}{2} \psi '(r)^2+ \frac{g'(r)}{r g(r)}+\frac{q^2 \psi (r)^2 \phi (r)^2}{2 g(r)^2 h(r)}+\frac{\phi '(r)^2}{g(r) h(r)}-\frac{1}{r^2 g(r)}+\frac{1}{r^2} = 0,\\
h'(r) -r h(r) \psi '(r)^2  -\frac{r \, q^2 \psi (r)^2 \phi (r)^2}{g(r)^2 }= 0,\\
\phi ''(r)+\frac{2 \phi '(r)}{r}-\frac{h'(r) \phi '(r)}{2 h(r)}  -\frac{q^2 \psi (r)^2 \phi (r)}{2 g(r)} = 0,\\
\psi ''(r)+\frac{g'(r) \psi '(r)}{g(r)}+\frac{h'(r) \psi '(r)}{2 h(r)}+ \frac{2 \psi '(r)}{r} +\frac{q^2 \psi (r) \phi (r)^2}{g(r)^2 h(r)} = 0.
\eea 
If we want look at asymptotically flat space solution, we can expand the fields $g, h, \phi$ and $\psi$ around $r\rightarrow\infty$ in powers of $1/r$. Now plugging these solutions back into the equations of motion, and solving the equations order by order, we will get that all the coefficients in the expansion for $\psi$ will be forced to zero, and we will end up with RN-Black hole as the general solution. This is the way in which the no-hair theorem manifests itself in our set up.

But if the manifold has a boundary at $r=\rb $, we can again perform a series expansion of the four fields around $r=\rb $ and plug it back into the equations of motion, and solve the coefficients order by order. This gives the boundary functions in terms of $\psi_{0}^{b}=\psi(\rb),\psi_{1}^{b}=\psi'(\rb),\phi_{0}^{b}=\phi(\rb),\phi_{1}^{b}=\phi'(\rb),g_{0}^{b} = g(\rb),h_{0}^{b}=h(\rb)$, as
\bea
&\psi(r) = \psi _0{}^b+\left(r-r_b\right) \psi _1{}^b+ \dots ,\\
&\phi(r) = \phi _0{}^b+\left(r-r_b\right) \phi _1{}^b+ \dots ,\\
&g(r) = g_0{}^b+\left(r-r_b\right) \left(\frac{1-g_0{}^b}{r_b}-\frac{r_b \left(2 \left(g_0{}^b\right){}^2 h_0{}^b \left(\psi _1{}^b\right){}^2+g_0{}^b \left(\phi _1{}^b\right){}^2+2 q^2 \left(\psi _0{}^b\right){}^2 \left(\phi _0{}^b\right){}^2\right)}{4 g_0{}^b h_0{}^b}\right)+ \dots ,\\
&h(r) =h_0{}^b+\left(r-r_b\right) r_b \left(\frac{q^2 \left(\psi _0{}^b\right){}^2 \left(\phi _0{}^b\right){}^2}{\left(g_0{}^b\right){}^2}+h_0{}^b \left(\psi _1{}^b\right){}^2\right)+ \dots .
\eea
The expansions at $r = 0$ or $r= \rh $ for the boson star and hairy black hole respectively, along with the boundary conditions, are discussed when we look at the specific solutions.

At this point, it seems relevant to discuss the some aspects of the scalar field. For the Einstein-Maxwell system, the information contained in a box is essentially the same as that in the asymptotic case. However, this is not the case for the scalar field. Taking the limit $\rb\rightarrow\infty$ is subtle, as the asymptotic space cannot support the scalar hair. In evaluating the free energy, this manifests as the Brown-York quasilocal energy definition being insufficient to capture the mass of the scalar. We will not try to propose an alternate definition for the quasilocal energy , instead we will evaluate the free energy using the on-shell action, $F = -T \log\mathcal{Z} = T S_{cl}$. We will discuss these points further in the conclusions.

We will now explicitly construct hairy solutions. There are two such classes of solutions, those without horizons and those with horizons. The former will be called a boson star (in analogy with similar solutions in AdS) and the latter is the hairy black hole.

\subsection{Boson Star}
The boson star is a a horizon-less configuration. At $r=0$, the derivatives of all the functions are set to zero. At $r=\rb$ we set Dirichlet boundary condition for the scalar, $\psibnd=0$. The expansions of the functions around $r=0$, such that they solve the equations of motion, are calculated to be
\bea
\psi(r) = \psizero - \dfrac{q^2 \phizero{}^2 \psizero  }{6 \hzero}r^2 + \dots ,\\
\phi(r) = \phizero + \dfrac{1}{3} q^2 \phizero \psizero{}^2  r^2 + \dots , \\
g(r) = 1 - \dfrac{q^2 \phizero^2 \psizero{}^2 }{6 \hzero}r^2 + \dots ,\\
h(r) = \hzero + \dfrac{1}{2} q^2 \phizero^2 \psizero^2  r^2 +\dots .
\eea
Here, we have $\psizero =\psi(0)$, $\phizero = \phi(0)$ and $\hzero = h(0)$, which parametrize the solutions, and all the six boundary parameters are determined from these three. The value of $\hzero$ can be arbitrary as we have to rescale the function $h(r)$ at the end to have the right boundary behaviour. The solutions are found by fixing a value for $\psizero$, setting $\hzero =\, $const., say 1, and choosing $\phizero$ such that $\psibnd$ is zero. 

The boson star configuration can have arbitrary temperature, and the value of chemical potential above which it can exist is controlled by $q$. This point of instability of the flat empty box to forming a boson star can be calculated analytically. At the point of instability, the scalar profile is not strong enough to cause any backreaction. Thus, we can take $\psi(r)\rightarrow \alpha\psi(r)$, where $\alpha\ll 1$, and look at the equations of motion upto linear order in $\alpha$. With the given boundary conditions, we will get the solution $g(r)=1, h(r)=1, \phi(r)=\mu $, and the scalar equation of motion gives
\bea
\psi''(r) + \dfrac{2}{r}\psi'(r) + \mu^{2}q^{2} \psi (r) =0.
\eea
Imposing the scalar boundary conditions, we get the solution
\bea\label{bbsi}
\psi(r) = \psizero \dfrac{\sin \mu q r}{r}, \ \text{with} \ \mu_{bsi} q = n \pi, \ n= 1,2,3,\dots \;.
\eea
We will be looking at the first eigenmode, i.e. $n=1$. 

\begin{figure}
	\centering
	\begin{subfigure}[b]{0.4\textwidth}
		\includegraphics[width=\textwidth]{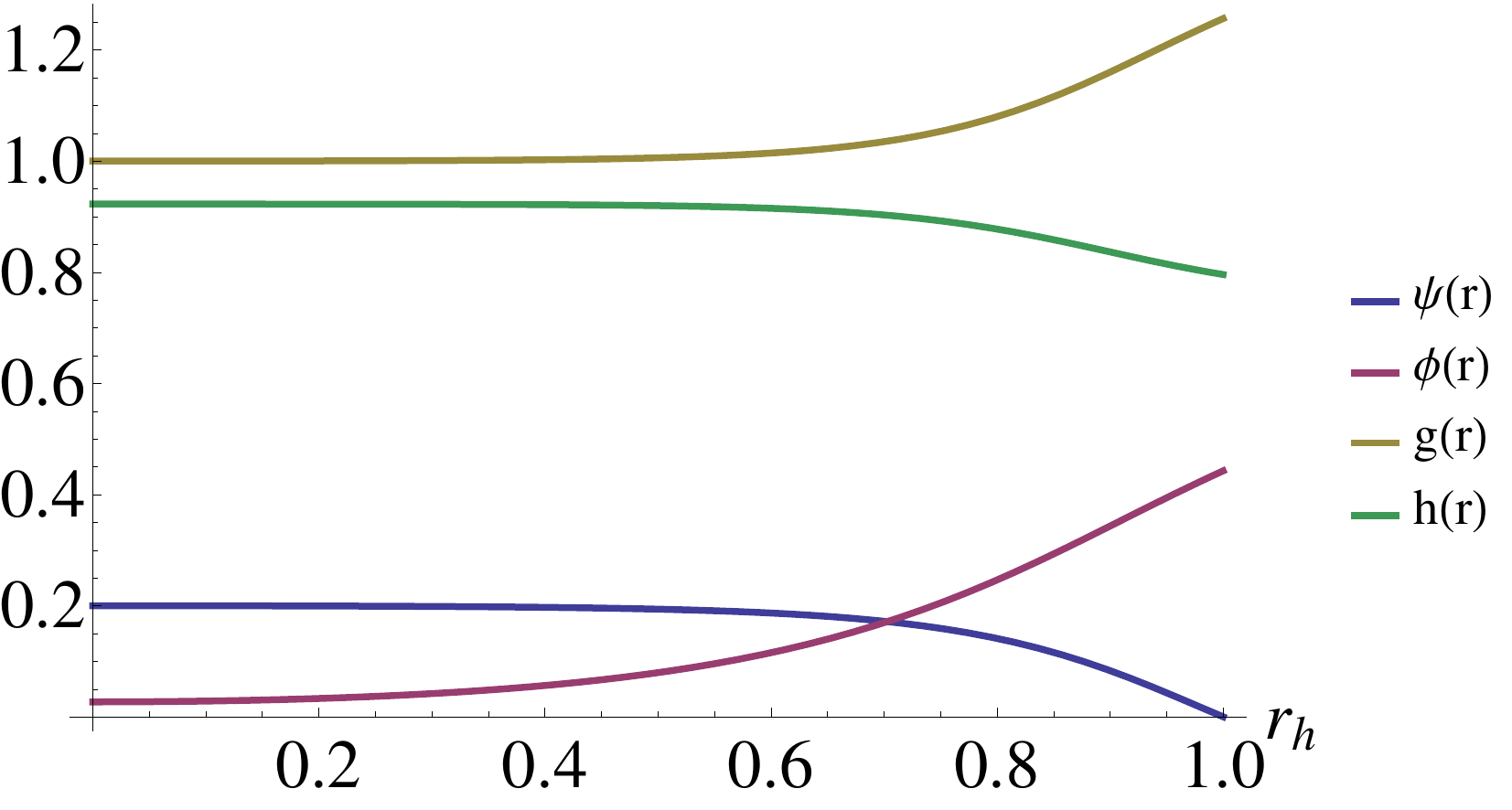}
		\caption{$q=40,\psi_{0} =0.2$}
	\end{subfigure}
	\begin{subfigure}[b]{0.4\textwidth}
		\includegraphics[width=\textwidth]{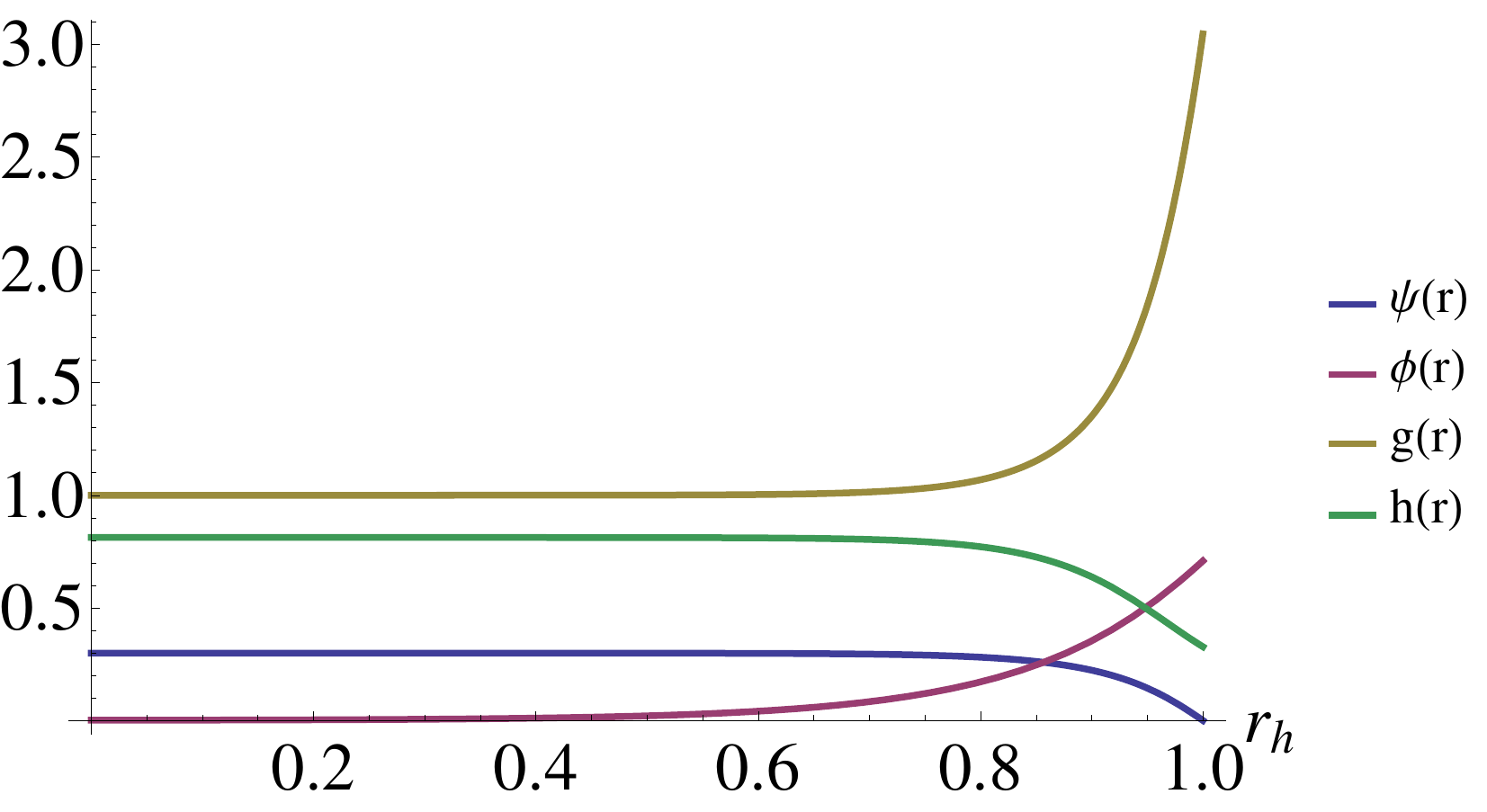}
		\caption{$q=40,\psi_{0} =0.3$}
	\end{subfigure}
	\caption{Sample profiles of $\psi(r), \phi(r),g(r)$ and $h(r)$ in the fully backreacted solution for boson star, with $\rb=1$. The quantities on the y-axis are labelled to the right of the figure.}\label{bbsplot}
\end{figure}

Of course, we can also construct fully backreacted solutions as well, numerically. In Fig.\ref{bbsplot}, we have shown the profiles of the functions for two fully backreacted solutions.

\subsection{Hairy Black Hole}

The hairy black hole is a system with a horizon and a non-trivial scalar profile. The existence of solutions in the box shows that the no-hair theorems of the asymptotic space do not apply when one is looking at a box. The boundary conditions at the horizon are $g(\rh)=0$ and $\phi(\rh)=0$, the latter ensures that the Maxwell field is regular at the horizon. Around $r=\rh $, the functions can be written as a series, such that they solve the equations of motion,
\bea
&\psi(r) = \psiin - \dfrac{ q^2  \rh^2 \hin \phiin^2 \psiin }{
	4 (\hin - \rh^2 \phiin^2 )^2} (r - \rh)^2 +\dots , \\
&\phi(r) = \phiin (r - rh) + \dfrac{ \phiin (8\rh^2 \hin \phiin^2  - 4\rh^4 \phiin^4  +  \hin^2 ( q^2 \rh^2 \psiin^2 -4 ))}{4 \rh (\hin - \phiin^2 \rh^2)^2}(r - \rh)^2  + \dots , \\
&g(r) =  (\dfrac{1}{\rh} - \dfrac{\phiin^2 \rh}{\hin}) (r - \rh)- \dfrac{4 \hin^2 + 8 \rh^4 \phiin^4 \rh^4 + 3\rh^2 \hin \phiin^2  (q^2 \rh^2 \psiin^2  -4 )}{ 4  \rh^2 \hin (\hin - \rh^2 \phiin^2 )}  (r - \rh)^2 + \dots ,  \nonumber \\
\\
&h(r) = \hin + \dfrac{\hin^2 \phiin^2 \psiin^2 q^2  \rh^3}{(\hin - 
	\phiin^2 \rh^2)^2}  (r - \rh)  %+  \dfrac{  \hin^2 \phiin^2 \psiin^2 q^2 \rh^2 (6 \phiin^4 \rh^4 +    \hin^2 (2 + \psiin^2 q^2 \rh^2) +   \hin \phiin^2 \rh^2 (-8 + 3 \psiin^2 q^2 \rh^2))}{ 4 (\hin - \phiin^2 \rh^2)^4} (r - rh)^2 
+ \dots ,
\eea
where $\psiin= \psi(\rh)$, $\phiin= \phi'(\rh)$ and $\hin= h(\rh)$. The value of the six parameters at the boundary are determined from the choice of these three parameters. The choice of $\hin$ is arbitrary as the solution is rescaled at the end to get the correct boundary metric. Thus, we set $\hin=1$, and tune $\phiin$ such that $\psibnd =0 $, for different values of $\psiin$, $q$ and $\rh$, and then appropriately rescale the functions $\phi$ and $h$.

\begin{figure}[h!]
	\centering
	\begin{subfigure}[b]{0.4\textwidth}
		\includegraphics[width=\textwidth]{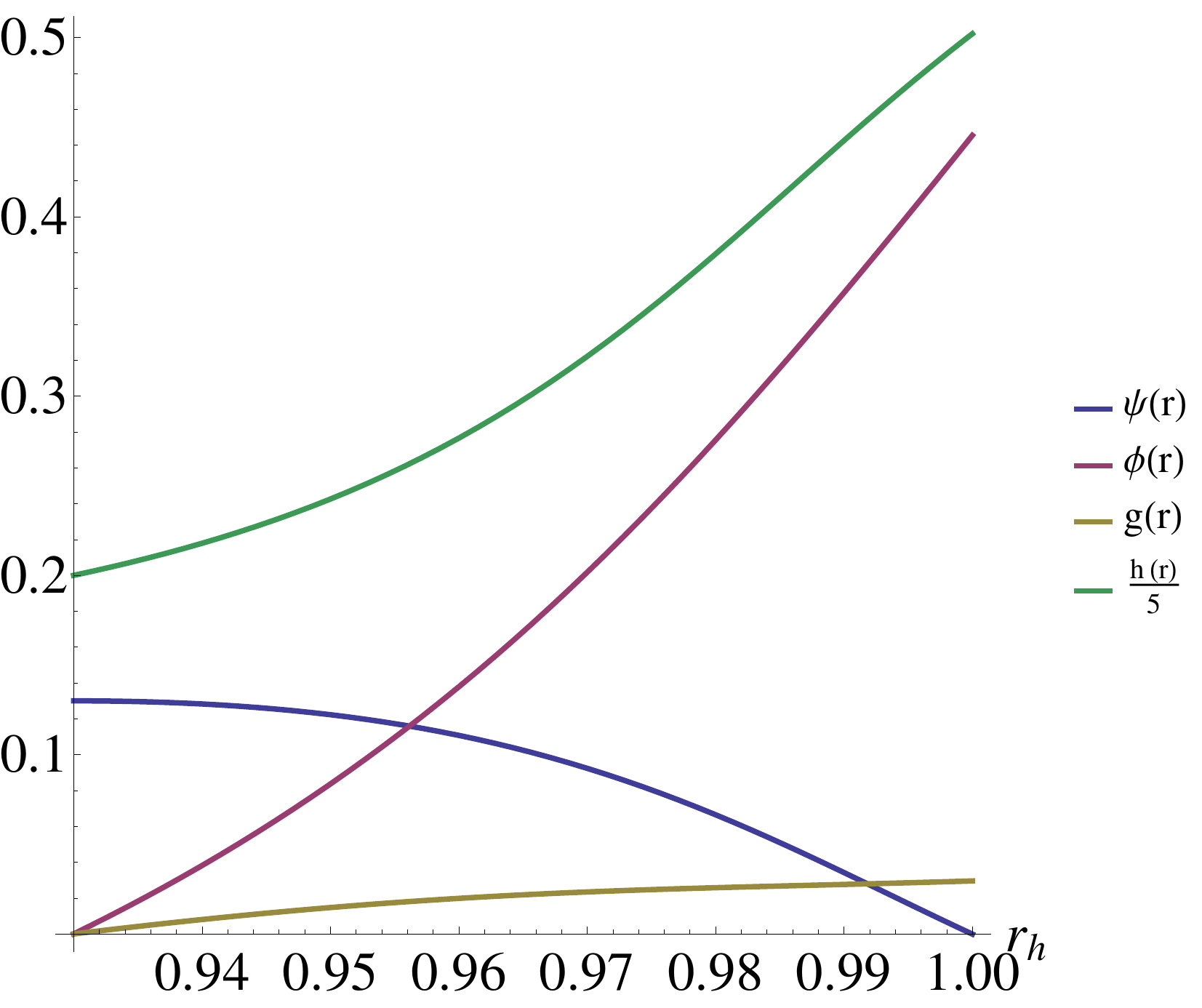}
		\caption{$q=40,\rh = 0.93, \psi_{0} =0.13$}
	\end{subfigure}
	\begin{subfigure}[b]{0.4\textwidth}
		\includegraphics[width=\textwidth]{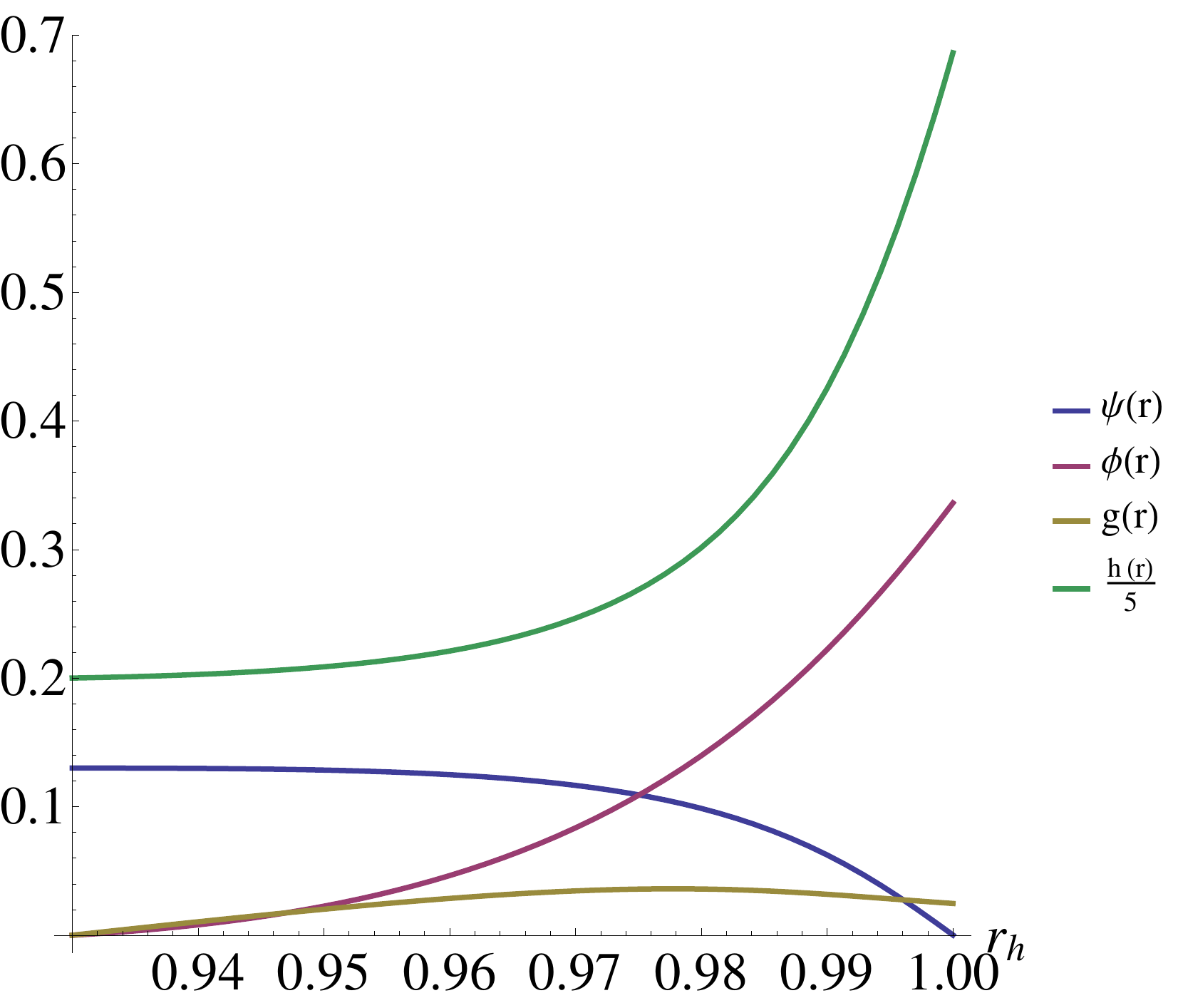}
		\caption{$q=100,\rh=0.93 ,\psi_{0} =0.13$}
	\end{subfigure}
	\caption{Sample profiles of $\psi(r), \phi(r)$, $g(r)$ and $\frac{h(r)}{5}$ in the fully backreacted solution for hairy black hole, with $\rb=1$. The quantities on the y-axis are labelled to the right of the figure.} \label{bbhprofiles}
\end{figure}

The instability of a RN black hole to develop hair is dependent on $q$ and $\rh$, and cannot be evaluated analytically. In the limit $\psi\rightarrow \alpha\psi(r)$ with $\alpha\ll 1$, and looking at upto terms linear in $\alpha$, we get the RN solution, and a homogeneous equation for $\psi(r)$, which can be solved numerically to find the first eigenmode,
\bea
g(r)= 1 -\dfrac{1}{r}\left(\rh+ \dfrac{Q^{2}}{\rh }\right), \ h(r) = \dfrac{1}{g(\rb)}, \ \text{and} \ \phi(r) = \dfrac{Q}{\sqrt{g(\rb)}} \left(\dfrac{1}{\rh} - \dfrac{1}{\rb }\right),\\
\psi ''(r) +\dfrac{\left(Q^2-2  \rh\,r+\rh^2\right)}{(r-\rh) \left(Q^2-\rh \,r \right)}  \psi '(r) + \dfrac{q^2 Q^2 r^2 }{\left(Q^2-\rh r\right)^2}\psi (r)= 0.
\eea

The profiles of two fully backreacted solutions are given in Fig.\ref{bbhprofiles}.

\section{The Phase Diagram}

As we discussed earlier, the free energy of the system when there is a non-trivial scalar profile present is done by evaluating the on-shell action. The full action is given by the sum of \eqref{baction} and \eqref{bscaction}. Since the boundary metric of all the systems are rescaled to be of the same form as the boundary metric of empty box, the temperatures of all the systems can be consistently compared. Using the equations of motions, we can rewrite the action (for details of a similar calculation see Appendix \ref{app_ch2})
\bea
F = \dfrac{S}{\beta} =\dfrac{1}{\sqrt{g(\rb)h(\rb)}} \left(-\dfrac{1}{2} \int_{\rh}^{\rb} \sqrt{h(r)}  \, dr -\dfrac{\rb}{2}\sqrt{h(\rb)} \left(g(\rb) - \dfrac{\rb}{2} g'(\rb)\right)- \dfrac{\rb^{2}}{4} \dfrac{g(\rb)}{\sqrt{h(\rb)}}h'(\rb) \right)\nonumber\\
-\left(-\rb + \dfrac{\rh }{2}\right).
\eea
The phase diagram is intricately dependent on $q$, which gives three distinct types of phase diagrams, which have two, three or four of the four possible solutions as thermodynamically acceptable solutions. We will look at each of theses cases in detail.

\subsection{$q_{1}$ < $q$ < $\infty$ }
\begin{figure}[h!]
	\centering
	\includegraphics[width=0.6\textwidth]{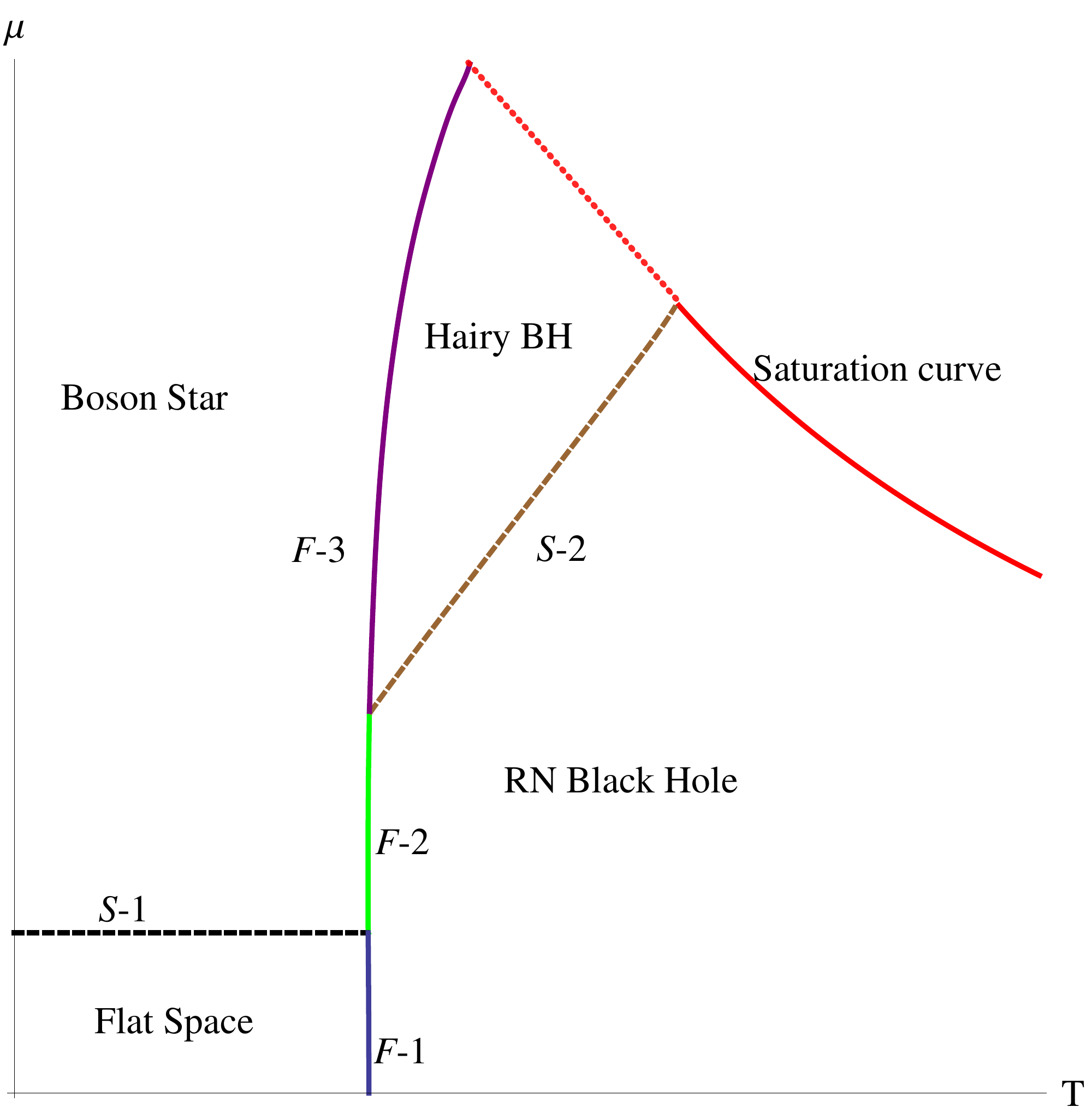}
	\caption{Schematic phase diagram for $\infty<q<q_{2}$.}
	\label{largeqsample}
\end{figure}
For this case all of the four solutions appear in the phase diagram in certain regions of the $ T-\mu $ plot. A representative diagram is given in Fig.\ref{largeqsample}. For values of  $\mu $ smaller than the boson star instability, $\mu_{bsi}$, for the given value of $q$ given by \ref{bbsi}, the phase boundary, $\mathit{F}$-1, separates the empty flat box from the RN black hole, which is a first order phase transition, and can be computed analytically. Above the value at which the boson star instability happens, say $\mu_{bsi}$, the favourable phase is a boson star, which is a second order phase transition from the flat empty box, indicated by $\mathit{S}$-1, and has lower free energy than the empty box. The phase transition between boson star and RN black hole is first order in nature and th phase boundary is indicated by$\mathit{F}$-2, which is computed semi-analytically by the following method. For a given value of $q$, the value of $\mu $ and $F$ can be determined for each value of $\psizero$, and we can do a fit to get free energy as a function of $\mu$, and then find the value of $T$ for the RN black hole with the same value of $\mu$ and $F$.

The curve $\mathit{F}$-2 comes to an end at the point where it intersects the hairy black hole instability curve, $\mathit{S}$-2. The RN black hole to hairy black hole transition is also a second order phase transition. One can check that for a given value of $T$ and $\mu$ (where the hairy black hole solution exists), the hairy black hole has a lower free energy than a RN black hole with the same $(T,\mu)$. The phase boundary between the boson star and hairy black hole is another first order transition, which is given by the curve $\mathit{F}$-3. This curve is slightly more difficult as the values of $(T,\mu,F)$ for both the competing phases are found by {\em numerically} solving the fully backreacted equations of motion of the respective systems. For the boson star case, we have the free energy as a function of $\mu$. In the case of the box, for a hairy black hole of a larger value of $\rh $ than the one of the black hole residing at the intersection of $\mathit{F}$-2 and $\mathit{S}$-2, say $r_{h}^{c}$ will go to a boson star phase as we increase $\mu$ past some critical point (as opposed to the case in global AdS). We start with a value of $\rh >\rh^{c}$, and keep increasing the value of $\psiin$ till the free energy of the hairy black hole becomes equal to the free energy of the boson star with the same $\mu $.
\begin{figure}[h!]
	\centering
	\begin{subfigure}[b]{0.4\textwidth}
		\includegraphics[width=\textwidth]{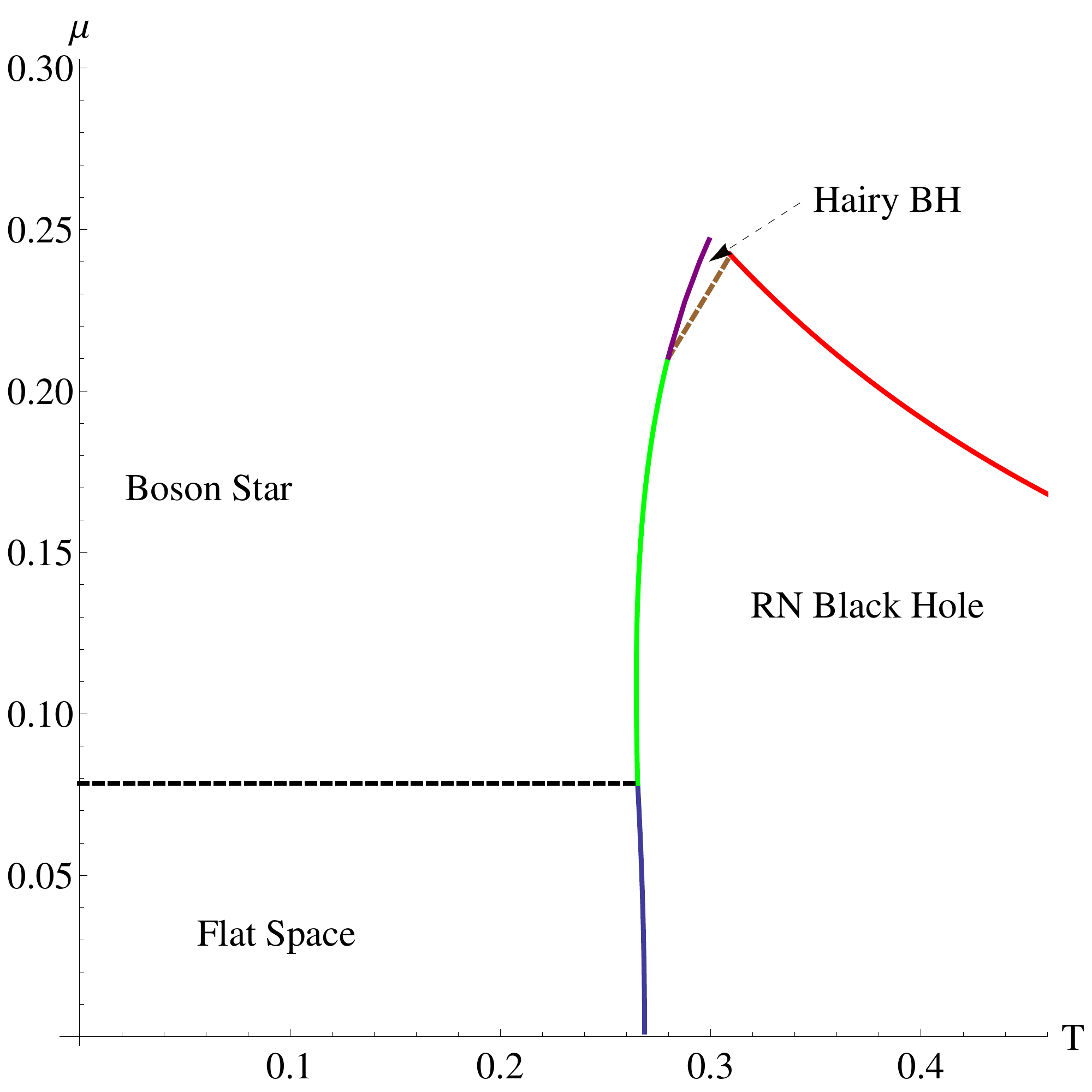}
		\caption{$q=40$}
	\end{subfigure}
	\begin{subfigure}[b]{0.4\textwidth}
		\includegraphics[width=\textwidth]{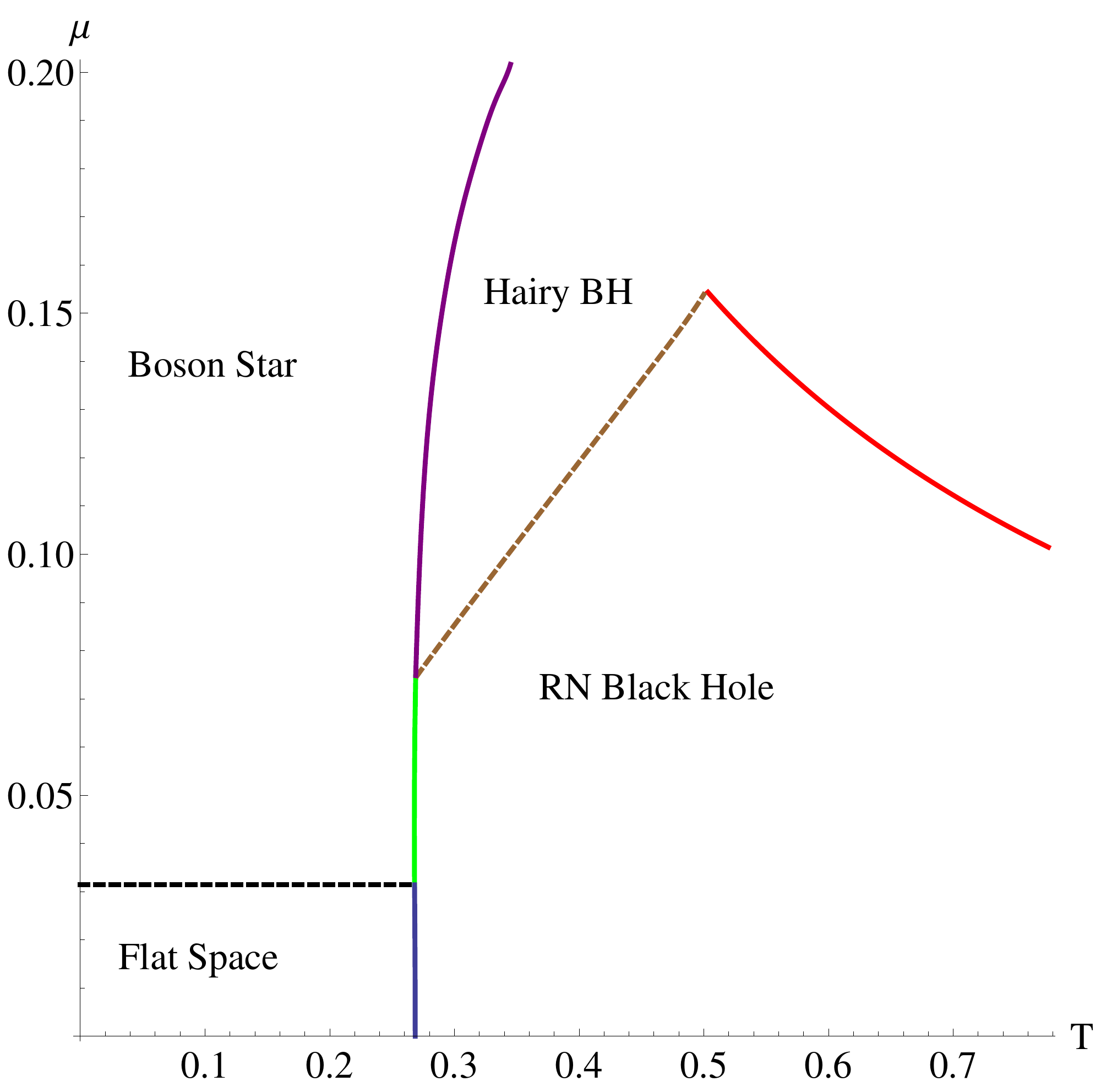}
		\caption{$q=100$}
	\end{subfigure}
	\caption{Phase diagrams for $q=40,100$.}\label{largeq}
\end{figure}
In principle the curve $\mathit{F}$-3 should be computable for all the values of $\rh$ till it becomes very close to $\rb $. However, the shooting procedure that we use to find the solutions becomes increasingly difficult as we take $\rh $ closer to $\rb$, depending on the value of $q$, For example, for $q=100$, we are able to obtain the curve upto around $\rh=0.962$, and for $q=20$, we can go to about $\rh=0.99$ (all with $\rb =1$). In other words, the exact structure of the red-dotted line in our schematic diagram Fig. \ref{largeqsample} cannot be precisely obtained with our current numerics. That the hairy black hole region has to be bounded is based on the fact that the black hole size is limited by the box. The precise form of the way the region closes as $\rh\rightarrow\rb $ does not alter the punchline that the hairy black hole is a thermodynamically favourable phase in some regions of the $(T,\mu)$-plane. So we will relegate that to future work.

The exact phase diagrams for $q=40,100$ are shown in Fig.\ref{largeq}. As we can see that region in which the hairy black hole can exist shrinks as we go to smaller values of $q$. Eventually, we end up with the case where $\rh^{c}\rightarrow\rb $, which happens at $q=q_{1}$. It is difficult to find the exact value of $q_{1}$ as the numerical value will have to be found by a tedious trial and error method. However, we have found that it happens for $q$ slightly greater than $36$.

\subsection{$q_{2}<q<q_{1}$}
\begin{figure}[h!]
	\centering
	\includegraphics[width=0.6\textwidth]{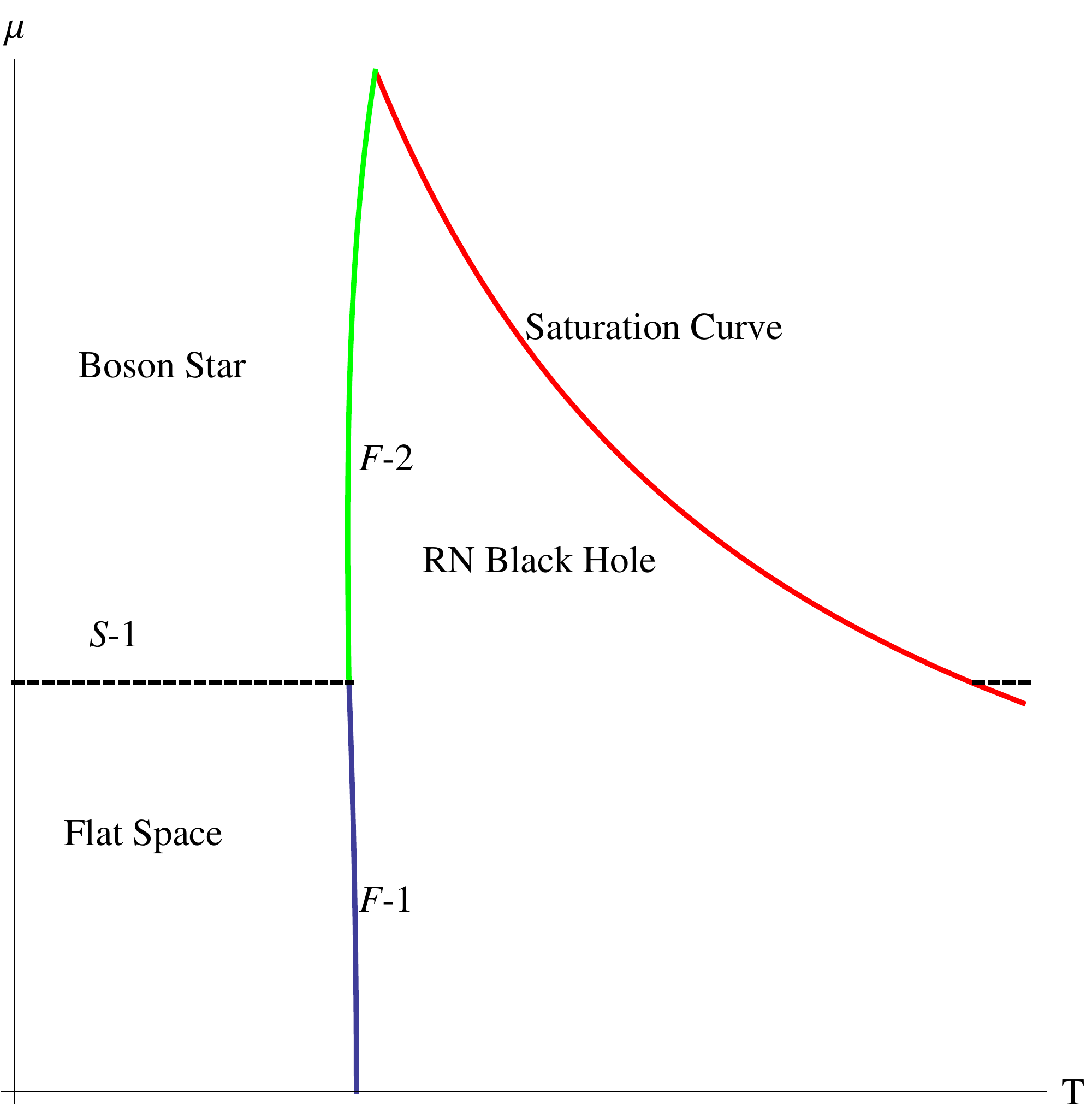}
	\caption{Schematic phase diagram for $q_{2}<q<q_{1}$.}
	\label{intersample}
\end{figure}
As we can see from the discussion in the previous case, this range of $q$ will give a phase diagram where only three of the four solutions can exist as thermodynamically favourable phases, namely flat empty box, RN black hole and boson star, see Fig.\ref{intersample}. The curves $\mathit{F}$-1, $\mathit{S}$-1 and $\mathit{F}$-2 are computed using the same procedure as mentioned for the large $q$ case. The RN black hole instability happens for values of $(T,\mu)$ where the RN black hole itself is not the favourable phase, whereby the second order transition does not happen, and the hairy black hole does not appear as a thermodynamically favourable phase. As a result, the curve $\mathit{F}$-2 ends when it intersects with the saturation curve.

\begin{figure}[h!]
	\centering
	\begin{subfigure}[b]{0.4\textwidth}
		\includegraphics[width=\textwidth]{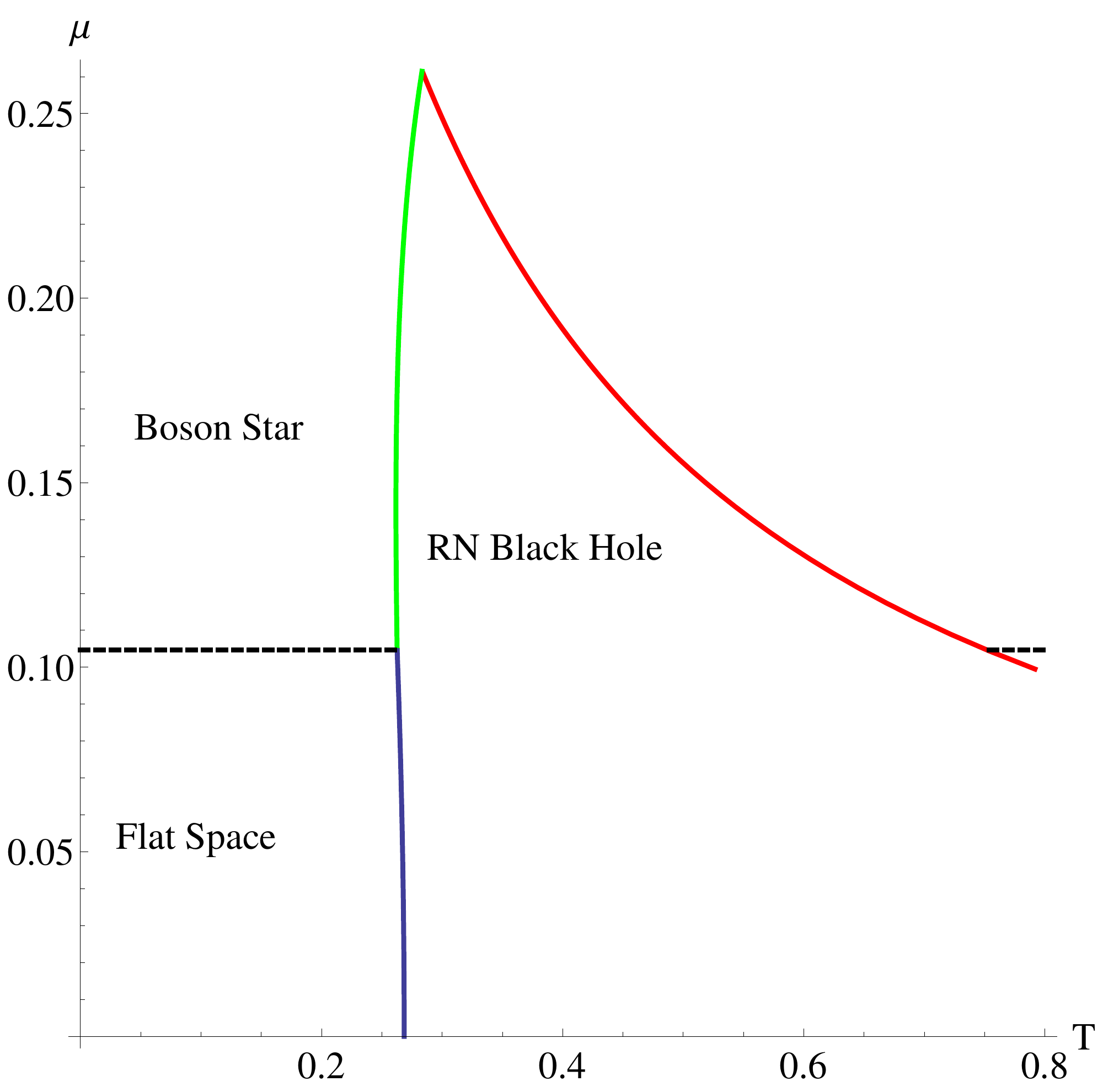}
		\caption{$q=30$}
	\end{subfigure}
	\begin{subfigure}[b]{0.4\textwidth}
		\includegraphics[width=\textwidth]{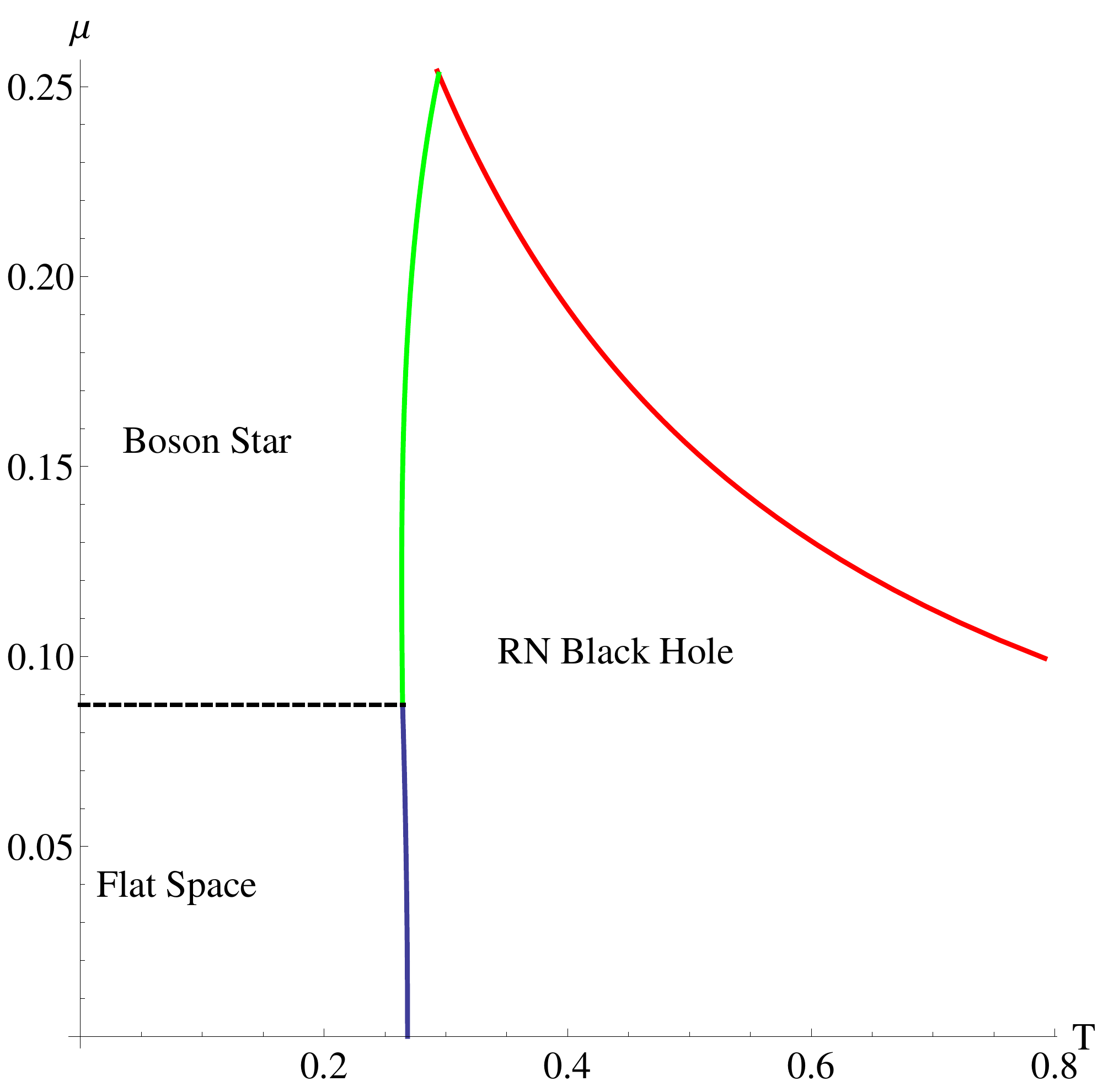}
		\caption{$q=36$}
	\end{subfigure}
	\caption{Phase diagrams for $q=30,36$.}\label{interq}
\end{figure}

We have plotted the exact phase diagram for $q=36,30$ in Fig.\ref{interq}. This type of phase diagram can exist only until the values of $q$, such that the value of $\mu_{bsi}<\frac{1}{3}$, or in other words $q<q_{2} = 3\pi$, below which there will be no phase boundary between RN black hole and boson star.

\subsection{$q<q_{2}$}

\begin{figure}[h!]
	\centering
	\includegraphics[width=0.6\textwidth]{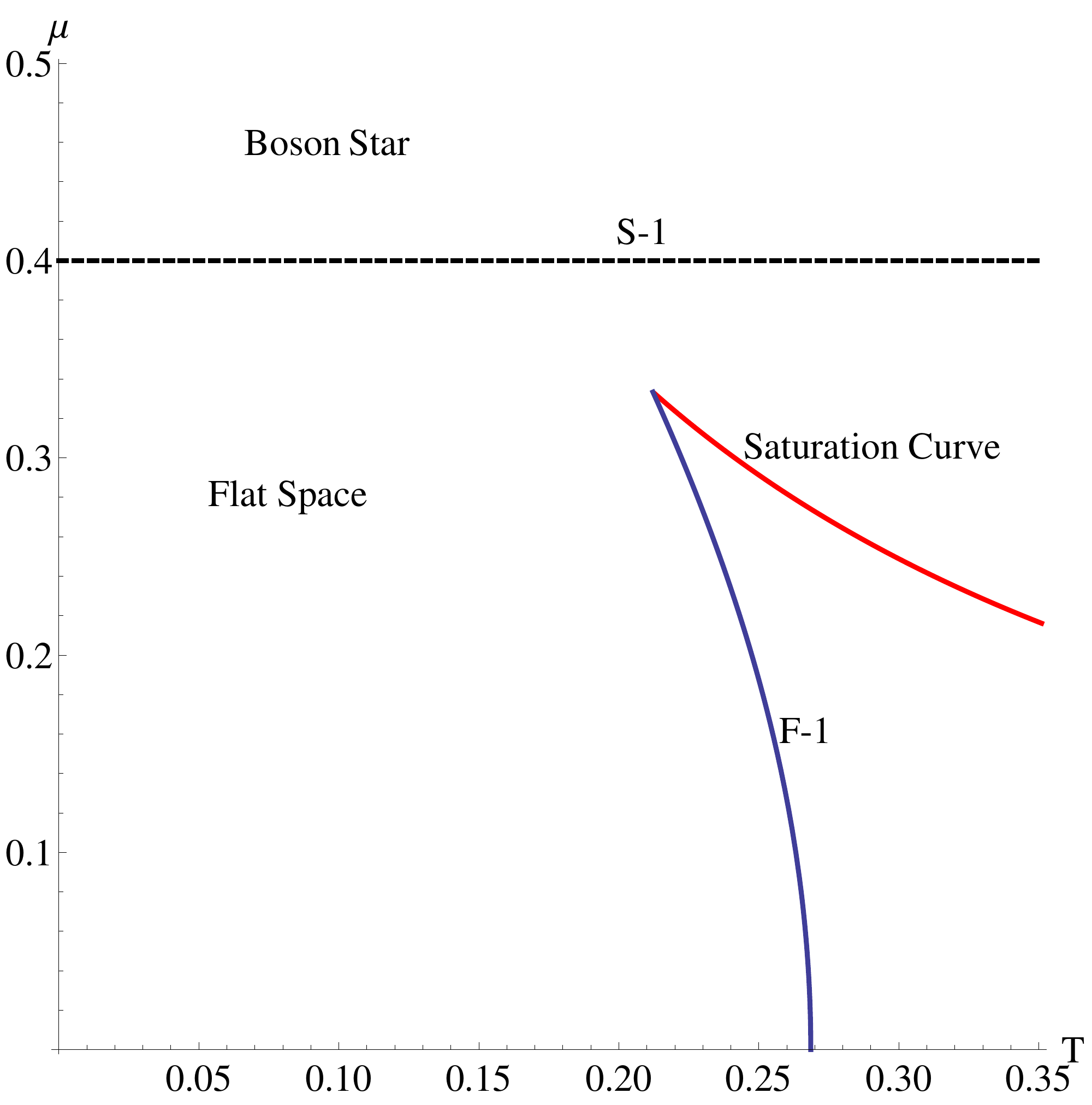}
	\caption{Phase diagram for $q=7.85$.}
	\label{smallsample}
\end{figure}

For values of $q<q_{2} = 3\pi$, the boson star instability happens at $\mu_{bsi}>\frac{1}{3}$, which in some sense leads to a trivial extention of the phase diagram of scalar-less case, see Fig.\ref{smallsample}. The boson star becomes the dominant phase above $\mu=\frac{\pi}{q}$. 

\section{Conclusions, Discussion and Future Directions}

We have charted out the phase diagram of the Einstein-Maxwell-scalar system in a box and demonstrated the existence of thermodynamically stable hairy solutions with and without horizons. The results that we find have close analogues in global AdS, but also some differences. These differences are closely tied to the fact that the box is a hard cut-off on the size of the black hole: extremal limit of thermodynamically stable black holes in AdS correspond to zero size black holes (both inner and outer horizon shrink to zero size), whereas in the box they correspond to the outer and inner horizon reaching the box size simultaneously. Our phase diagram is nearly complete, but it will be nice to precisely chart the upper boundary of the hairy black hole solutions where the black hole size reaches that of the box. 

One interesting observation is that having a non-trivial scalar in the box means that in the limit where the box size goes to infinity, these solutions are {\em not} the within-box truncation of the asymptotically flat (and therefore hairless) solutions. One might wonder what makes the scalar case and the gauge field so different when they are considered in a box. The point is that holding the gauge field fixed at a finite radius contains essentially the same information as fixing it at infinity (once the $r_h$ and $r_b$ are given). There exists a simple limit for gauge field when the $r_b \rightarrow \infty$, where the gauge field remains non-vanishing. But this is {\em not} the case for scalar. The $r_b \rightarrow \infty$ limit is non-trivial for the scalar, because no hair theorems force the scalar to be zero at all orders in $1/r$ in that limit.

This has consequences for defining the quasi-local mass in a box in the sense of \cite{BrownYork}. Typically, the quasi-local mass just puts a box around some region to define the mass/charges in that regions {\em while} allowing the field itself to decay to its asymptotically flat values outside the box. But when we have a non-trivial scalar profile in the box the quasi-local definition of mass will not work because it is not just about putting a box around the region of interest, but also about changing the boundary conditions of the scalar at the box. (This is not the case for the gauge field, where non-trivial boundary conditions at the box are automatically obtained by fixing them at infinity.) That the quasi-local mass cannot work is straightforward to check, because thermodynamic relations of the form $F \sim E-TS -\mu Q$ etc do not work in the box, if we use the quasi-local definitions. Fortunately, we do not need explicit forms of these quantities to chart out the phase diagram, we just need the ability to compute the action aka free energy directly. That, together with the fact that the phase diagram has internal consistency (the various independent curves in it intersect consistently and the overall structure matches very closely with that of AdS) give us confidence that the results are correct. 

It should be possible to generalize the definitions of quasi-local quantities so that one can define a thermodynamically useful notion of the mass of a scalar with a non-trivial profile in a finite region. We hope to come back to this interesting problem in the future. It seems evident that the box boundary acts a compression cavity to hold the scalar in, and therefore a pressure-like term will have to be added to the total mass of the spacetime.

It will also be interesting to consider extremal solutions (not necessarily thermodynamically stable) and perhaps see the possibility of attractor behavior. Note however that attractor behavior is typically associated to uncharged scalars, so the flavor here is slightly different. One can also try to construct hairy solutions with other boundary conditions (see eg., \cite{Neumann,Neumann1,Neumann2,Krishnan:2017bte}). One other interesting line is to consider generalizations of these solutions to higher dimensions: there exists a large class of solutions in higher dimensions \cite{5d} with a rich phase structure (see also \cite{Lahiri} for a ``dual" analysis of phases) and it will be interesting to see how adding a scalar (in the box) changes these results. 

\chapter{Quantum Chaos and Holographic Tensor Models}\label{tensormodel}

\setcounter{footnote}{0}

%%%%%%%%%%%%%%%%%%%%%%%%%%%%%%%%%%%%%%%%%%%%%%%%%%%%%%%%%%%%%%%%%%%%%%%%%%%%%%%%%%%%%%%%%%%%%%
%%%%%%%%%%%%%%%%%%%%%%%%%%%%%%%%%%%%%%%%%%%%%%%%%%%%%%%%%%%%%%%%%%%%%%%%%%%%%%%%%%%%%%%%%%%%%%
\section{Motivation and Conclusions}
\noindent

A theory that has (a) solvability in the large-$N$ limit, (b) maximal chaos \cite{bound}, and (c) emergent conformal symmetry in the infrared would make an excellent candidate for a controllable holographic model for quantum black holes. The Sachdev-Ye-Kitaev(SYK) model \cite{SYK,Kitaev} satisfies all these criteria. 

The SYK model has ``quenched disorder'', which means that it is a theory whose correlation functions are to be considered {\em after}\footnote{Of course, one can also consider the theory where the couplings realize only a single element of the ensemble. Indeed, we will see that this could actually be interesting for our discussions, see also section 8 of \cite{crowdsourced}. But the exact solvability at large-$N$ of SYK is unfortunately and crucially tied to the ensemble average.} an average over an ensemble of couplings. This means that the SYK (ensemble-averaged) correlation functions cannot themselves be interpreted as those of a true quantum system, and therefore one might worry about the lessons one can extract about the quantum behavior of black holes by studying them. 

As an antidote to this, Witten proposed \cite{WittenSYK} a class of tensor models  (building on the work of Gurau and collaborators \cite{guraudump}) which have the same large-$N$ ``melonic'' behavior  \cite{Gurau} as the SYK model and therefore shares its nice features, but does not require a quench. We will call these models and their relatives \cite{klebanov} Holographic Tensor Models (HTM). In this chapter, we will explicitly solve the simplest\footnote{The effective $N$ for this model turns out to be 32, which makes it comparable to the $N=32$ version of the SYK model that already exhibits \cite{crowdsourced} many large-$N$ features.} non-trivial Gurau-Witten tensor model.

Our interest in this problem is directly motivated by the work of \cite{crowdsourced, garcia}, who studied spectral properties of the SYK model and showed that it exhibits various features that are characteristic of random matrices and quantum chaos \cite{Haake, Shukla}. In particular, \cite{crowdsourced} considered a specific function constructed from the spectrum of the theory\footnote{They call this function the Spectral Form Factor (SFF), and we will adopt this terminology. See \cite{sffpapers,1996chao.dyn..6010P,1997PhRvE..56..264B,Papadodimas:2015jra,Papadodimas:2015xma} for various previous discussions on the SFF.} and showed that a specific dip-ramp-plateau structure in its time-dependence is a signature also shared by random matrices in the appropriate ensembles. This statement is true without further qualifications for the SYK model after the ensemble average. But even for a single realization of SYK, this statement holds %at early times as it stands, and at late times 
after a running time average\footnote{See section 8 of \cite{crowdsourced} and our discussions later for a precise definition of the running time average.}.  In this chapter, we will show that striking qualitative similarities with this picture exist also in the Gurau-Witten tensor model (after the running time average to kill the late-time fluctuations). This is interesting because unlike in the (single realization of the) SYK model, the coupling here is a single (dimensionful) number, not ${\cal O}(N^4)$ numbers each chosen from a Gaussian distribution. This result is indicative that despite this, there is randomness and chaos in the system.

We will also see however that there are some interesting differences between the tensor model and SYK. One of the most striking features is that the tensor model has what looks \footnote{To within our numerical error, which we cannot entirely rule out: We have precision up to $\approx 10$  decimals.} like a huge degeneracy in the middle of the energy spectrum, as well as moderate degeneracy elsewhere. The ground state however, is unique. It is tempting to speculate that such a large degeneracy has to do with the entropy of black hole states in the theory  \cite{QFTBH}, and that it has something to say about the zero modes of the broken emergent reparametrization \cite{Maldacena} in the IR. But we emphasize that the true ground state is unique and {\em not} degenerate. See \cite{ads2} for discussions on the emergent reparametrization in the holographically dual AdS$_2$. Since the system is fermionic, it is plausible that the ``half-filled" state should be viewed as the Fermi surface and states above and below it are to be thought of as particles and holes. This is especially likely in light of the fact that the spectrum has a particle-hole-like mirror symmetry as we will discuss. 

It will be very interesting to understand these degeneracies in terms of some underlying symmetry of the Gurau-Witten model. Note that for some values of $N$ the SYK model also had degeneracies because of fermionic symmetries related to Bott periodicity. But these were degeneracies that affected every level. Here on the other hand, the degeneracies affect every state except the most positive and most negative energy states, but the actual degeneracies are different for each level. The maximum degeneracy occurs at $E=0$. But we also note that extremely finely spaced quasi-degeneracies near zero are known in some condensed matter systems \cite{Sambuddha}. Also, in related uncolored tensor models of \cite{klebanov}, at least in some cases we have checked \cite{Vamsi} that many of the same features we see here remain, but the degeneracy is lifted. See the footnote in our final section for some more comments on this.

Interestingly, once we remove the degeneracies and look at the (unfolded) level spacing distribution $P(s)$, we find distinct evidence that the system shows level repulsion at low $s$ indicative of chaotic dynamics. Another feature  we see is that the spectrum has gaps in it, especially close to the mid-levels of the energy. In this, and the fact that the spectrum has no (linear) spectral rigidity, the holographic tensor model is distinct from  SYK \cite{crowdsourced}. Spectral rigidity (see eg. \cite{Berry}) is a measure of how much the integrated density of states (sometimes called the spectral staircase) deviates from a linear fit: it seems from our figure 4 that our spectrum does not have linear spectral rigidity\footnote{We have checked that the spectrum in a single sample of SYK can be fit quite well with a linear fit.}. The lack of spectral rigidity is responsible for the difference between the ``ramp" parts of our plots of the SFF, see figures 5 and 6, and those in \cite{crowdsourced}. In the holographic tensor model, we find that the plot rises up quite quickly after the dip to a plateau, in other words the ramp (to the extent that it is well-defined) is quite steep. This is perhaps not surprising because in \cite{crowdsourced} it was shown that their slow ramp structure is related to rigidity of the spectrum. We emphasize however that the statistics we have for the eigenvalues is relatively small, and that these claims should be taken with a pinch of salt. We note that the late-time plateau is also related to the level repulsion that wee see in the spacing distribution \cite{crowdsourced}. Quantum chaos, random matrix-like aspects and eigenstate thermalization in certain gapped systems has been studied in \cite{Rigol}.

Yet another striking feature of the spectrum is that it has a mirror symmetry, by which we mean that the energy levels come in pairs around the center as
\bea
(E_0+E_n, \ \ E_0-E_n).
\eea
The midpoint energy is $E_0=0$ and it is at that energy that we see the huge degeneracy. The presence of spectral mirror symmetry is an indication that the system has a discrete symmetry which we will discuss in detail later. We will see that it can be traced to the existence of a unitary operator that {\em anti-}commutes with the Hamiltonian \cite{WittenSYK}. We will explicitly construct this operator for our Gurau-Witten model.  Together with the presence of a Particle-Hole Symmetry operator which has already been identified for SYK and SYK-like models like ours \cite{crowdsourced,Ludwig,Fu:2016yrv}, this helps us fix the symmetry class of the theory. We will find that the symmetry class is the so-called BDI class in the 10-fold classification of Altland and Zirnbauer \cite{AZ}. This means that unlike the SYK models which were controlled (depending on the parity properties of $N$) by the Gaussian Unitary, Orthogonal and Symplectic ensembles of Dyson, the random matrix behavior of this model is likely to be controlled by the chiral Gaussian Orthogonal Ensemble. We leave a detailed study of these and numerous other interesting questions for future work, some of which we comment on in a final section.

This Chapter is based on \cite{Krishnan:2016bvg}.

\section{Aside on Quantum Chaos}

Before we delve into the details of the HTMs, we briefly review the idea of \emph{Quantum Chaos}, and how it gets related to Random Matrix Theory (RMT) via the SYK Model. In \cite{Shenker:2013pqa}, Shenker-Stanford introduced the idea of looking at quantum chaos by studying the Out-of Time Ordered Correlators (OTOCs). To illustrate this consider two operators $ V(0) $ and $ W(0) $ which are simple Hermitian operators with zero thermal one point functions, and represent $ \calo(1) $ degrees of freedom. Now, consider the following 
\bea
C(t) = -\langle [W(t),V(0) ]^{2} \rangle,
\eea
which measures the effect of perturbation by $ V $ at a later measurement of $ W $ and vice versa. The time at which the quantity $ C(t) $ becomes significant is called \emph{Scrambling time} $ t_{\star} $. There is also a short time scale called the \emph{dissipation time} $ t_{d} $, related to the exponential decay time of the two point functions like $ \bra V(0)V(t)\ket $. Considering the classical mechanics analogues $ V = p $ and $ W(t) = q(t) $, and looking at Poisson brackets, we can see that $ i\hbar\{q(t),p\} = i\hbar \de q(t)/\de q(0) $, represents the divergence of final positions with small changes in the initial conditions, indicative of classical chaos. With this understanding, we consider $ C(t) $ to be a diagnostic of Quantum Chaos, and identify the dissipation time with the Lyapunov exponent, $ t_{d}\sim 1/\lambda_{L} $. With this, we can also identify the scrambling time as $ t_{\star}\sim -\frac{1}{\lambda_{L}} \log \hbar $. In \cite{bound}, it was shown that the Lyapunov exponent is bounded by the relation
\bea
\lambda_{L} \leq \dfrac{2\pi}{\beta}.
\eea
The saturation of this bound shows that the system is maximally chaotic, which is a desirable property for describing black holes holographically.

The SYK Model saturates the chaos bound at large $ N $\cite{Kitaev,Maldacena}, hence making it a theory of interest for studying black holes holographically. In \cite{crowdsourced}, it was shown that the eigenvalue spectrum of finite $ N $ cases of the SYK Model agrees with the RMT predictions. One of the characteristics of RMTs is the existence of level repulsion due to the lack of degeneracies in the spectrum, which was observed in the case of SYK Model. They also find a precise match in the late time structure of the Spectral Form Factor (SFF), a quantity we will introduce later, between SYK Model and RMTs. With these in mind, we can expect a Quantum Chaotic system to exhibit behaviour analogous to an RMT.

\section{The Holographic Tensor Model}

The general Gurau-Witten tensor model contains $q=D+1$ real fermionic fields 
\bea
\psi_{a, i_{a0}...\slashed{i_{aa}}...i_{aD}}
\eea
where $a, b \in \{0,1,...,D\}$ are called colors, and each of the $i_{am}$'s run from $1,..,n$, where $n$ is independent of $D$. The notation $\slashed{i_{aa}}$ means that $i_{aa}$ is omitted in the indices. The transformation property of the index $i_{am}$ is what defines the symmetry group of the theory, and it is fixed as follows. First we define a group $G_{ab}=O(n)$ for each unordered pair $(a,b)$ of distinct elements in $\{0,1,...,D\}$. This means that upto an overall discrete group that we will not keep track of, the symmetry group of the theory is
\bea
G \sim O(n)^{D(D+1)/2}
\eea
Now the index $i_{am}$ is thought of as transforming in the vector represnetation of $G_{am}$ for each $m \neq a$. Since there are $D$ groups $G_{ab}$ with $a \neq b$ for a given $a$, each $\psi_a$ has $n^D$ components. Now the Gurau-Witten action is written as
\bea
S_{GW}=\int dt \Bigl(\frac{i}{2}\psi_i\partial_{t}\psi_i -  \frac{i^{(D+1)/2}J}{n^{D(D-1)/4}} \ \psi_{0} \psi_{1} .. \psi_{D} \Bigr) \label{GWD}
\eea
where we have suppressed the contractions in the interaction term. Since $a$ runs from 0 to $D$, the total number of real fermions in the theory is $N=(D+1)n^D$. This is the $N$ that is relevant for large $N$, in the sense of comparison to SYK: remember the $q$ in SYK is $(D+1)$ here. The sum over $i$ in the kinetic term is from 1 to $N$. It should be clear that because the index structure of each $\psi_a$ is explicitly constructed to reflect the rest of the fields in the theory, the contraction structure when explicitly written out is a bit of a mess; see eg. \cite{Gurau} for the explicit form of the action. We will only discuss the simplest Gurau-Witten theories where it will be straightforward to write down the contractions by inspection. We also note that the scaling in the coupling $J$ is introduced so that we have well-defined large-$N$ limit. We will often set this $J$ to unity, taking advantage of the fact that it is dimension-ful.

Lets start with the simplest theories, where $D=1$. In this case, we have two sets of fields: $\psi_0$ transforming as a vector under $G_{01}=O(n)$ and $\psi_1$ transforming as a vector under $G_{10}=G_{01}$. This means that the theory is an $O(n)$ theory and explicitly we have
\bea
S_{GW}^{D=1}=\int dt \Bigl(\frac{i}{2}\psi_a^i\partial_{t}\psi_a^i -  {i J} \ \psi_{0}^i\psi_1^i \Bigr) \label{GW1}
\eea
where all indices are explicit and repeated indices are summed over their appropriate ranges. This theory is trivially solved for any value of $n$ because it is free after an appropriate diagonalization in field space: we will not present the details. Essentially identical discussions can be found in eg. \cite{crowdsourced,Maldacena} in the context of SYK.

Since the Lagrangian has to be a boson, the next simplest example corresponds to $D=3$.  
Some index chasing and being careful about the locations of contractions shows that the explicit action is given by\footnote{It is worth mentioning \cite{CK} that quartic couplings have also appeared recently in another context in AdS: this is in the context of AdS instability \cite{instability,Basu:2014sia,Basu:2015efa,Evnin:2015gma}. There, the dynamics is classical and bosonic and exhibits connections to integrable systems, here it is quantum, fermionic and has connections to chaos and random matrices. It will be interesting to study the interpolation between these two types of systems. A paper that has some connections to these questions has very recently appeared \cite{OlegNew}.}
\bea
S_{GW}^{D=3}= \int dt \Bigl(\frac{i}{2}\psi_{a}^{ijk}\partial_{t}\psi_{a}^{ijk} + \frac{J}{n^{3/2}} \psi_{0}^{ijk} \psi_{1}^{ilm} \psi_{2}^{njm} \psi_{3}^{nlk} \Bigr) \label{GW3}
\eea
The specific contraction structure that we follow here follows a similar contraction in \cite{klebanov}. But one can check (and we have, explicitly) that other consistent contractions also lead to identical eigenvalue spectra: this is expected because this just affects the ordering of the assignment of gamma matrices to the spinors (see next section). 
The theory has an $O(n)^6$ symmetry group, and the number of fermions in the theory is $4 n^3$. The case $n=2$ will be the subject matter of most of our discussions. 

\section{The $D=3, n=2$ Gurau-Witten Hamiltonian} 

Our goal in this chapter is to diagonalize the Hamiltonian corresponding to \eqref{GW3} and use it to investigate whether the system exhibits any features of chaos/random-matrix behavior. 

The canonical anti-commutation relations of the theory immediately lead to the Clifford algebra
\bea
\{\psi_a^{ijk}, \psi_b^{lmn} \} = \delta_{ab}\delta^{il}\delta^{jm}\delta^{kn}.
\eea
This means that we can realize the fermion operators in 0+1 dimensions as Euclidean Gamma matrices\footnote{The nomenclature here in the condensed matter literature is a bit confusing to the high energy theorist. To emphasize the obvious: there are no genuine spinors in 0+1 D. What is meant by a fermion in 0+1 dimensional quantum mechanics is an operator that satisfies the Clifford algebra, in other words a gamma matrix. The dimensionality of the Clifford representation is a choice one has the freedom to make, independent of the spacetime dimension which is of course 0+1. In the SYK model for instance, this choice of $N$ gets interpreted as the number of lattice sites.} of $SO(N)=SO((D+1) n^D)$. The dimension of the spinors on which they act grow exponentially fast in $N$, so if we want to have any chance of solving these on a computer, we need to stick to low values for $D$ and $n$: the upper limit for $N$ that is tractable on a computer is about 32, 34, ... from what we see in papers on the subject. Quite fortunately, we find that the first non-trivial value for $N$ in the Gurau-Witten model corresponds to $n=2$ which yields $N=32$. This is the model we will solve. 

Note that we got lucky: the next lowest GW model is computationally inaccessible and requires too much RAM to store the matrices (at least by our resources and skills in computing), as we will discuss later. It is also fortuitous that the solvable $N$ is not too low! If it were, we could not legitimately hope to reasonably claim that we are seeing hints of any large-$N$ physics. As it happens, $N=32$ happens to fall in the right range, and it also happens to be around the upper boundary of $N$ considered in the work of \cite{crowdsourced}.

\subsection{Friendly and Really-Real Spinor Representations}

The gamma matrices we will need are those of $SO(32)$ which means they are going to be 65536 $\times$ 65536 matrices. To solve them with our computing resources, we found it best to work not with the standard representation of gamma matrices which are complex, but instead with a real symmetric representation. The fact that such a representation exists is guaranteed in $N=0 \mod 8$ dimensions. We will use the so-called friendly representation of gamma matrices \cite{sugra} where the gamma matrices are "really real" in $N=0 \mod 8$ dimensions. To construct them systematically, we adopt the following recipe. We first construct Euclidean gamma matrices $E_i$ in $N=8$ 
\bea
E_{1} &=& \sgm{1} \otimes \idd \otimes \idd\otimes \idd, \nonumber\\
E_{2} &=& \sgm{3} \otimes \idd \otimes \idd\otimes \idd, \nonumber\\
E_{3} &=& \sgm{2} \otimes \sgm{2} \otimes \sgm{1}\otimes \idd, \nonumber\\
E_{4} &=& \sgm{2} \otimes \sgm{2} \otimes \sgm{3}\otimes \idd, \nonumber\\
E_{5} &=& \sgm{2} \otimes \sgm{1} \otimes \idd \otimes \sgm{2}, \nonumber\\
E_{6} &=& \sgm{2} \otimes \sgm{3} \otimes \idd \otimes \sgm{2}, \nonumber\\
E_{7} &=& \sgm{2} \otimes \idd \otimes \sgm{2} \otimes \sgm{1}, \nonumber\\
E_{8} &=& \sgm{2} \otimes \idd \otimes \sgm{2} \otimes \sgm{3}.
\eea
These can be explicitly checked to satisfy the Clifford algebra. Together with the definition
\bea
E_{*} &=& E_{1}\dots E_{8} = \sgm{2} \otimes \sgm{2} \otimes \sgm{2} \otimes \sgm{2},
\eea 
now we can follow the recipe \cite{sugra} 
\bea
\gamma^{\mu} &=& \tilde{\gamma}^{\mu} \otimes E_{*}, \ \ \mu = 0,1,\dots, D-1,\nonumber \\ 
\gamma^{D-1+i} &=& \idd \otimes E_{i} ,\ \ \quad i=1,2\dots,8.
\eea
to construct gamma matrices in $D+8$ dimensions starting from those in $D$. Starting from eight dimensions and doing this three times we get from $N=$8 to 16 to 24 to 32, which is the case we want. These gamma matrices are real and symmetric. 

\subsection{Hamiltonian in Terms of Gamma Matrices}

\begin{figure}[H]
\centering
\includegraphics[width=0.4\textwidth]{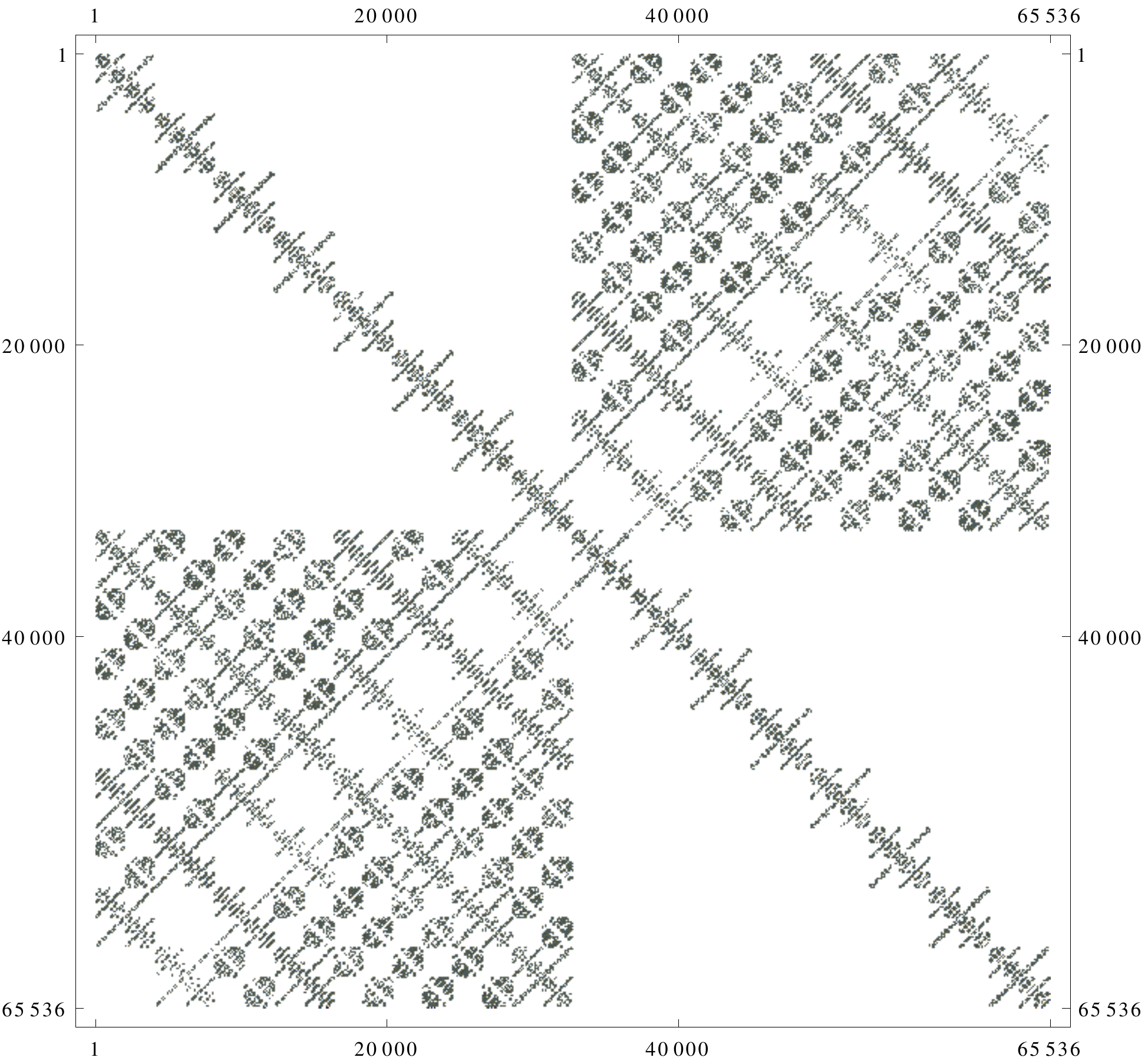}
\caption{The MatrixPlot of Hamiltonian \eqref{Ham}.}
\label{hamplot}
\end{figure}

\begin{figure}[H]
\centering
\includegraphics[width=0.4\textwidth]{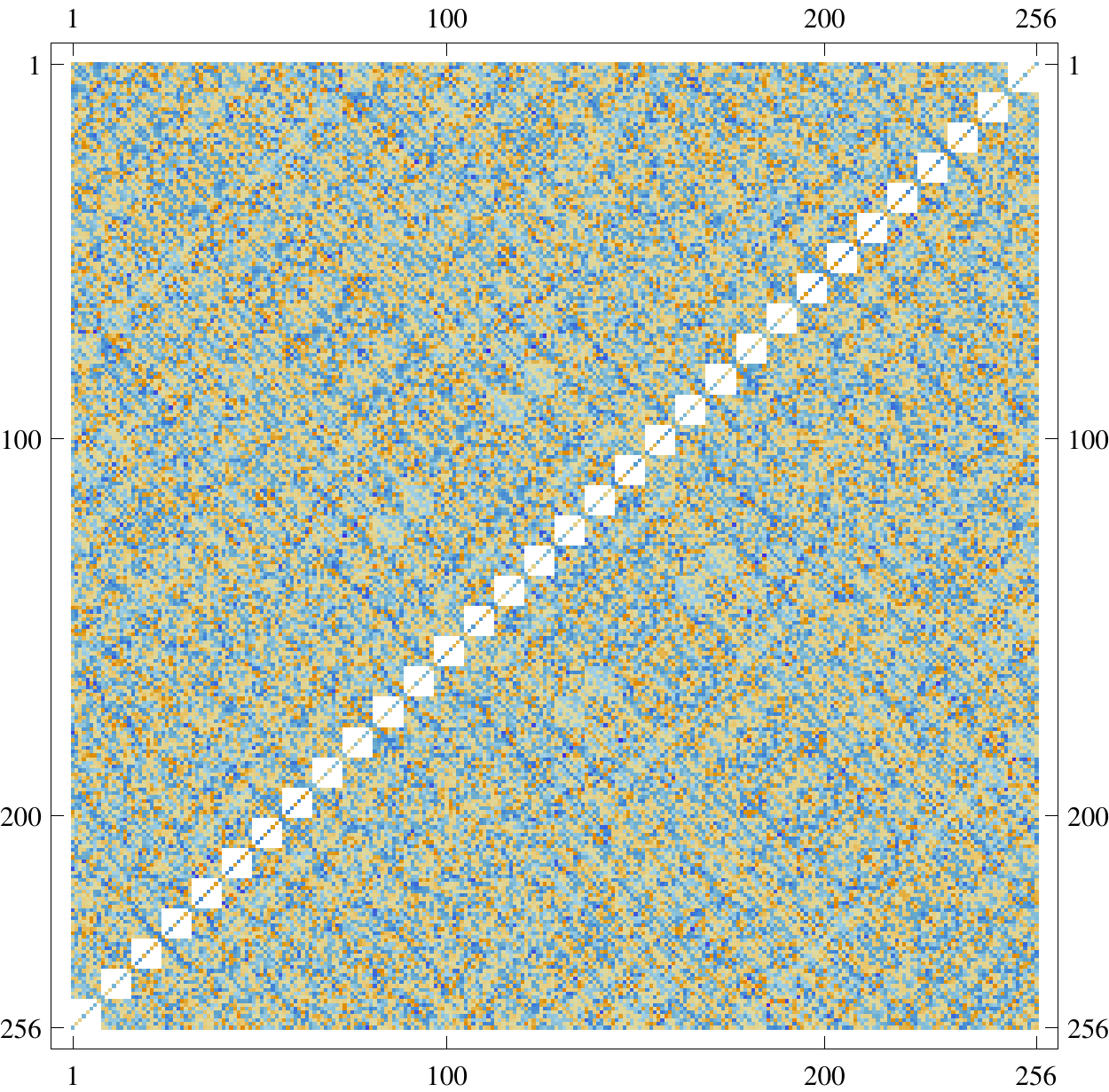}
\caption{The MatrixPlot of an SYK Hamiltonian for a single draw from the ensemble. We are considering the case $N=16$, with really real Gamma matrices.}
\label{hamplotsyk}
\end{figure}

Using these gamma matrices as our definition of the fermions, we can explicitly write out the Gurau-Witten Hamiltonian in terms of the $SO(32)$ gamma matrices. The result is a bit cumbersome:
\bea
H &=& \frac{J}{\sqrt{8}}\Bigl(\gamma_{1}\gamma_{9}\gamma_{17}\gamma_{25}+\gamma_{1}\gamma_{9}\gamma_{21}\gamma_{29}+\gamma_{1}\gamma_{10}\gamma_{18}\gamma_{25}+\gamma_{1}\gamma_{10}\gamma_{22}\gamma_{29}+\gamma_{1}\gamma_{11}\gamma_{17}\gamma_{27} \nonumber \\
&&+\gamma_{1}\gamma_{11}\gamma_{21}\gamma_{31}+\gamma_{1}\gamma_{12}\gamma_{18}\gamma_{27}+\gamma_{1}\gamma_{12}\gamma_{22}\gamma_{31}+\gamma_{2}\gamma_{9}\gamma_{17}\gamma_{26}+\gamma_{2}\gamma_{9}\gamma_{21}\gamma_{30}\nonumber \\
&&+\gamma_{2}\gamma_{10}\gamma_{18}\gamma_{26}+\gamma_{2}\gamma_{10}\gamma_{22}\gamma_{30}+\gamma_{2}\gamma_{11}\gamma_{17}\gamma_{28}+\gamma_{2}\gamma_{11}\gamma_{21}\gamma_{32}+\gamma_{2}\gamma_{12}\gamma_{18}\gamma_{28}\nonumber \\
&&+\gamma_{2}\gamma_{12}\gamma_{22}\gamma_{32}+\gamma_{3}\gamma_{9}\gamma_{19}\gamma_{25}+\gamma_{3}\gamma_{9}\gamma_{23}\gamma_{29}+\gamma_{3}\gamma_{10}\gamma_{20}\gamma_{25}+\gamma_{3}\gamma_{10}\gamma_{24}\gamma_{29}\nonumber \\
&&+\gamma_{3}\gamma_{11}\gamma_{19}\gamma_{27}+\gamma_{3}\gamma_{11}\gamma_{23}\gamma_{31}+\gamma_{3}\gamma_{12}\gamma_{20}\gamma_{27}+\gamma_{3}\gamma_{12}\gamma_{24}\gamma_{31}+\gamma_{4}\gamma_{9}\gamma_{19}\gamma_{26}\nonumber \\
&&+\gamma_{4}\gamma_{9}\gamma_{23}\gamma_{30}+\gamma_{4}\gamma_{10}\gamma_{20}\gamma_{26}+\gamma_{4}\gamma_{10}\gamma_{24}\gamma_{30}+\gamma_{4}\gamma_{11}\gamma_{19}\gamma_{28}+\gamma_{4}\gamma_{11}\gamma_{23}\gamma_{32}\nonumber \\
&&+\gamma_{4}\gamma_{12}\gamma_{20}\gamma_{28}+\gamma_{4}\gamma_{12}\gamma_{24}\gamma_{32}+\gamma_{5}\gamma_{13}\gamma_{17}\gamma_{25}+\gamma_{5}\gamma_{13}\gamma_{21}\gamma_{29}+\gamma_{5}\gamma_{14}\gamma_{18}\gamma_{25}\nonumber \\
&&+\gamma_{5}\gamma_{14}\gamma_{22}\gamma_{29}+\gamma_{5}\gamma_{15}\gamma_{17}\gamma_{27}+\gamma_{5}\gamma_{15}\gamma_{21}\gamma_{31}+\gamma_{5}\gamma_{16}\gamma_{18}\gamma_{27}+\gamma_{5}\gamma_{16}\gamma_{22}\gamma_{31}\nonumber \\
&&+\gamma_{6}\gamma_{13}\gamma_{17}\gamma_{26}+\gamma_{6}\gamma_{13}\gamma_{21}\gamma_{30}+\gamma_{6}\gamma_{14}\gamma_{18}\gamma_{26}+\gamma_{6}\gamma_{14}\gamma_{22}\gamma_{30}+\gamma_{6}\gamma_{15}\gamma_{17}\gamma_{28}\nonumber \\
&&+\gamma_{6}\gamma_{15}\gamma_{21}\gamma_{32}+\gamma_{6}\gamma_{16}\gamma_{18}\gamma_{28}+\gamma_{6}\gamma_{16}\gamma_{22}\gamma_{32}+\gamma_{7}\gamma_{13}\gamma_{19}\gamma_{25}+\gamma_{7}\gamma_{13}\gamma_{23}\gamma_{29}\nonumber \\
&&+\gamma_{7}\gamma_{14}\gamma_{20}\gamma_{25}+\gamma_{7}\gamma_{14}\gamma_{24}\gamma_{29}+\gamma_{7}\gamma_{15}\gamma_{19}\gamma_{27}+\gamma_{7}\gamma_{15}\gamma_{23}\gamma_{31}+\gamma_{7}\gamma_{16}\gamma_{20}\gamma_{27}\nonumber \\
&&+\gamma_{7}\gamma_{16}\gamma_{24}\gamma_{31}+\gamma_{8}\gamma_{13}\gamma_{19}\gamma_{26}+\gamma_{8}\gamma_{13}\gamma_{23}\gamma_{30}+\gamma_{8}\gamma_{14}\gamma_{20}\gamma_{26}+\gamma_{8}\gamma_{14}\gamma_{24}\gamma_{30}\nonumber \\
&&+\gamma_{8}\gamma_{15}\gamma_{19}\gamma_{28}+\gamma_{8}\gamma_{15}\gamma_{23}\gamma_{32}+\gamma_{8}\gamma_{16}\gamma_{20}\gamma_{28}+\gamma_{8}\gamma_{16}\gamma_{24}\gamma_{32}\Bigr) \label{Ham}
\eea
This object is what we will diagonalize and study in the upcoming sections. All its elements are either +1, -1 or zero. The matrix is largely sparse, and it is useful for some of our purposes later to have an idea about the distribution of its non-trivial matrix elements, so we plot it in Figure \ref{hamplot}. It is evident that it has some interesting (almost fractal-like) structure. It is also interesting to note that the result of a single draw of the SYK ensemble  (with the same really real Gamma matrices) results in a Hamiltonian which looks a lot more ``random" and less sparse in appearance. We present its sparseness structure in Figure \ref{hamplotsyk} for comparison. It is worth noting that the non-zero elements of such an SYK Hamiltonian are randomly distributed numbers, whereas the elements of the GW Hamiltonian are  +1, -1 or zero. And yet, we will see that it produces features of randomness. This is not unfamiliar in the case of condensed matter systems where eigenvalue spectra of adjacency matrices can give rise to randomness.

We have diagonalized the Hamiltonian above numerically, and we report on various aspects of the result in the next section.

\section{Numerical Results}

We first present the spectrum, and then in the subsequent subsections present qualitative comparisons to various spectral properties of the SYK model as well as to hints of random matrix-like behavior and chaos. We also mention the differences  from SYK.

\subsection{The Eigenvalue Spectrum}

\begin{figure}[H]
\centering
\includegraphics[width=0.5\textwidth]{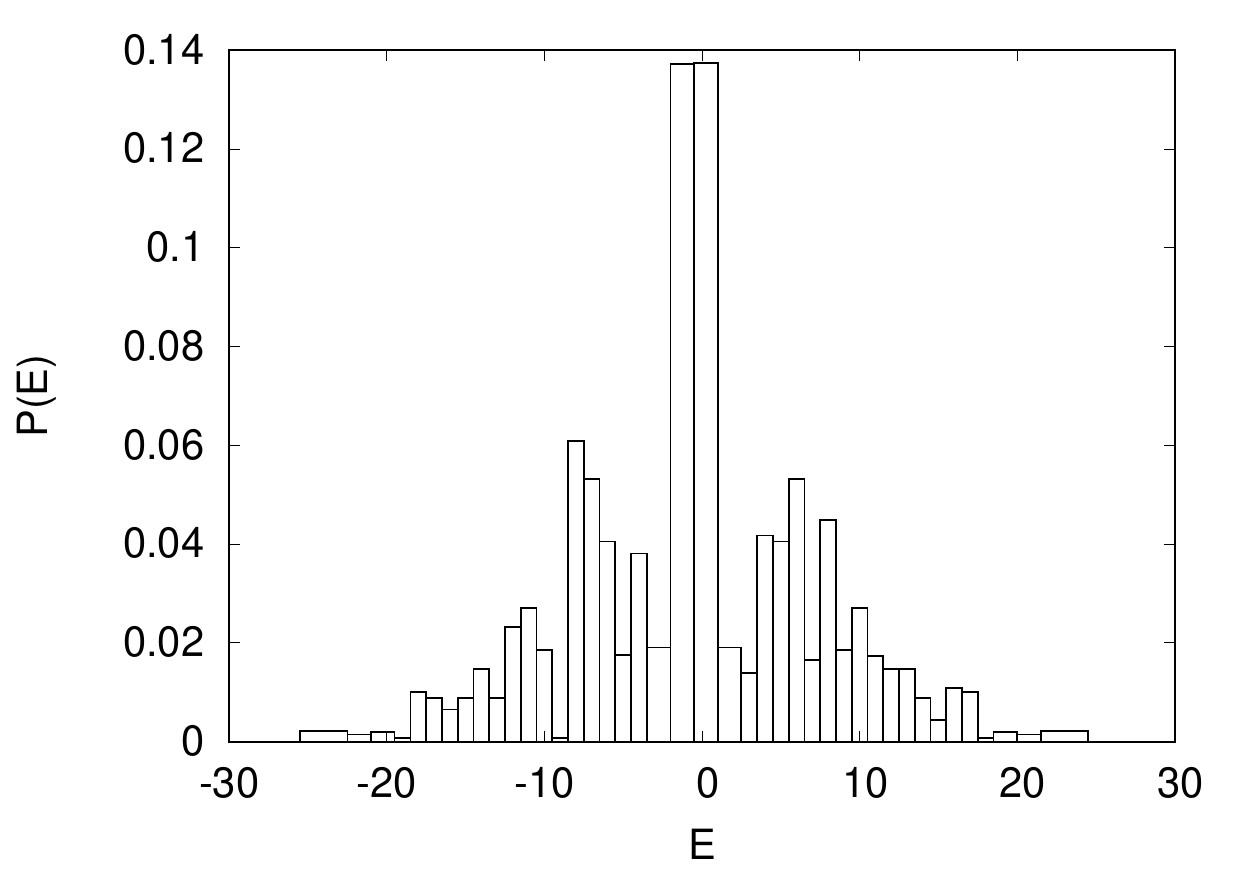}
\caption{The density of states. The d.o.s is symmetric: the slight asymmetry is an artifact of the binning of the eigenvalues.}
\label{dosplot}
\end{figure}

The density of states is plotted in Figure \ref{dosplot}.  It has a multi-peak structure that differs from the SYK single draw case \cite{Maldacena}.  
We also note that the spectrum is {\em exactly} symmetric around $E=0$. We will have more to say about this in the next section, but for now, we note that an {\em approximate} symmetry of this type existed also in (a single draw of) the SYK spectrum as well: see figure 13 in \cite{Maldacena}. We also note that the ground state is unique and has no degeneracies, but there is a huge degeneracy around $E=0$ (within our numerical precision).
\begin{figure}[H]
\centering
\includegraphics[width=0.5\textwidth]{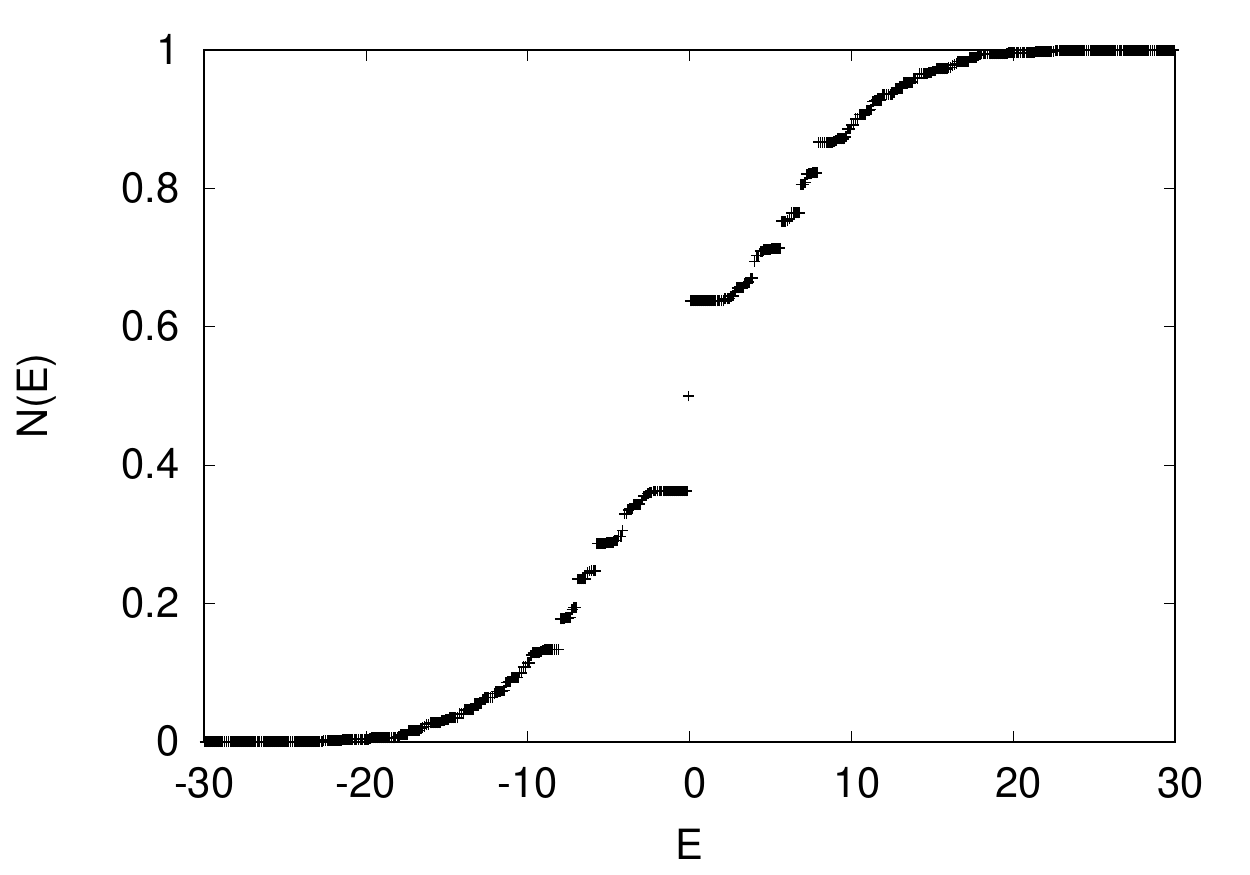}
\caption{The integrated density of states. The jump around zero is a result of the degeneracy at $E=0$.}
\label{idosfullplot}
\end{figure}

\subsection{Spectral Form Factor}

The plots of the spectral form factor, which is defined \cite{crowdsourced} as 
\bea
F_\beta(t)=\frac{|Z(\beta,t)|^2}{|Z(\beta)|^2}
\eea
with 
\bea
Z(\beta,t)\equiv {\rm Tr}\bigl(e^{-(\beta+it)H}\bigr)
\eea
was used as a measure of the random-matrix-like behavior of the SYK model. A dip-ramp-plateau structure in the theory was argued to be evidence for this. The work of \cite{crowdsourced} mostly focused on the ensemble-averaged case, but it was also noted that a running time average in the single draw case results in qualitatively similar features. 
\begin{figure}[H]
\centering
\includegraphics[width=0.5\textwidth]{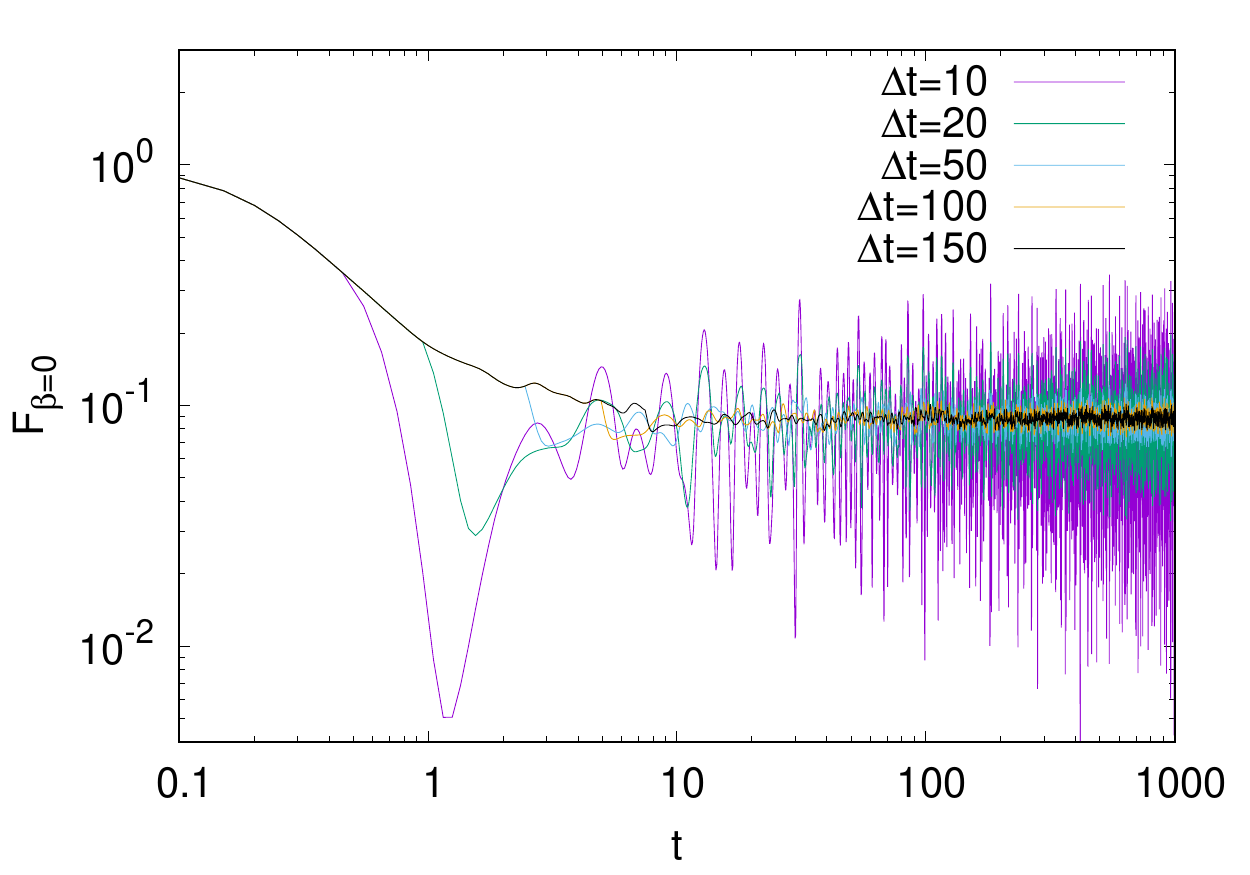}
\caption{The SFF for $\beta=0$}
\label{F0plot}
\end{figure}

\begin{figure}[H]
\centering
\includegraphics[width=0.5\textwidth]{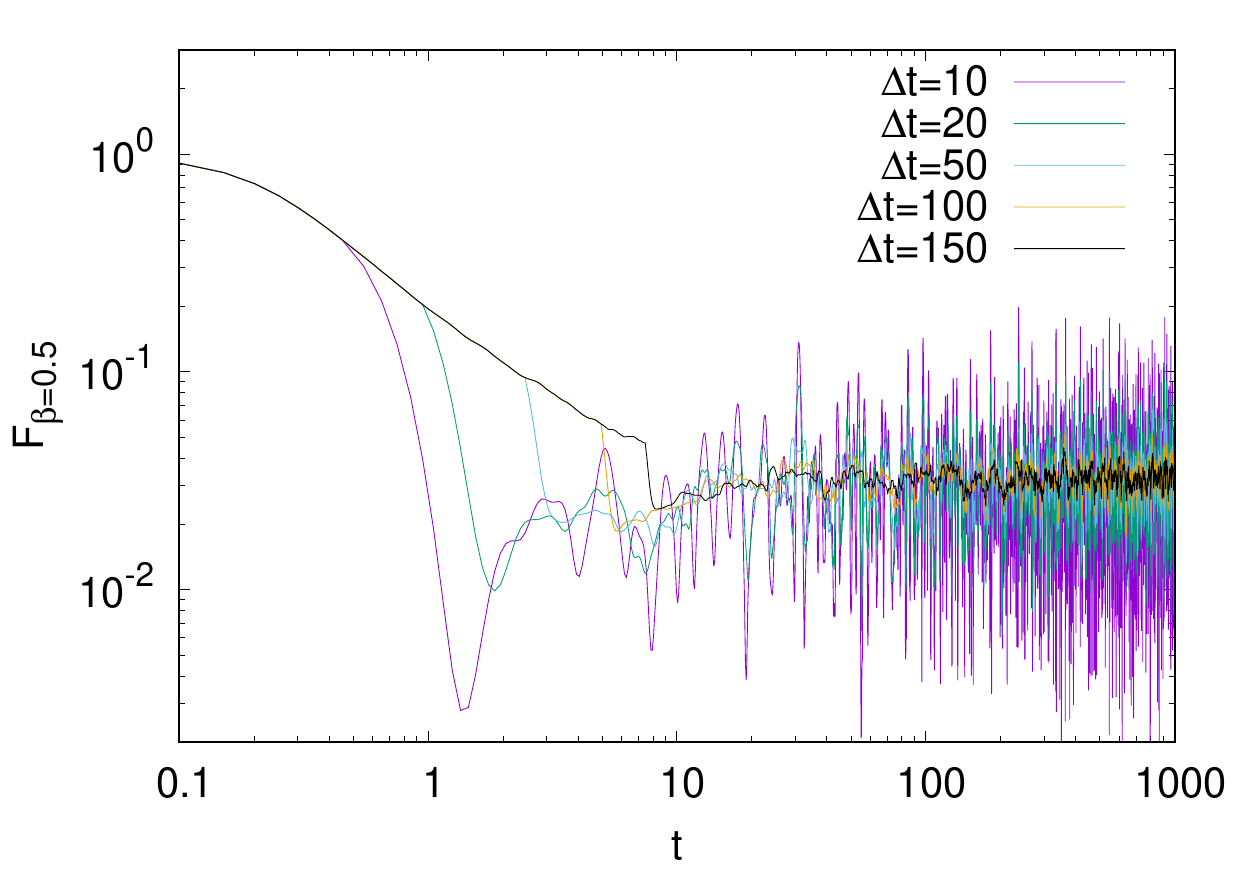}
\caption{The SFF for $\beta=0.5$}
\label{F2plot}
\end{figure}
We have computed the same quantity in the Gurau-Witten theory and we report the plots after a running time average. This means we plot a sliding window average with fixed time windows given by $\Delta t$. The averaging times $\Delta t$  are quoted in the figures. We see a pattern that is quite parallel to that found in \cite{crowdsourced}. Note also that our ramp is steeper than the one found there. We also note (as observed in \cite{Craps}) that there is some tension between increasing the averaging window and the existence of the ramp. 

\subsection{Level Repulsion} 

Once the degeneracies are removed (so that the delta function at the origin of the level spacing distribution goes away), we find that the level spacing distribution $P(s)$ shows distinct signs of level repulsion. 

To see this, we first have to unfold the spectrum (see \cite{garcia} and refernces therein). In integrable systems, the unfolded level spacing distribution typically shows a Poisson distribution steadily increasing as $s \rightarrow 0$. The absence of this, and a turnaround in the distribution close to zero is called level repulsion and is often taken as an indicator of chaotic behavior in the dynamics. In the plot \ref{unfold}, we see distinct evidence for this type of level repulsion\footnote{We have truncated the plot at large $s$ to avoid featuring edge effects: these are not relevant for seeing the level repulsion.}. 
\begin{figure}[h!]
\centering
\includegraphics[width=0.5\textwidth]{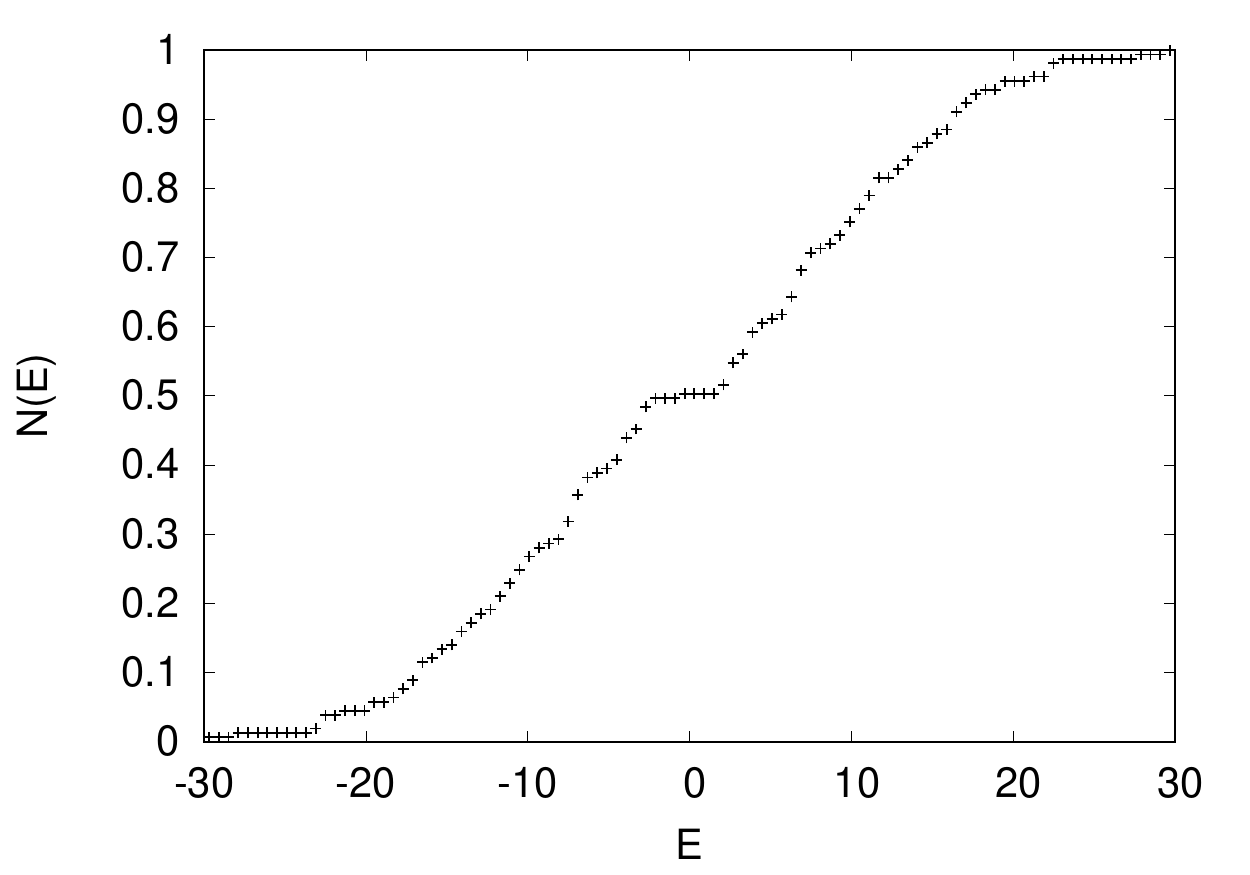}
\caption{The integrated d.o.s plot after degeneracies have been removed. This is the data that we use for doing the unfolding.}
\label{idosnondegplot}
\end{figure}

\begin{figure}[h!]
\centering
\includegraphics[width=0.5\textwidth]{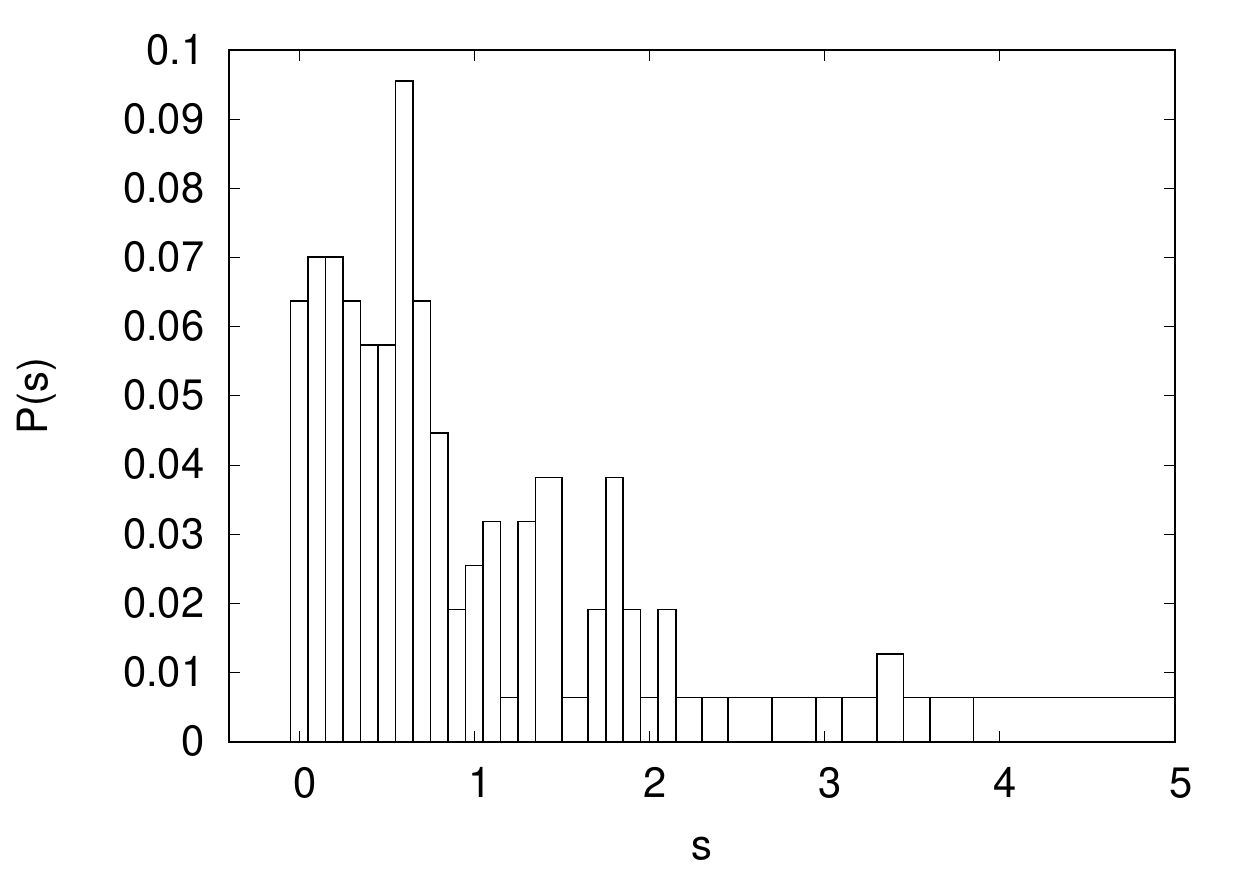}
\caption{Unfolded level spacing distribution showing level repulsion near $s\rightarrow 0$. The level repulsion is evident, but we emphasize that after the degeneracies are removed, the eigenvalues available are not many.}
\label{unfold}
\end{figure}

%\subsection{Thermodynamics}

%In this subsection we present plots of the thermodynamic quantities of the system in the canonical ensemble, taking advantage of the fact that we know its spectrum explicitly. The goal is merely to note that 

\section{Discrete Symmetries and the Choice of Ensemble}

From a glance at the spectrum, it becomes clear that the eigenvalues are exactly symmetrical around zero. Such a spectrum is said to exhibit {\em spectral mirror symmetry} \cite{Shukla}. In this section we will understand this symmetry in the spectrum in terms of an underlying discrete symmetry of the system. This will enable us to also identify the ensemble that is likely to control the random matrix-like behavior of the $D=3, n=2$ Gurau-Witten theory.

The basic observation here is simple. We note that flipping the sign of any one of the $\psi_a$'s in the theory changes the sign of the Hamiltonian: there is a unitary \cite{WittenSYK} operator under which the Hamiltonian is odd. Following the conventions of \cite{Shukla}, we will call this the $S$ operator. The statement then is that
\bea
S H S^\dagger= -H
\eea
What is this operator explicitly? It is straightforward to see this in the gamma matrix language. Flipping $\psi_0$ corresponds in this language to flipping the signs of all the $\gamma_i$'s in the range $i=1, ..8$ while retaining the signs of all the rest\footnote{Flipping the signs of any of the other $\psi_a$'s can be understood as a (signed) permutation of the $\psi_a$'s together with the $S$ operation, and the former is a symmetry of the theory, so these do not give rise to essentially new $S$ operators.}. This means that $S$ is defined by
\bea
S= \gamma_1 \gamma_2 ... \gamma_8
\eea
so that 
\bea
S \gamma_i S = -\gamma_i \  {\rm for} \ i=1,...,8 \\
S \gamma_i S = +\gamma_i \ {\rm for} \  i=9,...,32
\eea
Note also that in the really real representation that we are working with, the gamma's are real and symmetric and so the Clifford algebra guarantees that $S^2=S S^{\dagger}=S S^T=1$. So what we are left with is a unitary operator $S$ that anti-commutes with the Hamiltonian, and squares to 1.

Furthermore, it was noted in \cite{crowdsourced,Ludwig,Fu:2016yrv} that the theory has a symmetry $P$ that has been called a particle-hole symmetry\footnote{It is perhaps more usefully called a $T$ operator. We will adopt this terminology. It contains an anti-linear piece and is related to Kramer's degeneracy, see page 10 of \cite{Fu:2016yrv}.}. The same construction goes through in our case as well. For SYK with $N=0$ mod 8, as well as in our case, it is straightforward to check that it squares to 1. 

Together then, we have two discrete symmetries. An $S$ that squares to 1, and a $T$ that squares to 1. It turns out that these two symmetries are the defining features of the symmetry class BDI in the Altland-Zirnbauer 10-fold classification. It is also referred to as the chiral Gaussian Orthogonal Ensemble. This observation is a strong suggestion that unlike in the SYK cases, the random matrix ensembles corresponding to the Holographic Tensor Models need not be the Wigner-Dyson ensembles.   
\begin{figure}[H]
\centering
\includegraphics[width=0.5\textwidth]{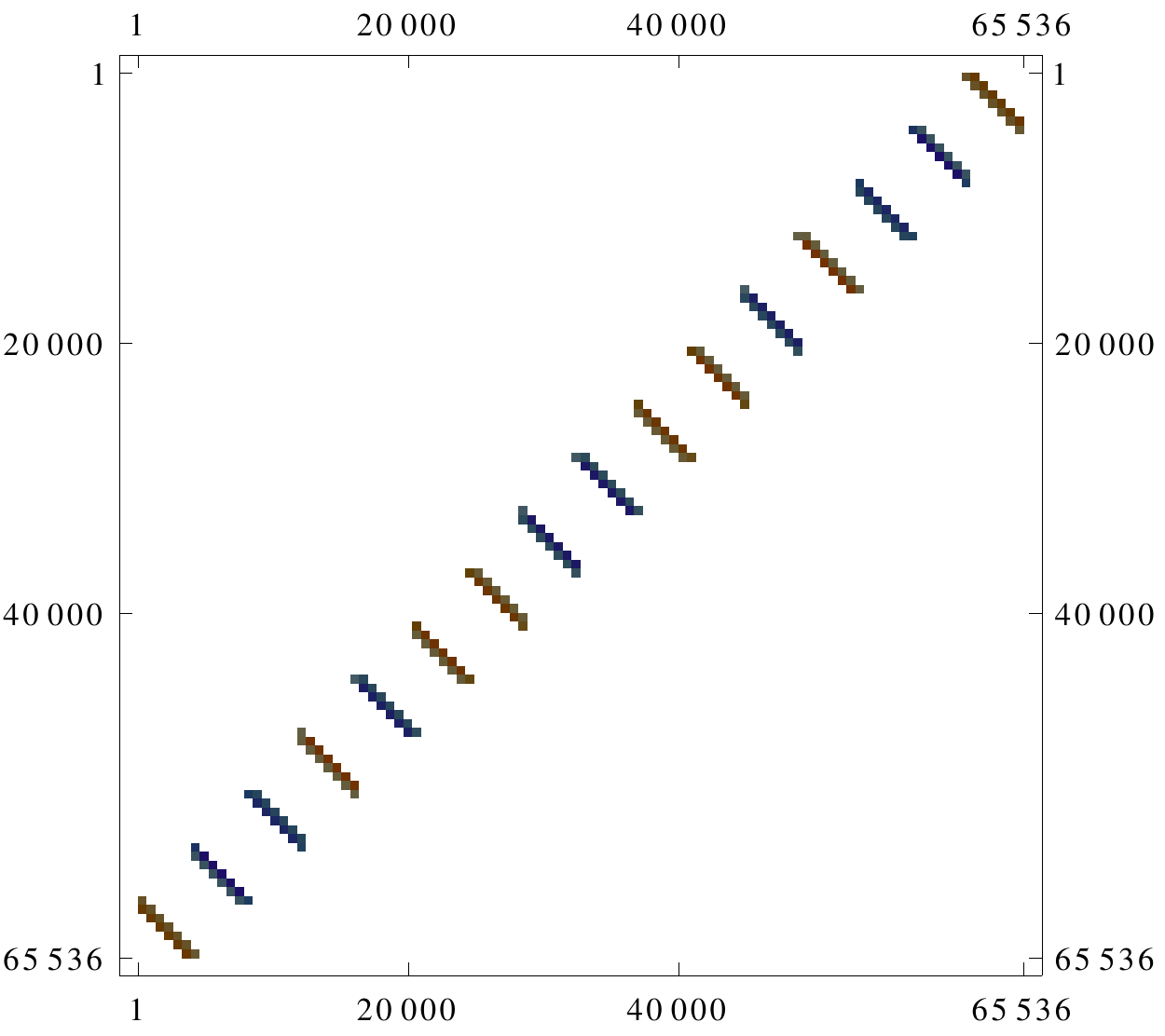}
\caption{The MatrixPlot structure of S.}
\label{Splot}
\end{figure}

We conclude this section with one brief comment. Note that Figure \ref{hamplot} is very suggestive of a Bogolubov-de Gennes (BdG) structure for the Hamitonian. This structure refers to Hamiltonians of the form
\bea
H = \begin{pmatrix}
\; A & B \;\\
\; B^{\dagger} & -A^{T}\; 
\end{pmatrix}
\eea
which are common in mesoscopic physics. One can in fact check explicitly that our Hamiltonian actually satisfies $A=A^T$.
Also since the Hamiltonian is real symmetric, we also have $B^\dagger=B^T$. But our Hamiltonian does {\em not} satisfy $B=\pm B^T$ which would have taken it to one of the other symmetry classes instead of BDI. Operationally this is because the $S$ operator in our case is {\em not} of the form 
\bea
\begin{pmatrix}
\; 0 & 1 \;\\
\; \pm 1 & 0\; 
\end{pmatrix},
\eea
see for example \cite{Shukla}. Explicit evaluations shows that its structure is as in Figure \ref{Splot} in the gamma matrix representation that we are working with.

\section{Comments}

Clearly, we have only considered the most basic features of a specific holographic tensor model. The results we find are a strong suggestion that there is a lot to be understood here. We only make some brief comments of immediate relevance.

It will be very interesting to understand the detailed level spacing distribution and other "random matrix-like" quantities of  HTMs with larger $N$: in our $N=32$ case we do not have too much statistics once the degeneracies are removed because the total number of eigenvalues of the Hamiltonian is merely 65536. The next simplest Gurau-Witten model however is at $D=3, n=3$ and $D=5, n=2$ which corresponds to $N=(D+1) n^D=108$ and $192$ which is computationally inaccessible via brute force\footnote{We are informed by J. Sonner that one can avoid dealing with explicit matrix assignments for gamma matrices, by treating operations involving them as logical operations on their matrix elements. This will reduce some of the demands on computing.}. Another possibility is to consider the model considered in \cite{klebanov}\footnote{We have completed a similar investigation there as well [19], and the results are quite parallel. In the $n=3, D=3$ case ($N=27$), there is spectral mirror symmetry and the plots of the various quantities are qualitatively similar. (Except for an overall 16-fold degeneracy that can be understood from symmetries.). The $n=2, D=3$ case ($N=8$) turns out to be too small to exhibit chaos. We have not been able to diagonalize the $n=2, D=5$ case ($N=32$) because the matrix is too dense for our rather simple-minded endeavors in computing.}, where the model is uncolored and therefore one gets a reduction in degrees of freedom by a factor of $D+1$. The $N$-dependence of the various features would be interesting to understand.

One thing we have not emphasized is the existence of the $\sim O(N)^{D(D+1)/2}$ symmetry in the Gurau-Witten theory, which should appropriately be thought of as gauged for holographic purposes. The spectrum of singlet states of GW Model has been studied in \cite{Krishnan:2018hhu} with the eigenvalues computed analytically and found to agree with the numerical results presented here (see also \cite{Krishnan:2017lra,Krishnan:2017txw,Krishnan:2017yyv,Krishnan:2018jsp,Klebanov:2018nfp}).

\chapter{On the KKLT Goldstino}\label{KKLT}

Controllably breaking supersymmetry  (SUSY) in supersymmetric theories is generally a difficult problem. This raises a challenge for (super)string theory, because the real world is non-supersymmetric and has a positive cosmological constant, which means that for string theory to be phenomenologically viable \cite{Willy}, it needs to admit (likely meta-stable\footnote{It is possible that metastable SUSY-breaking vacua \cite{ISS,Argurio:2006ny,Argurio:2007qk} are generic in supersymmetric theories with complicated potentials, even if they have supersymmetric vacua elsewhere in the potential landscape. Such vacua have also been found to be cosmologically viable \cite{WillyVadimCK}.}) de Sitter vacua. 

The first example of such a de Sitter vacuum in string theory was constructed by KKLT \cite{KKLT}. They did this by considering a fully moduli stabilized warped AdS compactification \cite{GiddingsKP} and proceeding to place a small number $p$ of anti-D3 branes in this warped geometry\footnote{This whole program relies on the existence of flux vacua. Our work does not have much to say directly about this point: our concern is with the nature of SUSY-breaking in them {\em assuming} they exist. This assumption is implicit in all of these works, but see the recent paper \cite{Sethi} which challenges the conventional wisdom.}. The idea is that this breaks supersymmetry and produces positive vacuum energy (which is hierarchically small because of the warping in the geometry) while having a fully stabilized compactification. 

In concrete discussions of KKLT, it is often useful to think of a non-compact Calabi-Yau geometry called the conifold, instead of a fully stabilized compact space. In this non-compact setting, one adds anti-D3 branes \cite{KPV} to the tip of the so-called warped deformed conifold geometry, which is known to be holographically dual to an ${\cal N}=1$ non-conformal gauge theory called the Klebanov-Strassler (KS) cascading\footnote{The duality cascade in turn can be understood \cite{Jarah} via the mechanism suggested in \cite{KPV}.} gauge theory \cite{KS}. The advantage of considering this set up is threefold. Firstly, it enables us to modularize the problem: one can address questions that are not tied to the technicalities of stabilizing the compactification in this more relaxed context, and then hope to ``attach'' the result to a fully stabilized compact Calabi-Yau. The fact that the conifold is an example of a generic Calabi-Yau singularity \cite{KW} makes this approach plausible. Secondly, the duality between the warped deformed conifold and the cascading gauge theory enables us to use powerful holographic techniques to address various bulk questions in the geometry. Indeed, this will be our primary strategy in this chapter. Thirdly, the Klebanov-Strassler theory gives us a concrete setting where we can do explicit calculations, but whose results are expected to have generic significance. 

For this approach to be of any use however, one needs to make sure that when one adds anti-D3 branes at the tip of the throat, the resulting bulk solution should be interpretable as a state in the dual cascading gauge theory\footnote{Note that it is not easy to determine from the bulk side alone if the cascading geometry with and without anti-D3 branes belongs to the same theory. As far as the supergravity is considered, anti-D3 brane sources are like a boundary condition in the IR, and are in some sense arbitrary. Whether they really belong to the spectrum of the gravity theory depends on the UV completion of the supergravity into string theory, and is something which we do not know well because we do not have full control on Klebanov-Strassler as a string background. What holography and the dual cascading theory does here, is to give us a non-perturbative definition of the theory so that we can in principle ask whether certain states belong to its spectrum. 
}. In particular, since the anti-branes break bulk supersymmetry, the corresponding state in the dual theory should be one where SUSY is spontaneously broken, which means that it should be characterized by a goldstino mode. Such a mode was indeed identified  in \cite{DKM} and later in \cite{Bertolini} within the context of a certain five dimensional $SU(2) \times SU(2) \times \Z_2\times U(1)_R$ truncation of IIB supergravity on $T^{1,1}$ using the holographic renormalization technology developed by \cite{Aharony}. 

In this chapter, our goal is to extend these results to an $SU(2) \times SU(2) \times \Z_2 \times \Z_{2R}$ truncation by including supergravity fields which are not neutral under the R-symmetry: the previous constructions \cite{DKM,Bertolini} were working with Klebanov-Tseytlin (KT) \cite{KT} asymptotics, whereas we will deal with the full Klebanov-Strassler.  Klebanov-Tseytlin geometry is singular in the IR and cannot incorporate the deformation of the conifold, while Klebanov-Strassler is a fully regular solution. The (implicit or explicit) hope of the calculations in \cite{DKM,Bertolini} was that since the deformation parameter is a supersymmetric perturbation, it is unlikely to destroy the claims about the SUSY-breaking perturbations. But it must be borne in mind that to discuss this question adequately, one must work with a fully consistent truncation that {\em allows} the deformation in the first place, and see whether (a) such a truncation allows for more SUSY-breaking parameters in the UV asymptotic solution\footnote{That is, a truncation that captures the conifold deformation parameter in the IR has more fields, and might allow more SUSY-breaking parameters in the UV. This cannot be settled merely by looking at the $U(1)$ truncation, one needs at least the $\Z_2$ truncation.}, and (b) whether the holographic Ward identities \cite{Bertolini,Argurio} get modified in any substantive way. We will answer both these questions in the negative, by working with the $SU(2) \times SU(2) \times \Z_2 \times \Z_{2R}$ truncation. 

The price we pay for working with a more realistic truncation is that there are extra fields in the system which make the problem more complicated. More conceptually, we will see that the extra fields that we turn on correspond to relevant sources, and that the mixing\footnote{Supergravity fields are not naturally diagonal in the field theory scaling dimension, and so we need to work with appropriate combinations of fields.} of fields that they cause on the supergravity side needs to be suitably taken care of. In the $U(1)$ case, only marginal sources were present and their mixing was dealt with \cite{Bertolini} by defining composite supergravity fields which were diagonal in the scaling dimensions. However, this construction is not always unique, and in the case of relevant sources, we find it more convenient to deal with the leading fall-offs of the would-be composite sources directly. We will see that this information is sufficient to compute the one- and two-point functions required for a holographic calculation of the Ward identities. 

In section \ref{sec:2}, we will review the details of our $\Z_2$ truncation and the resulting 5d effective supergravity action, starting from the 10d type IIB theory. Then in section \ref{ktsection}, we proceed to describe the Klebanov-Tseytlin and Klebanov-Strassler backgrounds in a 5d language and compare their UV asymptotics. In section \ref{sec:4}, we obtain and present the most general SUSY-breaking solution that asymptotes to the Klebanov-Strassler solution perturbatively in the UV. We show that despite the presence of extra supergravity fields w.r.t. the $U(1)$ truncation, here also there are only two such SUSY-breaking parameters. There are new SUSY-preserving parameters (apart from the conifold deformation parameter) that show up in the solution which we safely ignore since they do not contribute to SUSY breaking dynamics. In section \ref{sec:5} we give a holographic derivation of SUSY and trace Ward identities. We begin by setting up the gauge/gravity dictionary. We identify the holographic sources for dual operators (in particular, sources for marginal and relevant operators). This leads to some subtleties because (as we previously mentioned) the supergravity fields are not automatically diagonal to the field theory operators, so we need to consider appropriate combinations of them. Once these sources and their supersymmetric partners are identified in a useful form, we proceed to derive the SUSY Ward identities. We also derive the identities for the Weyl and super-Weyl transformations for completeness (and because we can). Since we are doing these calculations holographically we will be working with the local supersymmetries and diffeomorphisms of the bulk supergravity theory and derive the Ward identities by demanding that the variation of the renormalized on-shell action under these transformations is zero. To do this, we will need the transformations of the sources, which we compute following \cite{Bertolini}. Finally, in Section \ref{sec:6}, we conclude by checking these identities on the vacua dual to the SUSY-breaking solution found in Section \ref{sec:4} by explicitly calculating one-point functions that characterizes the scale of SUSY breaking. The results we find are consistent with the expectations of \cite{DKM} which were presented in the context of the $U(1)$ truncation. Since the deformation of the conifold is a SUSY-preserving parameter, it is not surprising that our results are consistent with those of \cite{DKM}. It is somewhat remarkable that even in this generalized setup, there are no more SUSY-breaking perturbations, on top of the ones found in the $U(1)$ case and that the number of SUSY-breaking parameters in the UV remains two. So in the end, we find that despite the complications involved in the relevant sources, the final Ward identities remain substantively unchanged. In Appendix \ref{app_ch4}, we give relevant details needed to reproduce the results in the main text. 

This Chapter is based on \cite{Krishnan:2018udc}.

\section{Dimensional Reduction of Type IIB SUGRA}\label{sec:2}
In this section we give a brief summary of dimensional reduction of type IIB supergravity theory on $T^{1,1}$ which gives rise to a particular ${\cal N}=4$, 5d gauged supergravity. We will truncate this theory to a particular ${\cal N}=2$ subsector that contains the Klebanov-Strassler solution. This truncation will be relevant for the rest of this chapter. The interested reader can find more details in \cite{Buchel,Cassani,Bena:2010pr,Halmagyi:2011yd,Liu:2011dw}. 

The type IIB supergravity in the Einstein frame, takes the form
\begin{align}
\begin{split}
S_{10} &= \dfrac{1}{2\kappa_{10}^{2}}\int_{\man_{10}} \left( R_{10} - \dfrac{1}{2} \left(d\phi\right)^2 - \dfrac{1}{2}e^{-\phi} H_{3}^2 - \dfrac{1}{2} e^{\phi} F_{3}^2 - \dfrac{1}{2} e^{2\phi} F_{1}^2- \dfrac{1}{4} F_{5}^2\right)\star1\\
&-\dfrac{1}{8\kappa\dtn^{2}}\int_{\man_{10}} (B_{2}\wedge dC_{2} - C_{2}\wedge dB_{2})\wedge dC_{4}~.
\end{split}
\end{align}
The ten dimensional space-time is denoted by $ \man\dtn $. $ \kappa\dtn $ is related to the ten dimensional Newton's constant. The field strengths satisfy the following Bianchi identities
\bea
dF_{1}= 0~,\ \ dH_{3} = 0~, \ \ dF_{3} = H_{3}\wedge F_{1}~, \ \ dF_{5} = H_{3}\wedge F_{3}~.
\eea
The equations of motion of the type IIB supergravity action have to be supplemented with the self-duality condition
\bea
\star\dtn F_{5} = F_{5}~.
\eea
We are interested in reductions of this theory on the coset $ T^{1,1} = (SU(2)\times SU(2))/U(1)$ with the $U(1)$ embedded in the two $SU(2)$'s diagonally. $T^{1,1}$ can be parametrized in terms of polar coordinates $ (\theta_{1},\phi_{1},\theta_{2},\phi_{2},\psi) $, with ranges $ 0\leq \theta_{1,2}< \pi, 0\leq \phi_{1,2}< 2\pi $ and $ 0\leq \psi< 4\pi $ in the following way
\begin{align}\label{oneforms1}
\begin{split}
&\ee^{1} = -\sin\theta_{1} \;d\phi_{1}~, \  \ \ \ee^{2}  = d\theta_{1}~, \\
& \ee^{3} = \cos\psi \sin\theta_{2} \;d\phi_{2} - \sin\psi \;d\theta_{2}~,\\
& \ee^{4} = \sin\psi \sin\theta_{2} \;d\phi_{2} + \cos\psi \;d\theta_{2}~,\\
& \ee^{5} = d\psi + \cos\theta_{1}\; d\phi_{1} +\cos\theta_{2}\;d\phi_{2}~.
\end{split}
\end{align}
The left-invariant 1- and 2-forms are given by \cite{Cassani}
\bea
\begin{split}
	\eta = -\dfrac{1}{3}\ee^{5}~, \qquad \Omega = \dfrac{1}{6} (\ee^{1} + i \ee^{2} )\wedge (\ee^{3} - i \ee^{4})~ ,\hspace{0.2in} \\
	J = \dfrac{1}{6}(\ee^{1}\wedge \ee^{2} - \ee^{3} \wedge \ee^{4})~,\qquad \Phi = \dfrac{1}{6} (\ee^{1}\wedge \ee^{2} + \ee^{3} \wedge \ee^{4})~.
	\label{BasisofT11}
\end{split}
\eea
The dimensional reduction proceeds by factoring the 10d spacetime ${\cal M}_{10}$ into the warped product space ${\cal M}_{10}={\cal M}_5\times_w T^{1,1}$ and expanding out the 10d form fields in the basis of the left invariant one forms \eqref{BasisofT11} (see Section 3.2 of \cite{Cassani}). The 10d scalars $\phi$ and $C_0$ and all the coefficients in this reduction ansatz are taken to be functions of the coordinates on ${\cal M}_5$. Non-trivial cycles of the internal manifold can allow for additional terms in the expansion for the field strengths. This ansatz retains all and only those modes of type IIB supergravity that are invariant under the action of the isometry group of $T^{1,1}$ which is $SU(2)\times SU(2)$ and automatically guarantees the consistency of the reduction. The resulting 5d effective action matches with the structure of 5dimensional $ \mathcal{N}=4 $ gauged supergravity. The field content of the dimensionally reduced theory, along with its type IIB origins, is reproduced from \cite{Cassani} in Table \ref{n=4trunc} below.

\begin{table}[h]
	\centering
	\begin{tabular}{c|c|c|c|c}
		\hline 
		IIB fields  & scalars & 1-forms & 2-forms & 5d metric \\ 
		\hline 
		10d metric & $u,v,w,t,\theta$ & $A$ &   & $g_{\mu\nu}$  \\		 
		
		$\phi$ &  $\phi $ &   &   &   \\ 
		
		$B_2$ & $b^{J},b^{\Phi},b^{\Omega}$ & $b_{1}$ & $b_{2}$ &  \\ 
		
		$C_{0}$ & $C_{0}$ & & & \\
		
		$C_{2}$ & $c^{J},c^{\Phi},c^{\Omega}$ & $c_{1}$ & $c_{2}$ &  \\
		
		$C_{4}$ & $a$ & $a_{1}^{J},a_{1}^{\phi},a_{1}^{\Omega}$ & $a_{2}^{\Omega}$ & \\
		\hline 
	\end{tabular} 
	\caption{5d fields along with their 10d origins.}
	\label{n=4trunc}
\end{table}
Apart from these 5d fields, there are also flux terms $p,q$ and $k$ that descends from the expansion of the field strengths with legs along the cohomologically non-trivial cycle $\Phi\wedge\eta$ and the volume form. These parameters appear explicitly in the scalar potential and characterizes the gauging. 

By consistently turning off the following 5d fields, one finds a further truncation to an ${\cal N} = 2$ gauged supergravity
\bea
b_{2} = c_{2} = a_{2}^{\Omega}=a_{1}^{\Omega} = b_{1} = c_{1} =b^{J} = c^{J} = 0~.
\eea
The so-called $\caln = 4$ Betti vector multiplet, consisting of $\{ a_{1}^{\phi}, w, \bphi, c^{\Phi}, t,\theta \}$, in the original reduction can be viewed as an $\caln = 2$ vector multiplet $\{ a_{1}^{\phi}, w \}$ together with a $\caln = 2$ hypermultiplet $\{\bphi,c^{\Phi}, t,\theta \} $. Setting either of them to zero is a consistent sub-truncation and gives rise to an ${\cal N}=2$ theory. Truncating out the vector multiplet gives rise to an ${\cal N}=2$ gauged supergravity coupled to three hypermultiplets and a vector multiplet which are invariant under a $\mathbb{Z}_2$ symmetry (not to be confused with the $\mathbb{Z}_{2R}$ symmetry associated to the gaugino condensation in the dual field theory).  This symmetry acts in the following way
\begin{itemize}
	\item $(\theta_1,\phi_1)~~\leftrightarrow ~~ (\theta_2,\phi_2)$ .
	\item Flip the signs of field strengths $H_3$ and $F_3$ (this corresponds to the action of $-\mathbb{I}$ of $SL(2,\mathbb{Z})$ duality group of Type IIB)~.
\end{itemize}
Here $\theta_i$ and $\phi_i$ are the coordinates on $T^{1,1}$. Under the above transformation, the scalar fields $b_J, c_J$ and $w$ flip sign\footnote{This is because in \cite{Cassani}, the 2-form $J$ is invariant and $\left(\ee^1\right)^2+\left(\ee^2\right)^2 \leftrightarrow \left(\ee^3\right)^2+\left(\ee^4\right)^2$
	under the transformation under $(\theta_1,\phi_1) \leftrightarrow (\theta_2,\phi_2)$.}. In the aforementioned sub-truncation these fields are not present. The fields that survive are presented in Table \ref{n=2trunc} below. We can refer to this sector as the $\mathbb{Z}_2$ truncation \cite{Aharony}. On top of this $\Z_2$ these fields also have an $\Z_{2R}$ symmetry\footnote{This can be found by looking at the 10d reduction ansatz. The complex 2-form $\Omega$ has an over multiplicative factor of $e^{ - i \psi }$. Since the coordinate $\psi$ is in the range $(0,4 \pi)$, to see the $U(1)_R$ it is convenient to define another coordinate (say) $\sigma = \psi / 2$. In terms of $\sigma$, $\Omega$ has the multiplicative factor $e ^{- 2 i\sigma}$. This means that $\Omega$ has $U(1)$ R-charge $-2$ which implies that $b^\Omega$ and $c^\Omega$ have R-charge $2$. 
	This means that the elements of the $U(1)_R$ which leave $\Omega$ invariant are $\sigma = 0, \pi$, which corresponds to the elements $1$ and $-1$ of the $U(1)_R$. Thus, $\Omega$ preserves a $\Z_{2R}$ subgroup of the full $U(1)_R$. 
	Consequently the fields $b^\Omega$ and $c^\Omega$ also preserve the $Z_{2R}$ subgroup of the full $U(1)_R$.
	An analogous analysis of the reduction ansatz of the metric leads to the fact that both $t$ and $\theta$ preserve a $\Z_{2R}$ subgroup of the $U(1)_R$.}.  Therefore the 5d modes appearing in the entire truncation in table 2 is invariant under an $SU(2) \times SU(2) \times \Z_2 \times \Z_{2R}$. As we will see in the next section, the Klebanov-Strassler solution can be embedded in this truncation \cite{Cassani}. 

\begin{table}[h]
	\centering
	\begin{tabular}{c|c|c|c|c}
		\hline 
		IIB fields & scalars & 1-forms & 2-forms & 5d metric \\ 
		\hline 
		10d metric & $u,v,t,\theta$ & $A$ &   & $g_{\mu\nu}$  \\		 
		
		$\phi$ &  $\phi $ &   &   &   \\ 
		
		B & $b^{\Phi},b^{\Omega}$ & & &  \\ 
		
		$C_{0}$ & $C_{0}$ & & & \\
		
		$C_{2}$ & $c^{\Phi},c^{\Omega}$ &  &  &  \\
		
		$C_{4}$ & $a$ & $a_{1}^{J}$ &  & \\
		\hline
	\end{tabular}
	\caption{Field content for the $ \mathcal{N}=2,$ $SU(2)\times SU(2)\times \mathbb{Z}_2$ truncation.}
	\label{n=2trunc}
\end{table}

In the Klebanov-Strassler solution, the flux parameter $p$ and the following fields are consistently set to zero
\bea\label{nonKSfields}
\{ \real[b^{\Omega}] , \imag[c^{\Omega}] , a, C_{0} , c^{\Phi} , \theta ,  A, a_{1}^{J} \}~.
\eea
We will not consider perturbations of the Klebanov-Strassler solution by the above fields.  The fields that remain have the discrete $\Z_2$ $R$-symmetry from before, and so are again part of an $SU(2)\times SU(2)\times\mathbb{Z}_2 \times \Z_{2R}$ truncation\footnote{This can also be understood as a sub-truncation of the Papadopoulos-Tseytlin ansatz \cite{PT}.}.  For brevity, we will often refer to it as the $\Z_2$-truncation as well: since this is the truncation we will work with exclusively, it should not cause any confusion with the full $\Z_2$ truncation of the previous paragraph. We will study perturbations of the KS solution by scalar fields which are already activated in the background. The action governing these perturbations is given by \cite{Cassani}
\begin{align}\label{z2action}
\begin{split}
S_{b} &= \dfrac{1}{2\kappa_{5}^{2}}\int\biggl[R -\dfrac{28}{3} du^{2} -\dfrac{4}{3}dv^{2} -\dfrac{8}{3}du\,dv - dt^{2} - e^{-4u-\phi} \cosh 2t \, (db^{\Phi})^{2} \\
&\qquad  - \dfrac{1}{2} d\phi^{2}  + 2 \, e^{-4u-\phi} \sinh 2t \, db^{\Phi}d\boi -e^{-4u-\phi} \cosh 2t \, (db_{\mathcal{I}}^{\Omega})^{2}   \\
&\qquad- e^{-4u +\phi} (dc_{\mathcal{I}}^{\Omega})^{2}   - 4 e^{-\frac{20}{3}u + \frac{4}{3}v} + 24 \cosh t \, e^{-\frac{14}{3}u -\frac{2}{3}v}- 9 \sinh^{2}t \, e^{-\frac{8}{3}u -\frac{8}{3}v}  \\
&\qquad - 9 e^{-\frac{20}{3}u -\frac{8}{3}v -\phi} (b_{\mathcal{I}}^{\Omega})^{2}  -2e^{-\frac{32}{3}u -\frac{8}{3}v} \left( 3b_{\mathcal{I}}^{\Omega} c_{\mathcal{R}}^{\Omega} -q \, b^{\Phi} +k \right)^{2} \\
&\qquad - e^{-\frac{20}{3}u -\frac{8}{3}v +\phi} \, \left(9 ( c_{\mathcal{R}}^{\Omega})^{2}\cosh 2t - 6 q \; \cor \sinh 2t + q^{2} \cosh 2t \right) \Biggr]\star 1 ~,
\end{split}
\end{align}
where $\kappa_5$ is related to $\kappa_{10}$ as follows
\bea
\kappa_{5}^{2}= \dfrac{\kappa_{10}^{2}}{V_{Y}}~,\  ~~~~\text{where}~~~~~ V_{Y}= \dfrac{1}{2}\int_{T^{1,1}} J\wedge J\wedge \eta=\frac{16\pi^3}{27}~.
\eea
$V_{Y}$ is the unit volume of $T^{1,1}$. For later convenience we write down the metric on the scalar manifold in the basis
\begin{equation}\label{scalbasis}
\varphi^{I} = \{u,v,t,\phi,\bphi,\boi,\cor \}~,
\end{equation}
as follows
\bea\label{scalmetric}
\calg_{IJ} &=& \left(
\begin{array}{ccccccc}
	\frac{28}{3} & \frac{4}{3} & 0 & 0 & 0 & 0 & 0 \\
	\frac{4}{3} & \frac{4}{3} & 0 & 0 & 0 & 0 & 0 \\
	0 & 0 & 1 & 0 & 0 & 0 & 0 \\
	0 & 0 & 0 & \frac{1}{2} & 0 & 0 & 0 \\
	0 & 0 & 0 & 0 & e^{-4 u-\phi } \cosh 2 t & -e^{-4 u-\phi } \sinh 2 t & 0 \\
	0 & 0 & 0 & 0 & -e^{-4 u-\phi } \sinh 2 t & e^{-4 u-\phi } \cosh 2 t & 0 \\
	0 & 0 & 0 & 0 & 0 & 0 & e^{ -4 u +\phi} \\
\end{array}
\right)~.
\eea
With these definitions we can write the bosonic action as
\bea\label{z2bosonic}
S_{b} =\dfrac{1}{2\kappa_{5}^{2}} \int d^{5}x \sqrt{-g} \left(R -  \calg_{IJ} \partial_{A}\varphi^{I} \partial^{A}\varphi^{J} + \calv (\varphi) \right)~,
\eea
where $ A,B $ are indices for the spacetime coordinates and $ I,J$ index the scalar fields. The scalar potential $\cal V$, given by, 
\begin{align}\label{scalarpotential}
\begin{split}
{\cal V}(\varphi)&=- 4 e^{-\frac{20}{3}u + \frac{4}{3}v} + 24 \cosh t \, e^{-\frac{14}{3}u -\frac{2}{3}v}- 9 \sinh^{2}t \, e^{-\frac{8}{3}u -\frac{8}{3}v}  \\
&\qquad - 9 e^{-\frac{20}{3}u -\frac{8}{3}v -\phi} (b_{\mathcal{I}}^{\Omega})^{2}  -2e^{-\frac{32}{3}u -\frac{8}{3}v} \left( 3b_{\mathcal{I}}^{\Omega} c_{\mathcal{R}}^{\Omega} -q \, b^{\Phi} +k \right)^{2} \\
&\qquad - e^{-\frac{20}{3}u -\frac{8}{3}v +\phi} \, \left(9 ( c_{\mathcal{R}}^{\Omega})^{2}\cosh 2t - 6 q \; \cor \sinh 2t + q^{2} \cosh 2t \right) ~,
\end{split}
\end{align}
can be written in terms of a superpotential $\cal W$, given by
\bea\label{superpotential}
\mathcal{W} = e^{-\frac43  (4u+v)}\left(3\,\boi \cor -q\,b^{\Phi} +k\right)+3 \cosh t \, e^{-\frac43(u+v)}+2 e^{-\frac23(5u-v)}~,
\eea
as follows
\bea
\calv(\varphi) &=& -\calg^{IJ}\partial_{I}\mathcal{W} \partial_{J}\mathcal{W}+ \dfrac{4}{3}\mathcal{W}^{2}~.
\eea

Supersymmetric solutions of this system are obtained by analyzing the vanishing of the fermionic variations. The dimensional reduction of the 10d fermionic SUSY variations was performed in \cite{Liu:2011dw}. After converting their formulas into the notation of Cassani and Faedo (see appendix \ref{appd}) we obtain the fermionic variations listed in Appendix \ref{appd1}.

Among the 5d fields in Table \ref{n=2trunc}, the complex scalars $b^\Omega, c^\Omega$ and $z=\tanh t ~e^{i\theta}$ have R-charge 2 under the $U(1)$ R-symmetry of the boundary theory\footnote{A different way to see the R-charges is as follows: The holomorphic (2,0)-form $\Omega$ has a non-zero charge $q=-3$ under the action of the Reeb vector $\xi=3\partial_\psi$ (where the coordinate $\psi$ is defined in \eqref{oneforms1}). For a tensor $X$ its charge $q$ under the action of the Reeb vector is defined as ${\cal L}_\xi X=iqX $ (see for instance \cite{Ashmore:2016oug}). The R-charge $r$ is related to $q$ by $q=3r/2$. Therefore $\Omega$ has R-charge $-2$ which implies that $b^\Omega$ and $c^\Omega$ have R-charge $+2$. Alternatively, one can also look at the gauge covariant derivative of $b^\Omega, c^\Omega,z$ and read off $q=+3$.}. Setting these scalars to zero consistently gives rise to a further truncation to $SU(2)\times SU(2)\times\mathbb{Z}_2 \times U(1)_R$ invariant modes. For later convenience, we will refer to this sector as the $U(1)$ truncation. The resulting model was considered in \cite{Bertolini,Aharony,Buchel}. The model is comparatively simpler and the action reads
\begin{align}\label{u1action}
\begin{split}
S= \dfrac{1}{2\kappa_{5}^{2}}\int\biggl[R -&\dfrac{28}{3} du^{2} -\dfrac{4}{3}dv^{2} -\dfrac{8}{3}du\,dv - e^{-4u-\phi} \, (db^{\Phi})^{2} - \dfrac{1}{2} d\phi^{2} - 4 e^{-\frac{20}{3}u+ \frac{4}{3}v} \hspace{1cm} \\
& +24 \, e^{-\frac{14}{3}u -\frac{2}{3}v} -2e^{-\frac{32}{3}u -\frac{8}{3}v} \left(-q \, b^{\Phi} +k \right)^{2} - e^{-\frac{20}{3}u -\frac{8}{3}v +\phi} \, q^{2} \Biggr]\star 1~.
\end{split}
\end{align}
The action reduces to the form considered in \cite{Bertolini} with the following identification
\bea\label{UVcomb}
U = 4u + v~,~~~~ V = u - v~.
\eea
The fields $U$ and $V$ have the geometric interpretation of the breathing and squashing mode of $T^{1,1}$ respectively. We will, at times, use these linear combinations to compare with the notations of \cite{Bertolini}.

\section{Klebanov-Tseytlin vs Klebanov-Strassler: UV Asymptotics}\label{ktsection}

In this section, we present the Klebanov-Tseytlin (KT) and Klebanov-Strassler (KS) solutions in terms of the fields of the five-dimensional gauged supergravity theory discussed in the previous section. Both KS and KT are supersymmetric solutions and preserves $1/2$ of the ${\cal N}=2$ supersymmetry of the supergravity theory. We present the BPS equations and the explicit form of the solutions. We end this section with a comparison of the UV asymptotics of the two solutions. 
\subsection{Klebanov-Tseytlin solution}

The KT solution is a $1/2$ BPS solution and can be embedded in the $U(1)$ truncation \eqref{u1action}. From 5d point of view, the KT solution is a flat domain-wall where the 5d metric takes the following form
\bea\label{ktmetric}
&& ds_{5}^{2} = \dfrac{1}{z^{2}}\Biggl( e^{2X(z)} dz^{2} + e^{2Y(z)} \eta_{\mu\nu} dx^{\mu}dx^{\nu}\Biggr),
\eea
and the scalars are functions of the radial coordinate $z$ only. In the above parametrization of the metric the boundary is at $ z=0 $. The indices $ \mu,\nu$ run over $0,1,2,3$.  On this ansatz, the BPS equations resulting from the fermionic variations in Appendix \ref{appd3} take the following gradient flow form
\bea\label{bpsu1}
e^{-X(z)}z \partial_{z}\phi^{I}-\calg^{IJ}\partial_{J}\mathcal{W}=0~, \ \ \ e^{-X(z)}z \partial_{z} \log \left(\dfrac{e^{Y(z)}}{z}\right) + \dfrac{1}{3}\mathcal{W} = 0~.
\eea
The KT solution, which solves this set of BPS equations, is given by
\begin{align}\label{ktsol}
\begin{split}
t &= 0~,\  \ b^{\Omega}_{\mathcal{I}} \ = \ 0~, \ \ c^{\Omega}_{\mathcal{R}} \ =\  0~, \\ 
\phi &= \log (g_{s})~, \ \ \ b^{\Phi} = -g_{s}q \, \log \biggl(\dfrac{z}{z_{0}}\biggr)~,\\ 
X &= \dfrac{2}{3} \log(h_{\tx{KT}})~, \ \ Y=\dfrac{1}{6} \log(h_{\tx{KT}})~,\\ 
u &=  \dfrac{1}{4} \log(h_{\tx{KT}})~, \ \  v \ = \ \dfrac{1}{4} \log(h_{\tx{KT}})~,\\ 
h_{\tx{KT}}(z) &= \dfrac{1}{8} \biggl[ -4k + g_{s}q^{2} -4 g_{s} q^{2} \log \biggl(\dfrac{z}{z_{0}}\biggr) \biggr]~,
\end{split}
\end{align}
where  $ g_{s} $ is an integration constant, which, upon uplifting to 10d string theory becomes the string coupling constant. The independent flux parameters $k$ and $q$ are related to the number of regular and fractional branes respectively in the uplifted theory. $ z_{0} $ is an arbitrary scale introduced to make the argument of the log dimensionless. 

\subsection{Klebanov-Strassler solution}

The KS solution is a $1/2$ BPS solution of \eqref{z2action}. Unlike the KT solution, the KS solution cannot be embedded in the $U(1)$ truncation \eqref{u1action} because the $U(1)$ charged fields $t, \boi, \cor$ are activated in the KS solution. The KT solution in \eqref{ktsol} has a naked singularity at $ z_{s} $, such that $ h(z_{s}) =0 $, and and therefore cannot capture the full dynamics of the dual field theory. On the other hand, in the full ten-dimensional spacetime, the KS solution (which asymptotically matches the KT solution) is smooth in the IR \footnote{The five-dimensional compactification of the KS solution turns out to be singular in the IR. This can be deduced by calculating the curvature invariants and probing the IR limit. However this singularity (which is an artefact of dimensional reduction) is an acceptable singularity since it satisfies Gubser's criterion of good singularities \cite{Gubser:2000nd}.}. From 5d point of view the KS solution can be seen as a flat domain-wall where the metric takes the following form
\bea\label{ksmetric}
&& ds_{5}^{2} = e^{2\mathfrak{X}(\tau)} d\tau^{2} + e^{2\mathcal{Y}(\tau)} \eta_{\mu\nu} dx^{\mu}dx^{\nu}~. 
\eea
In this parametrization the boundary is at $\tau=\infty$. On this ansatz, the BPS equations take the following form
\bea\label{bpsu2}
e^{-\mathfrak{X}(\tau)} \partial_{\tau}\phi^{I}+\calg^{IJ}\partial_{J}\mathcal{W}=0, \ \ \ e^{-\mathfrak{X}(\tau)}\partial_{\tau}\mathcal{Y}(\tau) - \dfrac{1}{3}\mathcal{W} = 0.
\eea
The seemingly different relative sign compared to that of \eqref{bpsu1} is due to the fact that the boundary is at $ \tau=\infty $.

The KS solution, for this choice of metric is given by
\begin{align}\label{kssol}
& \cor = \frac{q \; \tau}{3 \sinh (\tau)}~,~ t = -\log \left(\tanh \left(\frac{\tau}{2}\right)\right)~,~ e^{2u}={3\over 2}h^{1/2}\epsilon^{4/3}K~\sinh\tau~,\\
&  e^{2v}={3\over 2}\frac{h^{1/2}\epsilon^{4/3}}{K^2}~,~ b^{\Phi} = \dfrac{g_{s}\; q\; \coth (\tau )}{3}  \big(\tau  \coth (\tau )-1\big)~,~\boi =\dfrac{g_{s} q \big(\tau \cosh (\tau)-\sinh (\tau)\big)}{3 \sinh ^2(\tau)}~,\non\\
& e^{2\mathfrak{X}}={1\over4}h^{4/3}\epsilon^{32/9}\left({3\over2}\right)^{2/3}K^{-4/3}\sinh^{4/3}\tau~,~~~e^{2\mathcal{Y}}=h^{1/3}\epsilon^{20/9}\left({3\over2}\right)^{5/3}K^{2/3}\sinh^{4/3}\tau~,\non
\end{align}
where
\bea
&& K(\tau)=\frac{\left(\sinh(2\tau)-2\tau\right)^{1/3}}{2^{1/3}\sinh\tau}~,~~~~h'(\tau)=-\alpha\frac{l(\tau)}{K^2(\tau)\sinh^2\tau}~,
\eea
and the function $l(\tau)$ is given by
\bea
l(\tau)=\frac{\tau \coth \tau-1}{4\sinh^2\tau}(\sinh 2\tau-2\tau)~.
\eea
The dilaton is constant in this solution and is given by $ \phi = \log (g_{s})$. In these formulas $ \epsilon $ is the conifold deformation parameter and $\alpha=(16 g_s q^2)/(81 \epsilon ^{8/3})$. The function $h(\tau)$ is the integral $h(\tau) = \int_{\infty}^{\tau}h'(x)\; dx$ which cannot be evaluated in a closed form.

To compare the KS solution with the KT solution, we first find the asymptotic relation between the radial coordinate $\tau$ in the KS metric \eqref{ksmetric} and the radial coordinate $z$ in the KT metric \eqref{ktmetric}. This relation is found to be
\bea\label{relcoord}
z^{2}= \dfrac{2^{5/3}}{3}\epsilon^{-4/3}  e^{-2\tau/3}~.
\eea
Using this relation, we expand the KS solution to $ \calo(z^{4}) $ 
\bea\label{asymptks}
t &=& 2a^{3}z^{3} + \calo(z^{9}),\  \ b^{\Omega}_{\mathcal{I}} \ = -\dfrac{2}{3}g_{s}q \left(1+3\log(a\,z)\right) a^{3}z^{3}+ \calo(z^{9}) \ ,\\  
c^{\Omega}_{\mathcal{R}} &=& -2 q a^{3}z^{3}\log (a\,z) + \calo(z^{9}) , \ \ 
\phi \ = \ \log (g_{s}), \ \ \ b^{\Phi} = -\dfrac{g_{s}q}{3}-g_{s}q \, \log (a\, z) + \calo(z^{6}), \non\\ 
e^{2u}&=& h_{\tx{KS}}^{1/2}+\calo(z^{6})~, ~e^{2v}= h_{\tx{KS}}^{1/2}+\calo(z^{6})~,~ e^{2\mathfrak{X}} = \frac19 h_{\tx{KS}}^{4/3}+\calo(z^{6})~,~e^{2\mathcal{Y}}=\frac{1}{z^2} h_{\tx{KS}}^{1/3}+\calo(z^{4})~,\non
\eea
where
\bea\label{warpKS}
h_{\tx{KS}}(z)\ = \ -g_{s}q^{2}\biggl[\dfrac{1}{24}+\dfrac{1}{2}\log(a\, z) \biggr]+\calo(z^{6})~, \ \text{ and } \ a\ = \ \dfrac{3^{1/2}}{2^{5/6}} \epsilon^{2/3} .
\eea
The metric in \eqref{ksmetric}, under the coordinate change, is given by
\bea
ds_{5}^{2} &=& \dfrac{h_{\tx{KS}}^{4/3}}{z^{2}}\ dz^{2} +   \dfrac{h_{\tx{KS}}^{1/3}}{z^{2}} \ \eta_{\mu\nu}dx^{\mu} dx^{\nu} \equiv \dfrac{e^{2X}}{z^{2}} \ dz^{2} + \dfrac{e^{2Y}}{z^{2}} \ \eta_{\mu\nu}dx^{\mu} dx^{\nu},
\eea
where $ e^{2X} =h_{\tx{KS}}^{4/3} $ and $ e^{2Y} =h_{\tx{KS}}^{1/3} $. 

Plugging \eqref{relcoord} and the asymptotic expressions for $X$ and $Y$ found above in the KS metric \eqref{ksmetric}, we recover the form of the KT metric \eqref{ktmetric}. The comparison of $h_{\tx{KS}}$ with $h_{\tx{KT}}$ relates the flux parameter $k$ in terms of the flux parameter $q$, the conifold deformation parameter $\e$ and the scale $z_0$ introduced in \eqref{ktsol}. This relation, together with the $\mathbb{Z}_2$ symmetry, reflects the fact that the KS solution is dual to the symmetric point on the baryonic branch which exists only when $k$ is proportional to $q$. In contrast, the KT solution is more generic where $k$ and $q$ are independent parameters. On a related note, we furthermore see that although there is a smooth limit $(q\to 0)$ of the KT solution to the conformal Klebanov-Witten solution, there is no such limit for the the KS solution (since under $q\to 0$, $h_{\tx{KS}}\to0$). The baryonic branch of the deformed conifold has been discussed in \cite{GranaMinasian} and the mesonic branch in \cite{CKK2}.

\section{SUSY breaking perturbations of the KS solution}\label{sec:4}

In this section, we discuss the sub-leading perturbations of the KS solution by analyzing the full bosonic equations of motion and present the most general SUSY breaking deformation of KS upto order $z^4$. The equations of motion for the action \eqref{z2bosonic} are given by
\begin{align}\label{EOMs}
\begin{split}
& \dfrac{2}{\sqrt{-g}}\partial_{A} \left(\sqrt{-g} g^{AB} \calg_{IJ} \, \partial_{B} \varphi^{J}\right) + \dfrac{\partial \calv}{\partial \varphi^{I}} - \dfrac{\partial \calg_{JK}}{\partial \varphi^{I} } \partial_{A} \varphi^{J} \partial_{B} \varphi^{K} = 0~, \\
& R_{AB} = \calg_{IJ} \partial_{A} \varphi^{I} \partial_{B} \varphi^{J} - \dfrac{1}{3} g_{AB} \calv(\varphi)~.
\end{split}
\end{align}
We will take the flat domain-wall ansatz used in \eqref{ksmetric} for the metric which is supported by non-trivial profile for the seven scalars \eqref{scalbasis} along the radial direction. There are seven second order ordinary coupled differential equations coming from the scalar sector and two more from the $zz$ and $\m\n$ component of the Einstein's equation. We make the following ansatz for the asymptotic expansions\footnote{The particular parametrization for the scalars $U,V,X$ and $Y$ is motivated by a natural 10d uplift as explained in Appendix \ref{appa}.}
\begin{align}\label{UVExpAnsatz}
\begin{split}
&\varphi^{I}(z) = \varphi^{I}_{\tx{KS}} + \sum_{i=1}\left(C^{I}_{(i)} + D^{I}_{(i)} \log a z\right) z^{i}~,~~~~ \forall \ I \neq U,V ~, \\
& e^{U(z)} = h(z)^{\frac{5}{4}} \, h_{2}(z)\, h_{3}(z)^{4}~,~~~~  e^{V(z)} = h_{3}(z)\, h_{2}(z)^{-1}~,\\
& e^{X(z) } = h(z)^{\frac{2}{3}}\,h_{2}(z)^{\frac{1}{3}}h_{3}(z)^{\frac{4}{3}}~,~ \ e^{Y(z)} = h(z)^{\frac{1}{6}}\,h_{2}(z)^{\frac{1}{3}}h_{3}(z)^{\frac{4}{3}}~, 
\end{split}
\end{align}
where
\begin{align}\label{UVExpAnsatz1}
\begin{split}
&h(z) = h_{\tx{KS}} + \sum_{i=1}\left( h^{(1)}_{i} + h^{(2)}_{i} \log z \right)z^{i}~,\\
& h_{a}(z) = 1 + \sum_{i=1}\left( C^{a}_{(i)}  + D^{a}_{(i)} \log az \right) z^{i}, \ \ \ \ a=2,3~.
\end{split}
\end{align}
The subscript $\tx{KS}$ indicates the KS solution expanded around $ z=0 $ as given in \eqref{asymptks} and \eqref{warpKS}. 

Before presenting our asymptotic SUSY breaking solution, we make one technical comment about \eqref{UVExpAnsatz}. In setting up the power series ansatz, we have used a series expansion in $z$ (together with the logarithmic terms). If we were working with a conventional Fefferman-Graham gauge, only even powers of $z$ in the warp factors would be necessary. But since it is somewhat harder to capture the KS solution in the conventional Fefferman-Graham coordinate system, we prefer to keep both $X(z)$ and $Y(z)$ in the metric ansatz. For such a choice for the 5d metric, the series expansion with only even powers of $z$ was considered in \cite{Aharony}. We do not know of an argument why this is correct {\em a priori}, so we have kept the full expansion in all powers of $z$. If we work with terms that involve such odd powers of $z$, we will get solutions which are supported by coefficients appearing at linear order in the ansatz for the warp factors (one such coefficient is $ C^{(2)}_{(1)} $ appearing in \eqref{UVExpAnsatz1}). However, we find that such solutions are unphysical and can be gauged away by a redefinition of the radial coordinate\footnote{In Appendix \ref{appb}, we show this explicitly in pure AdS by showing that this mode can be gauged away by a redefinition of the radial coordinate.}. Apart from this gauge mode, all the terms that appear up to $\calo(z^4)$ are even powers (consistent with \cite{Aharony}) and have physical interpretations (either as parameters in the KS solution, or as SUSY-breaking parameters appearing in \cite{DKM, Bertolini, Kuperstein}).

Upon substituting the series expansions into the equations of motion and solving them order by order in the radial coordinate $z$, we find the solution presented in \eqref{KSSUSYb} below. 
\begin{subequations}\label{KSSUSYb}
	\begin{align}
	\phi &= \log g_{s}+ \left(\varphi + 3\cals \log az\right)a^4z^{4} + \calo(z^{6})~,\\
	\bphi &= -\dfrac{1}{3}g_{s}q-g_{s}q \log az + \frac{g_{s}q}{16} \Big(7\cals -4\varphi -24 \cals\log az\Big)  a^4z^{4} +\calo(z^{6})~,\\
	\boi &= -\left(\dfrac{2}{3}g_{s}q + 2 g_{s}q \log a z \right)  a^{3}z^{3} + \calo(z^{6})~,\\
	\cor &=  -2 q \log a z \, a^{3}z^{3} + \calo(z^{6})~,\\
	t &= 2a^{3}z^{3} + \calo(z^{6})~,\\
	h &= -\dfrac{\gs q^{2}}{24} (1+12 \log az) + \dfrac{\gs q^{2}}{192}\Big(35 \cals - 12\varphi -48 \cals \log az \Big)a^4z^{4} + \calo(z^{6})~,\\
	h_{2} &= 1+ \dfrac{1}{2}\cals \, a^4z^{4}+ \calo(z^{6})~,\\
	h_{3} &= 1+ \calo(z^{6})~.
	\end{align}
\end{subequations}
Up to order $ z^{4} $ and $z^4 \log z$, the solution is determined by two independent, SUSY-breaking, integration constants ${\cal S}$ and $\varphi$. There are no new SUSY-breaking integration constants with respect to SUSY-breaking deformations of the KT solution studied in \cite{DKM,Bertolini}. The authors of \cite{BGGHM} found the most general deformation of the KS solution by considering the $SU(2)\times SU(2)\times \Z_2$ invariant Papadopoulos-Tseytlin ansatz in the Type IIB supergravity. Our finding is consistent with their result in that the subleading perturbations are characterized by a two parameter family of SUSY-breaking integration constants. We also find a number of SUSY-preserving integration constants. However we have set them to zero as they do not play any role in subsequent sections\footnote{Some of the additional integration constants are related to reparametrization of the radial coordinate (see Appendix \ref{appb} for illustration in pure AdS). Therefore discussion about its SUSY might seem unnecessarily pedantic. But the principle of setting SUSY-preserving parameters to zero is a more generally a useful idea. In Appendix \ref{appa}, we discuss the details of a more general ansatz and count the number of SUSY-preserving/breaking parameters in them.}. 

\section{Holographic Ward identities}\label{sec:5}

We would like to associate the SUSY breaking solution found in the previous section with a SUSY breaking vacua of the Klebanov-Strassler gauge theory. Since the KS theory is an ${\cal N}=1$ supersymmetric gauge theory, supersymmetry Ward identities should hold in any of its vacua. In this section we will derive the SUSY ward identities holographically and check them against the solution found in \eqref{KSSUSYb}. We we also derive other operator identities involving the trace of the energy-momentum tensor and $\gamma$-trace of the supercurrent. As we will see, these identities can be derived from relations between one-point functions of operators at generic sources \cite{Bertolini}. Therefore, we begin by identifying the holographic sources for dual operator and defining the one-point functions.

\subsection{Sources and dual operators}

In order to find the sources for the operators of the dual gauge theory, we study the equations of motion linearized around the asymptotic KS solution \eqref{asymptks}. In the superconformal Klebanov-Witten theory, the usual AdS/CFT correspondence dictates that fields of a certain mass $m$ in the bulk are dual to gauge invariant operators in the CFT of a certain conformal dimension $\Delta$. The mass-dimension relation depends upon the spin of the fields/operators. In Table \ref{massSpectrum} below we present the mass/dimension of fields/operators that are present in the $SU(2)\times SU(2) \times \Z_2$ truncation. 

\begin{table}[h]
	\begin{center}
		$	\begin{array}{lcccccccl}\hline
		\mathcal N=2\: {\rm multiplet} && {\rm field\: fluctuations} && \hskip-0.3cm\text{AdS mass}   &&  {\rm spin} && \Delta \\ \hline
		\rule{0pt}{3ex}
		{\rm gravity} && \begin{array}{c}(A+2a_1^J)_A\\ \Psi_{A} \\ g_{AB} \end{array} && \hskip-0.3cm\begin{array}{l} m^2=0 \\m={3\over 2} \\ m^2=0 \end{array}  &&  \begin{array}{l} 1 \\ {3\over 2} \\ 2 \end{array} && \begin{array}{l} 3 \\{7\over 2} \\ 4 \end{array}  \\ \hline
		{\rm universal \;hyper} && \begin{array}{c} b^\Omega + i \,c^\Omega \\ \z_\f\\ \tau=C_0+ie^{-\f} \end{array} && \begin{array}{l} m^2=-3 \\m=-{3\over 2}\\ m^2=0 \end{array} &&  \begin{array}{l} 0 \\ {1\over 2} \\ 1 \end{array} && \begin{array}{l} 3 \\{7\over 2} \\ 4 \end{array}  \\ \hline
		{\rm Betti \;hyper} &&	\begin{array}{c}	 t \,e^{i\theta}\\ \z_b\\ b^\F,\;\;c^\F \end{array} && \begin{array}{l}m^2= -3 \\m=  -{3\over 2}\\m^2= 0   \end{array} &&  \begin{array}{l} 0 \\ {1\over 2} \\ 1 \end{array} && \begin{array}{l} 3 \\{7\over 2} \\ 4 \end{array}  \\ \hline
		{\rm massive \;vector} && \begin{array}{c} V\\ \z_V\\left(A-a_1^J)_A\\ b^\Omega - i \,c^\Omega \\ \z_U\\ U \end{array} && \begin{array}{l} m^2=12\\m={9\over2} \\m^2= 24 \\m^2= 21\\m=-{11\over2}\\m^2= 32  \end{array}   &&  \begin{array}{l} 0 \\ {1\over 2} \\ 1 \\0\\\frac12\\0\end{array} && \begin{array}{l} 6 \\{13\over 2} \\ 7\\7\\{15\over 2}\\8 \end{array}  \\ \hline
		\end{array}$
		\caption{Mass spectrum of bosons and fermions in the ${\cal N}=2,~\Z_2$ truncation of \cite{Cassani} around the supersymmetric $AdS_5$. In our conventions, setting $k= -2$ leads to a unit AdS radius (5d indices  are dubbed $A,B$).}
		\label{massSpectrum}
	\end{center}
\end{table}
All fields in this table are organized in multiplets of $5d$, ${\cal N}=2$ supersymmetry.  All the fermions are in the Dirac representation. The gravity multiplet contains the metric, a $U(1)$ vector field and the gravitino which comprises 8 bosonic and 8 fermionic on-shell real degrees of freedom. The hypermultiplets contains four real scalars and a Dirac fermion which comprises of 4 bosonic and 4 fermionic on-shell real degrees of freedom. The massive vector multiplet can be though as a massless vector that has undergone a Higgs mechanism by eating up an entire hypermultiplet. It contains 8 bosonic and 8 fermionic on-shell real degrees of freedom. To sum up, the matter content of the $\Z_2$ truncation can be seen as consisting of one vector multiplet and three hypermultiplets (splitting the massive vector into a massless vector and a hypermultiplet is convenient for writing down supersymmetry transformation rules). 
\subsubsection{Bosonic sector}
For bulk scalar fields which lie outside the double quantization window (as it is the case here), the non-normalizable mode is interpreted as a source for the dual operator. Since the bosonic scalar operators in the Table \ref{massSpectrum} have integer scaling dimensions, the linearized equations of motion for these fields around pure $AdS_5$ is solved by integer power law solutions. When we move to the KT/KS background, these power law solutions will get corrected by logarithmic terms (which capture the log running of the gauge coupling in the dual theory). With this in mind we start with an ansatz dictated by the pure AdS solution and add to it logarithmic terms. For convenience in the linearization procedure, we introduce a book-keeping parameter $ \varepsilon $ in the following ($n$ is not summed over in the following formulas)
\begin{align}
\begin{split}
&\varphi^{I}(z) = \varphi^{I}_{\tx{KS}} +\varepsilon\left(\delta\varphi^{I}_{(n)} + \delta\tilde{\varphi}^{I}_{(n)} \log a z\right) z^{n}~, ~~~~ \forall \ I \neq U,V~, \,\\
& e^{X(z) } = h(z)^{\frac{2}{3}}\,h_{2}(z)^{\frac{1}{3}}h_{3}(z)^{\frac{4}{3}}~, ~~ e^{Y(z)} = h(z)^{\frac{1}{6}}\,h_{2}(z)^{\frac{1}{3}}h_{3}(z)^{\frac{4}{3}}~, \\
& e^{U(z)} = h(z)^{\frac{5}{4}} \, h_{2}(z)\, h_{3}(z)^{4}~,  ~~~ \ e^{V(z)} = h_{3}(z)\, h_{2}(z)^{-1}~,
\end{split}
\end{align}
where
\begin{align}
\begin{split}
& h(z) = h_{\tx{KS}}^{0} + \varepsilon\left( \delta h_{(n)} + \delta \tilde{h}_{(n)} \log az \right)z^{n}~,\\
& h_{a}(z) = 1 + \varepsilon \left(\delta h^{(a)}_{(n)}  + \delta \tilde{h}^{(a)}_{(n)} \log az \right) z^{n}~, \ \ \ \ a=2,3~.
\end{split}
\end{align}
We will be interested in the following values of $n$:  $-4,-3,-2,0,1$. $n=1$ corresponds to the most relevant scalar operator of dimension three and $n=-4$ corresponds to the most irrelevant scalar operator of dimension eight in the theory. We solve the system separately for each $n$. In the presence of irrelevant operators, finding a solution to the full non-linear equations involving all the sources is an ill-defined problem \cite{Aharony}.

The solution presented below corresponds to sources for all scalar operators of a given dimension turned on one at a time.
\begin{subequations}\label{holsources}
	\bea
	(i)&& \delta h =\delta h_{(-4)}(x)z^{-4}~,\\
	(ii)&& \delta\boi =\delta\boi {}_{(-3)}(x)z^{-3}~, \ \ \delta\cor = -g_s^{-1}\delta\boi {}_{(-3)}(x)z^{-3}~, \\
	(iii) && \delta h = \frac12\gs q^{2}\; \delta h^{(2)}_{(-2)}(x)z^{-2}~, \ \ \delta h_{2} = \delta h^{(2)}_{(-2)}(x)z^{-2}~, \ \ \delta h_{3} = -\frac14 \delta h^{(2)}_{(-2)}(x)z^{-2}~,\non\\
	&& \delta\bphi = -\frac12\gs q \; \delta h^{(2)}_{(-2)}(x)z^{-2}~, \\
	(iv)&&  \delta\bphi =  \delta\bphi_{(0)}(x) -g_{s}q  \delta\phi_{(0)}(x) \log az~,  \ \ \delta\phi =  \delta\phi_{(0)}(x)~, \ \\
	&& \delta h = \dfrac{1}{8} \left(4 q \delta\bphi_{(0)}(x)+g_{s}q^{2} \delta\phi_{(0)}(x) -4 g_{s}q^{2}\delta\phi_{(0)}(x) \log az \right)~,\non \\
	(v) && \delta t= \dfrac{1}{\gs q}\left(\delta b_{(1)}+\delta t_{(1)} \log a z\right)z~,~~\delta \cor = \frac{ 1}{24\gs} \left[18 \delta b_{(1)} + \delta t_{(1)} \left(1+12 \log az\right)\right]z~,\non\\
	&&\delta \boi = \frac{1}{12} \left( 3 \delta b_{(1)} + 2\delta t_{(1)} \right)z~.
	\eea
\end{subequations}

In the solutions listed above, the first three correspond to sources for irrelevant operators of dimensions eight, seven and six respectively. The solution in $ (iv) $ contains the source for the two marginal scalar operators (that corresponds to the sum and the difference of the gauge couplings). The solutions in $ (v) $ contain sources for operators of dimension three which are new in the $\Z_2$ truncation. These sources corresponds to gaugino mass terms and therefore break supersymmetry explicitly.

In the metric sector, we have the transverse-traceless fluctuations of the metric induced on a finite radial cut-off surface, which in the boundary limit, sources the energy-momentum tensor of the boundary theory. The induced metric at a finite radial cut-off is given by
\bea
\gamma_{\mu\nu} &=& e^{2Y}\ol{\gamma}_{\mu\nu}, \ \ \text{where} \ \ \ol{\gamma}_{\mu\nu} = \dfrac{\eta_{\mu\nu}}{z^{2}}~.
\eea
The independent source from the metric which decouples from the rest of the sources is then given by
\bea\label{metricsource}
\delta \ol{\gamma}_{\mu\nu } = \dfrac{\delta h_{\mu\nu}(x)}{z^{2}}~.
\eea

Having obtained the sources, we now give the field operator map. The $SU(2)\times SU(2)\times \Z_2$ invariant sector of gauge invariant operators in the Klebanov-Strassler theory are, in general, dual to bulk fields which are composite. The two marginal operators ${\cal O}_+\equiv \Tr\left(F_{(1)}^2+F_{(2)}^2\right)$ and ${\cal O}_-\equiv\Tr\left(F_{(1)}^2-F_{(2)}^2\right)$ (that correspond to the sum and the difference of the gauge coupling) are dual to $e^{-\phi}$ and $e^{-\phi}b^\Phi$ respectively \cite{KS} whereas the two relevant operators ${\cal Q}_+\equiv \Tr\left(W_{(1)}^2+W_{(2)}^2\right)$ and ${\cal Q}_-\equiv \Tr\left(W_{(1)}^2-W_{(2)}^2\right)$ (that correspond to mass terms of the gaugino bilinears) are dual to the combination $b^\Omega+ig_s c^\Omega$ and $t$ respectively \cite{Loewy:2001pq,Kuperstein:2003yt}\footnote{In their convention, the linear combination dual to $ {\cal Q}_+ $ is $ b^\Omega - ig_s c^\Omega $.}. The sources in \eqref{holsources} and \eqref{metricsource} corresponds to these operators as follows
\begin{equation}\label{sourceOp}
{\cal O}_+ \leftrightarrow \delta\phi_{(0)}~,~~~~{\cal O}_- \leftrightarrow \delta\bphi_{(0)}~,~~~~{\cal Q}_+ \leftrightarrow \delta b_{(1)}~,~~~~{\cal Q}_- \leftrightarrow \delta t_{(1)}~,~~~~T_{\m\n} \leftrightarrow \delta h_{\mu\nu}~.
\end{equation}

The sources obtained in \eqref{holsources} are not diagonal by which we mean that a mode for one field can simultaneously turn on multiple fields. For example a non-zero $ \delta\phi_{(0)} $ results in turning on $ \delta\phi $ and $ \delta\bphi $. On the other hand, the composite field $e^{-\phi}b^\Phi$ is not affected by $ \delta\phi_{(0)} $ (it is turned on by $\delta\bphi_{(0)}$ only). This, however, is not true for the sources of dimension three operators. Regardless, we find it convenient to define combinations which are diagonal in the sources as it will be relevant later when we consider supersymmetry transformation of the sources.
\begin{align}\label{diagSources}
\begin{split}
&\delta \ol{\phi} = \delta \phi~,~~~~~~ \delta \ol{b}^\Phi = \delta\bphi + \gs q~\delta \phi \log a z ~,\quad \quad\delta \ol{t} =  \dfrac{24}{5-12 \log az}\Big(\delta B_+ -\gs q \delta t\Big) ~,\\
&\delta \ol{B}_{+} = \dfrac{1}{5-12 \log az}\Big( -24 \log az \delta B_+ +\gs q (5+12 \log az)\delta t\Big) ~,
\end{split}
\end{align}
where we have defined $ \delta B_{+} = \delta\boi+\gs \delta\cor $. All the hatted  fields are sourced independently\footnote{In \cite{Bertolini}, where fields dual to marginal operators only mattered, analogous relations were written down for the composite fields by re-writing the explicit $z$-dependencies on the right hand sides in terms of bulk fields.} i.e., 
\begin{align}\label{diagSources1}
\begin{split}
\delta\ol{\phi}  &= \delta \phi_{(0)}~,~~~~~~ \delta\ol{b}^\Phi  = \delta \bphi_{(0)}~,\\
\delta \ol{B}_{+} &=  ~ \delta b_{(1)} z ~,~~~~ \delta \ol{t} = ~ \delta t_{(1)} z  ~.
\end{split}
\end{align}
The holographically renormalized one point functions of the marginal operators in \eqref{sourceOp} was first obtained in \cite{Aharony} by functionally differentiating the on-shell renormalized action w.r.t. the corresponding sources in \eqref{sourceOp}. The renormalized one-point functions are then obtained by taking appropriate boundary limits. In the following we give an independent derivation of these one-point functions (including the dimension three operators which are new) by taking a slightly different approach where, the renormalized one point functions are obtained by functionally differentiating the on-shell renormalized action w.r.t. to the hatted (composite induced) fields and taking appropriate limits. The two procedure are equivalent in AAdS background but as we will see later in the derivation of the Ward identities (section \ref{sec5.2}), the latter definition is crucial in the KS background. We have 
\begin{align}\label{oneptfn1}
\begin{split}
\lag T_{\m\n}\rag &=\frac{2}{\sqrt{-\hat{\gamma}}} \frac{\delta S_{\tx{ren}}}{\delta h^{\m\n} }= \frac{2}{\sqrt{-\hat{\gamma}}} \frac{\delta S_{\tx{ren}}}{\delta \gamma_{\rho\sigma} }\frac{\delta \gamma_{\rho\sigma}}{\delta h^{\m\n} }~,\\
%%%%%%%%%%%%%%%%%%%%%%%%%%%%%%%%%%%%
\lag {\cal O}_+\rag &= \frac{1}{2\sqrt{-\hat{\gamma}}}\frac{\delta S_{\tx{ren}}}{\delta\hat{\phi} } =\frac{1}{2\sqrt{-\hat{\gamma}}}\frac{\delta S_{\tx{ren}}}{\delta\phi_{(0)} } \\
&= \frac{1}{2\sqrt{-\hat{\gamma}}}\bigg[\frac{\delta S_{\tx{ren}}}{\delta\phi }\frac{\delta \phi}{\delta\phi_{(0)} }+\frac{\delta S_{\tx{ren}}}{\delta\bphi }\frac{\delta \bphi}{\delta\phi_{(0)} }+\frac{\delta S_{\tx{ren}}}{\delta U }\frac{\delta U}{\delta\phi_{(0)} }+\frac{\delta S_{\tx{ren}}}{\delta \gamma_{\m\n} }\frac{\delta \gamma_{\m\n}}{\delta\phi_{(0)} }\bigg]~,\\
%%%%%%%%%%%%%%%%%%%%%%%%%%%%%%%%%%%%
\lag {\cal O}_-\rag &= \frac{1}{2\sqrt{-\hat{\gamma}}}\frac{\delta S_{\tx{ren}}}{\delta\hat{b}^\Phi}  =\frac{1}{2\sqrt{-\hat{\gamma}}}\frac{\delta S_{\tx{ren}}}{\delta\bphi_{(0)} }=  \frac{1}{2\sqrt{-\hat{\gamma}}}\bigg[\frac{\delta S_{\tx{ren}}}{\delta\bphi }\frac{\delta \bphi}{\delta\bphi_{(0)} }+\frac{\delta S_{\tx{ren}}}{\delta U }\frac{\delta U}{\delta\bphi_{(0)} }+\frac{\delta S_{\tx{ren}}}{\delta \gamma_{\m\n} }\frac{\delta \gamma_{\m\n}}{\delta\bphi_{(0)} }\bigg]~,\\
%%%%%%%%%%%%%%%%%%%%%%%%%%%%%%%%%%%%
\lag {\cal Q}_+\rag &=\frac{1}{2\sqrt{-\hat{\gamma}}}\frac{\delta S_{\tx{ren}}}{\delta\hat{B}_{+} } =\frac{1}{2\sqrt{-\hat{\gamma}}}\frac{\delta S_{\tx{ren}}}{\delta b_{(1)}z }= \frac{1}{2\sqrt{-\hat{\gamma}}}\bigg[\frac{\delta S_{\tx{ren}}}{\delta t }\frac{\delta t}{\delta b_{(1)} }+\frac{\delta S_{\tx{ren}}}{\delta \boi }\frac{\delta \boi}{\delta b_{(1)} }+\frac{\delta S_{\tx{ren}}}{\delta \cor }\frac{\delta \cor}{\delta b_{(1)} }\bigg]~,\\
%%%%%%%%%%%%%%%%%%%%%%%%%%%%%%%%%%%%
\lag {\cal Q}_-\rag &=\frac{1}{2\sqrt{-\hat{\gamma}}}\frac{\delta S_{\tx{ren}}}{\delta\hat{t} } =\frac{1}{2\sqrt{-\hat{\gamma}}}\frac{\delta S_{\tx{ren}}}{\delta t_{(1)} z }= \frac{1}{2\sqrt{-\hat{\gamma}}}\bigg[\frac{\delta S_{\tx{ren}}}{\delta t }\frac{\delta t}{\delta t_{(1)} }+\frac{\delta S_{\tx{ren}}}{\delta \boi }\frac{\delta \boi}{\delta t_{(1)} }+\frac{\delta S_{\tx{ren}}}{\delta \cor }\frac{\delta \cor}{\delta t_{(1)} }\bigg]~.
\end{split}
\end{align}
In these formulas, $S_{\tx{ren}}$ is the renormalized on-shell action given by $S_{\tx{ren}}=S_{\tx{reg}}+S_{\tx{ct}}$, where $S_{\tx{reg}}$ is the regulated action computed at a finite radial cut-off and $S_{\tx{ct}}$ is the counterterm action. Using \eqref{holsources}, these expressions can be simplified to the following
\begin{align}\label{oneptfn2}
\begin{split}
\lag T_{\mu\nu} \rag &= \dfrac{2}{\sqrt{-\ol{\gamma}}} \,  \dfrac{\de S_{\tx{ren}}}{\de \gamma^{\mu\nu}}\hks^{1/3}~,\\
%%%%%%%%%%%%%%%%%%%%%%%%%%%%%%%%%%%%
\lag {\cal O}_+\rag &= \dfrac{1}{2\sqrt{-\ol{\gamma}}} \bigg[ \dfrac{\de S_{\tx{ren}}}{\de \phi} - \gs q \la \dfrac{\de S_{\tx{ren}}}{\de \bphi} + \Big( 1+ \dfrac{\gs q^{2}}{6\hks}\Big)\bigg( \dfrac{5}{4} \dfrac{\de S_{\tx{ren}}}{\de U} + \dfrac{1}{3} \gamma_{\mu\nu}\dfrac{\de S_{\tx{ren}}}{\de \gamma_{\mu\nu}} \bigg)\bigg]~,\\
%%%%%%%%%%%%%%%%%%%%%%%%%%%%%%%%%%%%%%%
\lag {\cal O}_-\rag &= \dfrac{1}{2\sqrt{-\ol{\gamma}}} \bigg[ \dfrac{\de S_{\tx{ren}}}{\de \bphi} + \dfrac{q}{2\hks}\bigg( \dfrac{5}{4} \dfrac{\de S_{\tx{ren}}}{\de U} + \dfrac{1}{3} \gamma_{\mu\nu}\dfrac{\de S_{\tx{ren}}}{\de \gamma_{\mu\nu}} \bigg) \bigg]~,\\
%%%%%%%%%%%%%%%%%%%%%%%%%%%%%%%%%%%%%%%
\lag {\cal Q}_+\rag &= \dfrac{1}{2\sqrt{-\ol{\gamma}}} \bigg[\frac{1}{4} \dfrac{\de S_{\tx{ren}}}{\de \boi} +\frac{3}{4\gs}  \dfrac{\de S_{\tx{ren}}}{\de \cor}  + \frac{1}{\gs q} \dfrac{\de S_{\tx{ren}} }{\de t} \bigg]~,\\
%%%%%%%%%%%%%%%%%%%%%%%%%%%%%%%%%%%%%%%
\lag {\cal Q}_-\rag &= \dfrac{1}{2\sqrt{-\ol{\gamma}}} \bigg[\frac{1}{6} \dfrac{\de S_{\tx{ren}}}{\de \boi} +\frac{1}{24\gs}\left(1+12 \la\right)  \dfrac{\de S_{\tx{ren}}}{\de \cor}  +\frac{1}{\gs q } \la \dfrac{\de S_{\tx{ren}} }{\de t} \bigg]~.
%%%%%%%%%%%%%%%%%%%%%%%%%%%%%%%%%%%%%%%
\end{split}
\end{align}
The renormalized QFT one-point functions of these operators are obtained by taking the following limits
\begin{align}\label{oneptfn3}
\begin{split}
\lag T^\m_{~\n}\rag_{\tx{QFT}}= &\lim\limits_{z\rightarrow 0 } z^{-4}~\lag T^\m_{~\n}\rag~,~~~~\lag {\cal O}_+\rag_{\tx{QFT}}= \lim\limits_{z\rightarrow 0 } z^{-4}~ \lag {\cal O}_+\rag~,~~~~\lag {\cal O}_-\rag_{\tx{QFT}}= \lim\limits_{z\rightarrow 0 } z^{-4}~ \lag {\cal O}_-\rag~,\\
&~~~~\lag {\cal Q}_+\rag_{\tx{QFT}}= \lim\limits_{z\rightarrow 0 } z^{-3}~ \lag {\cal Q}_+\rag~,~~~~\lag {\cal Q}_-\rag_{\tx{QFT}}= \lim\limits_{z\rightarrow 0 }z^{-3}~ \lag {\cal Q}_-\rag~.
\end{split}
\end{align}

\subsubsection{Fermionic sector}

The fermionic content of the full $ SU(2) \times SU(2) \times \Z_2$ truncation of $ \caln=2 $ supergravity is made up of a gravitino $\Psi_\m$, three hyperinos $\zeta^A$ ($A=1,2,3$) and a gaugino\footnote{$u_3$ is the scalar that appears in the `massless' vector multiplet. See discussion above \eqref{vsmMetric} and Eq. \eqref{bertolinimap} for a clarification on the notation used here.} $\lambda^{u_3}$. A detailed discussion of the fermionic sector and the supersymmetry of the $SU(2) \times SU(2) \times \Z_2$ truncation is given in Appendix \ref{appd}, including the mapping of notations used in \cite{Bertolini, Halmagyi:2011yd,Liu:2011dw}.

The equations of motion for the fermions and the gravitino were originally obtained in \cite{Liu:2011dw}. To obtain the sources of the dual fermionic operators, we first project the fermions onto definite chirality (which is well-defined at a given radial surface) and then solve the equations in the asymptotic KS background given in \eqref{asymptks}. We can make a crucial observation at this stage, by looking at the equations of motion in \cite{Liu:2011dw}: if we repackage the bosonic background in the equations of motion by powers of $ z $, the leading terms are sensitive only to the $ \calo(1) $ and $ \calo(\log az) $ terms of the bosonic fields. What this means is that the leading order terms in the first order differential equations are identical to that one finds in the KT background, with an appropriate identification of the parameters $ k $ and $ a $. The solutions to the equations of motion in the KT background have been found by \cite{Bertolini},
\begin{subequations}
	\bea
	\zeta_{\phi}^{-} &=& \sqrt{z}\, \hks(z)^{-\frac{1}{12}}\, \psi_{1}^{-}(x) + \calo(z^{\frac{3}{2}}),\\
	\zeta_{b}^{-} &=& \dfrac{\sqrt{z} \hks(z)^{-\frac{1}{12}}}{20 q} (24\hks(z) - 5g_{s}q^{2}) \psi_{1}^{-}(x) +  \sqrt{z} \hks(z)^{-\frac{3}{4}}\psi_{2}^{-}(x) + \calo(z^{\frac{3}{2}}),\\
	\zeta_{U}^{-} &=& \dfrac{3}{4}\sqrt{z}\hks(z)^{-\frac{1}{12}}\psi_{1}^{-}(x) +\dfrac{5q}{8} \sqrt{z} \hks(z)^{-\frac{7}{4}} \psi_{2}^{-}(x) + \calo(z^{\frac{3}{2}}), \\
	\zeta_{V}^{+} &=&  \calo(z^{\frac{3}{2}}) ,\\
	\Psi_{\mu}^{+} &=& \dfrac{\hks(z)^{\frac{1}{12}}}{\sqrt{z}} \psi^{+}_{\mu}(x) + i \dfrac{\hks(z)^{\frac{1}{6}}}{g_{s}q^{2}\sqrt{z}}\gamma_{\mu}  \bigg(-\dfrac{4}{5}\hks(z)^{\frac{11}{12}}\psi_{1}^{-}(x)  \non\\
	&&\qquad\qquad\qquad\qquad\qquad + q \hks(z)^{-\frac{7}{4}}\left(\hks(z)+\dfrac{g_{s}q^{2}}{12}\right)\psi_{2}^{-}(x)\bigg)+ \calo(z^{\frac{3}{2}}).\qquad\qquad
	\eea
\end{subequations}
For completeness, we restate some crucial comments regarding these solutions below. In pure AdS $ \zeta_{\phi} $ and $ \zeta_{b} $ have masses $ m_{\phi,b}=-3/2 $, $ \zeta_{U} $ has mass $ m_{U} = -15/2 $, $ \zeta_{V} $ has mass $ m_{V} =11/2 $ and the gravitino has a mass $ m_\Psi =3/2 $. One basic idea behind solving the equations of motion for Dirac fields in $AdS_5$ is that a Dirac spinor in five dimensions has the same number of components as a Dirac spinor in four dimensions. However a Dirac spinor in 5d is irreducible while in 4d it is reducible and the minimal spinors are Weyl spinors which contain half as many physical degrees of freedom. Since the boundary operators are of definite chirality, it is imperative to decompose the 4d projection of the 5d spinors onto a definite chirality. The two chiralities at a given radial slice have different UV fall-offs and is determined by the sign of the fermion mass term (see \cite{Henneaux} for a detailed discussion). Following this, the leading chirality of $ \zeta_{\phi},\zeta_{b}$ are negative, while for $ \Psi_{\mu} $ is positive. We don't consider the sources for the irrelevant operators dual to $ \zeta_{U}^{-} $ and $ \zeta_{V}^{+} $.

Now, we will focus our attention on finding the fermionic superpartners of the composite bosonic fields. We will only be focusing on the sources $ \hat{\phi} $, $ \hat{b}^\Phi $ the metric fluctuation $ h_{\mu\nu} $ and their fermionic superpartners, as these are the only inputs required in the SUSY Ward identity computation\footnote{This point is further elaborated with reasons in the subsection where we compute the Ward Identities.}.  To this effort, we can use the following relation
\bea
\delta_{\epsilon}\hat{\varphi} = \dfrac{\de \hat{\varphi}}{\de \varphi^{I}} \delta_{\epsilon} \varphi^{I},
\eea
and use the SUSY variations of the bosonic fields given in \eqref{bosonSUSYvar}. These relations will be useful in computing the SUSY Ward identities. We also need the following relations, where this relation is evaluated in the KS background, 
\bea
\delta_{\epsilon}\hat{\phi} &=& \dep\phi, \\
\dep \hat{b}^\Phi &=& \dep\bphi+ \gs q \log az \, \dep\phi,\\
\dep \hat{e}^{a}_{\mu} &=& \hks^{-1/6}\bigg(\dep e^{a}_{\mu} - \dfrac{1}{48\hks}e^{a}_{\mu}(4q \dep \bphi + \gs q^{2}\dep \phi)  \bigg)
\eea

Using the KS background and the bosonic SUSY variations in \eqref{bosonSUSYvar}, along with \eqref{bertolinimap}, we can write
\begin{subequations}
	\bea\label{compferm}
	\dep \hat{\phi} &\equiv& \dfrac{i}{2} \big( \bar{\epsilon}\hzd{\phi} -\hzdb{\phi} \epsilon \big) \ = \ \dfrac{i}{2}\big(\bar{\epsilon} \zeta_{\phi} - \bar{\zeta}_{\phi} \, \epsilon \big),\\
	\dep \hat{b}^\Phi &\equiv& \dfrac{i}{2} \big( \bar{\epsilon}\hzd{b} -\hzdb{b} \epsilon \big) \ = \ \dfrac{i}{2} \big( \bar{\epsilon} \zeta_{b} - \bar{\zeta}_{b} \, \epsilon  \big) +  \dfrac{i}{2}\gs q \log az\big(\bar{\epsilon} \zeta_{\phi} - \bar{\zeta}_{\phi} \, \epsilon \big) +  \dots,\\
	\label{compgraviton}\dep \hat{e}^{a}_{\mu} &\equiv& \dfrac{1}{2} \bar{\epsilon} \gamma^{a}\ol{\Psi}_{\mu} +\text{ h.c.} \ = \ \dfrac{\hks^{-1/6}}{2}\bar{\epsilon}\gamma^{a}\Psi_{\mu}  + \text{h.c.} + \dots,
	\eea
\end{subequations}
where the dots indicate subleading terms, which are suppressed by powers of $ z $ or by factors of $ \log az $ in the denominator. In the rest of this chapter, we will use the hatted fermions to indicate the combinations, to leading order, of the original supergravity fermions that are defined above. The subleading terms do not contribute when we take the $ z\rightarrow 0 $ limit.

Although we do not require the explicit form of the fermionic action for the purpose of computing the SUSY Ward Identities, we need the formal prescription for computing the one point functions, and taking the boundary limit. The one point functions of the fermions dual to the composite bosons and their boundary limits are defined as follows
\bea\label{foneptfundef}
\begin{split}
	\lr{\bar{S}^{\nu -}} &=& \dfrac{-2i}{\sqrt{-\ol{\gamma}}} \dfrac{\delta S_{f,\tx{ren}}}{\delta \hat{\Psi}^{+}_{\nu}} , \qquad \lr{\bar{S}^{\nu -}}\qft = \lim\limits_{z\rightarrow0} z^{-\frac{9}{2}}\hks^{-\frac{1}{12}}\, \lr{\bar{S}^{\nu -}}\\
	\lr{\bar{\mathcal{O}}^{+}_{\hat{\zeta}_{\ol{\phi}}}} &=& \dfrac{i}{\sqrt{2}\sqrt{-\ol{\gamma}}} \dfrac{\delta S_{f,\tx{ren}}}{\delta \hat{\zeta}^{-}_{\ol{\phi}}} , \quad \exvalb{+}{\hzd{\phi}}\qft = \lim\limits_{z\rightarrow0} z^{-\frac{7}{2}}\hks^{-\frac{1}{12}}\, \exvalb{+}{\hzd{\phi}} ,\\
	\lr{\bar{\mathcal{O}}^{+}_{\hat{\zeta}_{\hat{b}^\Phi}}} &=& \dfrac{i}{\sqrt{2}\sqrt{-\ol{\gamma}}} \dfrac{\delta S_{f,\tx{ren}}}{\delta \hat{\zeta}^{-}_{\ol{\bphi}}} , \quad \exvalb{+}{\hat{\zeta}_{\hat{b}^\Phi}}\qft = \lim\limits_{z\rightarrow0} z^{-\frac{7}{2}}\hks^{-\frac{1}{12}}\, \exvalb{+}{\hzd{\bphi}},
\end{split}
\eea
where $ S_{f,\tx{ren}} $ is the renormalized fermionic action.

\subsection{Diffeomorphisms and Local SUSY}

Our calculations in this sub-section and the next section are parallel to that given in \cite{Bertolini}, except for the fact that we have not explicitly introduced composite fields, but instead work with the diagonalized sources we have defined in \eqref{diagSources}. As mentioned in the previous subsection, our interest will remain with the bosonic sources $ \hat{\phi} $, $ \hat{b}^\Phi $, $ h_{\mu\nu} $ and their fermionic superpartners $ \hzd{\phi},\hzd{b} $ and $ \hat{\Psi}_{\mu} $

To get rid of the non-dynamical components in the metric and gravitino fields, we will choose the following gauge for the metric and gravitino 
\bea\label{gaugechoice}
ds^{2} =  dr^{2} + \gamma_{\mu\nu}(r,x) dx^{\mu}dx^{\nu}~,~~~~  \Psi_{r} = 0.
\eea
Since the calculations of the sources in the previous section are in a slightly more relaxed gauge (for metric and gravitino), we need to relate the two in explicit calculations. The metric ansatz chosen in \eqref{ktmetric} can be related via the identifications
$dr  = -\dfrac{e^{X}}{z}dz$, and  $\gamma_{\mu\nu} = \dfrac{e^{2Y}}{z^{2}} \eta_{\mu\nu}$.

In the remainder of this subsection we will look at bulk diffeomorphisms and supersymmetry transformations that preserve the gauge choice we have taken. In doing so, we will assume that the fields take the form in \eqref{asymptks}. That is, we will ignore the corrections that come at subleading orders to \eqref{asymptks}, with the understanding that in the asymptotic limit ($ z\rightarrow 0 $), where the QFT is defined, all the other contributions vanish sufficiently fast. 

\subsubsection{Weyl}

The set of bulk diffeomorphisms that preserve the gauge choice of the metric can be found by solving the Killing vector equations. The 5d Killing vector equations translate to
\bea
\partial_{r} \xi^{r} = 0, \ \ \  \partial_{r}\xi^{\mu } + \gamma^{\mu\nu} \partial_{\mu} \xi^{r} = 0.
\eea
The $\xi^{\mu}$ correspond to boundary diffeomorphisms and their Ward identities, which we are not interested in. The $\xi^r$ on the other hand can be interpreted as a Weyl transformation and is solved by $\xi^{r} = \sigma(x)$, which we will calculate. The action of the Weyl transformations on the bosonic fields is given by
\begin{align}
\begin{split}
\delta_{\sigma} \gamma_{\mu\nu} &= \sigma \partial_{r}\gamma_{\mu\nu} \ = \ - \sigma e^{-X}z \partial_{z}\gamma_{\mu\nu} \ = \ 2 \sigma \hks^{-\frac{2}{3}}\, \gamma_{\mu\nu} +\dots,\\
\delta_{\sigma} \phi &= - \sigma e^{-X}z \partial_{z}\phi \ = \ 0 +\dots ,\\
\delta_{\sigma} \bphi &= - \sigma e^{-X}z \partial_{z}\bphi \ = \ \gs q\sigma \hks^{-\frac{2}{3} } +\dots ~.
\end{split}
\end{align}
The variations of the hatted fields can be computed using the above
\bea
\des \ol{\phi} = 0 +\dots ~,~~~~\des \ol{\bphi} = \sigma \gs q \hks^{-\frac{2}{3}} +\dots ~,~~~~\des \ol{\gamma}_{\mu\nu} = 2\sigma \hks ^{-2/3}\ol{\gamma}_{\mu\nu}+\dots
\eea
For the fermionic fields, the action of the Weyl transformations is given by 
\begin{align}
\begin{split}
\des \zeta_{\phi}^{-} &= -\sigma e^{-X}z \de_{z}\zeta_{\phi}^{-} \ = \ -\dfrac{1}{2}\sigma\hks^{-\frac{2}{3}} \bigg(1+ \dfrac{1}{12\hks}\gs q^{2} \bigg)\zeta_{\phi}^{-} + \dots~,\\
\des \zeta_{b}^{-} &= -\sigma e^{-X}z \de_{z}\zeta_{b}^{-}~, \\
&=  -\dfrac{1}{2}\sigma\hks^{-\frac{2}{3}} \bigg(1+\dfrac{3}{8\hks} \gs q^{2}  \bigg)\zeta^{-}_{b} +\dfrac{1}{12}\sigma \hks^{-\frac{5}{3}} \Big(12-\dfrac{1}{\hks}\gs q^{2} \Big) \zeta^{-}_{\phi} + \dots,\\
\des \Psi_{\mu}^{+} &= -\sigma e^{-X}z \de_{z}\Psi^{+}_{\mu}\\
&= \dfrac{1}{2}\sigma\hks^{-\frac{2}{3}} \bigg(1+\dfrac{1}{12\hks} \gs q^{2}  \bigg)\Psi^{+}_{\mu} -\dfrac{i \gamma_{\mu}}{3}\sigma\bigg(q \hks^{-\frac{3}{2}} \Big[1+\dfrac{5}{24\hks}\gs q^{2}\Big]\zeta_{b}^{-}+\dfrac{5}{96\hks^{\frac{5}{2}}} \gs q^{4} \zeta_{\phi}^{-} \bigg) +\dots
\end{split}
\end{align}
We can use the above results and the definitions for the fermionic superpartners of the composite fields, to find their Weyl transformations. In the following, we have kept only the leading order terms of the powers of $\hks $ (the reason being $ \hks^{-\frac{2}{3}} |_{z\rightarrow 0} \sim \frac{1}{(\log z)^{2/3}} $ and anything subleading to $\hks^{-\frac{2}{3}}  $ will be falling off much faster, and will not contribute to the Ward identity computations)
\begin{align}
\begin{split}
\des \hzd{\phi}^{-} &= -\dfrac{1}{2}\sigma \,\hks^{-\frac{2}{3}}\,  \hzd{\phi}^{-}  + \dots~,\\
\des \hzd{b}^{-} &= -\dfrac{1}{2}\sigma \,\hks^{-\frac{2}{3}}\, \Big(\hzd{b}^{-} + \gs q \hzd{\phi}^{-} \Big) + \dots~,\\
\des \hat{\Psi}_{\mu}^{+} &= \dfrac{\sigma}{2}\hks^{-\frac{2}{3}}\hat{\Psi}_{\mu}^{+} + \dots~.
\end{split}
\end{align}

\subsubsection{Local Supersymmetry}

The fact that we have gauge-fixed the gravitino means that the gravitino SUSY variation in \eqref{fermionSUSYvar} gives rise to a differential equation for the supersymmetry parameter
\bea
\delta_{\epsilon}\Psi_{r} = \left(\nabla_{r} + \dfrac{1}{6}\calw \Gamma_{r} \right)\epsilon + \calo(z^{3}) = 0.
\eea
By projecting out the two chiralities (see discussions about spinors in the Appendix) with $ \Gamma_{r} \epsilon^{\pm} = \mp \epsilon^{\pm} $ and looking at the leading order terms in $ z $, we get
\bea
\partial_{r} \epsilon^{\pm} \mp \dfrac{1}{6}\calw \epsilon^{\pm} = 0, \ \ 
\Rightarrow \ \  \epsilon^{\pm}(z,x) = z^{\mp1/2}\hks(z)^{\pm1/12} \epsilon^{\pm}_{0} + \dots \ .
\eea
For the transverse coordinates of the gravitino
\bea
\delta_{\epsilon}\Psi_{\mu} =\nabla_{\mu}\epsilon + \dfrac{1}{6}\calw \Gamma_{\mu} \epsilon + \calo(z^{3}) 
= \partial_{\mu}\epsilon + \dfrac{1}{2} \omega_{\mu}^{zi}\, \gamma_{zi}\epsilon + \dfrac{e^{Y}}{6z} \calw \,  \delta_{\mu}^{i}\gamma_{i}\epsilon ++ \calo(z^{3})  .
\eea
Since we need only the leading asymptotics, we can project to the positive chirality of the gravitino, and using the on-shell values of $ \omega_{\mu}^{zi} $ and $ \calw $, we get
\bea
\delta_{\epsilon}\Psi_{\mu}^{+}&=& \partial_{\mu}\epsilon^{+} + \dfrac{1}{3} \calw \Gamma_{\mu} \epsilon^{-} + \calo(z^{3}) .
\eea
For the composite gravitino, using the above relation with \eqref{compgraviton} and \eqref{epsmu}, we get
\bea
\dep \hat{\Psi}_{\mu}^{+} = \hks^{-\frac{1}{6}}\de_{\mu}\epsilon^{+} + \hks^{-\frac{2}{3}} \hat{\Gamma}_{\mu} \epsilon^{-}+\dots
\eea

Now we turn to the SUSY-variations of the scalars and fermions. We can find the SUSY variations of the fermionic fields by evaluating to the relations given in \eqref{fermionSUSYvar} in the KS background. For the purpose of finding the SUSY Ward Identities, we need only the leading order results, given by
\begin{align}\label{epsmu}
\begin{split}
\depm \zeta_\phi^{-} &=  0 +\dots ~,~~~~~~\depm  \zeta_{b}^{-} = -i \gs q \hks^{-\frac{2}{3} }\,\epsilon^{-} +\dots  ~.\\
\end{split}
\end{align}
The supersymmetry transformations of the composite fields can be computed using the above results,
\bea
\depm \hzd{\phi}^{-} = 0 +\dots ~,~~~~~~\depm \hzd{b}^{-} =-i\gs q \hks^{-\frac{2}{3}} \,\epsilon^{-}+\dots ~.
\eea
For the bosonic fields of interest, we get the SUSY variations from \eqref{compferm} to be
\begin{align}
\begin{split}
\delta_{\epsilon^{+}} \hat{\phi} &=  \dfrac{i}{2}\Big( \bar{\epsilon}^{\, +} \hzd{\phi}^{-} - \hzdb{\phi} ^{\, -} \epsilon^{+} \Big) =  \dfrac{i}{2}\Big( \bar{\epsilon}^{\, +} \zeta_{\phi}^{-} - \bar{\zeta}_{\phi} ^{\, -} \epsilon^{+} \Big)\\
\delta_{\epsilon^{+}} \hat{b}^\Phi &= \dfrac{i}{2} \big( \bar{\epsilon}^{\, +}\hzd{b}^{-} -\hzdb{b}^{\, -} \epsilon^{+} \big) \ = \ \dfrac{i}{2} \big( \bar{\epsilon}^{\, +} \zeta_{b}^{-} - \bar{\zeta}_{b}^{-} \, \epsilon^{+}  \big) +  \dfrac{i}{2}\gs q \log az\big(\bar{\epsilon}^{\, +} \zeta_{\phi}^{-} - \bar{\zeta}_{\phi}^{-} \, \epsilon^{+} \big) +  \dots ,
\end{split}
\end{align}
where only the $ \epsilon^{+} $ variations are considered. This is because both $ \zeta_{\phi} $ and $ \zeta_{\bphi} $ are sourced by the negative chirality, and we are only interested in looking at the SUSY variations of the sources in this section.

We state this to emphasize the fact that we do not need an explicit form of the covariant sources as composite fields as was done in \cite{Bertolini}: we can derive all the necessary results we need in the computation of the 1-point functions using the above facts because only linear parts of variations show up in these calculations.

Finally, we turn to the SUSY variations of the metric. Using the supersymmetry transformation of the vielbein given in \eqref{bosonSUSYvar}, we can write the supersymmetry transformation of the boundary metric as
\bea
\begin{split}
	\delta_{\epsilon}\gamma_{\mu\nu} &=& \delta_{\epsilon} \big(e_{\mu}^{a}e_{\nu}^{b}\eta_{ab} \big) = \dfrac{1}{2} \Big(\bar{\epsilon} \, \Gamma_{\mu} \Psi_{\nu} + \bar{\epsilon} \, \Gamma_{\nu}\Psi_{\mu} \Big) + \text{h.c.}  \\
	&=& \bar{\epsilon}^{\,+} \, \Gamma_{(\mu} \Psi_{\nu)}^{+} +\bar{\epsilon}^{\, -} \, \Gamma_{(\mu} \Psi_{\nu)}^{-}+ \text{h.c.}.
\end{split}
\eea
The symmetrization here contains the factor of $ 1/2 $. We can drop the $ \Psi_{\nu}^{-} $ owing to the fact that it is subleading (and therefore does not corresponds to a source), and we get
\bea
\begin{split}
	\delta_{\epsilon^{+}}\gamma_{\mu\nu} &=& \bar{\epsilon}^{+} \, \Gamma_{(\mu} \Psi_{\nu)}^{+} + \text{h.c.} + \dots,\\
	\Rightarrow \delta_{\epsilon^{+}} \ol{\gamma}_{\mu\nu} &=&  \bar{\epsilon}^{+} \, \hat{\Gamma}_{(\mu} \hat{\Psi}_{\nu)}^{+} + \text{h.c.} + \dots
\end{split}
\eea

\subsection{Derivation of SUSY and Trace Ward Identities}\label{sec5.2}

Now we can put all these ingredients together to compute the SUSY and trace Ward identities. We directly present the results: the approach is parallel to that in \cite{Bertolini}. In order to compute the SUSY Ward identities, we only turn on those sources that do not break SUSY explicitly. As noted in the description of the linearized sources for the bosons, the relevant sources that are present in the \zz theory explicitly break SUSY. Hence, to compute the SUSY Ward Identities, we take the action to be a functional of the SUSY preserving sources, 
\bea
S_{ren} \equiv S_{ren} \Big[\hat{\gamma}_{\mu\nu},\hat{\phi},\hat{b}^\Phi,\hat{\Psi}_{\mu},\hzd{\phi}^{-},\hzd{b}^{-} \Big]~.
\eea
The action we use to compute the Ward Identities is to be understood as the full $ \caln =2 $ renormalized supergravity action, with both the bosonic and fermionic fields. However, we do not need the explicit form of the action to carry out the computations.

In computing the SUSY Ward/Trace Identities, the set of sources used here are the one that appear in the $ U(1) $ truncation as well. However, the presence of more fields does change the SUSY variations of substantially. The reason we go through this section (and Appendix \ref{appd}) in such detail is to ensure that all the falloffs that go in to the Ward/Trace identity computations are under control.
\subsubsection{SUSY Ward Identities}

We consider $\epsilon^{+}, \epsilon^{-}$ and $\sigma$ in turn. We use the results from the previous sub-section, where we found the action of $ \sigma,\epsilon^{\pm} $ on the sources, to compute the Ward Identities. First, we will look at the $ \epsilon^{+} $ variation which will give rise to SUSY Ward identities in the boundary QFT. We have
\begin{align}
\delta_{\epsilon^{+} }S_{\tx{ren}} &= \int d^{4}x\sqrt{-\ol{\gamma}} \Big(\dfrac{i}{2} \lr{\bar{S}^{\mu-}}\delta_{\epsilon^{+}}\hat{\Psi}^{+}_{\mu} + \dfrac{1}{2}\lr{T^{\mu\nu}} \delta_{\epsilon^{+}}\ol{\gamma}_{\mu\nu} +2 \exval{+} \delta_{\epsilon^{+}}\ol{\phi}  + 2 \exval{-} \delta_{\epsilon^{+}}\ol{\bphi}  +\text{h.c.}\Big)\non \\
&= \int d^{4}x\sqrt{-\ol{\gamma}} \bigg[ -\dfrac{i}{2}\lr{\de_{\mu} \bar{S}^{\mu-}}\hks^{-\frac{2}{3}} -\dfrac{1}{2} \lr{T^{\mu\nu} } \hat{\Gamma}_{\mu}\hat{\Psi}^{+}_{\mu} -i\exval{+} \hzdb{\phi}^{-} -i\exval{-} \hzdb{b}^{-} \bigg]\epsilon^{+} +\text{h.c.} 
\end{align}
In these formulas (and formulas in subsequent subsections), we have used \eqref{foneptfundef} for the definition of fermionic one-point functions at non-zero source. For one-point function of the stress-tensor and other bosonic operators, we use the definition in \eqref{oneptfn1}. By setting $ \delta_{\epsilon^{+}}S = 0 $, we get the following operator relation at a finite radial cut-off surface at non-zero sources
\bea
\dfrac{i}{2}\hks^{-\frac{1}{6}}\, \lr{\de_{\mu} \bar{S}^{\mu-}}  &=& -  \dfrac{1}{2} \lr{T^{\mu\nu} } \overline{\hat{\Psi}}^{+}_{\mu}\hat{\Gamma}_{\nu} -i\exval{+} \hzdb{\phi} -i\exval{-} \hzdb{b}^{-} 
\eea
Taking the functional derivatives of this w.r.t. the different fermionic sources gives rise to the following identities
\begin{subequations}\label{SUSYWI}
	\bea
	\hks^{-\frac{1}{6}}\,\lr{\de_{\mu} \bar{S}^{\mu-}(x)\bar{S}^{\nu-}(0)}  &=& 2i\hat{\Gamma}_{\mu}\lr{T^{\mu\nu}}\delta^{4}(x,0)~,\\
	\hks^{-\frac{1}{6}}\,\lr{\de_{\mu} \bar{S}^{\mu-}(x)\calo^{+}_{\hzd{\phi}}(0)}  &=& \sqrt{2}\exval{+}\delta^{4}(x,0)~,\\
	\hks^{-\frac{1}{6}}\,\lr{\de_{\mu} \bar{S}^{\mu-}(x)\calo^{+}_{\hzd{b}}(0)}  &=& \sqrt{2}\exval{-}\delta^{4}(x,0)~,
	\eea
\end{subequations}
where $ \delta^{4}(x,y) = \sqrt{-\ol{\gamma}}\,\delta^{4}(x-y) $ . Now, using the definitions for the QFT one-point functions, we finally get SUSY Ward identities
\begin{subequations}
	\bea\label{sWI}
	\lr{\de_{\mu} \bar{S}^{\mu-}(x)\bar{S}^{\nu-}(0)}\qft  &=& 2i\gamma_{i}\delta^{i}_{\mu}\lr{T^{\mu\nu}}\qft \delta^{4}(x),\\
	\label{OpWI}
	\lr{\de_{\mu} \bar{S}^{\mu-}(x)\calo^{+}_{\hzd{\phi}}(0)}\qft  &=& \sqrt{2}\exval{+}\qft\delta^{4}(x),\\
	\label{OmWI}
	\lr{\de_{\mu} \bar{S}^{\mu-}(x)\calo^{+}_{\hzd{b}}(0)}\qft  &=& \sqrt{2}\exval{-}\qft\delta^{4}(x).
	\eea
\end{subequations}
Eqn. \eqref{sWI}, is the Ward identity involving, the supercurrent and the stress tensor, which sit in the supercurrent multiplet. Eqns. \eqref{OpWI},\eqref{OmWI} are Ward identities for operator sitting in a chiral supermultiplet in which the highest component operators are ${\cal O}_{\pm}$. Therefore a non-zero vev for ${\cal O}_{\pm}$ would correspond to a supersymmetry broken vacuum. However, this is not the sole criterion for spontaneous supersymmetry breaking, as we will see in the next subsection.
%%%%%%%%%%%%%%%%%%%%%%%%%%%%%%%%
\subsubsection{Trace Identities}
Looking at the variation of $S_{\tx{ren}}$ under $\epsilon^{-}$ gives us
\bea
\begin{split}
	\delta_{\epsilon^{-}} S_{\tx{ren}}  &=& \int d^{4}x \sqrt{-\ol{\gamma}} \, \bigg[ \dfrac{i}{2}\lr{\bar{S}^{\mu-}} \delta_{\epsilon^{-}}\hat{\Psi}_{\mu}^{+} -i\sqrt{2} \exvalb{+}{\hzd{b}}  \delta_{\epsilon^{-}}\hzd{b}^{-}  +\text{h.c}  \bigg]~,\\
	&=& \int d^{4}x \sqrt{-\ol{\gamma}} \, \bigg[\dfrac{i}{2}\hks^{-2/3}\lr{\bar{S}^{\mu-}} \hat{\Gamma}_{\mu} -\sqrt{2} \gs q\, \hks^{-2/3} \exvalb{+}{\hzd{b}} \bigg]\epsilon^{-} +\tx{h.c.}
\end{split}
\eea
Setting $ \delta_{\epsilon^{-}}S_{\tx{ren}} = 0 $, we get
\bea
\dfrac{i}{2}\lr{\bar{S}^{\mu-}} \hat{\Gamma}_{\mu} &=& \sqrt{2}  \gs q\exvalb{+}{\hzd{b}} ~.
\eea
Finally, taking the boundary limit, we get the following QFT operatorial relation between the $\gamma$-trace of the supercurrent and the fermionic superpartner of ${\cal O} _-$ (upto potential anomaly terms\footnote{To calculate the anomaly terms we need to know the explicit form of the bosonic and fermionic counterterm. For a systematic derivation of these terms in $4d$, ${\cal N}=1$ and $3d$, ${\cal N}=2$ superconformal theories on an arbitrary curved background, see \cite{Papadimitriou:2017kzw,Papadimitriou:2004rz,An:2017ihs}.}).
\bea\label{superWeylWI}
\dfrac{i}{2}\lr{\bar{S}^{\mu-} \gamma_{i}\delta^{i}_{\mu}}\qft &=& \sqrt{2} \gs q\exvalb{+}{\hzd{b}}\qft~.
\eea
Finally we consider the invariance of the $S_{\tx{ren}}$ under rescalings of the radial coordinate. We have
\bea
\des S_{\tx{ren}}  &=& \int d^{4x}\sqrt{-\ol{\gamma}}\bigg[ \dfrac{1}{2}\lr{T^{\mu\nu}}\des \ol{\gamma}_{\mu\nu} + 2\exval{-} \des \ol{\bphi}  +\dfrac{i}{2}\lr{\bar{S}^{\mu-}}\des \hat{\Psi}_{\mu}^{+} \non\\
&& \qquad\qquad\qquad -\sqrt{2}i\Big(\exvalb{+}{\hzd{\phi}}\des \hzd{\phi}^{-} +\exvalb{+}{\hzd{b}}\des \hzd{b}^{-} \Big)+ \text{h.c.}\bigg]~.
\eea
Using the $ \sigma $-variations and turning off the fermionic sources, we get
\bea
\lr{T^{\mu}_{\mu}} &=& -2 \gs q ~\exval{-}~.
\eea
By taking the boundary limit, we get the following relation between the trace of the stress-tensor and the operator ${\cal O}_-$
\bea\label{WeylWI}
\lr{T^{\mu}_{\mu}}\qft &=& -2 \gs q~ \exval{-}\qft ~ ,
\eea
upto potential anomaly terms which do not appear here since we have taken the boundary metric to be Minkowskian. The relation \eqref{WeylWI} is the bosonic counterpart of the fermionic relation in \eqref{superWeylWI} and the two results are in perfect agreement. In the next section we will check these Ward identities  on a vacua of the KS theory dual to the two-parameter SUSY breaking solution in \eqref{KSSUSYb} by explicitly calculating the one-point functions. This will allows us to comment upon the nature of supersymmetry breaking.

%%%%%%%%%%%%%%%%%%%%%%%%%%%%%%%%%%%%%%%%%
\section{One-point Functions and the Goldstino Pole}\label{sec:6}

To obtain the QFT one point functions, we evaluate the functional derivatives of the renormalized on-shell action appearing \eqref{oneptfn2} and take the limits in \eqref{oneptfn3}. The regulated action in $S_{\tx{reg}}$ is given by
\bea
S_{\tx{reg}} = S_{\tx{5D}} + S_{\tx{GH}}~,
\eea
where $S_{\tx{5D}}$ the boundary contribution coming from the five-dimensional gauged supergravity action and $S_{\tx{GH}}$ is the Gibbons-Hawking term. Correlation functions computed from $S_{\tx{reg}}$ are typically divergent because of the infinite volume of spacetime. Finite quantities can be obtained through the standard procedure of holographic renormalization where we first identify the divergences of the regularized on-shell action and then add appropriate local covariant counterterms to kill these divergence \cite{deHaro,Papadimitriou:2004ap,Neumann2}. The renormalized action thus obtained is finite when the cut-off surface is taken to the boundary. However there are scheme ambiguities associated to finite terms which may be required to preserve supersymmetry. For flat-domain wall solution the superpotential ${\cal W}$ in \eqref{superpotential} has all the necessary finite terms to render $ S_{\tx{ren}}=S_{\tx{reg}}+S_{\tx{ct}} =0 $ on supersymmetric configurations \cite{Bianchi}. Therefore we take the following as out bosonic counter term 
\bea\label{counterterm}
S_{\tx{ct}} = -\int d^{4}x\sqrt{-\gamma } ~2\calw~.
\eea
This, along with the fact that counter terms have to be universal for any solution to the equations of motion for the given potential, fixes them once and for all, regardless of the bulk solution being supersymmetric or not. The calculation of the one point functions for the marginal operators proceed as in \cite{Bertolini} and we find no further subtleties.
\begin{subequations}
	\begin{align}
	\label{optT}
	\lag T^\m_{~\n}\rag_{\tx{QFT}}&= -3 {\cal S}a^4 \delta^\m_{~\n}~,\\
	\label{optOp}
	\lag {\cal O}_+\rag_{\tx{QFT}}&= \frac12\left(3{\cal S}+4\varphi\right)a^4~,\\
	\label{optOm}
	\lag {\cal O}_-\rag_{\tx{QFT}}&= \frac{6{\cal S}}{g_s q}a^4~,
	\end{align}
	These expressions were first obtained in \cite{DKM} and were later independently derived in \cite{Bertolini}. Here we find that even in the full KS theory, these one-point functions remain unaffected (upto a trivial modification by the conifold deformation parameter $a$ defined in \eqref{warpKS}). 
	
	We now expand around this result. Integrating the SUSY ward identity in \eqref{sWI} in $x_\mu$ gives us the two point function of the supercurrent\cite{Argurio,Argurio0}. The right hand side contains a massless fermionic pole provided the vev of the stress-tensor (that corresponds to the vacuum of a QFT state) is non-zero. This massless pole is hallmark signature of the presence of a goldstino which is associated to the spontaneous breaking of supersymmetry. Since from \eqref{optT} we have that the one-point function of the stress-tensor is non-zero and gets contribution from the parameter ${\cal S}$ only, we conclude that ${\cal S}$ corresponds to spontaneous supersymmetry breaking. Furthermore, the one-point functions in \eqref{optT},\eqref{optOm} satisfy the trace identity derived in \eqref{WeylWI}. We see that there is no contribution to the vacuum energy from the vev of ${\cal O}_+$. On the other hand the parameter $\varphi$ corresponds to explicit breaking of supersymmetry and does not corresponds to a vacua of the KS gauge theory. This is because in this SUSY breaking solution (where ${\cal S}=0$), the vacuum energy vanishes ($\lag T^\mu_{~\mu}\rag=0$) and therefore the residue of the Goldstino pole vanishes. Despite the technical complications, we find that even in the full theory, which captures the conifold deformation parameter, all of the aforementioned results are identical to those obtained via the $U(1)$-truncated supergravity action considered in \cite{DKM,Bertolini}.

	Finally let us comment upon the one-point functions of the gaugino bilinear operators ${\cal Q}_\pm$ which do not participate in any of the Ward identities (since we have not turned on sources for these operators). Without the inclusion of the counterterm in \eqref{counterterm} we find that the vev of ${\cal Q}_+$ is finite and its value is 
	\begin{align}
	\lag {\cal Q}_+\rag_{\tx{QFT}}&= \dfrac{18 a^3}{\gs q}~.
	\end{align}
	Hence the one-point function of ${\cal Q}_+$ does not require renormalization. The counterterm gives the same finite contribution but with opposite sign. Therefore if we include the counterterm contribution we will not be able see the vev which is actually non-zero. The inverse dependence of the vev on $q$ can be attributed to the normalization of the operator ${\cal Q}_+$. In \cite{Loewy:2001pq} the operator ${\cal Q}_+$ (as defined in \eqref{sourceOp}) was identified with the gaugino bilinear that condense in the Klebanov-Strassler gauge theory. Here we find that even in the presence of the SUSY breaking perturbations, this vev is uneffected. On the other hand the vev of ${\cal Q}_-$ is divergent, and therefore needs renormalization. Upon adding the counterterm \eqref{counterterm}, we do not find any non-vanishing finite contributions. Therefore we conclude that 
	\begin{align}
	\lag {\cal Q}_-\rag_{\tx{QFT}}&= 0~.
	\end{align}
\end{subequations}

It would be interesting to generalize our analysis to the full $SU(2) \times SU(2)\times \Z_2$ truncation by turning on fields in \eqref{nonKSfields} to see if there are more SUSY-breaking parameters and study if there exists spontaneously supersymmetry breaking vacua. Our minimal goal here, namely to derive the SUSY Ward identities in a truncation of Type IIB SUGRA, that admits the deformation of the conifold parameter, has been accomplished. Since the parameter $\mathcal{S}$ is known to be triggered by anti-D3 branes on the tip of the throat \cite{DKM}, this shows that {\em if} the KKLT construction is (meta-) stable\footnote{See discussions on some aspects of this issue in \cite{Sethi, Halmagyi, Polchinski, Dasgupta}}, it is a spontaneously broken (and therefore bonafide) vacuum of string theory. The goldstino on the worldvolume of the anti-D3 brane has been noted in previous work in \cite{Wrase}.

\chapter{Summary \& Conclusions}
In this thesis, we have dealt with different systems unified by a central theme of Holography. We summarize the methods and findings of our investigation below.

\section{Hairy Black Holes in Global AdS}
We studied the fully backreacted Einstein-Maxwell-Scalar system with a negative cosmological constant and conformally coupled scalar, in 3+1-dimensional spacetime. The solutions we looked at asymptote to \emph{global AdS}, which allows for four distinct possibilities: Empty global AdS, Reissner-Nordstrom (RN-AdS) Black Hole, Boson star and Hairy Black Hole. The boson star solution corresponds to a horizon-less spacetime, where the scalar has a non-trivial profile and so does the gauge field. The Hairy Black Hole, on the other hand, is a spacetime with a horizon and non-trivial profiles for the scalar and gauge field. The scalar-less system already is an interesting theory as it has two different phases (global AdS and RN-AdS) separated by a \emph{First Order} phase boundary, and the phase transition is called the \emph{Hawking-Page} phase transition. With the addition of scalars, the phase structure of the theory becomes even more interesting, as there are four phases competing against each other. Since the boson star and hairy black hole solutions are condensate formations over the global-AdS and RN-AdS spaces respectively, the system also has second order phase transitions, adding to the complexity of the phase diagram. The phase diagrams, as we have found, depend crucially on the charge of the scalar $ q $, and we study the system for different ranges of $ q $.

For $ q>1 $, the boson star instability occurs at $ \mu<1 $. This implies that all the four phases will be present in the phase diagram. The global AdS/boson star transition is a second order transition, and so is the RN-AdS/hairy black hole transition. These phase boundaries are easy to evaluate, as the phase boundaries sit at the point where the instability to forming a non-trivial scalar profile occurs. The $ (T,\mu) $ values of the phase boundaries can thus be evaluated by doing a probe computation on the respective backgrounds. The phase transitions between global AdS or boson star to the RN-AdS or hairy black hole will be a first order phase transition, and the phase boundary between boson star/hairy black hole becomes particularly challenging to find as they are both numerical solutions. We find a method to carry out this computation efficiently, and chart out the phase diagrams for $ q=5,3,1.2 $.

The case of $ q<1 $ allows for only three of the four phases to exist, as the boson star instability happens at $ \mu>1 $ where already the global AdS phase is thermodynamically unstable. The phase diagram has one first order phase boundary corresponding to the Hawking-Page transition, and the RN-AdS/hairy black hole transition, both of which are straightforward to evaluate.

\section{Hairy Black Holes in a Box}
It is common lore that global AdS is like a \emph{Box}, which leads to the natural question: Does the Einstein-Maxwell-Scalar system in a Flat Box also allows for an interesiting phase diagram, and more importantly, can \emph{No-Hair Theorems} of asymptotically flat spaces be evaded by imposing a radial cutoff for the space-time (i.e. Box)? We answer both of these questions in the affirmative, and chart out the phase diagram for different values of charge of the scalar. To begin with, the scalar-less system already is interesting inside a box, as there exists a Hawking-Page-like first order phase transition. However, the fact that the black hole has to sit inside the box results in a very peculiar phase diagram already in the scalar-less case. For the case with the scalar turned on, we find that there are again four possible solutions: empty Box, RN Black Hole, Boson Star and Hairy Black Hole. We find that depending on the value of $ q $, there exists \emph{three} distinct types of phase diagrams.

For $ q_{1}(\sim 36)<q<\infty $, all four phases are present, and we chart out the full phase diagram using (semi)-analytic and numerical methods, similar to what we used in the global AdS computation. We report the exact figures for $ q=40,100 $. This proves that in addition to having Hairy Black Holes solutions in a box, they are also thermodynamically favourable phases in appropriate regions of the parameter space. The Hairy Black Hole solution ceases to exist as a thermodynamically favourable phase for $ q<q_{1} $, although the solution still exists. Thus for $ q<q_{1} $, the phase diagrams are an interplay of three different phases. However, the phase boundary between boson star and RN black hole is only present for $ q_{2}<q<q_{1} $, and disappears for $ q<q_{2}(=3\pi) $, leading to a very simple phase diagram.

The study however provides conclusive proof that No-Hair Theorems of asymptotically flat space can be evaded successfully in a box, and Hairy Black Hole solutions exist as thermodynamically stable solutions in some regions of the parameter space. In doing so, we also find that the definition of quasi-local stress tensor is not general enough to capture the presence of scalar fields in the box.

\section{Quantum Chaos and Holographic Tensor Models}
We investigated the finite $ N $ behaviour of the Gurau-Witten Model. Gurau-Witten (GW) Model is a tensor model constructed out of fermionic tensors, the large-$ N $ behaviour of which is \emph{melonic}, same as that of the Sachdev-Ye-Kitaev(SYK) Model. The SYK and GW models are solvable at large-$ N $, have emergent conformal symmetry in the IR and saturate the \emph{Chaos bound} at large-$ N $, suggesting that they are maximally chaotic. This makes them tentative models of quantum black holes. The SYK model is constructed out of $ q $-fermion interaction, with the coupling drawn out of a Gaussian distribution. The computation of the correlation functions have to be, at the end, disorder averaged over the ensemble of couplings; this means that the correlation functions of the system are not \emph{truly quantum mechanical}. GW Model on the other hand is purely a quantum mechanical theory, as there is no disorder averaging involved.

We study the first non-trivial GW Model with $ N =32 $, which also happens to be an upper bound on computational accessibility. However, signatures of quantum chaos already exist in this particular model in spite of the smallness of $ N $. We find that the Spectral form factor does resemble a single sample result of the SYK Model, and in doing a running-time-average the spectral form factor exhibits a dip-ramp-plateau structure. We also studied the unfolded level spacing distribution $ P(s) $, and we can see that the spectrum $ s $ exhibits level repulsion and that it does not have linear spectral rigidity. We also find that the ground state is unique and that there is a large degeneracy at the mid-level. Furthermore, the system has mirror symmetry, which means that the energy levels come in pairs around the central value of 0. This can be traced back to the existence of a unitary operator that anti-commutes with the Hamiltonian. Along with this we find that the GW Model with $ N=32 $ corresponds to the BDI class in the Andreev-Altland-Zirnbauer 10-fold classification of Random Matrix Ensembles.

\section{KKLT Goldstino}
We study the $ SU(2)\times SU(2)\times \Z_2 \times Z_{2R} $ truncation of Type IIB supergravity, a truncation that is general enough to capture the deformed conifold solution, i.e. Klebanov-Strassler(KS) solution. The system is of interest as the conifold is a generic singularity of 6-d Calabi-Yau(CY) manifolds, and deformation of the conifold is one of the ways to get rid of the singularity. The deep IR physics of the deformed-conifold geometry is relevant to understand the nature of anti-D3 branes in a fully stabilized 6-d CY compactification. 

We choose an ansatz that has a good interpretation in the Type IIB uplift, and find the solution that asymptotes to Klebanov-Strassler in the UV. This solution has two modes, $ \cals $ and $ \varphi $, that deforms the space from KS. Using the BPS equation, it is clear that both of these two modes result in supersymmetry breaking. In order to find the way in which SUSY is broken, spontaneously or explicitly, we have to find the Ward Identity corresponding to the supercurrent two-point function. For this purpose, first we have to find the sources to dual operators in the boundary gauge theory. The mass-eigenstates of the supergravity theory mix non-trivially in the KS background, which requires us to find combination that result in a good falloff at the asymptotic infinity. We can also find the action of supersymmetry and bulk diffeomorphisms on these diagonalized sources, which will be necessary to find the Ward/Trace Identities. We also find the fermions dual to these diagonalized sources, which are combinations of the four fermions in the theory. Using the action of SUSY and bulk diffeomorphisms on the fields, we find the Ward/Trace Identities by requiring the on-shell action to be stationary under the action of these symmetries. We compute the one-point functions of operators dual to the diagonalized sources, which along with the Ward Identities indicate that the perturbation $ \cals $ corresponds to spontaneous SUSY-breaking and the $ \varphi $ perturbation corresponds to explicit SUSY-breaking. We also identify and evaluate the one-point function which corresponds to the gaugino condensate. 

The mode $ \cals $ has been identified previously to capture the presence of smeared anti-D3-branes placed at the tip of the warped throat. Since the supercurrent Ward Identity is related to $ \cals $, we identify this as the spontaneous SUSY-breaking parameter. What this means is that if KKLT solution is metastable, then it is indeed a bonafide vacuum of string theory with the SUSY being broken spontaneously by the anti-D3 branes.

\newpage

\appendix
\chapter{Appendix to Chapter \ref{gads}}\label{app_ch2}
\section{Free Energy computation by background subtraction}
The free energy of a system can be computed by evaluating the classical action directly. However, the classical energy computed this way is divergent, and has to be made finite by using counter-terms or by using background subtraction, where we subtract out the classical action of the global AdS solution after matching the temperatures of the two configurations at the asymptotic region. 

Let us take the equations of motion to be of the form $R_{\mu\nu} - \frac{1}{2}g_{\mu\nu} R = \tilde{T}_{\mu\nu}$. It can be easily checked for our ansatz that
\bea
\mathcal{L} - R = \dfrac{2}{r^{2}}\tilde{T}_{33} = \dfrac{2}{r^{2}\sin^{2}\theta}\tilde{T}_{44}.
\eea

Now using $G_{\mu\nu}= R_{\mu\nu} - \dfrac{1}{2}g_{\mu\nu} R$, and $G^{\mu}_{\mu} = - R $, the classical action will simplify to the following form, in the Euclidean signature
\bea
S_{Euc} = -  \dfrac{1}{16\pi} 2\;{\rm Vol}_{3}\left(\int_{r_{0}}^{\infty}dr \sqrt{h(r)} - r g(r) \sqrt{h(r)}\Biggr|_{r\rightarrow\infty} \right),
\eea
where ${\rm Vol}_{3}= 4\pi \beta$. Also, $r_{0}=0$ for any solution without a horizon, namely global AdS and boson star, and $r_{0} = \rh $ for the RNAdS black hole and the hairy black hole solutions. The action calculated this way will be divergent and we can use holographic renormalization or background subtraction to regularize the action. We will be using background subtraction using global AdS to regularize and we get
\bea
S_{reg} = -  \dfrac{1}{2} \beta \left(\int_{r_{0}}^{\infty}dr \sqrt{h(r)} - r g(r) \sqrt{h(r)}\Biggr|_{r\rightarrow\infty} \right) + \beta_{0}\dfrac{1}{2}\left(\rb - \rb \;(1+\rb^{2}) \right)\Biggr|_{\rb\rightarrow\infty},
\eea
where $\beta$ and $\beta_{0}$ are the periodicities of the $t$ integrals of the geometry that we are interested in and of global AdS, respectively. Now, we have to adjust $\beta_{0}$ such that the geometry at the hypersurface $r=\rb $ of both the spaces are the same. This gives the relation
\bea
\dfrac{\beta}{\beta_{0}} &= &\sqrt{\dfrac{g_{tt}^{gAdS}(\rb)}{g_{tt}(\rb)}} = \sqrt{\dfrac{1+\rb^{2}}{g_{tt}(\rb)}},\\
\Rightarrow F &= &\dfrac{S_{reg}}{\beta} = -  \dfrac{1}{2} \left(\int_{r_{0}}^{\infty}dr \sqrt{h(r)} - r g(r) \sqrt{h(r)}\Biggr|_{r\rightarrow\infty} \right) + \dfrac{1}{2}\sqrt{\dfrac{g_{tt}(\rb)}{1+\rb^{2}}}\left(\rb - \rb \;(1+\rb^{2}) \right)\Biggr|_{\rb\rightarrow\infty}.
\eea

One can check that the classical action computed using this equation matches with that computed using \eqref{gtherm} analytically for RNAdS black hole, and numerically for the hairy solutions.

The evaluation of thermodynamic variables in the case of numerical hairy solutions is done by fitting the curves of $g(r), \phi(r)$ and we take $E$ to be $(-\frac{1}{2})$ times the coefficient of $\frac{1}{r}$ term of $g(r)$, and chemical potential($\mu $) and charge $(Q)$ from the falloff of $\phi(r)$, using the relation $\phi(r\rightarrow\infty) \approx 2 \mu - \frac{2 Q}{r}$.

\section{Extremal Black Hole instability}

In the full set of possible solutions, there are two systems with zero temperature and scalar condensate, boson star and extremal hairy black hole. %We would like to compare the instability points of the two systems, for arbitrary $q$, to get an understanding of how the chemical potential required to form the condensate in the two cases compare for different values of $q$. 

The near-horizon instability of the extremal black hole is analytically tractable. The extremal RNAdS black hole has a charge, in terms of $\rh$ (with $L=1$),
\bea
Q = \pm 	2 \rh \sqrt{3 \rh^2 + 1}.
\eea
The chemical potential for the instability here is calculated as follows. We expand the scalar equation of motion in the extremal RNAdS background, in the near-horizon region. The leading order terms multiplying $\psi(r),\psi'(r),\psi''(r)$, together gives the equation of motion for a scalar in $AdS_{2}\times S^{2}$, which is the near-horizon geometry of $4$-d RNAdS. The instability point for the $4$-d extremal RNAdS is taken to be the value of $\mu$ that saturates the Breitolehner-Freedman bound for the $AdS_{2}$. This gives
\bea
-M_{(2)}^{2}L_{(2)}^{2}  =\dfrac{1}{4}=\dfrac{4q^2\mu^2 (\mu^2-1)}{3(2 \mu^2-1)^2}+\frac{2(\mu^2-1)}{3(2\mu^2-1)}.
%\frac{2}{\frac{12}{\mu ^2-4}+6}+\frac{4 \left(\frac{1}{48} \left(\mu ^2-4\right)^2+\frac{1}{12} \left(\mu ^2-4\right)\right) q^2}{\left(\frac{1}{2} \left(\mu ^2-4\right)+1\right)^2}.
\eea
We can solve for $\mu$ for a given value of $q$ for which the instability sets in and it is given by
\bea
\mu =  \sqrt{\frac{2 q^2}{4 q^2+1}+\frac{3}{2(4 q^2+1)}+\frac{ \sqrt{4 q^4+q^2+1}}{4 q^2+1}}.
\eea
%{\bf What fixes the sign of the inside square root?}
The sign in the inside square-root has been fixed by noting that $\mu^2=5/2$ when $q=0$, which can be directly read off from the previous equation. For any value of $q$ (including $q=0$ !) this shows that for sufficiently large $\mu$ there is a near-horizon instability. 
 
We can use the above expression to compare the values of $\mu$ at which instability is triggered in the extremal and the boson star cases (the latter happens at $\mu =1/q$ as already noted). %We plot the two curves together and get Fig.\ref{compareq}.
%It can be seen that the boson star instability happens, for a given $q$, at a smaller value of $\mu$ as compared to that of the extremal black hole instability. 
We can see that for large values of $q$, the boson star instability happens at a smaller value of $\mu$ than that of the extremal-RNAdS instability, and vice versa. We can also evaluate the $q$ at which the two curves intersect and it is found to be around $\approx 0.9$. % Since $q$ is less than 1, this means that the boson star is anyway not present as a dominant phase here, so the relevant instability is 
%The boson star instability should be compared to the superradiant instability discussed in \cite{Min}, but o
We can think of the boson star instability as a proxy for the bulk instability of AdS black holes when the black hole is small (this is sometimes called superradiant instability, \cite{Min1, Min2}). It will be interesting to see  if the interplay between these two types of instabilities leads to a quantum critical phase transition for the hairy extremal solutions.

\chapter{Appendix to Chapter \ref{KKLT}}\label{app_ch4}

\section{Truncations, Ansatzes and Uplifts}\label{appa}

The KT solution in the Type IIB setting and the linearized SUSY breaking perturbations that asymptote to KT were discussed in \cite{DKM} and in terms of 5d Supergravity in \cite{Bertolini} (where they use the notations of \cite{Buchel,Cassani}). We will discuss some salient points in the uplift of 5d Supergravity solutions to the 10d Type IIB. This will serve to both establish the correspondence with the notations in various previous papers, as well as to emphasize some subtleties.

In the notation of \cite{DKM}, the 10d metric for the $ U(1) $ truncation is given by
\bea\label{DKMmetric}
ds^{2}_{10\,(DKM)} = r^{2}e^{2a(r)}\eta_{\mu\nu}dx^{\mu}dx^{\nu} + \dfrac{e^{-2a(r)}}{r^{2}} dr^{2} +  \dfrac{1}{6}e^{2(c(r)-a(r))}\sum_{a=1}^{4} (\ee^{a})^{2} + \dfrac{1}{9}e^{2(b(r)-a(r))} (\ee^{5})^{2}~.\non\\
\eea
The two scalar fields coming from the dilaton and B-field of IIB are denoted by $ \Phi(r) $ and $ k(r) $ in \cite{DKM}, which are denoted by $ \phi(z) $ and $ b^{\phi}(z) $, respectively, in our discussions, and the radial coordinates are related as $ r=1/z $. The linearized solution to the equations of motion around the KT background allows for perturbations of the fields $ \{a,b,c,k,\Phi\} $. 

The 10d metric in the notation of \cite{Buchel}, keeping only the fields corresponding to the $ U(1) $ truncation is given by
\bea
ds_{10}^{2} = e^{-\frac{2}{3}(4u+v)} ds^{2}_{5} + \dfrac{1}{6} e^{2u} \sum_{a=1}^{4} (\ee^{a})^{2} + \dfrac{1}{9}e^{2v} (\ee^{5})^{2},
\eea
where $ ds^{2}_{5} = g_{AB} dx^{A}dx^{B} $ is the 5d metric. In \cite{Bertolini} the 5d metric is taken to be of the form
\bea
ds^{2}_{5} = \dfrac{1}{z^{2}} \left(e^{2X} dz^{2} + e^{2Y} \eta_{\mu\nu} dx^{\mu} dx^{\nu}\right).
\eea
The equations of motion are solved using the parametrization
\bea
e^{X(z) } = h(z)^{\frac{2}{3}}\,h_{2}(z)^{\frac{1}{4}},&& \ e^{Y(z)} = h(z)^{\frac{1}{6}}\, h_{2}(z)^{\frac{1}{4}}\,h_{3}(z)^{\frac{1}{4}} \non\\
e^{U(z)} = h(z)^{\frac{5}{4}} \, h_{2}(z)^{\frac{3}{4}}, && \ e^{V(z)} = h_{2}(z)^{-\frac{3}{4}},
\eea
where $ U=4u+v $ and $ V=u-v $. On uplifting this ansatz to 10d, this is in a slightly different gauge for the radial coordinate comapred to \cite{DKM}: %However, at the level of solutions to equation of motion, ansatz of \cite{Bertolini} is equivalent to that of \cite{DKM} up to  $ z^{4} $ and $ z^{4}\log z $.
\bea
ds_{10}^{2} = \dfrac{h(z)^{-\frac{1}{2}}h_{3}(z)^{\frac{1}{2}}}{z^{2}} \eta_{\mu\nu}dx^{\mu}dx^{\nu} + \dfrac{h(z)^{\frac{1}{2}}}{z^{2}} dz^{2} +  \dfrac{1}{6} h(z)^{\frac{1}{2}}\sum_{a=1}^{4} (\ee^{a})^{2} + \dfrac{1}{9}h(z)^{\frac{1}{2}} h_{2}(z)^{\frac{3}{2}} (\ee^{5})^{2}.
\eea
After a coordinate change to $ r=1/z $ and defining $ H^{\frac{1}{2}} = r^{-2} h^{\frac{1}{2}} h_{3}^{-\frac{1}{2}} $, we get
\bea
ds_{10}^{2} = H^{-\frac{1}{2}} \eta_{\mu\nu}dx^{\mu}dx^{\nu} + H^{\frac{1}{2}} \bigg( h_{3}^{\frac{1}{2}}dr^{2} +  \dfrac{1}{6} r^{2}h_{3}^{\frac{1}{2}}\sum_{a=1}^{4} (\ee^{a})^{2} + \dfrac{1}{9}r^{2}h_{3}^{\frac{1}{2}} h_{2}^{\frac{3}{2}} (\ee^{5})^{2}\bigg)
\eea

The most general parametrization of the functions can be taken in the form
\bea
e^{X(z) } =h_{X}^{\frac{2}{3}}(z),&& \ e^{Y(z)} =  h_{Y}^{\frac{1}{6}}(z)\non\\
e^{U(z)} =  h_{U}^{\frac{5}{4}}(z), && \ e^{V(z)} = h_{V}(z).
\eea
where $h_{X},\, h_{Y} $ and $ h_{U} $ at leading order is given by $ \hks $ and $ h_{V} =1 $ at leading order. The  functions are each a double series in $ z^{n} $ and $ z^{n}\log z $.This metric uplifts to
\bea
ds_{10}^{2} &=& \dfrac{h_{U}^{-\frac{5}{6}}\, h_{Y}^{\frac{1}{3}}}{z^{2}}\eta_{\mu\nu}dx^{\mu}dx^{\nu} + \dfrac{ h_{U}^{-\frac{5}{6}}\, h_{X}^{\frac{4}{3}}}{z^{2}} dz^{2} \non\\
&& +  \dfrac{1}{6} h_{U}^{\frac{1}{2}}\, h_{V}^{\frac{2}{5}}\sum_{a=1}^{4} (\ee^{a})^{2} + \dfrac{1}{9}  h_{U}^{\frac{1}{2}}\, h_{V}^{-\frac{8}{5}} (\ee^{5})^{2}.
\eea
\begin{comment}
\bea
e^{X(z) } = h_{KS}^{\frac{2}{3}}(z)\,h_{X}^{\frac{2}{3}}(z),&& \ e^{Y(z)} = h_{KS}^{\frac{1}{2}}(z)\, h_{Y}^{\frac{1}{6}}(z)\non\\
e^{U(z)} = h_{KS}^{\frac{5}{4}}(z) \, h_{U}^{\frac{5}{4}}(z), && \ e^{V(z)} = h_{V}(z).
\eea
where $ h_{KS}(z) $ is fixed by the KS function\footnote{We extract this fixed function to make things easier for Mathematica.}, while the other functions are each a double series in $ z^{n} $ and $ z^{n}\log z $.This metric uplifts to
\bea
ds_{10}^{2} &=& \dfrac{\hks^{-\frac{1}{2}}\,h_{U}^{-\frac{5}{6}}\, h_{Y}^{\frac{1}{3}}}{z^{2}}\eta_{\mu\nu}dx^{\mu}dx^{\nu} + \dfrac{\hks^{\frac{1}{2}}\, h_{U}^{-\frac{5}{6}}\, h_{X}^{\frac{4}{3}}}{z^{2}} dz^{2} \non\\
&& +  \dfrac{1}{6}\hks^{\frac{1}{2}}\, h_{U}^{\frac{1}{2}}\, h_{V}^{\frac{2}{5}}\sum_{a=1}^{4} (\ee^{a})^{2} + \dfrac{1}{9}\hks^{\frac{1}{2}}\, h_{U}^{\frac{1}{2}}\, h_{V}^{-\frac{8}{5}} (\ee^{5})^{2}.
\eea
\end{comment}
The equations of motion can be solved order by order for this ansatz (we also include the other fields in the $\Z_2$ truncation to do this, obviously), and we find that there are a total of 4 independent (SUSY-preserving) parameters on top of the SUSY-breaking ones. 

Note however that the above ansatz is not the most convenient for a few reasons. Firstly, we have not fixed the gauge freedom (this in particular means that we cannot be sure that all the perturbations we found are physical), and secondly, we find it (slightly) better to work with an ansatz that is more naturally adapted to a 10d brane ansatz form in the spirit of \cite{DKM}. A (partial) gauge fixing that accomplishes this is the ansatz we use in the main body of Chapter \ref{KKLT}:
\bea
e^{X(z) } = h(z)^{\frac{2}{3}}\,h_{2}(z)^{\frac{1}{3}}h_{3}(z)^{\frac{4}{3}},&& \ e^{Y(z)} = h(z)^{\frac{1}{6}}\,h_{2}(z)^{\frac{1}{3}}h_{3}(z)^{\frac{4}{3}}  \non\\
e^{U(z)} = h(z)^{\frac{5}{4}} \, h_{2}(z)\, h_{3}(z)^{4}, && \ e^{V(z)} = h_{3}(z)\, h_{2}(z)^{-1}, \label{ouransatz}
\eea
which in the $U(1)$ case, when uplifted to 10d takes the form
\bea
ds_{10}^{2} = \dfrac{h(z)^{-\frac{1}{2}}}{z^{2}}\eta_{\mu\nu}dx^{\mu}dx^{\nu} + \dfrac{h(z)^{\frac{1}{2}}}{z^{2}} dz^{2} +  \dfrac{1}{6} h(z)^{\frac{1}{2}}h_{3}(z)^{2}\sum_{a=1}^{4} (\ee^{a})^{2} + \dfrac{1}{9}h(z)^{\frac{1}{2}}h_{2}(z)^{2} (\ee^{5})^{2}.
\eea
It is straightforward to see that this metric and the metric in \eqref{DKMmetric} are the same form upto renaming of functions, with the identification $ r = 1/z $. For the \zz truncation, the same ansatz lifts to a metric of the form
\bea
ds_{10}^{2} &=& \dfrac{h(z)^{-\frac{1}{2}}}{z^{2}}\eta_{\mu\nu}dx^{\mu}dx^{\nu} + \dfrac{h(z)^{\frac{1}{2}}}{z^{2}} dz^{2} +  \dfrac{\cosh t}{6} h(z)^{\frac{1}{2}}h_{3}(z)^{2}\sum_{a=1}^{4} (\ee^{a})^{2} \non\\
&& \qquad + \dfrac{\sinh t}{3}h(z)^{\frac{1}{2}}h_{3}(z)^{2} \Big( \ee^{1} \ee^{3} + \ee^{2} \ee^{4} \Big)  + \dfrac{1}{9} h(z)^{\frac{1}{2}}h_{2}(z)^{2} (\ee^{5})^{2}.
\eea
The advantage of this ansatz, which is the one we use, is that it removes all the SUSY-preserving perturbations except for one (which we argue in the next Appendix is a gauge mode).

\section{Gauge Freedom in the 10d metric}\label{appb}

In this appendix, we will show that a specific perturbation that arises in the class of 10d metrics from the previous section when expanded around Klebnov-Witten, is a coordinate redefinition. The reason for our interest in this perturbation is that within the ansatzes that we work with\footnote{By which we mean the forms \eqref{ouransatz} as well as the combined expansions in $z^n$ and $z^n \ln z$ with $n$ not restricted to be even. If $z$ is restricted to be even as in \cite{Aharony} this term does not arise and this appendix can be skipped.}, this is the only perturbation (SUSY-preserving) that shows up around the KS background other than the parameters in KS and the SUSY-breaking perturbations. The fact that precisely this perturbation arises also around KW, and there it can be understood as a gauge artefact will be taken as motivation to believe that it is a gauge artefact around KS as well. We will work with the $U(1)$ truncation to keep the notation slightly cleaner, but the arguments go through precisely analogously in the $\Z_2$ case as well.

Let us start with the 10d metric 
\bea
ds_{10}^{2} = \dfrac{h(z)^{-\frac{1}{2}}}{z^{2}}\eta_{\mu\nu}dx^{\mu}dx^{\nu} + \dfrac{h(z)^{\frac{1}{2}}}{z^{2}} dz^{2} +  \dfrac{1}{6} h(z)^{\frac{1}{2}}h_{3}(z)^{2}\sum_{a=1}^{4} (\ee^{a})^{2} + \dfrac{1}{9}h(z)^{\frac{1}{2}}h_{2}(z)^{2} (\ee^{5})^{2}.
\eea
The KW solution is given by $ h(z) = h_{2}(z) = h_{3}(z)=1 $. Now, let us look at small arbitrary perturbation around this background. The metric becomes
\bea
ds_{10}^{2} &=& \dfrac{ (1+\delta  h(z))^{-\frac{1}{2}}}{z^{2}} \eta_{\mu\nu}dx^{\mu}dx^{\nu} + \dfrac{ (1+\delta  h(z))^{\frac{1}{2}}}{z^{2}} dz^{2}  +  \dfrac{1}{6}(1+\delta  h(z))^{\frac{1}{2}} (1+\delta  h_{3}(z))^{2}\sum_{a=1}^{4} (\ee^{a})^{2} \non\\
&& \qquad\qquad + \dfrac{1}{9}(1+\delta  h(z))^{\frac{1}{2}} (1+\delta  h_{2}(z))^{2} (\ee^{5})^{2}.
\eea
We can redefine the $ z $-coordinate to $ y $ in the following way
\bea
\dfrac{z^{2}}{ (1+\delta  h(z))^{-\frac{1}{2}}} &=& y^{2}\qquad \Rightarrow\quad y^{2} \ \simeq \  z^{2} \Big(1+\dfrac{1}{2}\delta h(z)\Big) \\
2y dy &=& \Big[2z\Big(1+\dfrac{1}{2}\delta h(z)\Big) +\dfrac{z^{2}}{2} \delta h'(z) \Big] dz.
\eea
We will only need the perturbation upto linear order, so these approximations will turn out to be consistent for our purposes.
Using this we get
\bea
\dfrac{ (1+\delta  h(z))^{\frac{1}{2}}}{z^{2}} dz^{2} &\approx& \dfrac{4y^{2}\Big(1+ \frac{1}{2}\delta h(z)\Big) dy^{2} }{y^{2}\Big(1-\frac{1}{2}\delta h(z)\Big) \Big[2z\Big(1+\frac{1}{2}\delta h(z)\Big) +\frac{z^{2}}{2} \delta h'(z) \Big]^2}\non\\
&\approx& \dfrac{4 dy^{2}}{\Big(1-\delta h(z)\Big) \Big( 4 z^{2}(1+ \delta h(z)) + 2 z^{3}  \delta h'(z) \Big)}\non\\
&\approx& \dfrac{dy^{2}}{z^{2}\Big(1+\frac{1}{2}z\, \delta h'(z)\Big)}.
\eea
If we now set $ \delta h(z) = \e z $, the denominator in the last line can be rewritten as $ z^{2}\Big(1+ \frac{1}{2} \e z\Big) = z^{2}\Big(1+ \frac{1}{2} \delta h(z)\Big) \simeq y^{2} $. Thus, we get
\bea
\dfrac{ (1+\delta  h(z))^{\frac{1}{2}}}{z^{2}} dz^{2} &\simeq& \dfrac{dy^{2}}{y^{2}}. 
\eea
In order to have the full metric unchanged under this redefinition, we need
\bea
(1+\delta  h(z))^{\frac{1}{2}} (1+\delta  h_{2}(z))^{2} = (1+\delta  h(z))^{\frac{1}{2}} (1+\delta  h_{3}(z))^{2} =1.
\eea
Altogether these conditions read
\bea
\delta h_{2}(z) = \delta h_{3}(z) = - \dfrac{1}{4} \delta h(z)=  -\dfrac{1}{4}\e z.
\eea

The reason we care about this, is because the 10d metric we started with, when expanded around KW has precisely this as a perturbation at $\mathcal{O}(z)$ when we demand that the equations of motion hold. This means that that particular perturbation can be viewed as a gauge artefact.

\section{Fermions in AdS: A mini review}

The spin-1/2 fermions in 5d are Dirac fermions. The gamma matrices are given by
\bea
\Gamma^{A} &=& e^{A}_{a}\gamma^{a},
\eea
where $ e^{A}_{a} $ are the vielbeins corresponding to the 5d metric. The $ \gamma^{a} $'s can be grouped into the gamma matrices of the boundary 4-d space $ \gamma^{i}, i=0,1,2,3 $ and $ \gamma^{z} $ of the radial direction
\bea
\{\gamma^{i},\gamma^{j}\} &=& 2\eta^{ij}, \ \ \ \gamma^{0\dagger} \ = \ -\gamma^{0}, \ \ \ \gamma^{i \dagger} \ = \ \gamma^{0}\gamma^{i}\gamma^{0},\\
\{\gamma^{z},\gamma^{j}\} &=& 0, \ \ \ \ \ \gamma^{z}{}^{2} \ = \  1,  \ \ \ \gamma^{z\dagger} \ = \ \gamma^{z}.
\eea
The conjugate spinor is defined as
\bea
\overline{\psi} &=& \psi^{\dagger}\,i \gamma^{0}.
\eea

One basic idea in solving fermionic fields in AdS is that a spinor in the bulk, being a 5d spinor has the same number of components as a 4D Dirac spinor. But the minimal spinors on the boundary are (4D) Weyl spinors and contain half as many degrees of freedom. When we want to use them as boundary data for solving the bulk (spinor) equations, the two possible chirality choices separate out. This is good: because unlike in the bosonic cases, the bulk spinor equations are first order. So it is good that the two chiralities on the boundary can yield a natural interpretation as source and condensate - as they do in the bosonic case for the field and its derivative (roughly).

Lets see how this works out in the case of Rarita-Schwinger fields and spin-1/2 fermions. The latter discussion we follow the very clear presentation in \cite{Henneaux}.

\subsection{Rarita-Schwinger field in AdS}

The Rarita-Schwinger equation, in the AdS background, for a gravitino of mass $ m=\frac{3}{2} $, is given by
\bea \label{rsrhoeq}
(\delta^{\rho}_{j}\delta^{\mu}_{i}\gamma^{j}\gamma^{i} - \eta^{\rho\mu})\bigg(-z^{3}\gamma^{z}\,\de_{z} \Psi_{\mu}(z,x) + \dfrac{z^{2}}{2}  (2\gamma^{z}-3) \Psi_{\mu}(z,x)\bigg)  &&\non\\
+z^{3}(\delta^{\rho}_{j}\delta^{\nu}_{k}\delta^{\mu}_{i}\gamma^{j} \gamma^{k}\gamma^{i} - \eta^{\rho\nu} \delta^{\mu}_{i}\gamma^{i} -  \eta^{\mu\nu} \delta^{\rho}_{j}\gamma^{j} + \eta^{\rho\mu} \delta^{\nu}_{k}\gamma^{k} )\,\de_{\nu} \Psi_{\mu}(z,x) \ = \ 0 ,&&\\
\Rightarrow z^{3} (\delta^{\nu}_{j}\delta^{\mu}_{i} \gamma^{j}\gamma^{i}- \eta^{\mu\nu}) \de_{\nu}\Psi_{\mu}(z,x) + \label{rszeq}\dfrac{3 z^{2}}{2} \delta^{\mu}_{i}\gamma^{i} (1-\gamma^{z}) \Psi_{\mu}(z,x) \ = \ 0.&&
\eea
We can solve these equations near the boundary $ z=0 $, using the Frobenius method. We substitute the series expansion
\bea
\Psi_{\mu}(z,x) = z^{\Delta}\sum_{l=0} c_{\mu(i)}(x) z^{l},
\eea
in \eqref{rsrhoeq}, and set the coefficients of each of the $ z $ powers to zero. The leading equation is 
\bea
\Big(-\Delta \gamma^{z} +\gamma^{z} -\dfrac{3}{2} \Big) c_{\mu(0)}(x) = 0,
\eea
which is solved by
\bea
\Delta = \begin{cases}
	-\frac{1}{2} \qquad &\text{with}\ \ \gamma^{z}c_{\mu(0)}(x) = c_{\mu(0)}(x) \\
	\frac{5}{2} \qquad &\text{with}\ \ \gamma^{z}c_{\mu(0)}(x) = -c_{\mu(0)}(x).
\end{cases}
\eea
These two are the two independent boundary fields that fix the full gravitino solution in the bulk.

We stress here that this discussion is for the AdS background and not the KS background, where the fermions and gravitino are non-trivially coupled. 

\subsection{Spinors in AdS}

For simplicity, we will consider a single fermion of mass $ m $ in the AdS background. The equation of motion for the fermion is 
\bea\label{fermads}
z \gamma^{z}\de_{z} \zeta(z,x) + z \delta^{\mu}_{i}\gamma^{i} \de_{\mu}\zeta(z,x) -2\gamma^{z} \zeta(z,x) + m \zeta(z,x) = 0.
\eea
We can again use the Frobenius method near the boundary at $ z=0 $. We take the solution to be a series expansion in $ z $ of the form
\bea
\zeta(z,x)  = z^{\Delta} \sum_{l=0} c_{(l)}z^{l}.
\eea
Substituting this in \eqref{fermads} and from the leading order coefficient we get
\bea
\Delta = \begin{cases}
	2 - m ,\qquad &\text{with} \ \ \gamma^{z} c_{(0)}(x) = c_{(0)}(x)\\
	2 + m ,\qquad &\text{with} \ \ \gamma^{z} c_{(0)}(x) = -c_{(0)}(x).
\end{cases}
\eea
The most general solution can be written as, assuming $ m $ is positive,
\bea
\zeta(z,x) = c^{+}_{(0)}\,z^{2-m} +\dots + z^{2+m}\Big(c^{-}_{(0)} + c^{+}_{(2+m)} \log z \Big) + \dots
\eea
where the presence of the $ \log z $-term depends on the mass and is non-generic -- we will not need it in our discussions. 

\begin{comment}

{\bf appears when $2-m $ is integer}, and is determined in terms of $ c^{+}_{(0)} $. The entire solution is determined in terms of the parameters $ c^{\pm}_{(0)} $. However, if we are looking at supersymmetric theories, there is an additional constraint that we can impose, namely the residual gauge freedom in the supersymmetry transformation rules. Using this, we can fix the subleading coefficient (here $ c^{-}_{(0)} $) in terms of the leading order coefficient\cite{Bertolini1}. This straightforwardly generalizes to the fermions of the supergravity theory that is of interest here, in the AdS background.
\end{comment}

The above discussion focusses on empty AdS background, where all the scalars are set to zero. This simplifies the discussion as the fermions and gravitino are all decoupled. We could in principle perform a similar analysis in the KS background, but the non-trivial couplings complicates the analysis substantially, and this is what we have done perturbatively in the main text.

\section{Supersymmetry of ${\cal N}=2$, $SU(2)\times SU(2)\times \mathbb{Z}_2$ truncation}\label{appd}
In this appendix we map the consistent truncation ansatz used in Liu-Szepietowski \cite{Liu:2011dw} (henceforth LS) to that of Cassani-Faedo \cite{Cassani} (henceforth CF). We then use this map to write down the fermionic SUSY variations in the notations of CF from which we then extract the BPS equations. We begin by defining the following one-forms
\begin{align}
\begin{split}
\sigma_1=c_{\psi/2}d\theta_1+s_{\psi/2}s_{\theta_1}d\phi_1~,~~~~&\Sigma_1=c_{\psi/2}d\theta_2+s_{\psi/2}s_{\theta_2}d\phi_2~,\\
\sigma_2=s_{\psi/2}d\theta_1-c_{\psi/2}s_{\theta_1}d\phi_1~,~~~~&\Sigma_2=s_{\psi/2}d\theta_2-c_{\psi/2}s_{\theta_2}d\phi_2~,\\
\sigma_3={1\over 2} d\psi+c_{\theta_1}d\phi_1~,~~~~&\Sigma_3={1\over 2} d\psi+c_{\theta_2}d\phi_2~.
\end{split}
\end{align}
where $ s_{\bullet} =\sin(\bullet) $ and $ c_{\bullet} =\cos(\bullet)$. These one-forms satisfy the $SU(2)\times SU(2)$ structure equations
\bea
d\sigma_i={1\over 2}\epsilon_{ijk}\sigma_i\wedge\sigma_j,~~~~~~~~d\Sigma_i={1\over 2}\epsilon_{ijk}\Sigma_i\wedge\Sigma_j~.
\eea
Using these forms we can endow a K\"{a}hler structure on $T^{1,1}$ as follows. We first define the following complex one-forms
\bea
\begin{split}
	E_1={1\over\sqrt{6}}\left(\sigma_1+i\sigma_2\right)
\end{split}
\qquad  \qquad
\begin{split}
	E_2={1\over\sqrt{6}}\left(\Sigma_1+i\Sigma_2\right).
\end{split}
\eea
Using these two complex one-forms we now define a basis of left-invariant forms on $T^{1,1}$ used in LS
\bea
J_{1} = \dfrac{i}{12} E_{1}\times \bar{E}_{1}~,\qquad J_{2} = \dfrac{i}{12} E_{2}\times \bar{E}_{2}~, \qquad \Omega = \dfrac{1}{6} E_{1} \times E_{2}~ , \qquad \eta = \dfrac{1}{3}\ee^{5}~,
\eea
where $\ee^5$ is defined in \eqref{oneforms1}. To compare with the notation of CF, we define $J_{\pm} = J_{1} \pm J_{2}$. The conversion now reads as follows
\bea
\eta_{\tx{LS}} = -\eta_{\tx{CF}}~,\qquad J_{+\, \tx{LS}} = -J_{\tx{CF}}~,\qquad  J_{-\, \tx{LS}} = -\Phi_{\tx{CF}}~,\qquad \Omega_{\tx{LS}}=\Omega_{\tx{CF}}~.
\eea
The LS metric is parametrized in the following way
\bea\label{LSzMetric}
ds_{LS}^{2} = e^{2A}ds_{5}^{2} + \dfrac{1}{6}e^{2B_{1}} E_{1}\bar{E}_{1} + \dfrac{1}{6}e^{2B_{2}} \hat{E}_{2}\hat{\bar{E}}_{2} + \dfrac{1}{9}e^{2C} (\eta+3A)^{2},
\eea
where $ \hat{E}_{2} = E_{2} + \alpha \bar{E}_{1}  $, $ \alpha $ being a complex scalar. In order to compare with the CF metric
\bea\label{CFMetric}
ds_{\tx{CF}}^{2} &=& e^{-\frac{8u-2v}{3}} ds_{5}^{2} + \dfrac{e^{2u}}{6} \cosh t \Big(e^{2w}(e_{1}^{2}+e_{2}^{2}) + e^{-2w}(e_{3}^{2}+e_{4}^{2})\Big) + \dfrac{e^{2v}}{9} (\eta+ 3A)^{2} \non\\
&& + \dfrac{e^{2u}}{3}\sinh t\Big( \cos\theta (e_{1}e_{3}+e_{2}e_{4}) + \sin\theta ( e_{1}e_{4} - e_{2}e_{3})\Big)~,
\eea
\eqref{LSzMetric} can be expanded in terms of $ \ee^{i} $'s defined in \eqref{oneforms1}. Upon comparing we obtain
\bea
A_{\tx{LS}} = -\dfrac{4u+v}{3} \bigg|_{\tx{CF}},\qquad \alpha_{\tx{LS}} = e^{2w} \tanh t ~e^{i\theta}|_{\tx{CF}} ,\qquad C_{\tx{LS}} = v |_{\tx{CF}}, \non\\
B_{1} = u+w -\dfrac{1}{2}\log\cosh t, \qquad B_{1} = u - w +\dfrac{1}{2}\log\cosh t~.
\eea
Similarly, from the expansion ansatz of the two form potentials, we get
\bea
e^{1}_{0} = -\bphi~, \ \ j_{0}^{2} = q~, \ \ b^{1}_{0} = \dfrac{1}{2}\overline{b^{\Omega}}~, \ \ b^{2}_{0} = \dfrac{1}{2} \overline{c^{\Omega}}~.
\eea
In the above equation we have written down the map only for fields turned on in the Klebanov-Strassler solution. Other relevant relations are as follows
\begin{align}
\begin{split}
&h_1^1=-d\bphi,\ \ f_0^1= \frac32 i~ \overline{b^{\Omega}}, \ \ f_0^2= \frac32 i~ \overline{c^{\Omega}}~,\\
&f_1^1= \frac12  d\overline{b^{\Omega}}, \ \ f_1^2= \frac12  d\overline{c^{\Omega}}~,\\
&\hat{f}_1^1= \frac12  d\overline{b^{\Omega}} +\frac i2 \tanh t~ d\bphi,\ \ \hat{f}_1^2= \frac12  d\overline{c^{\Omega}}~,\\
&\hat{f}_0^1 = \frac32 i~ \overline{b^{\Omega}}, \ \ \hat{f}_0^2 = \frac32 i~ \overline{c^{\Omega}}-\frac i2 q \tanh t ~,\\
&\hat{\cal F}_1^1= \frac i2 \sinh t~  \left( - d\boi +\tanh t~ d\bphi  \right), \ \ \hat{\cal F}_1^2= \frac12 \tanh t ~ d \cor ~,\\
&\hat{\cal F}_0^1= \frac 32 \boi \tanh t, \ \ \hat{\cal F}_0^2= \frac i2 \sinh t \left( 3 \cor -q \tanh t\right) ~.
\end{split}
\end{align}
One has to remember that $ \Re[b^{\Omega}]  = \Im[c^{\Omega}] = 0$. However, since many computations involve taking absolute values or the real and imaginary parts of products of functions, it is better to set this condition after making sure all such functions have been evaluated. Or one could set it and then be careful not to miss the $ i $ coming from $ b^{\Omega} = i\boi $. 
From the five form we get
\bea
-\dfrac{1}{2}(4+ \phi_{0}) = (k -q\bphi + 3 \boi\cor),
\eea
The notation for the axio-dilaton is $\tau=\tau_1+i\tau_2=C_{0}+i e^{-\phi}$. It will be convenient to write down the $SL(2,\mathbb{R})$ vielbein 
\bea
v_{1} = -(C_0e^{\phi/2}+i e^{-\phi/2})~,\qquad v_{2} = e^{\phi/2}~,
\eea
that appears explicitly in the SUSY variations of the fermions.
\subsection{SUSY variation of Fermions}\label{appd1}
\begin{subequations}\label{fermionSUSYvar}
	\bea
	\delta \zeta^{1} &= & \bigg[-\dfrac{i}{2\sqrt{2}}\Gamma.\de\phi \bigg]\epsilon -\dfrac{i}{2\sqrt{2}}e^{-2u-\frac{\phi}{2}}\Bigg[\Gamma\cdot\Big(e^{\phi}\de\cor + \cosh t \de \boi - \sinh t \de\bphi\Big) \non\\
	&&\hspace{3cm} +e^{-\frac{4}{3}(u+v)}\Big(3\boi+3e^{\phi}\cosh t  \cor -qe^{\phi} \sinh t\Big)\bigg] \epsilon^{c}\\
	%%%%%%%%%%%%%%%%%%%%%%%%%%%%%%%%%%%%%%%%%%%
	\delta \zeta^{2} &=& -\dfrac{i}{2}e^{-2u-\frac{\phi}{2}}\cosh t\bigg[\Gamma\cdot\Big(\de \bphi -\tanh t \de \boi\Big) - e^{-\frac{4}{3}(u+v)+\phi}(3\tanh t \cor -q)\bigg]\epsilon\qquad\non\\
	&&+ \frac i2 \bigg[\Gamma\cdot\de t + 3 \sinh t e^{-\frac{4}{3}(u+v)}\bigg]\epsilon^c\\
	%%%%%%%%%%%%%%%%%%%%%%%%%%%%%%%%%%%%%%%%%%%
	\delta \zeta^{3} &=& 2\sqrt{2}\bigg[-\dfrac{i}{2} \Gamma\cdot\de u -\dfrac{i}{2}e^{-\frac{2}{3}(5u-v)} +\dfrac{i}{8} e^{-\frac{4}{3}(4u+v)} (4+\phi_{0})\bigg] \epsilon \non\\
	&& +\dfrac{i}{2\sqrt{2}}e^{-2u-\frac{\phi}{2}}\bigg[ \Gamma\cdot\Big(e^{\phi} \de\cor-\cosh t\de \boi + \sinh t \de\bphi  \Big) \non\\
	&& \qquad\qquad +e^{-\frac{4}{3}(u+v)} \Big(3  \boi-3e^{\phi} \cosh t \cor + q e^{\phi}\sinh t\Big) \bigg] \epsilon^{c}\\
	%%%%%%%%%%%%%%%%%%%%%%%%%%%%%%%%%%%%%%%%%%%
	\delta \lambda^{u_3} &=& -\bigg[ -\dfrac{i}{6} \Gamma\cdot\de(u+v) +\dfrac{i}{6} e^{-\frac{2}{3}(5u-v)} -\dfrac{i}{2}\cosh t e^{-\frac{4}{3}(u+v)} +\dfrac{i}{12}e^{-\frac{4}{3}(4u+v)}(4+\phi_{0})\bigg]\epsilon\non\\
	&&-\frac{i}{6} e^{-\frac{2}{3}(5u+2v) -\frac{\phi}{2}}\Big(3\boi -3e^{\phi}\cosh t\cor +q e^{\phi} \sinh t  \Big)\epsilon^{c}\\
	%%%%%%%%%%%%%%%%%%%%%%%%%%%%%%%%%%%%%%%%%%%
	\delta \Psi_{\mu} &=& \bigg[D_{\mu}+\dfrac{1}{6}\Gamma_{\mu}\calw \bigg]\epsilon +\bigg[\dfrac{\Gamma_{\mu}}{6}e^{-\frac{2}{3}(5u+2v)-\frac{\phi}{2}}\Big(3\boi -3e^{\phi}\cosh t \cor+ q e^{\phi}\sinh t\Big)\non\\
	&&\qquad\qquad +\dfrac{1}{2}e^{-2u-\frac{\phi}{2}}\Big(\cosh t \de_{\mu}\boi - e^{\phi} \de_{\mu} \cor - \sinh t \de_{\mu}\bphi \Big)\bigg]\epsilon^{c}
	\eea
\end{subequations}
The above SUSY transformation are taken from Eq. (102) of \cite{Halmagyi:2011yd} with the following definitions
\begin{align}
\begin{split}
\zeta^1_{\tx{here}}= \frac{1}{\sqrt{2}}\zeta^1_{\tx{there}}~,~~~~\zeta^2_{\tx{here}}= -\left(\zeta^2_{\tx{there}}\right)^c~,~~~~\zeta^3_{\tx{here}}= 2\sqrt{2}\zeta^3_{\tx{there}}~,~~~~\lambda^{u_3}_{\tx{here}}= -\xi^1_{\tx{there}}~.
\end{split}
\end{align}
The above field redefinitions are needed to extract the correctly normalized vielbeins of the scalar manifold such that they give rise to the metric ${\cal G}_{IJ}$ in \eqref{scalmetric}. The full scalar manifold can be seen as a direct product ${\cal Q} \otimes {\cal S}$ where ${\cal S}$ is a one dimensional very especial manifold and ${\cal Q}$ is twelve (real) dimensional quaternionic K\"ahler manifold.

Upon comparing with the notation of \cite{Ceresole:2000jd} one can extract the vielbeins and the SUSY variations of the scalars fields. In what follows we report this supergravity data. In writing down \eqref{fermionSUSYvar}, we have fixed some typos in \cite{Halmagyi:2011yd} which do not affect the BPS equations but do affect the metric on the scalar manifold .
\subsection{SUSY variation of Bosons}
The generic form of the SUSY variation of hyperino and gaugino in matter coupled ${\cal N}=2, D=5$ gauged supergravity is \cite{Ceresole:2000jd}
\bea\label{genferSUSY}
\begin{split}
	\delta \zeta^A=-\frac i2 f^A_{iX} \slashed{\de}q^X \e^i+...\\
	\delta \lambda^{x}_i=-\frac i2  \slashed{\de}\phi^x \e_i+...
\end{split}
\eea
where the dots denote the terms proportional to the gauging. All the fermions in the above formula are in the Symplectic-Majorana representation\footnote{In 5 dimensions, the minimal spinor is Dirac, so one cannot define a reality condition by relating the two minimal Weyl representations as in 4 dimensions. Instead, one takes two copies of Dirac to impose a complex conjugation condition relating them. The result is called a symplectic Majorana spinor. }. In the above formulas the index $i$ transforms in the fundamental representation of $SU(2)_R$ R-symmetry group, the index $A$ transforms in the fundamental representation of $USp(2n)$ (where $n$ is the number of hypermultiplets which in our case is three)\footnote{In \cite{Bertolini}, the fermionic sector was written in a sigma model form. The index carried by the fermions were treated on a similar footing as those of the scalars. While this notation allowed to write the fermionic Lagrangian and supersymmetry transformations (in the $U(1)$ truncation) compactly in terms of geometric quantities on the scalar manifold, it is not suitable for studying supersymmetry of the theory. It is not clear if a sigma model-type notation can be used for writing down the fermionic sector of the entire $\Z_2$ truncation.}.  The index $X$ labels coordinates on ${\cal Q}$ and the index $x$ labels coordinates on ${\cal S}$. To extract the vielbeins $f^A_{iX}$ on the quaternionic manifold, we first write down the symplectic-Majorana conditions for the fermions $\zeta^A$ that appears in the three hypermultiplets (here $A=1,2,3,4,5,6$) 
\bea\label{SMtoD}
\zeta^4=\left(\zeta^1\right)^c~,~~~~\zeta^5=\left(\zeta^2\right)^c~,~~~~\zeta^6=\left(\zeta^3\right)^c~.
\eea  
Here $\zeta^1,\zeta^2, \zeta^3$ are Dirac fermions that appear in \eqref{fermionSUSYvar}. The charge conjugation operation is defined as
\bea
\psi^c=\gamma_0C \psi^*~,
\eea
where $C$ is the charge conjugation matrix that satisfies the following properties
\bea
\begin{split}
	&&C=-C^\dagger=-C^T=-C^{-1}=C^*~,\\
	&&C^{-1}\gamma_\mu C =\gamma_\mu^T~.
\end{split}
\eea
Equation \eqref{genferSUSY} and \eqref{SMtoD} together imply the following relation between the vielbeins 
\begin{align}
\begin{split}
f^1_{1X}=f^4_{2X}~,~~~~f^1_{2X}=-f^4_{1X}~,\\
f^2_{1X}=f^5_{2X}~,~~~~f^2_{2X}=-f^5_{1X}~,\\
f^3_{1X}=f^6_{2X}~,~~~~f^3_{2X}=-f^6_{1X}~.
\end{split}
\end{align}
Upon comparing \eqref{genferSUSY} with \eqref{fermionSUSYvar} we get the following non-vanishing vielbeins of the quaternionic manifold ${\cal Q}$
\begin{align}\label{scalViel}
\begin{split}
&f^1_{1\phi}=\frac{1}{\sqrt{2}}~,~~~~f^1_{2\cor}=\frac{1}{\sqrt{2}}e^{-2u+\frac{\phi}{2}}~,\\
&f^1_{2\boi}=\frac{1}{\sqrt{2}}e^{-2u-\frac{\phi}{2}}\cosh t~,~~~~f^1_{2\bphi}=-\frac{1}{\sqrt{2}}e^{-2u-\frac{\phi}{2}}\sinh t~,\\
&f^2_{1\bphi}=e^{-2u-\frac{\phi}{2}}\cosh t~,~~~~f^2_{1\boi}=-e^{-2u-\frac{\phi}{2}}\sinh t~,~~~~f^2_{2t}=-1~,\\
&f^3_{1u}=2\sqrt{2}~,~~~~f^3_{2\cor}=-\frac{1}{\sqrt{2}}e^{-2u+\frac{\phi}{2}}~,\\
&f^3_{2\boi}=\frac{1}{\sqrt{2}}e^{-2u-\frac{\phi}{2}}\cosh t~,~~~~f^3_{2\bphi}=-\frac{1}{\sqrt{2}}e^{-2u-\frac{\phi}{2}}\sinh t~.
\end{split}
\end{align}
As a check of this result one can verify that with \eqref{scalViel}, one indeed reproduces the quaternionic metric in \eqref{scalmetric} via the following relation \cite{Ceresole:2001wi}
\bea\label{quatMetric}
g_{XY}\equiv C_{AB}\varepsilon^{ij} f^A_{iX}f^B_{jY}= f^{iA}_Xf_{YiA}~,
\eea
where $C_{AB}$ is the $USp(6)$ invariant tensor which in our convention reads
\begin{equation}C_{AB}=
\left(
\begin{array}{cc}
0 & \mathbb{I}_3 \\
-\mathbb{I}_3 & 0 \\
\end{array}
\right)~.
\end{equation}
In making this check we have to keep in mind the the metric $g_{XY}$ in \cite{Ceresole:2000jd} is defined upto a factor or 2 (see their Eq. (5.1)).

The metric on the very special manifold, ${\cal S}$, parametrized by the scalar $u_3=-\frac13(u+v)$ in the vector multiplet, can be obtained by the following relations \cite{Halmagyi:2011yd}
\begin{align}\label{vsmMetric}
\begin{split}
G_{{\ci}{\cj}}&=X_\ci X_\cj-C_{\ci\cj\ck}X^\ck~,\\
X_\ci &=\frac12C_{\ci\cj\ck}X^\cj X^\ck~,\\
g_{xy}&= \de_x X^\ci \de_y X^\cj G_{\ci\cj}~,
\end{split}
\end{align}
where $X^\ci(\phi^x)$ are the embedding coordinates of the very special manifold that satisfies the following constraint
\bea
\frac 16 c_{\ci\cj\ck} X^\ci X^\cj X^\ck =1~.
\eea
For the supergravity model under consideration we have
\bea
X^0=e^{4u_3}~,~~~~X^1=e^{-2u_3}~,~~~~C_{011}=2~,
\eea
which, using \eqref{vsmMetric}, gives $g_{u+v,u+v}=\frac83$. Combining with $g_{uu}$ from \eqref{quatMetric}, we recover the ${\cal G}_{uu},{\cal G}_{vv},{\cal G}_{uv}$ components in \eqref{scalmetric}

We are now in a position to write down the bosonic SUSY variations. From \cite{Ceresole:2000jd} the generic form of the SUSY variation of the scalars in the hyper and vector multiplet is
\begin{align}\label{genbosSUSY}
\begin{split}
\delta q^X&=-i\bar{\epsilon}^i \zeta^A f_{iA}^X~,\\
\delta \phi^{x}&=\frac i2 \bar{\e}^i\lambda_i^x ~.
\end{split}
\end{align}
We remark that the index $X$ in $f^A_{iX}$ is raised an lowered by the metric $g_{XY}$ in \eqref{quatMetric} and not the metric in \eqref{scalmetric} which differs by a factor of two.
Using the vielbeins \eqref{scalViel} we find the following SUSY variations for the bosonic fields
\begin{subequations}\label{bosonSUSYvar}
	\begin{align}
	\delta e^a_\mu & = \frac12 \left(\bar{\e} \gamma^a \Psi_\mu-\bar{\Psi}_\mu \gamma^a \e \right)~,\\
	\delta\phi &=\frac{i}{\sqrt{2}}\bar{\e}\zeta^1-\frac{i}{\sqrt{2}}\bar{\zeta^1}\e~,\\
	\delta \bphi &=-\frac{i}{2}e^{2u+\frac{\phi}{2}} \left[\cosh t\left(\overline{\zeta^2}\e-\bar{\e}\zeta^2\right)+\sinh t \left(\overline{\chi_+}\e-\bar{\e}\chi_+\right)\right]~,\\
	\delta \boi &=-\frac{i}{2}e^{2u+\frac{\phi}{2}} \left[-\sinh t\left(\overline{\zeta^2}\e-\bar{\e}\zeta^2\right)+\cosh t \left(\overline{\chi_+}\e-\bar{\e}\chi_+\right)\right]~,\\
	\delta \cor &=-\frac i2 e^{2u-\frac{\phi}{2}}\big[\bar{\e}\chi_- -\overline{\chi_-}\e\big]~,\\
	\delta t&= -\frac i2 \left[\overline{\e^c}\zeta^2-\bar{\e}\left(\zeta^2\right)^c\right]~, \\
	\delta u & =-\frac{i}{4\sqrt{2}}\left[\overline{\zeta^3}\e-\bar{\e}\zeta^3\right]~, \\
	\delta(u+v)&=-\frac{3i}{2}\left[\bar{\e}\lambda^{u_3}-\overline{\lambda^{u_3}}\e\right]~.
	\end{align}
\end{subequations}
In the above equations we have used a new spinor $\chi_\pm$ which is defined as follows
\bea
\chi_\pm=-\frac{1}{\sqrt{2}}\left(\zeta^4\pm\zeta^6\right)~,~~~~\chi_\pm^c=\frac{1}{\sqrt{2}}\left(\zeta^1\pm\zeta^3\right)~.
\eea
These bosonic variations reduce to those of the $U(1)$ consistent truncation of \cite{Bertolini} upon using the following identification
\bea\label{bertolinimap}
\zeta_\phi=\sqrt{2}\zeta^1~,~~~~\zeta_b= e^{2u+\frac{\phi}{2}} \zeta^2~,~~~~\zeta_U= \frac{3}{2\sqrt{2}}\zeta^3-3\lambda^{u_3}~,~~~~\zeta_V= \frac{1}{\sqrt{2}}\zeta^3+3\lambda^{u_3}~.
\eea
This is the basis that we use in the main text.
\subsection{BPS equations from the fermionic variations}\label{appd3}
In this section, we extract from the fermionic variations, the BPS equations for flat domain walls where the metric takes the form \eqref{ksmetric} and all the scalars are function of the radial coordinate $\tau$.  The BPS equations take the form of a gradient flow \eqref{bpsu2} in terms of the superpotential $\cal W$ in \eqref{superpotential}. On supersymmetric configurations, $\delta \tx{(fermions)}=0$. We begin by splitting the SUSY variation parameter $\e$ appearing in \eqref{fermionSUSYvar} as follows:
\bea
\e=\e_+ +\e_-~,
\eea
with the property that 
\bea
\g ^5 \e_\pm =\pm \e_\pm~,~~~~\g ^5 \e_\pm^c =\mp \e_\pm^c~.
\eea
The $\g^5$ above the tangent space gamma matrix and is related to the curved space gamma matrix by $\g^a= e^a_{~\m}\G^\m$, where $e^a_{~\m}$ are the vielbeins of the five-dimensional spacetime. From the gauge fixed form of the metric in \eqref{ksmetric} we read that
\bea
\g^5= e^X \G^\tau~.
\eea
We now construct the projector
\bea
P_\pm =\frac12 \left(1\pm \g^5\right)~,
\eea
which satisfies
\bea
P_\pm^2=P_\pm~,~~~~P_+P_-=P_-P_+=0~.
\eea
Therefore we can write 
\bea
P_\pm \e=\e_{\pm}~,~~~~P_\pm \e^c=\e_{\mp}^c~.
\eea
Setting either $\e_+$ or $\e_-$ to zero kills half of the supersymmetries because $P_\pm$ is a half rank matrix. The choice is arbitrary and we choose to set $\e_+=\e_+^c=0$. Hitting the fermionic SUSY variations in \eqref{fermionSUSYvar} by  $P_\pm$ we extract the BPS equations. The system simplifies considerably if we start with the variation of $\lambda^{u_3}$. From the term proportional to $\e^c$ we find a constraint
\bea\label{constraint1}
3\boi -3e^{\phi}\cosh t\cor +q e^{\phi} \sinh t=0~.
\eea
Next we move to the $\tau$ component of the gravitino variation. Since we have guage fixed $\Psi_\tau=0$, we find upon using \eqref{constraint1} another equation from the term proportional to $\e^c$ 
\bea\label{constraint2}
\cosh t \de_{\tau}\boi - e^{\phi} \de_{\tau} \cor - \sinh t \de_{\tau}\bphi =0~.
\eea
Using \eqref{constraint1} and \eqref{constraint2} we see that in variation of $\zeta^3$ the entire piece proportional to $\e^c$ vanish. From the remaining equations we get
\begin{subequations}
	\begin{align}
	&\de_\tau \f=0~,\\
	&e^{-X}\Big(e^{\phi}\de_\tau\cor + \cosh t \de_\tau \boi - \sinh t \de_\tau\bphi\Big) +e^{-\frac{4}{3}(u+v)}\Big(3\boi+3e^{\phi}\cosh t  \cor -qe^{\phi} \sinh t\Big)=0~,\\
	&e^{-X}\Big(\de_\tau \bphi -\tanh t \de_\tau \boi\Big) + e^{-\frac{4}{3}(u+v)+\phi}(3\tanh t \cor -q)=0~,\\
	&e^{-X}\de_\tau t + 3 \sinh t e^{-\frac{4}{3}(u+v)}=0~,\\
	&\dfrac{1}{2} e^ {-X}\de_\tau u -\dfrac{1}{2}e^{-\frac{2}{3}(5u-v)} +\dfrac{1}{8} e^{-\frac{4}{3}(4u+v)} (4+\phi_{0})=0~,\\
	&\dfrac{1}{6} e^{-X}\de_{\tau}(u+v) +\dfrac{1}{6} e^{-\frac{2}{3}(5u-v)} -\dfrac{1}{2}\cosh t e^{-\frac{4}{3}(u+v)} +\dfrac{1}{12}e^{-\frac{4}{3}(4u+v)}(4+\phi_{0})=0~.
	\end{align}
\end{subequations}
Upon using the constraints \eqref{constraint1} and \eqref{constraint2} it is straightforward to show that these equation reduce to \eqref{bpsu2}.

\bibliography{ref}
\bibliographystyle{utphys}
 
\end{document}